\documentclass[journal,10pt,doublecolumns]{IEEEtran}
\usepackage{amsfonts}
\usepackage{amsmath}
\usepackage{amsthm}
\usepackage{amssymb}
\usepackage{amsfonts}
\usepackage{amsmath}
\usepackage{amsthm}
\usepackage{graphicx}
\usepackage{epstopdf}
\usepackage{subcaption}
\usepackage{xcolor}
\usepackage[linesnumbered,ruled,lined]{algorithm2e}
\usepackage{multirow}
\usepackage{diagbox}
\usepackage{mathtools}
\usepackage{adjustbox}
\newtheorem{remark}{Remark}
\usepackage[font=footnotesize]{caption}
\usepackage{tabularx}
\newcolumntype{L}[1]{>{\raggedright\let\newline\\\arraybackslash\hspace{0pt}}m{#1}}
\newcolumntype{P}[1]{>{\centering\arraybackslash}p{#1}}
\newcolumntype{C}[1]{>{\centering\arraybackslash}m{#1}}
\usepackage{cite}
\usepackage{enumerate}

\usepackage{tikz}
\newcommand*\circled[1]{\tikz[baseline=(char.base)]{
		\node[shape=circle,draw,inner sep=1pt] (char) {#1};}}
\usetikzlibrary{arrows, calc, decorations.markings, positioning}

\newdimen\satlevel
\newdimen\satdiameter
\newcommand{\satisfaction}[2][]{%
	\satdiameter=1.9ex\relax
	\ifcase#2\relax
	\satlevel=0pt\relax
	\or
	\satlevel=0.125\satdiameter
	\or
	\satlevel=0.25\satdiameter
	\or
	\satlevel=0.375\satdiameter
	\or
	\satlevel=0.5\satdiameter
	\fi
	\tikz[baseline=-0.3\satdiameter]{%
		\draw[#1] (0,0) circle (0.5\satdiameter);
		\fill[#1] (0,0) circle (\satlevel);
	}%
}

\makeatletter
\newenvironment{timeline}[6]{%
	\newcommand{\startyear}{#1}
	\newcommand{\tlendyear}{#2}
	
	\newcommand{\yearcolumnwidth}{#3}
	\newcommand{\rulecolumnwidth}{#4}
	\newcommand{\entrycolumnwidth}{#5}
	\newcommand{\timelineheight}{#6}
	
	\newcommand{\templength}{}
	
	\newcommand{\entrycounter}{0}
	
	\long\def\ifnodedefined##1##2##3{%
		\@ifundefined{pgf@sh@ns@##1}{##3}{##2}%
	}
	
	\newcommand{\ifnodeundefined}[2]{%
		\ifnodedefined{##1}{}{##2}
	}
	
	\newcommand{\drawtimeline}{%
		\draw[timelinerule] (\yearcolumnwidth+5pt, 0pt) -- (\yearcolumnwidth+5pt, -\timelineheight);
		\draw (\yearcolumnwidth+0pt, -10pt) -- (\yearcolumnwidth+10pt, -10pt);
		\draw (\yearcolumnwidth+0pt, -\timelineheight+15pt) -- (\yearcolumnwidth+10pt, -\timelineheight+15pt);
		
		\pgfmathsetlengthmacro{\templength}{neg(add(multiply(subtract(\startyear, \startyear), divide(subtract(\timelineheight, 25), subtract(\tlendyear, \startyear))), 10))}
		\node[year] (year-\startyear) at (\yearcolumnwidth, \templength) {\startyear};
		
		\pgfmathsetlengthmacro{\templength}{neg(add(multiply(subtract(\tlendyear, \startyear), divide(subtract(\timelineheight, 25), subtract(\tlendyear, \startyear))), 10))}
		\node[year] (year-\tlendyear) at (\yearcolumnwidth, \templength) {\tlendyear};
	}
	
	\newcommand{\entry}[2]{%
		\pgfmathtruncatemacro{\lastentrycount}{\entrycounter}
		\pgfmathtruncatemacro{\entrycounter}{\entrycounter + 1}
		
		\ifdim \lastentrycount pt > 0 pt%
		\node[entry] (entry-\entrycounter) [below of=entry-\lastentrycount] {##2};
		\else%
		\pgfmathsetlengthmacro{\templength}{neg(add(multiply(subtract(\startyear, \startyear), divide(subtract(\timelineheight, 25), subtract(\tlendyear, \startyear))), 10))}
		\node[entry] (entry-\entrycounter) at (\yearcolumnwidth+\rulecolumnwidth+10pt, \templength) {##2};
		\fi
		
		\ifnodeundefined{year-##1}{%
			\pgfmathsetlengthmacro{\templength}{neg(add(multiply(subtract(##1, \startyear), divide(subtract(\timelineheight, 25), subtract(\tlendyear, \startyear))), 10))}
			\draw (\yearcolumnwidth+2.5pt, \templength) -- (\yearcolumnwidth+7.5pt, \templength);
			\node[year] (year-##1) at (\yearcolumnwidth, \templength) {##1};
		}
		
		\draw ($(year-##1.east)+(2.5pt, 0pt)$) -- ($(year-##1.east)+(7.5pt, 0pt)$) -- ($(entry-\entrycounter.west)-(5pt,0)$) -- (entry-\entrycounter.west);
	}
	
	\newcommand{\plainentry}[2]{% plainentry won't print date in the timeline
		\pgfmathtruncatemacro{\lastentrycount}{\entrycounter}
		\pgfmathtruncatemacro{\entrycounter}{\entrycounter + 1}
		
		\ifdim \lastentrycount pt > 0 pt%
		\node[entry] (entry-\entrycounter) [below of=entry-\lastentrycount] {##2};
		\else%
		\pgfmathsetlengthmacro{\templength}{neg(add(multiply(subtract(\startyear, \startyear), divide(subtract(\timelineheight, 25), subtract(\tlendyear, \startyear))), 10))}
		\node[entry] (entry-\entrycounter) at (\yearcolumnwidth+\rulecolumnwidth+10pt, \templength) {##2};
		\fi
		
		\ifnodeundefined{invisible-year-##1}{%
			\pgfmathsetlengthmacro{\templength}{neg(add(multiply(subtract(##1, \startyear), divide(subtract(\timelineheight, 25), subtract(\tlendyear, \startyear))), 10))}
			\draw (\yearcolumnwidth+2.5pt, \templength) -- (\yearcolumnwidth+7.5pt, \templength);
			\node[year] (invisible-year-##1) at (\yearcolumnwidth, \templength) {};
		}
		
		\draw ($(invisible-year-##1.east)+(2.5pt, 0pt)$) -- ($(invisible-year-##1.east)+(7.5pt, 0pt)$) -- ($(entry-\entrycounter.west)-(5pt,0)$) -- (entry-\entrycounter.west);
	}
	
	\begin{tikzpicture}
		\tikzstyle{entry} = [%
		align=left,%
		text width=\entrycolumnwidth,%
		node distance=12mm,% change column height
		anchor=west]
		\tikzstyle{year} = [anchor=east]
		\tikzstyle{timelinerule} = [%
		draw,%
		decoration={markings, mark=at position 1 with {\arrow[scale=1.5]{latex'}}},%
		postaction={decorate},%
		shorten >=0.4pt]
		
		\drawtimeline
	}
	{
	\end{tikzpicture}
	\let\startyear\@undefined
	\let\tlendyear\@undefined
	\let\yearcolumnwidth\@undefined
	\let\rulecolumnwidth\@undefined
	\let\entrycolumnwidth\@undefined
	\let\timelineheight\@undefined
	\let\entrycounter\@undefined
	\let\ifnodedefined\@undefined
	\let\ifnodeundefined\@undefined
	\let\drawtimeline\@undefined
	\let\entry\@undefined
}
\makeatother

\newcommand{\labelsubseccounter}[1]{
	\renewcommand\thesubsection{\Alph{subsection}}
	\addtocounter{subsection}{-1}
	\refstepcounter{subsection}
	\label{#1}
	\renewcommand\thesubsection{\thesection.\Alph{subsection}}
}
\begin{document}
	%%%
	\title{Rate-Splitting Multiple Access: Fundamentals, Survey, and Future Research Trends}
	
	\author{\IEEEauthorblockN{Yijie~Mao, \IEEEmembership{Member, IEEE}, Onur~Dizdar, \IEEEmembership{Member, IEEE}, Bruno~Clerckx, \IEEEmembership{Fellow, IEEE}, Robert~Schober,
			\IEEEmembership{Fellow, IEEE}, Petar Popovski, 
			\IEEEmembership{Fellow, IEEE}, and~H. Vincent Poor, \IEEEmembership{Life Fellow, IEEE}}
		
		\thanks{ \par This work has been partially supported by the U.K. Engineering and Physical
			Sciences Research Council (EPSRC) under grant EP/N015312/1, EP/R511547/1;  partially supported by Shanghai Sailing Program under Grant 22YF1428400. The work of Petar Popovski was partly supported by the Villum Investigator Grant “WATER” from the Velux Foundation, Denmark. Robert Schober's work was partly supported by the Federal Ministry of Education and Research of Germany under the programme of “Souveran. Digital. Vernetzt.” joint project 6G-RIC, project identification number: PIN 16KISK023. \textit{(Corresponding author: Bruno Clerckx)}
			\par Y. Mao is with the School of Information Science and Technology,
			ShanghaiTech University, Shanghai 201210, China (e-mail: maoyj@shanghaitech.edu.cn).
			\par O. Dizdar and B. Clerckx are with Department of Electrical and Electronic Engineering, Imperial College London, London SW7 2AZ, U.K (email: \{o.dizdar, b.clerckx\}@imperial.ac.uk).
			\par R. Schober is with the Institute of Digital Communications, Friedrich-Alexander University Erlangen-Nuremberg, 91054 Erlangen, Germany (e-mail: robert.schober@fau.de).
			\par P. Popovski is with the Department of Electronic Systems, Aalborg University, 9220 Aalborg, Denmark (e-mail: petarp@es.aau.dk).
			\par H. Vincent Poor is with the Department of Electrical and Computer Engineering, Princeton University, Princeton, NJ 08544 USA (e-mail: poor@princeton.edu).
	} }

	%\thanks{Department of Electrical and Electronic Engineering, Imperial College London,
	% United Kingdom, e-mail: \{e.piovano15, hamdi.joudeh10, b.clerckx\}@imperial.ac.uk.}% <-this % stops a space
	%%%%%%%%%%%%%%%%%%%
	\maketitle
	%\let\endthebibliography\endlist
	%%%%%%%%%%%%%%%%%%%%
	\begin{abstract}
		
		\par 
		%%%
		Rate-splitting multiple access (RSMA) has emerged as a novel, general, and powerful framework for the design and optimization of non-orthogonal transmission, multiple access (MA), and interference management strategies for future wireless networks. 
		%%%
		By exploiting splitting of user messages as well as non-orthogonal transmission of common messages decoded by multiple users and private messages decoded by their corresponding users, RSMA can softly bridge and therefore reconcile the two extreme interference management strategies of fully decoding interference and treating interference as noise.
		%%%
		RSMA has been shown to generalize and subsume as special cases four existing MA schemes, namely, orthogonal multiple access (OMA), physical-layer multicasting, space division multiple access (SDMA) based on linear precoding (currently used in the fifth generation wireless network--5G), and  non-orthogonal multiple access (NOMA) based on linearly precoded superposition coding with successive interference cancellation (SIC).
		%%%
		Through information and communication theoretic analysis, RSMA has been shown to be optimal (from a Degrees-of-Freedom region perspective) in several transmission scenarios.
		%%%
		Compared to the conventional MA strategies used in 5G, RSMA enables spectral efficiency (SE), energy efficiency (EE), coverage, user fairness, reliability, and quality of service (QoS) enhancements for a wide range of network loads (including both underloaded and overloaded regimes) and user channel conditions. Furthermore, it enjoys a higher robustness against imperfect channel state information at the transmitter (CSIT) and entails lower feedback overhead and complexity. 
		%%%
		Despite its great potential to fundamentally change the physical (PHY) layer and  media access control (MAC) layer of wireless communication networks, RSMA is still confronted with many challenges on the road towards standardization. 
		%%%
		In this paper, we present the first  comprehensive tutorial on RSMA by providing a survey of the pertinent state-of-the-art research, detailing its architecture, taxonomy, and various appealing applications, as well as comparing with existing MA schemes in terms of their overall frameworks, performance, and complexities.
		%%%
		An in-depth discussion of future RSMA research challenges is also provided to inspire future research on RSMA-aided wireless communication for beyond  5G systems.
		%%%
	\end{abstract}
	%%%%%%%%%%%%%%%%%%%%
	
	%%%
	
	\begin{table*}
		\caption{List of abbreviations.}
		\label{tab:Acronyms}
		\centering
		\begin{tabular}{|l|l||l|l|}
			\hline
			AMC   & Adaptive Modulation and Coding               & mmWave   & millimeter-Wave                                           \\
			BC    & Broadcast Channel                            & MU--LP   & Multi-User Linear Precoding                               \\
			BD    & Block Diagonalization                        & MU--MIMO & Multi-User Multiple-Input Multiple-Output                 \\
			CDMA  & Code Division Multiple Access                & NOMA     & Non-Orthogonal Multiple Access                            \\
			CoMP  & Coordinated Multi-Point                      & NOUM     & Non-Orthogonal Unicast and Multicast                      \\
			CSI   & Channel State  Information                   & NR       & New Radio                                                 \\
			CSIT  & Channel State Information at the Transmitter & OFDMA    & Orthogonal Frequency Division Multiple Access             \\
			C-RAN & Cloud-Radio  Access  Networks                & OMA      & Orthogonal Multiple Access                                \\
			DoF   & Degree-of-Freedom                            & PHY      & Physical                                                  \\
			DPC   & Dirty Paper Coding                           & QoS      & Quality of Service                                        \\
			DPCRS & Dirty Paper Coded Rate-Splitting             & RF       & Radio Frequency                                           \\
			D2D   & Device-to-Device                             & RS       & Rate-Splitting                                            \\
			EE    & Energy Efficiency                            & RSMA     & Rate-Splitting Multiple Access                            \\
			eMBB  & enhanced Mobile Broadband Service            & RS-CMD   & Rate-Splitting  and  Common  Message Decoding             \\
			ER    & Ergodic Rate                                 & RZF      & Regularized Zero-Forcing Beamforming                      \\
			ESR   & Ergodic Sum Rate                             & SAGIN    & Space-Air-Ground Integrated Networks                      \\
			FDD   & Frequency  Division  Duplex                  & SC       & Superposition Coding                                      \\
			FDMA  & Frequency Division Multiple Access           & SCA      & Successive Convex Approximation                           \\
			FBL   & Finite Blocklength                           & SDMA     & Space Division Multiple Access                            \\
			F-RAN & Fog-Radio Access Networks                    & SE       & Spectral Efficiency                                       \\
			GDoF  & Generalized Degree-of-Freedom                & SIC      & Successive Interference Cancellation                      \\
			HARQ  & Hybrid Automatic Repeat Request              & SISO     & Single-Input Single-Output                                \\
			HK    & Han  and  Kobayashi                          & SNR      & Signal-to-Noise Ratio                                     \\
			HRS   & Hierarchical RS                              & SVD      & Singular Value Decomposition                              \\
			IC    & Interference Channel                         & SWIPT    & Simultaneous  Wireless  Information  and  Power  Transfer \\
			IRS   & Intelligent  Reconfigurable   Surface        & TDD      & Time  Division  Duplex                                    \\
			JD    & Joint  Detection                             & TDMA     & Time-Division Multiple Access                             \\
			KPIs  & Key Performance Indicators                   & THP      & Tomlinson-Harashima Precoding                             \\
			LLS   & Link-Level Simulation                        & THPRS    & Tomlinson-Harashima Precoded Rate-Splitting               \\
			MA    & Multiple Access                              & UAV      & Unmanned Aerial Vehicles                                  \\
			MAC   & Multiple Access Channel                      & URLLC    & Ultra-Reliable  Low-Latency  Communication                \\
			MBF   & Matched Beamforming                          & VLC      & Visible Light Communication                               \\
			MIMO  & Multiple-Input Multiple-Output               & V2X      & Vehicle-to-Everything                                     \\
			MISO  & Multiple-Input Single-Output                 & WMMSE    & Weighted Minimum Mean Square Error                        \\
			MMF   & Max-Min  Fairness                            & WSR      & Weighted Sum Rate                                         \\
			MMSE  & Minimum Mean Square Error                    & ZFBF     & Zero-Forcing Beamforming                                  \\
			mMTC  & massive Machine-Type Communication           &          &                                                          \\ \hline
		\end{tabular}
	\end{table*}
	
	\section{Introduction}
	\label{sec:intro}
	The sixth generation mobile communications (6G) system is attracting significant attention from academia and industry. 
	%%%
	It is envisioned that 6G will enable the Internet of Everything, provide services with higher throughput, ultra reliability, heterogeneous quality of service (QoS), massive connectivity, and support the convergence of communications, sensing, localization, computing, and control.
	%%%
	6G therefore requires a more efficient use of the wireless resources and more powerful means to manage interference, which has triggered a rethinking and redesign of physical (PHY) layer and multiple access (MA) techniques for wireless communication systems \cite{6Gwhitepaper2021RSMA}. 
	%%%
	In this section, we first  briefly introduce a candidate MA scheme for 6G, namely, rate-splitting multiple access (RSMA), and then discuss the PHY layer challenges for the emerging MA schemes. After that, we specify the requirements of the  fifth generation (5G) and 6G of mobile communication and motivate the need for RSMA.
	%%%
	The major contributions of this treatise are summarized at the end of the section.
	%%%
	Table \ref{tab:Acronyms} details the main abbreviations used throughout this work.
	%%%
	\begin{figure}[t!]
		\centering
		%	\hspace*{0.2cm}
		\begin{subfigure}[b]{0.4\textwidth}
			\centering
			\includegraphics[width=0.98\textwidth]{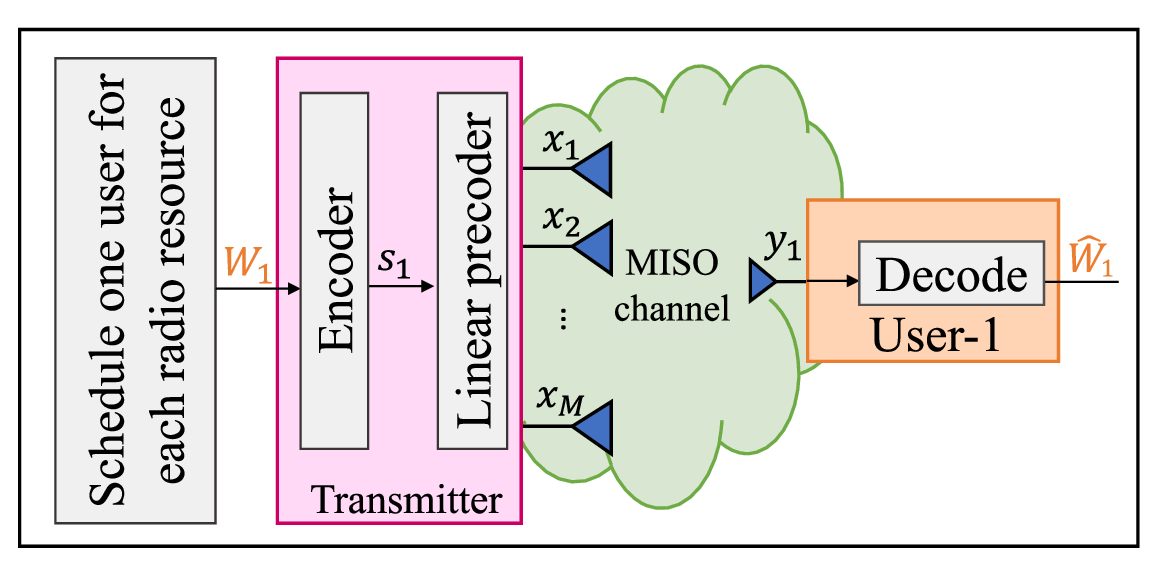}%
			\vspace{-2mm}
			\caption{OMA}
			\vspace{1mm}
		\end{subfigure}%
		\\
		\begin{subfigure}[b]{0.4\textwidth}
			\centering
			\includegraphics[width=\textwidth]{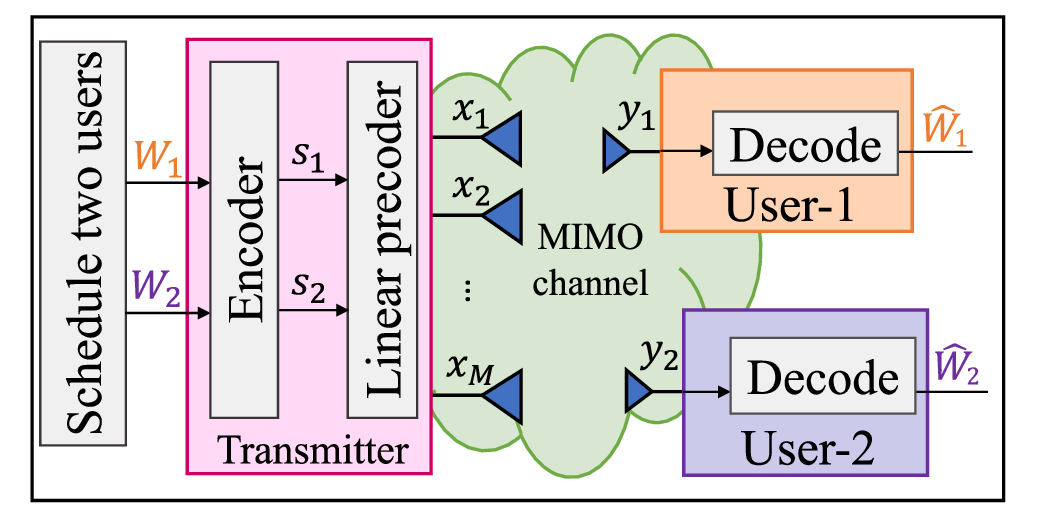}%
			\vspace{-2mm}
			\caption{SDMA}
			\vspace{2mm}
		\end{subfigure}%
		\\
		\begin{subfigure}[b]{0.43\textwidth}
			\centering
			\includegraphics[width=\textwidth]{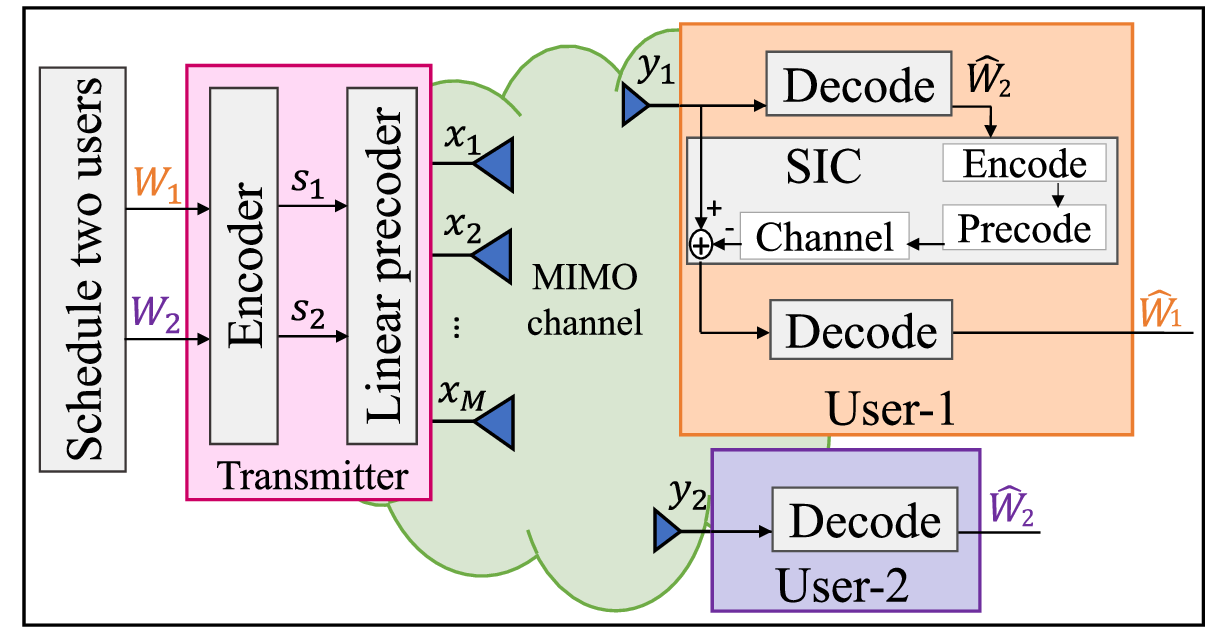}%
			\vspace{-1mm}
			\caption{NOMA (SC--SIC)}
			\vspace{2mm}
		\end{subfigure}%
		\\%%%
		\begin{subfigure}[b]{0.48\textwidth}
			\centering
			\includegraphics[width=\textwidth]{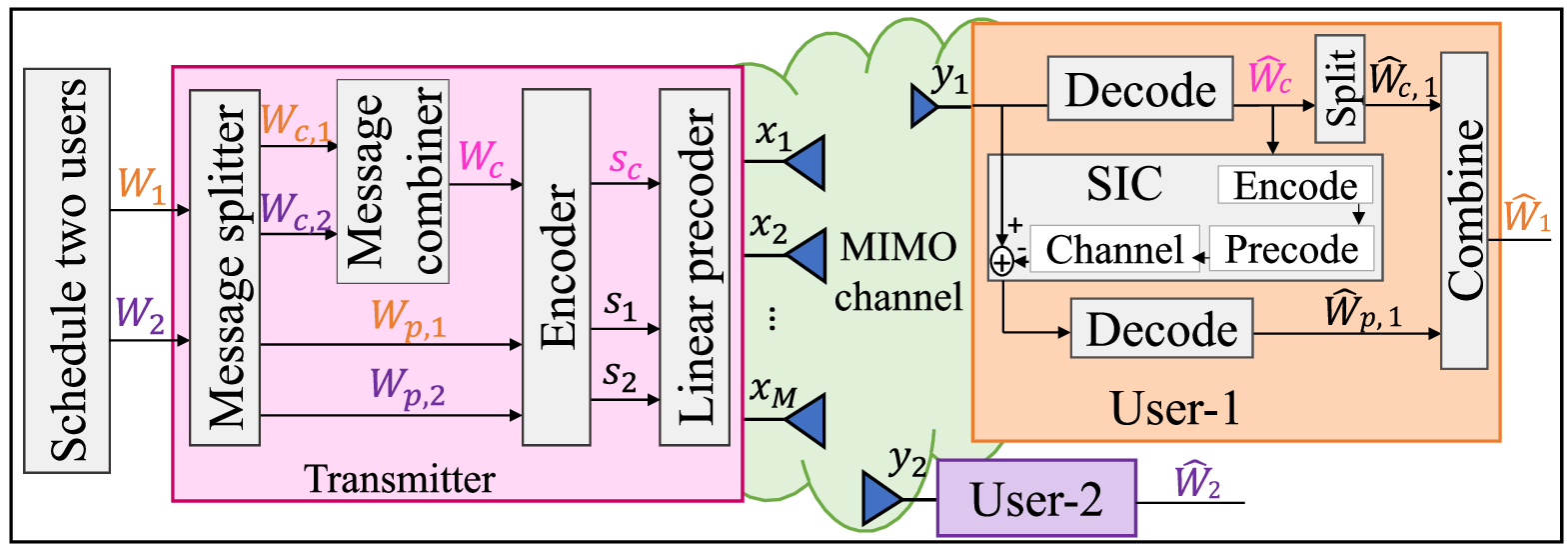}%
			\vspace{-1mm}
			\caption{RSMA (1-layer RS)}
			\vspace{1mm}
		\end{subfigure}%
		\caption{Two-user OMA, SDMA, NOMA, RSMA for MISO BC (within one radio resource). }
		\label{fig:twoUserBaseline}
	\end{figure}
	\par
	\subsection{Rate-Splitting Multiple Access}
	\labelsubseccounter{I-A}
	Over the past few decades, the evolution of cellular networks from the first generation (1G) to 5G has been reflected in the concurrent development of new MA techniques. 
	%%%
	From 1G to the fourth generation (4G) wireless networks, MA has evolved with one common goal, that is to allocate orthogonal radio resources  to the users so as to avoid multi-user interference. 
	%%%
	Orthogonal multiple access (OMA), where users are scheduled in orthogonal domains (e.g., the time domain as in time division multiple access--TDMA, the frequency domain as in frequency division multiple access--FDMA, the code domain as in code division multiple access--CDMA, or both the time and frequency domains as in orthogonal frequency division multiple access--OFDMA) is therefore adopted.
	%%%
	In 4G and 5G, the unprecedentedly exploding demand for wireless capacity and the scarcity of the spectrum resources have motivated the adoption of multiple-input multiple-output (MIMO) communication, where  multiple antennas are deployed at all access points.
	%%%
	MIMO networks create a spatial dimension, which opens the door for the application of space-division multiple access (SDMA). 
	%%%
	By properly utilizing  the spatial resources and multi-antenna processing at the transmitter,  SDMA is capable of  serving multiple users in the same time-frequency resource, and each user can directly decode the intended data streams by treating any residual interference as noise. 
	%%%
	To further enhance the spectral efficiency (SE), MA has progressed towards the direction of non-orthogonal multiple access (NOMA), where users are superposed in the same time-frequency resources via the power domain (e.g., power-domain NOMA) or code domain (e.g., sparse code multiple access) \cite{book2021Mao}. 
	%%%
	Power-domain NOMA\footnote{This treatise focuses on  power-domain NOMA. In the rest of the paper, ``power-domain NOMA" is referred to as ``NOMA"  for simplicity.} relies on superposition coding (SC) at the transmitter  to superpose user messages in the power domain and successive interference cancellation (SIC) at the receivers (a.k.a. SC--SIC) \cite{NOMA2013YSaito,NOMAsurvey2015,NOMAsurvey2017Ding,wshin2017RSNOMA}.  
	%%%
	Such approach manages multi-user interference by forcing (at least) one user to successfully decode messages (and remove interference) of other users. 
	%%%
	\par 
	Recently, RSMA, built upon the concept of rate-splitting (RS), has been recognized as a promising PHY-layer transmission paradigm for non-orthogonal transmission, interference management and MA strategies in 6G. 
	%%%
	The main idea behind RSMA is to split user messages into common and private parts, and enable the capability of \textit{partially decoding the interference and partially treating the interference as noise}, which contrasts with the extreme interference management strategies used in SDMA and NOMA. 
	The flexible nature of RSMA allows it to perform well for all levels of interference. 
	%%%
	When the interference is weak or strong, RSMA automatically reduces to SDMA or NOMA by tuning the powers and contents of the common and private streams. RSMA naturally  bridges SDMA and NOMA, including any hard switching between SDMA and NOMA \cite{mao2017rate}. 
	%%%
	\par 
	To capture the difference among the aforementioned MA schemes, we illustrate a two-user toy example for one radio resource in Fig. \ref{fig:twoUserBaseline} where the transmitter is equipped with $M$ antennas and serves two users.  For OMA, one user (i.e., user-1) is selected  to occupy the entire radio resource while the other user  (i.e., user-2) has to wait for access. For SDMA, the two user messages are independently encoded into two streams and linearly precoded at the transmitter. Each user directly decodes its intended stream by treating any residual interference from the other stream as noise. For NOMA, the user messages are independently encoded into streams, which  are superposed at the transmitter and broadcast to the users. The data stream of one user (i.e., user-2) has to be decoded by both users.
	%%%
	For RSMA, the transmitter splits the  message $W_k$ of each user-$k$  into  a common part $W_{c,k}$ and a private part $W_{p,k}$ and combines $W_{c,1}, W_{c,2}$ into a common message $W_c$. The three messages $W_c, W_{p,1}$, and $W_{p,2}$ generated from $W_{1}$ and $W_{2}$ are independently encoded and linearly precoded at the  transmitter. Each user first decodes the common stream $s_c$ by treating all private streams as noise. After removing the decoded common stream from the received signal, each user decodes the intended private stream $s_k$ by treating the  private stream of the other user as noise. 
	%%%
	This allows flexible management of the interference by {partially decoding interference} through the common stream decoding and partially treating the interference as noise when decoding the intended private stream at each user \cite{RSintro16bruno}.
	%%%
	By turning off the common stream, $s_c$, such that $W_k$ is encoded into private stream $s_k$ directly, RSMA reduces to SDMA in Fig. \ref{fig:twoUserBaseline}(b) and all interference between $s_1$ and $s_2$ is treated as noise. Similarly, RSMA reduces to NOMA in Fig. \ref{fig:twoUserBaseline}(c) by turning off $s_2$, encoding $W_2$ into $s_c$, and encoding $W_1$ into $s_1$.
	%%%
	\par
	%%%
	RSMA therefore unifies the existing MA schemes, and constitutes a promising PHY-layer transmission paradigm for non-orthogonal transmission, interference management, and multi-user communications. Next, we will explain the major PHY layer challenges for the existing MA schemes and why it is imperative to employ this new paradigm in the design of modern and future wireless networks.
	%%%
	\par 
	\subsection{PHY Layer Challenges for the Multiple Access Schemes}
	\label{sec:motivations}
	\labelsubseccounter{I-B}
	\par In this subsection, we summarize the properties of the state-of-the-art MA schemes including OMA, SDMA, NOMA along with their advantages and disadvantages.
	%%%
	
	\subsubsection{Orthogonal multiple access}
	%%%
	In 1G, FDMA was employed where the available spectrum was partitioned into non-overlapped frequency bands, each accommodating one user.
	%%%
	The second generation (2G) standard systems adopted TDMA where  time was partitioned into time slots allocated to different users. 
	%%%
	The third generation (3G) systems exploited a third dimension via CDMA. In particular, by utilizing orthogonal, user-specific codes to spread the modulated user symbols, CDMA serves multiple users simultaneously in the same time-frequency resources without causing multi-user interference (under ideal propagation conditions).
	%%%
	In 4G, OFDMA was deployed by  dividing the frequency and time resources into narrow subcarriers and time slots, which were grouped into  resource units and allocated to the users. 
	%%%
	FDMA, TDMA, CDMA, and OFDMA  are collectively referred to as OMA.
	%%%
	\par
	The advantages and disadvantages of OMA can be summarized as follows:
	\begin{itemize}
		\item \textit{Advantages:} The benefits of  OMA include a simple  transceiver design and the avoidance of multi-user interference.
		%%%
		\item \textit{Disadvantages:}  As each orthogonal radio resource in OMA is dedicated to a single user, the number of  users simultaneously supported is restricted by the total number of available radio resources, which therefore limits SE. 
		%%%
		Another issue of OMA is that a low-rate user (such as an Internet of Things--IoT sensor) needing only a small amount of resources may still occupy a full resource block by itself, which further leads to an inefficient use of spectrum. 
		%%%
		Moreover,  well-designed user scheduling, which entails a high signaling overhead, is required to guarantee the system performance for OMA.
		%%%
	\end{itemize}
	
	%%%
	\subsubsection{Space division multiple access}
	\label{sec: SDMA}
	MIMO has become one of the most essential and indispensable technologies for current  wireless networks, and is included in virtually all  high-rate wireless standards (e.g., 5G New Radio--NR,  4G Long Term Evolution--LTE, IEEE 802.11n, WiMAX). 
	%%%
	For the MIMO/multiple-input single-output (MISO) Gaussian broadcast channel (BC), dirty paper coding (DPC) is the only known strategy for achieving the capacity region \cite{capacityRegion2006HW} when the channel state information at the transmitter (CSIT) is perfect.  
	%%%
	Although appealing from an information-theoretic point of view,  DPC is impractical due to the high computational burden of the encoding process. 
	%%%
	Alternative non-linear precoding techniques such as Tomlinson-Harashima precoding (THP) and vector perturbation precoding  have therefore been proposed. They achieve a performance close to that of DPC, but are still relatively complex.
	%%%
	A more practical precoding technique for the multi-antenna BC (including the MIMO and MISO BC) is  multi-user linear precoding (MU--LP) relying on linear precoding at the transmitter and treating multi-user interference as noise at the receivers  \cite{clerckx2013mimo}. 
	%%%
	Although MU--LP is suboptimal for the multi-antenna BC, it achieves near-capacity performance when the CSIT is perfect and the user channels are nearly orthogonal with similar channel strengths or similar long-term signal-to-noise ratios (SNRs). 
	%%%
	SDMA based on MU--LP\footnote{In the rest of the paper, for simplicity, we use ``SDMA" when we want to refer to ``SDMA based on MU--LP".}  is therefore an integral part of numerous 4G and 5G  transmission schemes such as  multi-user MIMO (MU--MIMO), networked MIMO, coordinated multi-point (CoMP), massive MIMO, and millimeter-wave (mmWave) MIMO.
	%%%
	\par
	The advantages and disadvantages of SDMA can be summarized as follows:
	\begin{itemize}
		\item \textit{Advantages:} When the CSIT is perfect and the network  is underloaded (i.e., when the number of  antennas deployed at the transmitter is larger than the  number of receive antennas at all users),  SDMA  can successfully suppress (or eliminate) multi-user interference if the user channels are not aligned, and it achieves  the maximum  degrees-of-freedom (DoFs)\footnote{DoF, also known as (a.k.a.) spatial multiplexing gain, quantitatively captures  how well the   spatial dimension is exploited by a given communication strategy \cite{bruno2021MISONOMA}. The DoF of a user is the fraction or number of independent data streams that can be transmitted to that user. Its mathematical definition is detailed in Section \ref{sec:performanceCompare}.} of the underloaded multi-antenna BC \cite{DPCrateRegion03Goldsmith}. 
		%%%
		Moreover, the transmitter and receiver complexities of SDMA are low as the transmitter employs linear precoding and each receiver directly decodes the intended message by fully treating interference as noise.
		%%%
		\item \textit{Disadvantages:}  The major limitations of SDMA  are summarized into the following three points. 
		%%%
		\begin{itemize}
			\item \textit{First}, SDMA is sensitive to the network load.
			%%%
			It is only suitable for underloaded systems and the performance drops significantly when the network becomes overloaded (i.e., when the total number of streams, or equivalently the number of served users for the MISO BC, is larger than the number of antennas deployed at the transmitter) as having enough transmit antennas is a prerequisite for successful interference management based on using SDMA. 
			%%%
			A common method to handle overloaded settings is to separate users into different groups, schedule user groups via OMA, and perform SDMA in each user group, which however, reduces the QoS and increases latency.	
			%%%
			\item \textit{Second}, SDMA is sensitive to the user deployment (including the angles and strengths of the user channels), which therefore imposes a stricter requirement on the scheduler. 
			%%%
			SDMA requires the scheduler to pair  users with  nearly orthogonal channels and  relatively similar channel strengths.
			%%%
			Although suboptimal and low-complexity scheduling and user-pairing algorithms exist, the scheduler complexity rises swiftly when an exhaustive search is conducted to maximize the SE \cite{clerckx2013mimo}. 
			%%%
			\item \textit{Third}, SDMA is sensitive to CSIT inaccuracy. In contrast to its good performance in  the perfect CSIT setting, SDMA cannot achieve the maximal DoFs  when CSIT is imperfect \cite{RSintro16bruno}. In fact, its performance decreases dramatically in the presence of imperfect CSIT \cite{NJindalMIMO2006}. This is due to the fact that SDMA is designed  for perfect CSIT. Applying a  framework motivated by perfect CSIT under  imperfect CSIT conditions results in residual multi-user interference caused by the imprecise interference mitigation at the transmitter (via imperfect linear precoding)  \cite{RSintro16bruno}. 
		\end{itemize}
	\end{itemize}
	
	\subsubsection{Non-Orthogonal Multiple Access}
	\label{sec: NOMA}
	\par  
	%%%
	%\par
	The study of NOMA can be traced back several decades to  the information theory and wireless communications literature on the single-input single-output (SISO) BC and MAC. It is well-known that SC--SIC is the capacity-achieving strategy for the SISO BC, and it is capacity-achieving with time-sharing for the SISO MAC \cite{Tcover1972,tsefundamentalWC2005}.
	%the SISO (Gaussian) BC and  MAC 
	% SC--SIC is well-recognized as the capacity-achieving scheme of SISO (Gaussian) BC \cite{Tcover1972}, and SIC at receivers  is  capacity-achieving  in SISO MAC \cite{tsefundamentalWC2005}.
	%%%
	%In the SISO BC,  the superiority of NOMA  over OMA  is pronounced when users experience channel strength differences \cite{tsefundamentalWC2005}.
	%%%
	%However, when users experience identical channel strengths, OMA is sufficient to achieve the capacity region of the SISO BC and the advantage of NOMA disappears  \cite{tsefundamentalWC2005}.
	%%%
	% In the SISO MAC with $K$ users, the capacity region has the shape of a $K$-dimensional polyhedron with the vertices on the boundary of the capacity region being achieved by NOMA and all points along the line segment between adjacent vertices being achieved by time sharing. 
	%%%
	% OMA in contrast only achieves a few points on the boundary of the capacity region when power control is applied.
	%%%
	\begin{figure}[t!]
		\centering
		\includegraphics[width=3.0in]{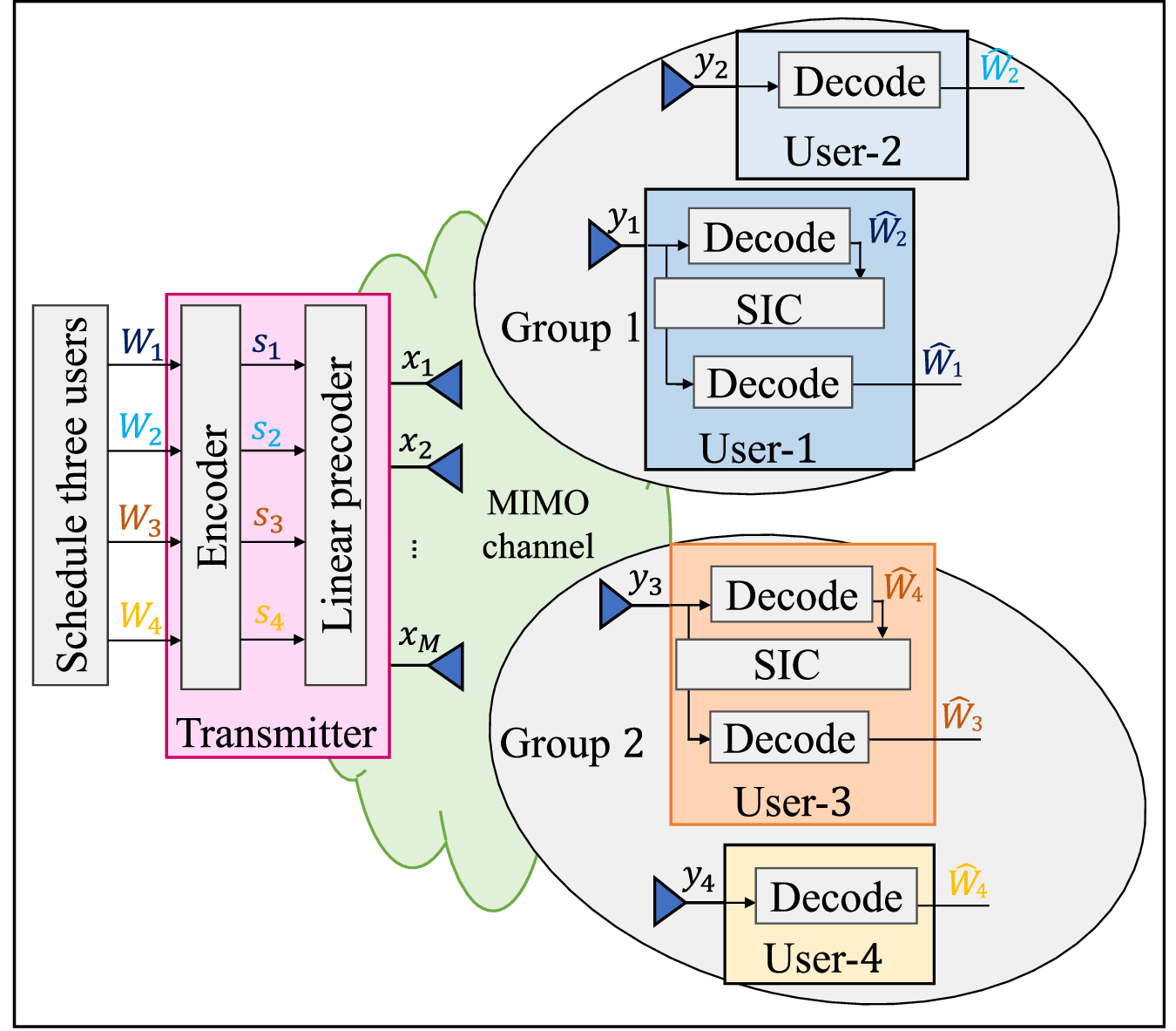}%
		\caption{Four-user multi-antenna NOMA with two user groups and two users in each group.}
		\label{fig:SCSICperGroup}
		\vspace{-2mm}
	\end{figure}
	%%%
	\par 
	Driven by the performance gain of NOMA over OMA in single-antenna networks, numerous attempts have been made to synergize NOMA and multi-antenna networks in recent years  \cite{Mojtaba2019MAbook,QSunNOMA2015,QZhangNOMA2016,NOMA2013YSaito,DingNOMA2016,yuanwei2021NGMA}. 
	%%%
	In the multi-antenna BC, MIMO/MISO NOMA combines SDMA and NOMA by separating $K$ users into $G$ ($1\leq G\leq K$) different groups. At each user, the interference from users within the same group is managed via  SC--SIC  while interference from users in different groups is managed via SDMA.
	%%%
	When $G=1$, all users are in a single group. This corresponds to a direct extension of single-antenna NOMA (SC--SIC) to the multi-antenna BC by ordering users according to their effective channel strengths (after linear precoding) and forcing users to successfully decode messages (and remove interference) \cite{QSunNOMA2015,QZhangNOMA2016}. 
	%%%
	When $G>1$, the transmitter uses linear precoding to decompose the multi-antenna BC into $G$ single-antenna NOMA channels and there is no interference between the decomposed NOMA channels if the transmitter has a sufficient number of transmit antennas and perfect CSIT knowledge
	\cite{NOMA2013YSaito,DingNOMA2016,yuanwei2021NGMA}. 
	%%%
	In the rest of the paper, the terms ``multi-antenna NOMA ($G=1$)" and ``multi-antenna NOMA ($G>1$)" are  used to denote the multi-antenna NOMA schemes with one user group and  multiple user groups, respectively. 
	%%%
	Fig. \ref{fig:SCSICperGroup} illustrates  a four-user multi-antenna NOMA model with $G=2$ user groups and two users in each group (i.e., user-1/2 in group 1 and user 3/4 in group 2). The inter-group interference managed by SDMA is treated as noise at all users, while the intra-group interference managed by NOMA is fully decoded at one user in each group.
	%%%
	For example, user-1 has to decode the message of user-2 before decoding its intended message, and it treats the interference from user-3/4 as noise.
	%%%
	%\par
	In the multi-antenna MAC, NOMA based on minimum mean square error (MMSE)--SIC receivers
	is known to achieve the vertices at the boundary of the capacity region, but time sharing is still required to achieve all points along the line segment between adjacent vertices.
	%%%
	\par
	The advantages and disadvantages of NOMA can be summarized as follows:
	\begin{itemize}
		\item \textit{Advantages:} The major advantage of NOMA is its potential of improving SE in severely overloaded scenarios by serving  users  with closely aligned channels  and diverse channel strengths  in the same time-frequency resources.
		%%%	
		\item \textit{Disadvantages:}  Though SC--SIC attains the  capacity region of the scalar Gaussian BC, as the number of users increases, the receiver complexity (i.e., the number of SIC layers) increases and  SIC error propagation is aggravated.
		%%%
		For the $K$-user SISO BC, the user with the strongest channel  requires $K-1$ layers of SIC to decode and remove the interference from the $K-1$  messages of all other co-scheduled users before being able to access its intended message.
		%%%
		One practical approach to address this issue is to restrict the number of SIC layers at each user by clustering users into smaller groups, applying SC--SIC in each group, and scheduling user groups via OMA. Such approach, however, may cause a loss in performance and increase latency. 
		%%%
		While forcing a user to  decode the entire message of another user is effective  in the degraded SISO BC, it is not an efficient approach in multi-antenna networks  for  most scenarios due to the inefficient use of the multiple antennas and the SIC receivers. 
		%%%
		As  revealed in \cite{bruno2021MISONOMA}, in multi-antenna networks, NOMA has several limitations, which are summarized in the following four points:
		\begin{itemize}
			%%%
			\item  \textit{First},  multi-antenna NOMA can suffer from a loss of DoF due to the inefficient use of SIC.
			%%%
			As per \cite[Table II]{bruno2021MISONOMA},
			for a MISO BC employing one $M$-antenna transmitter having perfect CSIT and simultaneously serving $K$ single-antenna users, the sum-DoF achieved by multi-antenna NOMA with $G=1$ is only 1, which is  equivalent to the DoF of OMA/single user-MISO transmission.
			%%%
			In comparison, SDMA achieves the optimal sum-DoF of $\min\{M, K\}$ in the same setting without using any SIC at the users.
			%%%
			Hence, in this case, the DoF loss of multi-antenna NOMA is accompanied by a dramatic  increase in the receiver complexity.
			%%%
			\item  \textit{Second}, multi-antenna NOMA can impose  significant computational burdens on the transmitter and the receivers. Besides the multiple SIC layers required at each user for decoding interference, the transmitter also requires a joint optimization of the precoders, user grouping, and decoding orders as  they are coupled with each other.
			%%%
			One commonly adopted approach for complexity reduction in multi-antenna NOMA is to fix the beamformers of users in the same group \cite{NOMA2013YSaito}. Such approach further restricts user performance because the  feasible region of the precoders shrinks accordingly.
			%%%
			\item \textit{Third}, multi-antenna NOMA can be sensitive to the  user deployment. 
			In particular, it is generally most suitable for highly overloaded scenarios where the user channels are nearly aligned in each user group and are relatively orthogonal  among different user groups. 
			%%%
			\item  \textit{Fourth}, multi-antenna NOMA can be  vulnerable to CSIT uncertainties as the inter-group interference  is managed in  the same manner as in SDMA. 
		\end{itemize}
	\end{itemize}
	
	%%%
	\par 
	Comparing  existing MA schemes, within one radio resource, OMA  only supports a single user while both SDMA and NOMA can accommodate multiple users.
	%%%
	In this sense, SDMA can also be regarded as a non-orthogonal transmission strategy that superposes users in the same radio resource.
	%%%
	In fact, whether the spectrum is orthogonally or non-orthogonally allocated among users  is not the fundamental problem.
	What is foremost is \textit{how  multi-user interference is managed}.
	%%%
	This is indeed the major difference between  OMA, SDMA, and NOMA.
	%%%
	Specifically, OMA avoids interference by allocating orthogonal radio resources among users. SDMA and NOMA use two distinct strategies to manage   interference, where   \textit{SDMA based on MU--LP  relies on fully treating any residual multi-user interference as noise, while NOMA based on SC--SIC relies on fully decoding and removing multi-user interference}. 
	%%%
	\par
	The aforementioned disadvantages of the existing MA schemes are consequences of how they manage interference.
	%%%
	As per \cite{Tse2008}, treating
	interference as pure noise is only promising when the multi-user interference is weak while fully decoding interference is only advisable when the interference is sufficiently strong. When the interference is medium (i.e., neither strong nor weak),  decoding the interference partially can significantly improve performance.
	%%%
	The lesson that can be learnt from \cite{Tse2008} for the design of future MA schemes is that  \textit{the amounts of interference to be decoded and treated as noise, respectively, should be adaptively changed based on the interference level.}
	%%%

	%%%
	\par
	Unfortunately, SDMA and NOMA are only well suited for weak and strong interference levels, respectively.
	%%%
	To achieve adaptive interference management, one intuitive approach is to dynamically switch between  SDMA  and NOMA \cite{quasidegrade2016}. In such an approach, however, either SDMA or NOMA is supported in the system, which does not work well for medium interference levels.
	%%%
	\subsection{5G/6G and the Need for RSMA}
	\label{sec:RSMAintro}
	\labelsubseccounter{I-C}
	5G New Radio (NR) has defined three core services, namely, ultra-reliable low-latency communication (URLLC), massive machine-type communication (mMTC), and enhanced mobile broadband (eMBB), where gigabit per second (Gbps)-level data rates are required for eMBB users, reliable transmissions with $10^{-5}$ block error rate (BLER) are required for URLLC users with a latency within 1 ms, and a high connection density and high energy efficiency (EE) are required for mMTC.
	%%%
	The above key performance indicators (KPIs) of NR are expected to be further tightened and enhanced in 6G.
	%%%
	Interference has been recognized as one of the major obstacles to achieving the aforementioned KPIs in NR, and is envisioned to become even more severe and  unpredictable in 6G.
	%%%
	If managed inappropriately,  multi-user interference will become dominant and degrade the system performance.
	%%%
	Therefore, there is an urgent need for a new paradigm for interference management in view of the limitations of existing MA schemes.
	%%%
	\par
	Moreover, in modern MIMO networks, one major source of multi-user interference is imperfect CSIT.
	%%%
	In the presence of imperfect CSIT, interference management is further impeded because CSIT imperfections imply that interference cannot be managed easily by the precoders at the transmitter anymore, e.g., interference cannot be eliminated since the channel is not known accurately.
	%%%
	Unfortunately, the acquisition of accurate CSIT is challenging due to many inevitable sources of  impairment. 
	%%%
	For example, in time division duplex (TDD) MIMO networks, due to the unavoidable pilot sequence reuse of the users in different cells, the issue of pilot contamination arises, which further contributes to  inter-cell interference as well as channel estimation errors \cite{PilotConta2011TWC}. 
	%%%
	In frequency division duplexing (FDD) MIMO networks, downlink channel estimation and uplink channel feedback usually lead to an overwhelming pilot signaling and feedback overhead, particularly when the dimension of the transmit antenna array is large, which therefore hampers CSIT acquisition \cite{FDD2015TCOM}. 
	%%%
	Moreover, feedback delay, user mobility,  inaccurate calibration of radio
	frequency (RF) chains, RF impairments (e.g., phase noise), and channel state information (CSI) estimated at the sub-band level (rather than the subcarrier level)  all prevent the CSIT to be accurate \cite{clerckx2013mimo}. 
	%%%
	According to the statistics revealed in \cite{5Gtrail2020},  the CSIT inaccuracy caused by an average user mobility of  13.2 km/h  leads to  at least 50\% cell average throughput loss in 5G massive MIMO.
	% According to the statistics  revealed in \cite{clerckx2013mimo},  the CSIT inaccuracy in MIMO networks leads to  at least 30\% cell average throughput loss and 42\% cell edge throughput  loss in 4G LTE-Advanced. 
	%%%
	It introduces additional multi-user interference and has become the primary performance bottleneck in MIMO networks.
	%%%
	Such severe  multi-user interference, however,  is simply treated as noise by both  SDMA and multi-antenna NOMA (with multiple user groups). 
	%%%
	The classical approach for dealing with this practical limitation of imperfect CSIT takes a ``robustification" stance where the precoders for SDMA and NOMA that have been designed under the assumption of perfect CSIT are tweaked to account for imperfect CSIT \cite{Member2007,clerckx2013mimo,QZhangNOMA2016}.
	%%%
	\setlength\extrarowheight{3pt}
	\begin{table*}[t!]
		\centering
		\caption{\label{tab:outline} Outline of the paper.}
		\begin{tabular}{|l l|}
			\hline
			\multicolumn{2}{|c|}{\textbf{Section \ref{sec:intro}. Introduction}} \\ 
			\ref{I-A}. Rate-Splitting Multiple Access          & \ref{I-B}. PHY Layer Challenges for the Multiple Access Schemes   \\   \ref{I-C}. 5G/6G and the Need for RSMA         
			&\ref{I-D}. Contributions  \\          \ref{I-E}. Organization and Notation               &                                                           \\ \hline \hline
			\multicolumn{2}{|c|}{\textbf{Section \ref{sec:principle}. Principles of RSMA}}                                               \\ 
			\ref{IV-A}. Definition and Design Principle                   & \ref{IV-B}. Downlink RSMA \\
			\ref{IV-C}. Uplink RSMA                   & \ref{IV-D}. Multi-cell RSMA \\ \hline \hline
			\multicolumn{2}{|c|}{\textbf{Section \ref{sec:literatureReview}.  Information-Theoretic Background and Milestones of RSMA }}                                                                  \\ 
			\ref{III-A}. RSMA in Single-Antenna Networks                        & \ref{III-B}. RSMA in Multi-Antenna Networks                                 \\ 
			\ref{III-C}. Milestones of RSMA                          &   \\ \hline \hline
			\multicolumn{2}{|c|}{\textbf{Section \ref{sec:resourceAllocation}. Communication-Theoretic Background of RSMA}}                                                                  \\ 
			\ref{V-A}. Resource Allocation for Single-Carrier RSMA                     &  \ref{V-B}. Resource Allocation for Multicarrier RSMA   \\ \hline \hline
			\multicolumn{2}{|c|}{\textbf{Section \ref{sec:RSMAphyLayer}. RSMA PHY Layer Design}}                                                  \\ 
			\ref{IV-A}. A Quick Introduction to RSMA PHY Layer                         &  \ref{IV-B}. RSMA PHY Layer Architecture                          \\ \hline \hline
			\multicolumn{2}{|c|}{\textbf{Section \ref{sec:RSMAvsOtherMA}. Comparison of Multiple Access Schemes}}                                                                  \\ 
			\ref{VI-A}. Framework Comparison                         & \ref{VI-B}. Complexity Comparison   \\                                                \ref{VI-C}. Performance Comparison                         & \ref{VI-D}. Advantages of RSMA                               \\
			\ref{VI-E}. Disadvantages of RSMA                        &     \\\hline \hline  
			\multicolumn{2}{|c|}{\textbf{Section \ref{sec:emergingApp}. Emerging Applications, Challenges, and Future Research Trends of RSMA}}                                                                            \\ 
			\ref{VIII-A}. Technical Aspects of RSMA -- The Road Ahead            & \ref{VIII-B}. RSMA for Enabling Technologies in 6G                                  \\ 
			\ref{VIII-C}. Standardization and Implementation of RSMA               &                                 \\\hline \hline
			\multicolumn{2}{|c|}{\textbf{Section \ref{sec:conclusion}. Conclusions}}                                                                                  \\ \hline
		\end{tabular}
	\end{table*}
	
	\par 
	On account of the aforementioned limitations of the existing MA schemes and their underlying interference management strategies, and the resulting imperfect CSIT bottleneck in multi-antenna networks,  the following three questions are raised and motivate the study of a new PHY layer:
	\begin{enumerate}
		\item   Is there an  MA scheme that is inherently \textit{robust} to  imperfect CSIT?
		\item  Is there an MA scheme that is  \textit{flexible} enough to adapt to the interference level rather than operate in the two extremes of fully treating interference as noise and fully decoding interference?
		\item Is there an MA scheme that is \textit{general} enough to encompass and outperform all existing MA schemes?
	\end{enumerate}
	%%%
	The three questions above can be answered with ``yes" by RSMA.
	%%%
	In the rest of this treatise, we will  dive into how RSMA addresses the above three questions.

	\subsection{Contributions}
	\label{sec:contri}
	\labelsubseccounter{I-D}
	\par 
	As summarized in the rest of this treatise,
	existing works on RSMA mainly focus on one narrow scenario and there is no review and tutorial paper which  pedagogically explains why, how, when it is beneficial to use RSMA in wireless communication networks. Meanwhile,  there is a lack of consensus in the RSMA literature   because of the recent explosive increase in research activity. Hence, it is timely to consolidate the existing literature.
	%%%
	\par
	The main goal of this treatise is to provide the first holistic tutorial overview on RSMA including a thorough review of its state-of-the-art, a detailed illustration of its design principles, a comprehensive summary of its merits as well as  a broad discussion of the related research challenges and future research directions.
	%%%
	Our major contributions can be summarized as follows:
	\begin{itemize}
		% \item We illustrate the evolution of MA, and summarize the pros and cons of existing MA including OMA, SDMA and NOMA. Specifically, we compare the limitations of the existing interference management approach adopted in each MA. 
		
		\item The principles and transmission frameworks  of downlink, uplink, and multi-cell RSMA are respectively elaborated including a quantitative complexity comparison between the existing downlink RSMA schemes. 
		
		\item An exhaustive survey of the state of the art of RSMA is provided from both the information-theoretic  and communication perspectives along with a recap of the pivotal historical milestones in the history of RSMA.

		\item The two prevalent precoder design strategies for RSMA, namely, precoder optimization and low-complexity precoder design, are comprehensively summarized.
		
		\item The PHY layer architecture design of RSMA is presented, with toy examples to gain a first intuition and detailed explanations.
		
		\item RSMA and existing MA schemes including SDMA, NOMA, OMA, and multicasting  are compared in terms of the achieved DoFs, complexity, and throughput performance by link-level simulations (LLS), which unveils the major advantages and disadvantages of RSMA.
		
		\item Emerging applications of RSMA and related research challenges  in numerous potential directions are summarized including many timely fields such as space-air-ground integrated networks (SAGIN), vehicle-to-everything (V2X) communications, and three-dimensional (3D) eMBB-URLLC-mMTC services.
		
		\item The pathways of RSMA in the context of the 3rd Generation Partnership Project (3GPP) are discussed, followed by a synopsis of RSMA standardization and implementation issues.
	\end{itemize}
	%%%
	We hope this tutorial will shed light on the fundamental RSMA concepts, contribute to the consolidation of the existing RSMA research, provide a useful reference on the use of RSMA in wireless communications, and ultimately serve as a platform to drive forward the research in this area.
	
	%%%
	
	\subsection{Organization and Notation}
	\label{sec:organi}
	\labelsubseccounter{I-E}
	The remainder of this paper is organized as follows.
	%%%
	%%%
	Section \ref{sec:principle} introduces the design principles of RSMA, and specifies the transmitter and receiver architectures for downlink, uplink, and multi-cell RSMA. 
	%%%
	Section \ref{sec:literatureReview} provides a thorough literature review on RSMA from an information-theoretic perspective followed by a summary of the milestones of RSMA research.
	%%%
	Section \ref{sec:resourceAllocation} discusses the communication-theoretic background of RSMA with a focus on RSMA resource allocation.
	%%%
	Section \ref{sec:RSMAphyLayer} presents examples for and details of the PHY layer architecture design for RSMA.
	%%%
	Section \ref{sec:RSMAvsOtherMA} compares RSMA with other MA schemes and unveils the major advantages and disadvantages of RSMA.
	%%%
	Section \ref{sec:emergingApp} presents emerging applications, research challenges, and future research trends for RSMA. 
	%%%
	Section \ref{sec:conclusion} concludes this tutorial overview. 
	%%%
	The outline of this paper is illustrated in Table \ref{tab:outline}.
	%%%
	
	\begin{table*}
		%% increase table row spacing, adjust to taste
		\renewcommand{\arraystretch}{0.9}
		% if using array.sty, it might be a good idea to tweak the value of
		% \extrarowheight as needed to properly center the text within the cells
		\caption{List of notation.}
		\label{tab:notations}
		\centering
		\begin{tabular}{|l|l||l|l|}
			\hline
			$\mathcal{A}$                & subset of the user set $\mathcal{K}$                                                                      & $R_k$                      & instantaneous rate of $s_k$ at user-$k$                    \\
			$C_k$                        & amount of $R_c$ allocated to user-$k$                                                                     & $R_c$                      & achievable rate of $s_c$                                   \\
			$\widehat{C}_k$              & amount of $\widehat{R}_{c}$ allocated to user-$k$                                                         & $R_{k,tot}$                & total achievable rate of user-$k$                          \\
			$\overline{C}_k$             & ergodic common rate allocated to user-$k$                                                                 & $R_k^{th}$                 & QoS rate constraint of user-$k$                            \\
			$\mathbf{c}$                 & common rate vector containing all $C_k$                                                                   & $\widehat{R}_{c,k}$        & worst-case achievable rate of $s_c$ at user-$k$            \\
			$\widehat{\mathbf{c}}$       & common rate vector containing all $\widehat{C}_k$                                                         & $\widehat{R}_{k}$          & worst-case achievable rate of $s_k$ at user-$k$            \\
			$\overline{\mathbf{c}}$      & ergodic common rate vector containing all $\overline{C}_k$                                                & $\widehat{R}_c$            & worst-case achievable rate of $s_c$ at all users           \\
			$d_k^{(j)}$                  & DoF of user-$k$ achieved by scheme $j$                                                                    & $\widehat{R}_{k,tot}$      & worst-case achievable rate of user-$k$                     \\
			$d_{\textnormal{s}}^{(j)}$   & sum-DoF of scheme $j$                                                                                     & $\overline{R}_{c,k}$       & ergodic rate of decoding $s_c$ at user-$k$                 \\
			$d_{\textnormal{mmf}}^{(j)}$ & MMF-DoF of scheme $j$                                                                                     & $\overline{R}_k$           & ergodic rate of decoding $s_k$ at user-$k$                 \\
			$\textrm{ER}_{k,tot}$        & ergodic achievable rate of user-$k$                                                                       & $\overline{R}_c$           & achievable ergodic rate of $s_c$                           \\
			$G$                          & number of user groups                                                                                     & $s_n$                      & data stream transmitted from the BS                        \\
			$\mathcal{G}$                & set of user groups                                                                                        & $s_c$                      & common stream for 1-layer RS                               \\
			$\mathbf{h}_k$               & channel vector between the BS and the user-$k$                                                            & $\mathbf{s}$               & data stream vector transmitted from the BS                 \\
			$\widehat{\mathbf{h}}_k$     & instantaneous channel estimate of $\mathbf{h}_k$                                                          & $t$                        & time instance                                              \\
			$\widetilde{\mathbf{h}}_k$   & instantaneous channel estimation error of $\mathbf{h}_k$                                                  & $u_k$                      & weight allocated to user-$k$                               \\
			$\mathbf{H}$                 & channel matrix containing $\mathbf{h}_k$ for all $k$                                                      & $U(\mathbf{P},\mathbf{c})$ & utility function of $\mathbf{P}$ and $\mathbf{c}$          \\
			$\widehat{\mathbf{H}}$       & channel estimate matrix containing  $\widehat{\mathbf{h}}_k$ for all $k$                                  & $W_k$                      & message of user-$k$                                        \\
			$\widetilde{\mathbf{H}}$     & channel estimation error matrix containing $\widetilde{\mathbf{h}}_k$ for all $k$                         & $W_k^l$                    & sub-message split from $W_k$                            \\
			$K$                          & number of users                                                                                           & $W_k^{\mathcal{K}_i}$      & inner-group common sub-messages for group $\mathcal{K}_i$  \\
			$\mathcal{K}$                & set of users                                                                                              & $W_n^*$                    & message transmitted from the BS                            \\
			$\mathcal{K}_i$              & set of users in group-$i$                                                                                 & $W_{c,k}$                  & common sub-message split from $W_k$                     \\
			$L$                          & number of sub-messages split from each message                                                         & $W_{p,k}$                  & private sub-message split from $W_k$                    \\
			$M$                          & number of transmit antennas at the transmitter                                                            & $\mathcal{W}$              & set of messages transmitted from the BS                    \\
			$N$                          & number of transmitted streams                                                                             & $\mathcal{W}_k$            & message set decoded at user-$k$                            \\
			$n_k$                        & additive white Gaussian noise of user-$l$                                                                 & $\alpha$                   & CSIT scaling factor                                        \\
			$P$                          & transmit power                                                                                            & $\eta$                     & power amplifier efficiency                                 \\
			$P_{\textrm{cir}}$           & circuit power consumption                                                                                 & $\epsilon$                 & time correlation coefficient                               \\
			$P_i$                        & power allocated to precoder $\mathbf{p}_i$                                                                & $\epsilon'$                & tolerance for the selection of MA schemes                  \\
			$\mathbf{P}$                 & precoding matrix for all data streams                                                                     & $\pi_k$                    & decoding order of user-$k$                                 \\
			$\mathbf{p}_i$               & precoding vector for $s_i$                                                                                & $\pi_{l,k}$                & decoding order at user-$k$ to decode the $l$-order streams \\
			$\bar{\mathbf{p}}_i$         & direction of the precoder $\mathbf{p}_i$                                                                  & $\sigma_k^2$               & variance of $\mathbf{h}_k$ or $\widehat{\mathbf{h}}_k$     \\
			$\mathcal{P}_1$              & precoder optimization problem illustrated in (\ref{eq:perfectCSIopt})   & $\sigma_{n,k}^2$           & variance of $n_k$                                          \\
			$\mathcal{P}_2$              & precoder optimization problem illustrated in (\ref{eq:imperfectCSIopt}) & $\tau$                     & power allocated to all private streams                     \\
			$\mathcal{P}_3$              & long-term precoder optimization problem obtained from  $\mathcal{P}_2$                                    & $\varpi_k$                 & weight allocated to user-$k$ for matched beamforming             \\
			$R_{c,k}$                    & instantaneous rate of $s_c$ at user-$k$                                                                   &                            &                                                            \\ \hline
		\end{tabular}
	\end{table*}
	%%%
	\par
	\textit{Notation:} Bold upper and lower case letters denote matrices and column vectors, respectively.
	{$(\cdot)^{T}$,   $(\cdot)^{H}$}, $|\cdot|$, $\|\cdot\|$, $\mathbb{E}\{\cdot\}$, and $\mathrm{tr}(\cdot)$ represent the transpose, Hermitian, absolute value, Euclidean norm, expectation, and trace  operators, respectively.  $\mathcal{CN}(0,\sigma^2 )$ denotes the circularly symmetric complex Gaussian (CSCG) distribution with zero mean and variance $\sigma^2$. The main notations used in this work are summarized in Table \ref{tab:notations}.
	%%%

	%%%
	\section{Principles of RSMA}
	\label{sec:principle}
	Starting from a rigorous definition of RSMA and its design principle, this section details the transmission framework for downlink, uplink, and multi-cell RSMA. 
	
	\subsection{Definition and Design Principle}
	\label{sec:RSMAdef}
\begin{figure}[t!]
		\centering
		\includegraphics[width=0.49\textwidth]{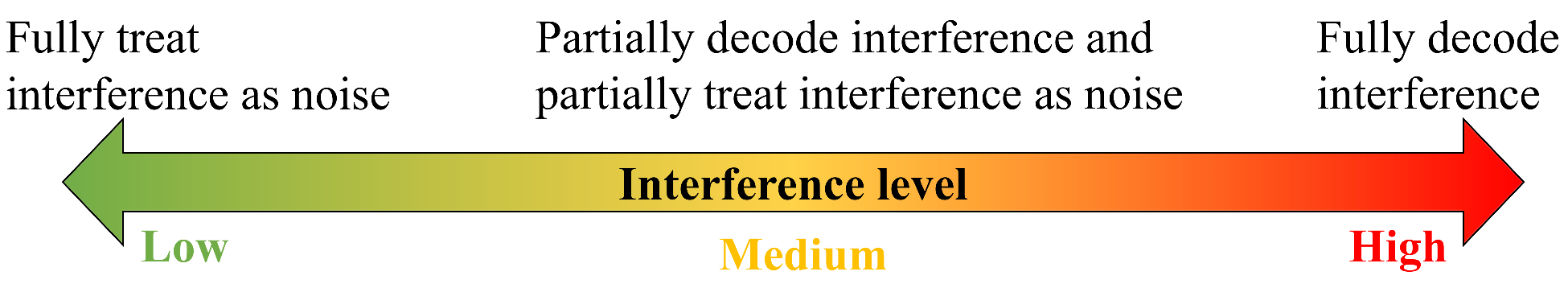}%
		\caption{Lesson learnt from \cite{Tse2008}: the amount of interference to be decoded and treated as noise should be adaptively changed with the interference level.}
		\label{fig:interferencelesson}
	\end{figure}

	\labelsubseccounter{IV-A}
	Due to the broadcast nature of wireless communication, interference has become an inevitable issue encountered in today's communication networks. As has been discussed in the introduction section,  the interference management approaches adopted in existing MA schemes can be categorized by the following taxonomy:
	\begin{itemize}
		\item Avoiding interference by allocating orthogonal radio resources to the users (as in OMA)
		\item Treating interference as noise (as in SDMA)
		\item Decoding interference (as in NOMA)
	\end{itemize}
	%The authors in  \cite{Tse2008} have shown that treating interference as noise and decoding interference are only suited in weak and strong interference regimes, respectively. 
	%%%
	%SDMA and NOMA would therefore hamper the system performance as they both use one of the above inflexible approaches to manage all levels of interference.
	%%%
	Next generation MA schemes are envisioned to possess a more powerful interference management capability, where the amount of interference to be decoded or treated as noise is adaptively changed with the level of interference, i.e., the interference is fully treated as noise (or fully decoded) when the level of interference is low (or high) while the interference is partially decoded and partially treated as noise when the interference level is medium, as illustrated in Fig. \ref{fig:interferencelesson}.
	%%%
	
	%%%
	\par 
	RSMA is a promising PHY layer strategy to serve multiple users based on the concept of RS\footnote{RSMA is commonly known as RS in previous works where the message of each user is split into two parts at most, same as in Fig. \ref{fig:twoUserBaseline}(d). Such model forms the basic building block of the RSMA framework for the multi-antenna BC \cite{mao2017rate}. In the following, we use ``RS" to denote this building block of RSMA so as to maintain consistency with the literature.}. 
	%%%
	The design principle of RSMA, which is \textit{to softly bridge the two extremes of fully decoding interference and fully treating interference as noise by partially decoding the interference and partially treating the interference as noise}, exactly caters to the requirements of future networks.
	%%%
	Therefore, RSMA constitutes a new paradigm for MA and non-orthogonal transmission in future wireless networks. It is also a new paradigm for interference management, as it is capable of harmonizing the existing interference management taxonomy, as illustrated in Table \ref{tab:lessonFromIT}.
	%%%
	\begin{table}[t!]
		\centering
		\caption{Comparison of preferable interference levels of different MA schemes.}
		\label{tab:lessonFromIT}
		\begin{tabular}{|c|c|c|c|}
			\hline 
			%  \diagbox{\textbf{MA schemes}}{ \begin{tabular}[c]{@{}c@{}}\textbf{Interference}\\ \textbf{levels}\end{tabular}}     
			\textbf{Interference levels}
			& \textbf{Low} & \textbf{Medium} & \textbf{High} \\ \hline \hline
			\textbf{SDMA} & $\surd$       & $\times$          & $\times$       \\ \hline
			\textbf{NOMA} & $\times$       & $\times$         & $\surd$       \\ \hline
			\textbf{RSMA} & $\surd$       & $\surd$         & $\surd$       \\ \hline
		\end{tabular}
		\vspace{0.1cm}
		
		{Notations: $\surd$: Suited. $\times$: Not well suited. }
	\end{table}
	%%%
	%%%
	\par In the following subsections, we illustrate why RSMA facilitates flexible and powerful interference management by elaborating  in detail the RSMA transmission frameworks for downlink, uplink, and multi-cell networks, respectively.
	%%%
	
	\subsection{Downlink RSMA}
	\label{sec:RSMADL}
	\labelsubseccounter{IV-B}
	Under the umbrella of downlink multi-user communication networks (a.k.a. broadcast channels--BCs), various RSMA schemes have been proposed, such as 1-layer RS \cite{RS2015bruno,RS2016hamdi,RS2016joudeh,enrico2016bruno,Lu2018MMSERS,mao2017rate,Medra2018SPAWC,mao2018EE,bruno2019wcl,hongzhi2020LLS,Onur2020LLS,bruno2020MUMIMO,Gui2020EESEtradeoff,mao2021IoT,bho2021globalEE,wonjae2021imperfectCSIR,longfei2021statisticalCSIT,onur2021mobility}, 2-layer hierarchical RS (HRS) \cite{Minbo2016MassiveMIMO,mao2017rate},  generalized RS \cite{mao2017rate,Zheng2020JSAC,alaa2020gRS}, RS and common
	message decoding (RS-CMD) \cite{Ala2019IEEEAccess,alaa2020EECRAN,alaa2020powerMini,alaa2020cranimperfectCSIT}, RSMA with non-linear THP (a.k.a. THPRS) \cite{Flores2018ISWCS,Andre2021THP},  and dirty paper coded RS (DPCRS) \cite{mao2019beyondDPC,mao2020DPCNOUM}, which  are detailed and compared in this subsection in terms of their transmitter and receiver designs. 
	%%%
	\subsubsection{Transmitter design}
	%%%
	\par
\begin{figure}[t!]
		\centering
		\includegraphics[width=0.45\textwidth]{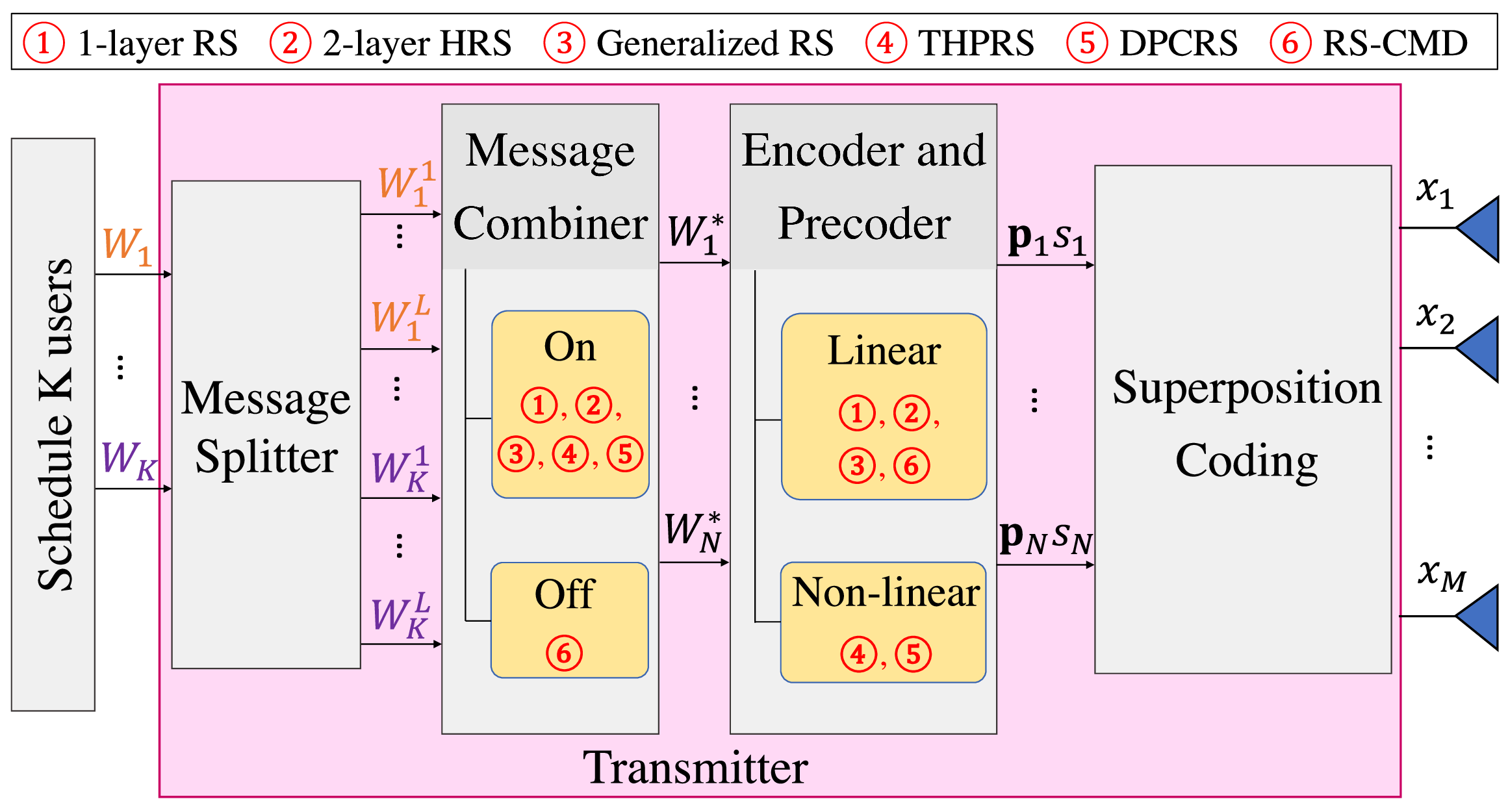}%
		\caption{$K$-user downlink RSMA transmitter. }
		\label{fig:RSMAtx}
	\end{figure}

\begin{figure}[t!]
		\centering
		\includegraphics[width=0.47\textwidth]{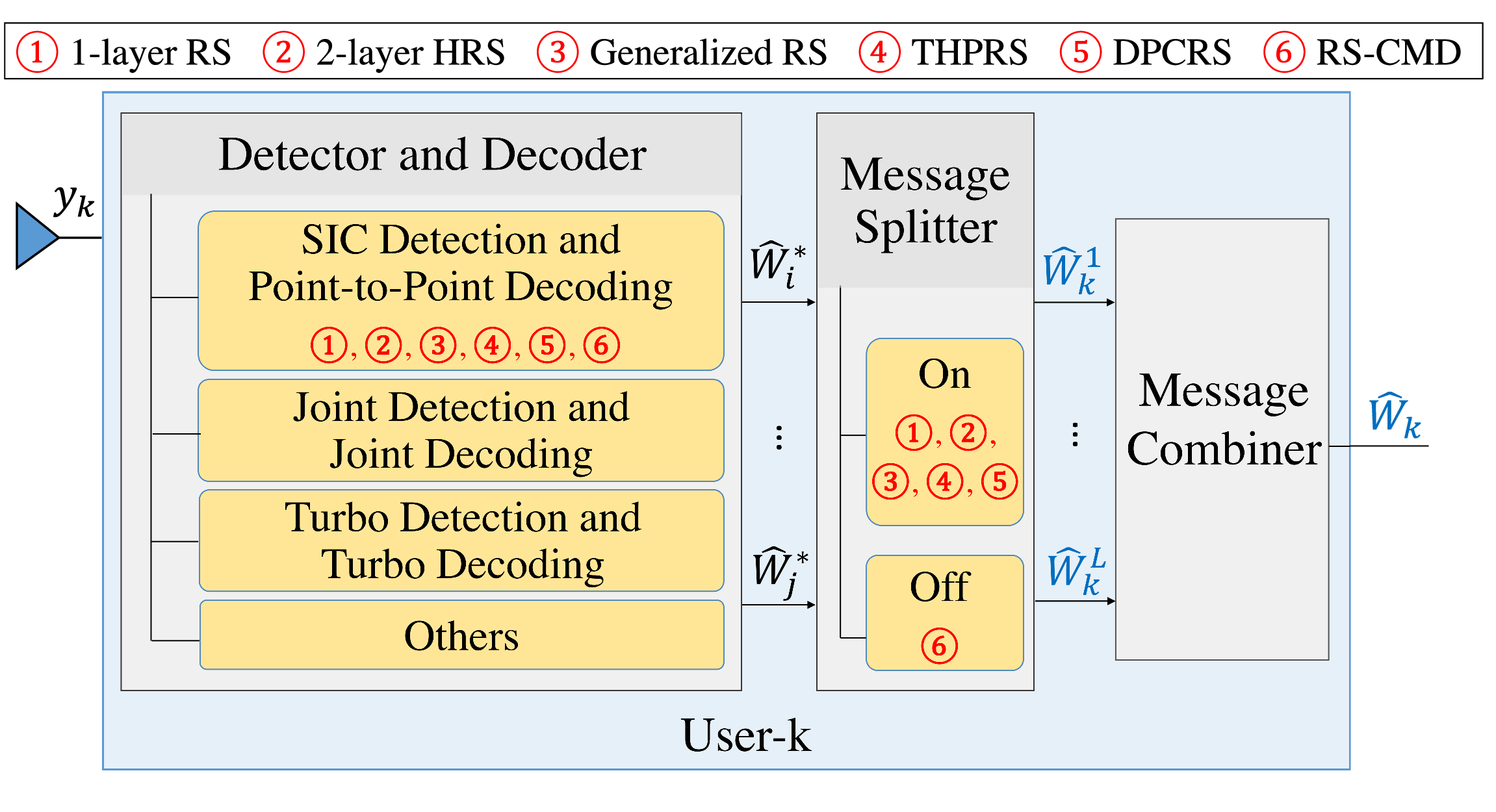}%
		\caption{Downlink RSMA receiver. }
		\label{fig:RSMArx}
	\end{figure}

 
	Generally, all downlink RSMA schemes can be represented in terms of the  universal   RSMA transmitter and receiver designs depicted in Fig. \ref{fig:RSMAtx} and Fig. \ref{fig:RSMArx}, respectively.
	%%%
	% Fig. \ref{fig:RSMAtx}  illustrates a universal $K$-user  RSMA transmitter design, which embraces all RSMA downlink schemes. 
	%%%
	The transmitter  is equipped with $M$ antennas  ($M\geq1$)  and serves $K$ single-antenna users\footnote{We limit the system model to single-antenna receivers for ease of explanation. Readers are referred to \cite{anup2021MIMO} for a detailed treatment of the extension to multi-antenna receivers.} indexed by $\mathcal{K}=\{1,\ldots,K\}$.  $K>1$ is assumed as RSMA is designed for better managing co-channel interference in multi-user transmission and it does not bring additional benefits for point-to-point transmission.  The proposed framework therefore applies to the SISO BC when $M=1$ and to the MISO BC when $M>1$.
	
	%%%
	\par 
	Compared with conventional frameworks, the fundamental difference of  RSMA-enabled transmitters compared with conventional MA transmitters is the use of  ``message splitter" for splitting user messages into  $L$ sub-messages so as to enable rate-splitting. 
	%%%
	For each user-$k$, its message $W_k$ is split into $\{W_k^1, W_k^2,\ldots, W_k^L\}$. 
	%%%
	The number of sub-messages $L$ into which the original message of each user is split depends on the particular RSMA scheme, as illustrated in Table \ref{tab:RSschemesCompare}. For 1-layer RS, THPRS, 1-DPCRS, and RS-CMD, the message of each user is split at most  into two sub-messages ($L=2$) while  2-layer HRS employs three sub-messages ($L=3$) for each user message. Generalized RS and M-DPCRS are more complex as  each user message is split into $L=2^{K-1}$ sub-messages, so as to encode the common streams into different layers and achieve more flexible interference management.   
	%%%
	%%%
	%%%
	\par
	After splitting, the ``message combiner" is turned on for some of the RSMA schemes (including 1-layer RS, 2-layer HRS, Generalized RS, THPRS and DPCRS) to combine the user sub-messages into $N$  messages $\mathcal{W}=\{W_1^*,\ldots, W_N^*\}$  depending on the  RSMA schemes  as illustrated in Table \ref{tab:RSschemesCompare}, where $N=K+1$ for 1-layer RS, THPRS and 1-DPCRS, $N=K+2$ for 2-layer HRS, and $N=2^{K}-1$ for generalized RS and M-DPCRS.
	%%%
	% The motivation of using the ``message combiner" at the transmitter is to reduce the receiver complexity as discussed in the receiver subsection.  
	%%%
	Different from other RSMA schemes, the ``message combiner" is turned off for RS-CMD and the total $LK$  sub-messages of  all users are directly encoded into $N=LK$ streams. 
	%%%
	Among the $N$ messages, a message to be decoded by multiple users is referred to as a \textit{common message}, while a message to be decoded by a single user is referred to as a \textit{private message}. 
	%%%
	Typically, a common message contains sub-messages of one or multiple users while a private message only contains a sub-message of a single user. 
	%%%
	Embedding the sub-messages of multiple users into one common message is also known as packet-level multiplexing \cite{packetLevelMutplex2001JSAC}. 
	%%%
	%%%
	\begin{table*}[t!]
		\centering
		\caption{A quantitative complexity comparison of the transceiver design of different RSMA schemes.}
		\label{tab:RSschemesCompare}
		%	\addtolength\tabcolsep{-2pt}
		\begin{tabular}{@{}|P{2.2cm}|P{1.3cm}|P{3.5cm}|P{1.3cm}|P{1.6cm}|P{2.6cm}|P{2.0cm}|@{}}
			\hline
			&
			\textbf{Proposed in Ref.} &
			\textbf{Number of sub-messages split from each message ($L$)} &
			\textbf{Message combiner} &
			\textbf{Precoding} &
			\textbf{Number of transmitted streams ($N$)} &
			\textbf{Layers of SIC at each user}\\ \hline
			\textbf{1-layer RS}     & \cite{RS2015bruno}          & $2$       & On & Linear     & $K+1$   & $1$    \\ \hline
			\textbf{2-layer HRS}    & \cite{Minbo2016MassiveMIMO} & $3$       & On & Linear     & $K+2$    & $2$   \\ \hline
			\textbf{Generalized RS} & \cite{mao2017rate}          & $2^{K-1}$ & On & Linear     & $2^{K}-1$  & $2^{K-1}-1$ \\ \hline
			\textbf{THPRS}          & \cite{Flores2018ISWCS}      & $2$       & On & Non-linear & $K+1$    & $1$   \\ \hline
			\textbf{$1$-DPCRS}      & \cite{mao2019beyondDPC}     & $2$       & On & Non-linear & $K+1$    & $1$   \\ \hline
			\textbf{M-DPCRS}      & \cite{mao2019beyondDPC}     & $2^{K-1}$ & On & Non-linear & $2^{K}-1$  & $2^{K-1}-1$ \\ \hline
			\textbf{RS-CMD}         & \cite{Ala2019IEEEAccess}    & $2$       & Off  & Linear     & $2K$    & $K$    \\ \hline
		\end{tabular}
	\end{table*}
	%%%
	\par 
	Each common message is then encoded into a common stream using a codebook shared by the intended users\footnote{In LTE/5G NR systems, all codebooks are  shared among the users as the same family of modulation and coding schemes (MCS) specified in the standard is used for all users \cite{bruno2021MISONOMA}. Therefore, using a shared codebook is not an issue in modern communication systems.} while each private message is encoded into a private stream using an independent codebook.  The  $N$ encoded streams $\mathbf{s}=[s_1,\ldots,s_N]^T$ are accordingly mapped via $N$ precoders $\mathbf{P}=[\mathbf{p}_1,\ldots,\mathbf{p}_N]^T$ onto $M$ transmit antennas. Assuming that $\mathbb{E}[\mathbf{s}\mathbf{s}^H]=\mathbf{I}$, we obtain the transmit power constraint as $\sum_{n=1}^N\left\|\mathbf{p}_{n}\right\|^2\leq P$.
	%%%
	The resulting transmit signal which superposes the precoded common and private streams is given as follows
	\begin{equation}
		\mathbf{x}=\sum_{n=1}^N \mathbf{p}_ns_n.
	\end{equation}
	%%%
	As illustrated in Fig. \ref{fig:RSMAcatogory}, due to the use of linear or non-linear precoding techniques,  RSMA can be classified into \textit{linearly-precoded RSMA} (including 1-layer RS, 2-layer HRS, generalized RS and RS-CMD) and \textit{non-linearly precoded RSMA} (including THPRS and DPCRS).
	%%%
	Specifically, the two non-linearly precoded schemes, THPRS and DPCRS,  respectively adopt THP and DPC precoding. 
	%%%
	Note that all linearly-precoded RSMA strategies can be extended to their THP/DPC or other non-linearly precoded counterparts by simply changing the precoding method.
	%%%
	\par 
	\begin{remark}
		\label{remark:multicast}
		We highlight here that from a message content perspective, a ``common message" in RSMA is fundamentally different from a  ``multicast message" that is originally intended for all users, though both of them are decoded by multiple users. 
		%%%
		A ``common message" in RSMA includes parts of the (unicast) messages for different users.
		%%%
		It is decoded by multiple users for interference management purposes, but the content of the message is not wholly needed by those users. 
		%%%
		Therefore, it can be considered in line with the packet-level multiplexing described in \cite{petar2017FBL}, which relies on obtaining new messages by concatenating multiple short messages intended for different users.
		%%%%
		In contrast, a multicast message is primarily intended for multiple  users and each user wants the whole content \cite{RSintro16bruno}.
		%%%
		But we should note that, from a PHY layer transmission perspective, the transmission of an encoded ``common message" is multicasted regardless of the content it conveys. 
		%%%
		This is common in satellite communications where the data intended for different users is bundled in the same frame and transmitted in a multicast fashion \cite{multicastSatellite2015twc,longfei2020multibeam}.
		%%%
		NOMA also has common messages/streams, though they are usually not referred to as such. 
		%%%
		As illustrated in Fig. \ref{fig:twoUserBaseline}(c), $W_2$ in NOMA is a common message. Though it only contains the information of user-2,  it is encoded into a common stream and decoded by both users.
		%We highlight here that a ``common message" in RSMA is fundamentally different from a  ``multicast message" that is originally intended for all users, though both of them are decoded by multiple users:
		%%%
		%%\[\textrm{a common message} \neq \textrm{a multicast message}.\]
		%%%
		%A ``common message" in RSMA includes parts of the (unicast) messages for different users.
		%%%
		%It is decoded by multiple users for interference management purposes, but is not wholly needed by those users. 
		%%%
		%In contrast, a multicast message is primarily intended for multiple  users and each user wants the whole multicast message \cite{RSintro16bruno}.
		%%%
		%NOMA also has common messages/streams, though they are usually not referred to as such. 
		%%%
		%As illustrated in Fig. \ref{fig:twoUserBaseline}(c), $W_2$ in NOMA is a common message. Though it only contains the information of user-2,  it is encoded into a common stream and decoded by both users.
	\end{remark}
	%%%
	\subsubsection{Receiver design}
	%%%
	\par The receiver designs of different RSMA  schemes are illustrated in Fig. \ref{fig:RSMArx}.
	%%%
	Using $\mathbf{h}_k\in \mathbb{C}^{M\times1}$ to denote the channel vector of user-$k$ and $n_k$ to denote the corresponding additive white Gaussian noise (AWGN) following distribution $\mathcal{CN}(0,\sigma_{n,k}^2)$, the signal received at user-$k$ can be modelled as follows 
	\begin{equation}
		y_k=\mathbf{h}_k^H\mathbf{x}+n_k, \forall k\in \mathcal{K}.
	\end{equation}
	%%%
	$y_k$ is passed through the decoder to obtain the output messages $\mathcal{W}_k=\{\widehat{W}_i^*,\ldots,\widehat{W}_j^*\}$. 
	%%%
	According to the different encoding rules of the different RSMA strategies, each user decodes only selected common streams and the remaining common streams are treated as noise.
	%%%
	The decoded message set $\mathcal{W}_k$  at each user is therefore a subset of the transmitted messages $\mathcal{W}$, i.e., $\mathcal{W}_k\subseteq \mathcal{W}$.   
	%%%
	Note that the  common messages $\mathcal{W}_k$ decoded at user-$k$ do not necessarily contain the intended sub-messages of this user.
	%%%
	\par
	As illustrated in Fig. \ref{fig:RSMArx}, RSMA receivers are not limited to SIC detectors (a.k.a. SIC receiver architectures). Other detectors, such as joint detectors (JD) and turbo detectors, can be used for RSMA receiver design as well \cite{Zheng2020JSAC}. 
	%%%
	For a given detector, different channel decoders can be applied, i.e., Vertical Bell Laboratories Layered Space-Time (V-BLAST) decoding  and Polar decoding have been respectively used with SIC detectors in RSMA receivers in \cite{anup2021MIMO} and \cite{Onur2020LLS}. 
	%%%
	The performance of different receivers are worth investigating in future works.
	% \par
	Mirroring the message splitter  and on/off message combiner  at the transmitter, each receiver contains an on/off message splitter for extracting the intended sub-messages as well as  a message combiner for recovering the original user message. 
	
	\par
\begin{figure}[t!]
		\centering
		\includegraphics[width=2.6in]{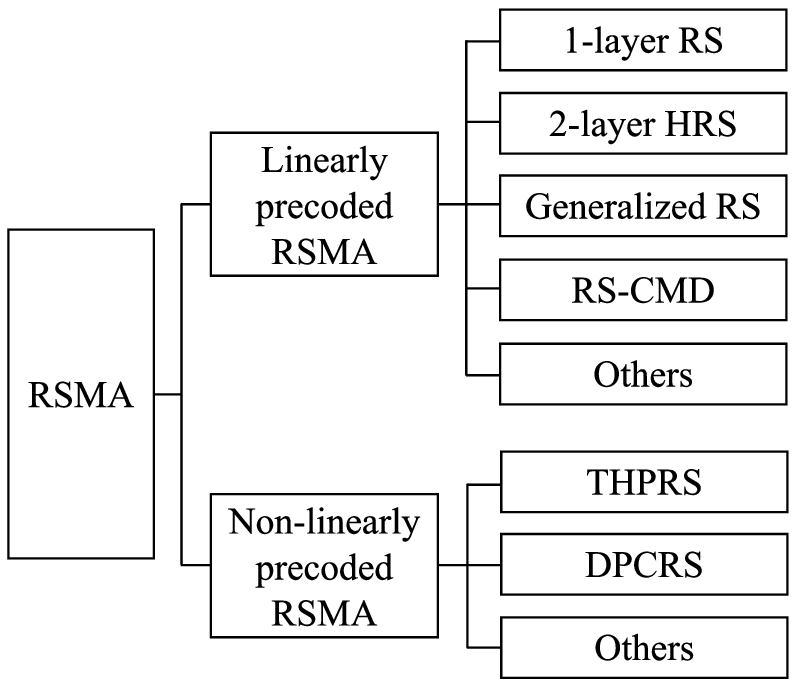}%
		\caption{Classification of RSMA strategies.}
		\label{fig:RSMAcatogory}
		% 	\vspace{-2mm}
	\end{figure}

 Next, we respectively detail the transceiver design of existing RSMA schemes in Fig. \ref{fig:RSMAcatogory} to further illustrate their differences. 
	%%%

	%%%
	\subsubsection{1-layer RS}
	\label{sec:1-layer RS}
	%%%
	\par
\begin{figure}[t!]
		\centering
		\includegraphics[width=3.4in]{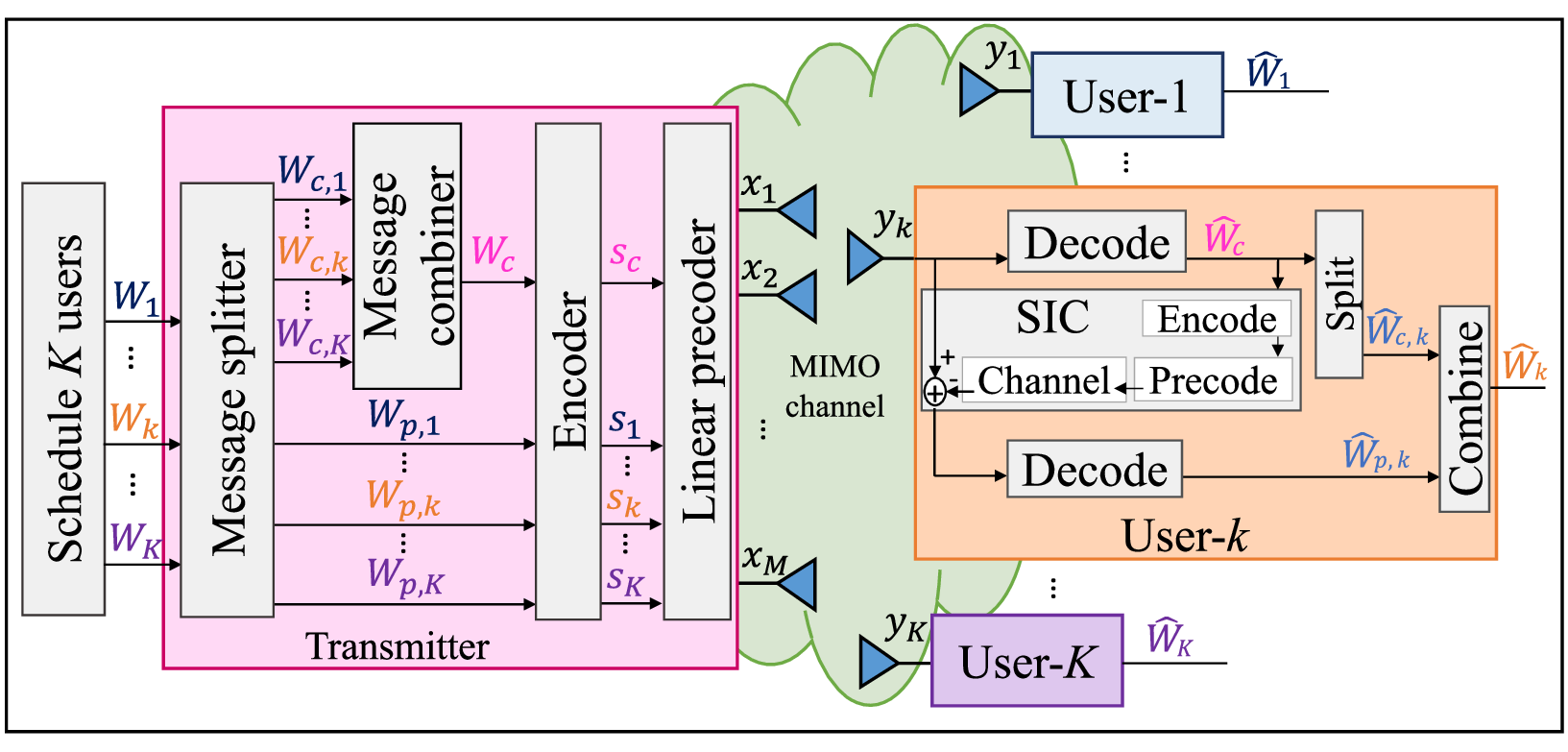}%
		\caption{Transceiver architecture of $K$-user 1-layer RS \cite{RSintro16bruno}.}
		\label{fig: OneLayerRS}
		\vspace{-2mm}
	\end{figure}

  The simplest and most practical RSMA scheme is 1-layer RS, which is the basic building block of almost all existing RSMA schemes. 1-layer RS has been widely studied for the multi-antenna BC with  perfect CSIT \cite{mao2017rate,mao2018EE,bruno2019wcl,hongzhi2020LLS,bruno2020MUMIMO,Gui2020EESEtradeoff,bho2021globalEE} and imperfect CSIT \cite{RS2015bruno,RS2016hamdi,RS2016joudeh,enrico2016bruno,Lu2018MMSERS,Medra2018SPAWC,Onur2020LLS,bruno2020MUMIMO,mao2021IoT,wonjae2021imperfectCSIR,longfei2021statisticalCSIT,onur2021mobility}. 
	%%%
	Fig. \ref{fig: OneLayerRS} illustrates the transceiver architecture of $K$-user 1-layer RS. 
	
	%%%
	\par 
	At the transmitter, the message combiner is turned on and linear precoding is adopted. Each message $W_k$ is split into two sub-messages ($L=2$)\footnote{Here, we consider the most general case when the messages of all users are split. 
		%%%
		Note that it is not always necessary to split all user messages. When maximizing the user sum rate without QoS rate constraint, it is sufficient to split one user message \cite{RS2016hamdi}.  However, when user fairness is considered as one of the design criteria (i.e., when maximizing the weighted sum rate--WSR or max-min rate or any other objective with individual QoS rate constraints), splitting the messages of all users becomes a prerequisite for the optimal solution \cite{mao2019maxmin,hamdi2017bruno,mao2017rate}.}, namely, one common sub-message $W_{c,k}$ and one private sub-message $W_{p,k}$.  
	%%%
	The common sub-messages of all users $W_{c,1},\ldots, W_{c,K}$ are combined into one common message $W_c$ and encoded into a common stream $s_c$ using a codebook shared by all users. $W_c$ has to be decoded by all users.
	%%%
	The private sub-messages of all users $W_{p,1},\ldots, W_{p,K}$ are  independently encoded into  private streams $s_1,\ldots,s_K$, which are decoded by the corresponding users only. 
	%%%
	Therefore, $K+1$ streams $\mathbf{{s}}=[
	s_{c},s_{1},\ldots, s_{K}]^{T}\in\mathbb{C}^{(K+1)\times1}$ are created from $K$ messages $W_{1},\ldots, W_{K}$. The streams are linearly precoded via  precoding matrix $\mathbf{{P}}=[\mathbf{{p}}_{c}, \mathbf{{p}}_{1},\ldots, \mathbf{{p}}_{K}]\in\mathbb{C}^{M\times (K+1)}$ with $\mathbf{p}_{c},\mathbf{{p}}_{k}\in \mathbb{C}^{M\times 1}$  such that the transmit signal is obtained as follows
	\begin{equation}
		\mathbf{{x}}=\mathbf{{p}}_{c}s_{c}+\sum_{k\in\mathcal{\mathcal{K}}}\mathbf{{p}}_{k}s_{k}.
	\end{equation}
	%%%
	%%%
	\par 
	The received signal  at  user-$k,  k\in\mathcal{K}$, is given by
	\begin{equation}
		\label{eq:signalReceiveHRS}
		y_k=\mathbf{{h}}_{k}^{H}\mathbf{{p}}_{c}s_{c}+\mathbf{{h}}_{k}^{H}\mathbf{{p}}_{k}s_{k}+\sum_{j\in\mathcal{\mathcal{K}},j\neq k}\mathbf{{h}}_{k}^{H}\mathbf{{p}}_{j}s_{j}+n_k.
	\end{equation}
	
	%%%
	\par Each user-$k$ first decodes the common stream $s_c$ into $\widehat{W}_c$ by treating the interference from all private streams as noise. Using SIC, $\widehat{W}_c$  is then re-encoded, precoded, and subtracted from the received signal such that user-$k$ decodes its private stream $s_k$ into $\widehat{W}_{p,k}$ by treating the remaining interference from the other private streams as noise\footnote{Here, we fix the decoding order at each user to  $s_c \rightarrow s_k$ as we follow the rule that streams intended for more users have a higher decoding priority \cite{NOUM2019Gunduz,Liu2017LDM} for the entire RSMA framework. The decoding order of $s_{c}$ and $s_{k}$ can be further optimized.}. User-$k$ reconstructs the original message by extracting $\widehat{W}_{c,k}$ from $\widehat{W}_{c}$, and combining $\widehat{W}_{c,k}$ with $\widehat{W}_{p,k}$ into $\widehat{W}_{k}$. 
	%%%
	Under the assumption of Gaussian signaling and infinite block length, the instantaneous rates for decoding the common and private streams at user-$k$ are given as follows
	\begin{equation}
		\label{eq:rate1RS}
		\begin{aligned}
			R_{c,k}&=\log_{2}\left(1+\frac{\left|\mathbf{{h}}_{k}^{H}\mathbf{{p}}_{c}\right|^{2}}{\sum_{j\in\mathcal{K}}\left|\mathbf{{h}}_{k}^{H}\mathbf{{p}}_{j}\right|^{2}+\sigma_{n,k}^2}\right),
			\\R_{k}&=\log_{2}\left(1+\frac{\left|\mathbf{{h}}_{k}^{H}\mathbf{{p}}_{k}\right|^{2}}{\sum_{j\in\mathcal{K},j\neq k}\left|\mathbf{{h}}_{k}^{H}\mathbf{{p}}_{j}\right|^{2}+\sigma_{n,k}^2}\right).
		\end{aligned}
	\end{equation}
	%%%
	To ensure $s_c$ is successfully decoded by all users, its rate can not exceed
	\begin{equation}
		\label{eq:commonstream1}
		R_{c}=\min\left\{R_{c,1},\ldots,R_{c,K}\right\}
	\end{equation}
	As $s_c$ contains sub-messages $W_{c,1},\ldots,W_{c,K}$ of $K$ users, the rate distribution of $R_c$ among the users is adapted to the amount of sub-messages that each user contributed. Let  $C_k$ denote the portion of rate $R_c$  allocated to user-$k$ for $W_{c,k}$. Then, we have
	\begin{equation}
		\label{eq:commonstream2}
		\sum_{k\in\mathcal{K}}C_k=R_{c}.
	\end{equation}
	The overall achievable rate of user-$k$, $k\in\mathcal{K}$, is
	\begin{equation}
		\label{eq: R_tot}
		R_{k,tot}=C_{k}+R_{k}.
	\end{equation}
	Apparently, the rate of each user is split into two parts, namely, the rate of $s_k$ (a.k.a. the private rate) and relevant part of the rate of $s_c$  (a.k.a. the common rate).
	
	\par Based on the system model of 1-layer RS, the precoding matrix $\mathbf{{P}}=[\mathbf{{p}}_{c}, \mathbf{{p}}_{1},\ldots, \mathbf{{p}}_{K}]\in\mathbb{C}^{M\times (K+1)}$ can be designed as low-complexity precoding, such as  zero-forcing beamforming (ZFBF) for the private streams and random beamforming for the common stream \cite{bruno2019wcl}, or via optimization under different objectives, i.e.,  maximizing the WSR \cite{RS2016hamdi,mao2017rate}, maximizing the worst-case user rate \cite{RS2016joudeh}, maximizing EE \cite{mao2018EE}, minimizing the transmit power \cite{Medra2018SPAWC}, etc. In Section \ref{sec:resourceAllocation}, the precoder design  for RSMA is discussed in detail. 
	%%%
	\subsubsection{2-layer HRS}
	%%%
	\par 2-layer HRS was initially proposed  for FDD massive MIMO \cite{Minbo2016MassiveMIMO} for enhancing the robustness to imperfect CSIT and boosting the achievable rate of all users.
	%%%
	The $K$ users are grouped into $G$ separate  groups indexed by $\mathcal{G}=\{1,\ldots,G\}$ and each group-$i$  contains $\mathcal{K}_i$ users with $\bigcup_{i\in\mathcal{G}}\mathcal{K}_i=\mathcal{K}$.
	%%%
	Each user-$k$ splits its message $W_k$ into three sub-messages ($L=3$), namely, an inter-group common sub-message $W_k^{\mathcal{K}}$, an inner-group common sub-message $W_k^{\mathcal{K}_i}$, and a private sub-message $W_k^k$.  The inter-group common sub-messages $\{W_k^{\mathcal{K}}|k\in\mathcal{K}\}$ are encapsulated into one  common message $W_{\mathcal{K}}$, which is encoded into a common stream $s_{\mathcal{K}}$ using a codebook shared by all users and is decoded by all users.
	%%%
	The inner-group common sub-messages $\{W_k^{\mathcal{K}_i}|k\in\mathcal{K}_i\}$ of the users in group-$i$ are merged into one common message $W_{\mathcal{K}_i}$ and encoded into an inner-group common  stream $s_{\mathcal{K}_i}$ using a codebook shared by the users in $\mathcal{K}_i$. $W_{\mathcal{K}_i}$ is therefore decoded by the users in group-$i$. 
	%%%
	The private sub-messages  $\{W_k^{k}|k\in\mathcal{K}\}$ are independently encoded into $K$ private streams $s_1,\ldots,s_K$, which are decoded by the corresponding users.
	%%%
	The overall encoded streams $\mathbf{s}=[s_{\mathcal{K}},s_{\mathcal{K}_1},\ldots,s_{\mathcal{K}_G},s_{1},\ldots,s_K]^T\in\mathbb{C}^{(K+G+1)\times 1}$ are linearly precoded with $\mathbf{{P}}=[
	\mathbf{{p}}_{\mathcal{K}}, \mathbf{{p}}_{\mathcal{K}_1},\ldots,\mathbf{{p}}_{\mathcal{K}_G},\mathbf{{p}}_{1},\ldots, \mathbf{{p}}_{K}]\in\mathbb{C}^{M\times (K+G+1)}$. The   signal sent from the transmitter is given as follows
	\begin{equation}
		\mathbf{{x}}=\mathbf{{p}}_{\mathcal{K}}s_{\mathcal{K}}+\sum_{i\in\mathcal{G}}\mathbf{{p}}_{\mathcal{K}_i}s_{\mathcal{K}_i}+\sum_{k\in\mathcal{\mathcal{K}}}\mathbf{{p}}_{k}s_{k}.
	\end{equation}
	%%%
	\par 
	The received signal at each user-$k,  k\in\mathcal{K}$, is given as follows
	\begin{equation}
		\label{eq:signalReceive1RS}
		y_k=\mathbf{{h}}_{k}^{H}\mathbf{{p}}_{\mathcal{K}}s_{\mathcal{K}}+\sum_{i\in\mathcal{G}}\mathbf{{h}}_{k}^{H}\mathbf{{p}}_{\mathcal{K}_i}s_{\mathcal{K}_i}+\sum_{j\in\mathcal{\mathcal{K}}}\mathbf{{h}}_{k}^{H}\mathbf{{p}}_{j}s_{j}+n_k.
	\end{equation}
	%%%
	Each user-$k$ ($k\in \mathcal{K}_i$) employs two layers of SIC to sequentially decode $s_{\mathcal{K}}$, $s_{\mathcal{K}_i}$, and $s_{k}$  with $s_{\mathcal{K}}$ being decoded first, $s_{\mathcal{K}_i}$ second, followed by  $s_{k}$.
	%%%
	Under the assumption of Gaussian signaling and infinite block  length, the rates for decoding streams $s_{\mathcal{K}}$, $s_{\mathcal{K}_i}$, and $s_{k}$ at user-$k$ are given as follows
	\begin{equation}
		%\small
		\resizebox{0.49\textwidth}{!}{$
			\begin{aligned}
				&R_{k}^{\mathcal{K}}=\log_{2}\left(1+\frac{\left|\mathbf{{h}}_{k}^{H}\mathbf{{p}}_{\mathcal{K}}\right|^{2}}{\sum\limits_{i\in\mathcal{G}}\left|\mathbf{{h}}_{k}^{H}\mathbf{{p}}_{\mathcal{K}_i}\right|^{2}+\sum\limits_{j\in\mathcal{K}}\left|\mathbf{{h}}_{k}^{H}\mathbf{{p}}_{j}\right|^{2}+\sigma_{n,k}^2}\right),
				\\
				&R_{k}^{\mathcal{K}_i}=\log_{2}\left(1+\frac{\left|\mathbf{{h}}_{k}^{H}\mathbf{{p}}_{\mathcal{K}_i}\right|^{2}}{\sum\limits_{i'\in\mathcal{G},i'\neq i}\left|\mathbf{{h}}_{k}^{H}\mathbf{{p}}_{\mathcal{K}_{i'}}\right|^{2}+\sum\limits_{j\in\mathcal{K}}\left|\mathbf{{h}}_{k}^{H}\mathbf{{p}}_{j}\right|^{2}+\sigma_{n,k}^2}\right),
				\\
				&R_{k}=\log_{2}\left(1+\frac{\left|\mathbf{{h}}_{k}^{H}\mathbf{{p}}_{k}\right|^{2}}{\sum\limits_{i'\in\mathcal{G},i'\neq i}\left|\mathbf{{h}}_{k}^{H}\mathbf{{p}}_{\mathcal{K}_{i'}}\right|^{2}+\sum\limits_{j\in\mathcal{K},j\neq k}\left|\mathbf{{h}}_{k}^{H}\mathbf{{p}}_{j}\right|^{2}+\sigma_{n,k}^2}\right).
			\end{aligned}
			$}
	\end{equation}
	%%%
	\par Following  (\ref{eq:commonstream1}) and  (\ref{eq:commonstream2}), we obtain the inter-group and inner-group common rates of  $s_{\mathcal{K}}$, $s_{\mathcal{K}_i}$, as follows
	\begin{equation}
		\begin{aligned}
			\sum_{k\in\mathcal{K}}C_k^{\mathcal{K}}&=\min\left\{R_{k}^{\mathcal{K}}\mid k\in \mathcal{K}\right\},\\
			\sum_{k\in\mathcal{K}_i}C_k^{\mathcal{K}_i}&=\min\left\{R_{k}^{\mathcal{K}_i} \mid k\in \mathcal{K}_i\right\}, \forall i\in\mathcal{G},
		\end{aligned}
	\end{equation}
	where  $C_k^{\mathcal{K}}$ and $C_k^{\mathcal{K}_i}$ are the respective rates allocated to user-$k$ for transmitting $W_k^{\mathcal{K}}$  and $W_k^{\mathcal{K}_i}$, respectively. The total achievable rate of user-$k, k\in \mathcal{K}_i$, is given by 
	\begin{equation}
		R_{k,tot}=C_{k}^{\mathcal{K}}+C_{k}^{\mathcal{K}_i}+R_{k}.
	\end{equation}
	
	%%%
	\par
\begin{figure}[t!]
		\centering
		\includegraphics[width=3.5in]{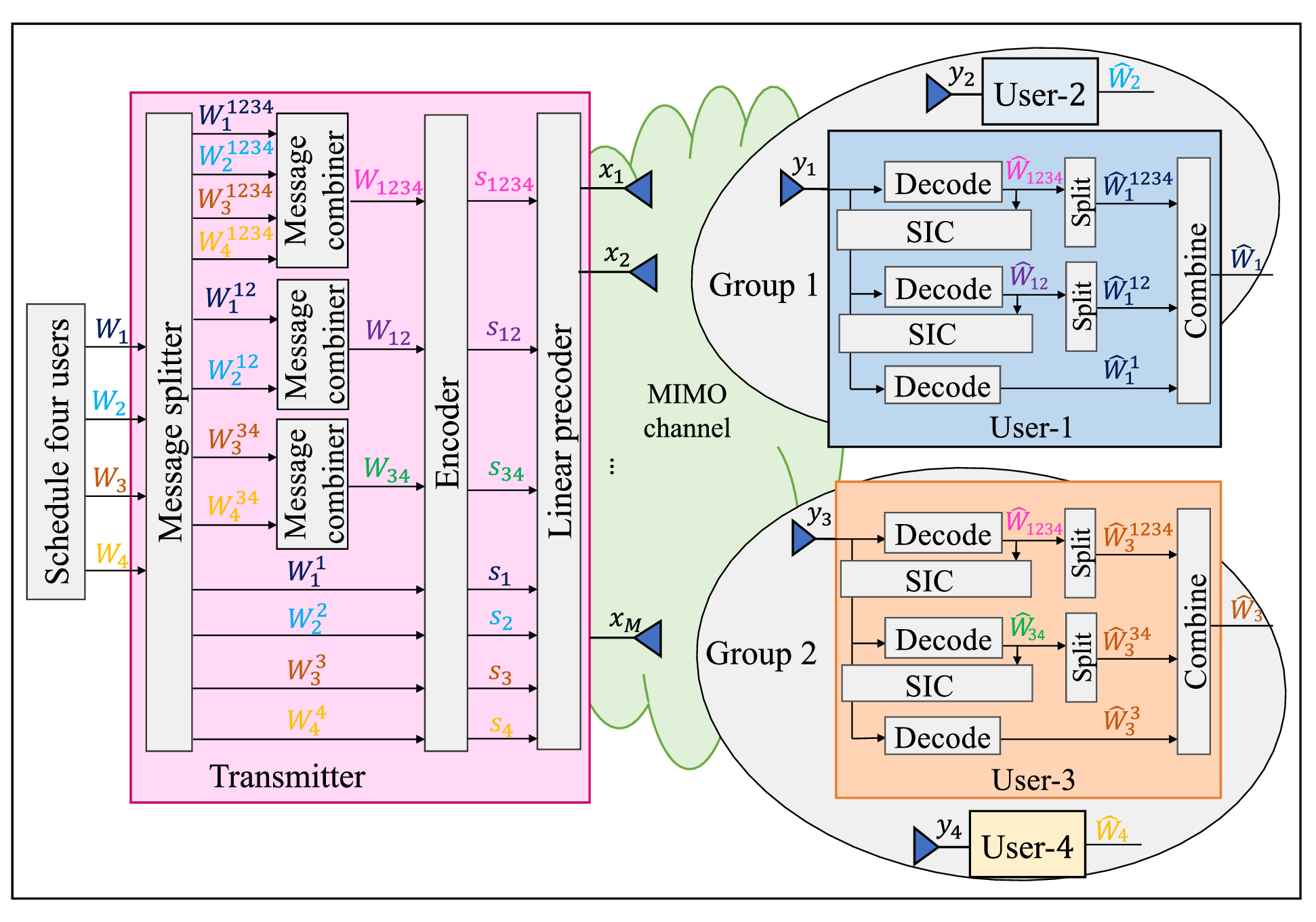}%
		\caption{Transceiver architecture of 4-user 2-layer HRS.}
		\label{fig: HRS}
		% 	\vspace{-2mm}
	\end{figure}

  A 2-layer HRS example for $K=4$ users is depicted in  Fig. \ref{fig: HRS} with user-$1$/$2$ in group-$1$, user-$3$/$4$  in group-$2$.
	%%%
	$s_{1234}$ is an inter-group common stream which is decoded by the four users. $s_{12}$ and $s_{34}$ are the inner-group common streams which are respectively decoded by the users in group-$1$ and group-$2$ only.  
	%%%
	By turning off $s_{12}$ and $s_{34}$, the four-user 2-layer HRS example in  Fig. \ref{fig: HRS} boils down to 1-layer RS where $s_{1234}$ plays the role of $s_{c}$ in Fig. \ref{fig: OneLayerRS}.
	% The receivers  of user-2 and user-4 resemble those of user-1 and user-2 by respectively changing the private streams $s_1$ and $s_3$ to be decoded to $s_2$ and $s_4$.
	%%%

	%%%
	\subsubsection{Generalized RS}
	%%%
	\label{sec: generalized RS}
	%%%
	\par
	Generalized RS, introduced in  \cite{mao2017rate}, is a generalized linearly precoded RS scheme aiming at  maximizing the achievable rate  and QoS of linearly precoded RS at the expense of a higher transceiver complexity. 
	%%%
	In contrast to 1-layer RS and 2-layer HRS, which maintain a constant number of message splits $L$ for each user message and a fixed number of SIC layers at each user independent of the number of users $K$, $K$-user generalized RS requires $L$ to be increased with $K$ as $L=2^{K-1}$ in order to synthesize different common streams, decoded by different user subsets of $\mathcal{K}$, and consequently, the number of SIC layers at each user increases with $K$ as $2^{K-1}-1$.
	%%%
	At the transmitter, message $W_k$ of user-$k$  is split as $\{ W_k^{\mathcal{A}'} | \mathcal{A}' \subseteq \mathcal{K}, k \in \mathcal{A}' \}$. For any user subset $\mathcal{A}\subseteq\mathcal{K}$, the BS loads sub-messages  $\{W_{k'}^{\mathcal{A}}|k'\in\mathcal{A}\}$ with the same superscript $\mathcal{A}$  onto data stream $s_{\mathcal{A}}$, which is decoded by all users in subset $\mathcal{A}$ and treated as noise by the other users.
	%%%
	The concept of \textit{stream order} is introduced in generalized RS to define the streams decoded by different numbers of users. 
	%%%
	Specifically, the streams decoded by $l$  users are defined as \textit{$l$-order streams} \cite{mao2017rate}. 
	% For instance,  the common stream $s_{\mathcal{K}}$ decoded by all users in $\mathcal{K}$ is a $K$-order stream and the private stream $s_{k}$ is a $1$-order stream to be decoded by a single user. 
	%%%
	Let $\{s_{\mathcal{A}'}|\mathcal{A}'\subseteq\mathcal{K},|\mathcal{A}'|=l\}$ denote the set of all $l$-order streams with ${K\choose l}$ elements, which forms the \textit{$l$-order data stream vector} $\mathbf{s}_l\in\mathbb{C}^{{K\choose l}\times 1}$.
	%%%
	As only  one $K$-order stream $s_{\mathcal{K}}$ exists, $\mathbf{s}_{K}$ is simplified to $s_{\mathcal{K}}$ when $l=K$.
	%%%
	$\mathbf{s}_l$ is linearly precoded by precoding matrix $\mathbf{P}_{l}$, which is composed of $ \{\mathbf{p}_{\mathcal{A}'}|\mathcal{A}'\subseteq\mathcal{K},|\mathcal{A}'|=l\}$. The transmit signal of $K$-user generalized RS is given as follows
	\begin{equation}
		\mathbf{{x}}=\sum_{l=1}^{K}\mathbf{{P}}_{{l}}\mathbf{{s}}_{{l}}=\sum_{l=1}^{K}\sum_{\mathcal{A}'\subseteq\mathcal{K},|\mathcal{A}'|=l}\mathbf{{p}}_{\mathcal{A}'}{{s}}_{\mathcal{A}'}.
	\end{equation}
	%%%
	%%%
	\begin{figure}[t!]
		\centering
		\includegraphics[width=3.5in]{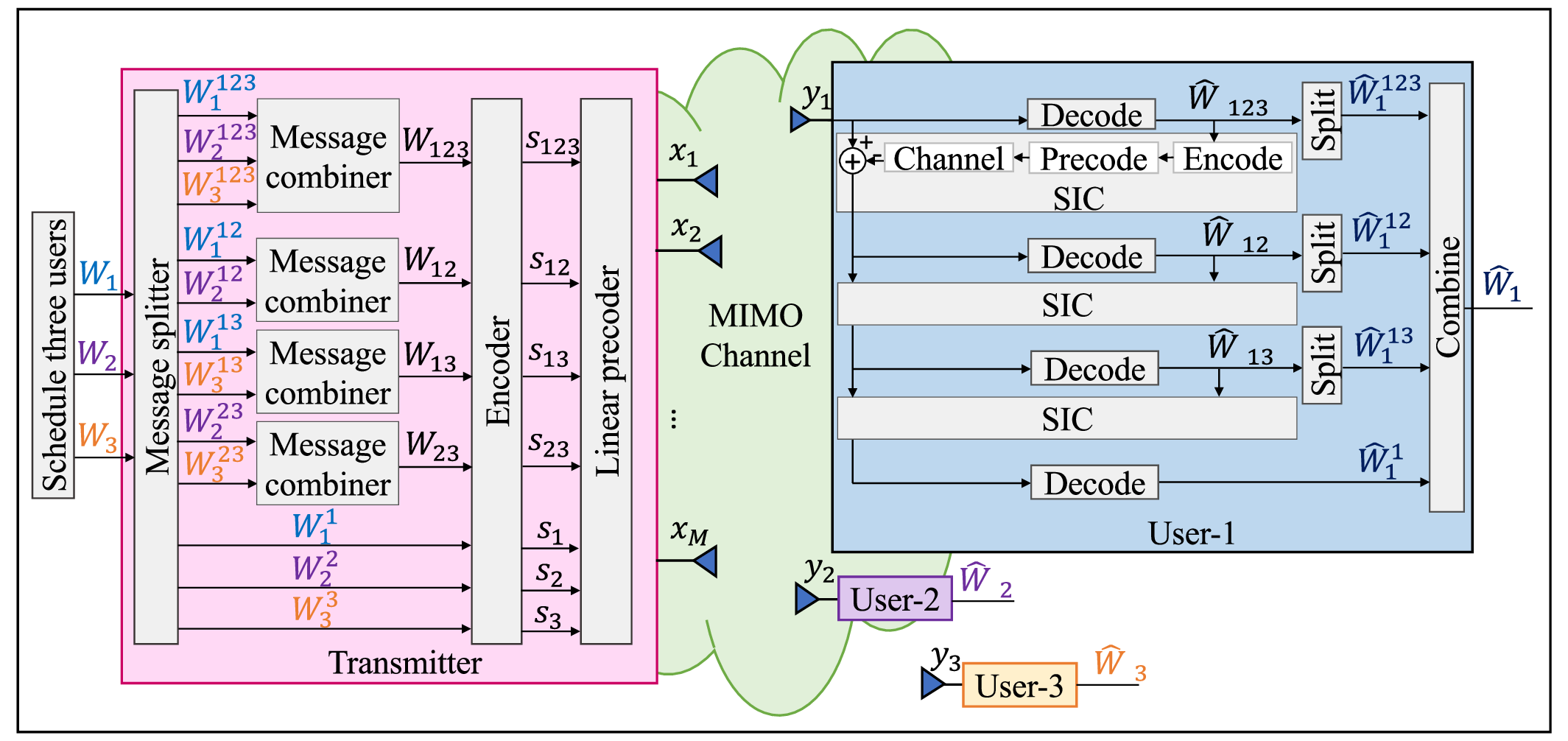}%
		\caption{Transceiver architecture of 3-user generalized RS \cite{mao2017rate}.}
		\label{fig: GRS}
		\vspace{-2mm}
	\end{figure}
	%%%
	\par Each user-$k$ employs $2^{K-1}-1$  SIC layers to decode the intended common streams and its private stream. Starting with the $K$-order stream  $s_{\mathcal{K}}$,   the decoding process goes downwards until the $1$-order private stream $s_{k}$ is reached.
	%%%
	The $l$-order streams  involved in $\mathcal{S}_{l,k}=\{s_{\mathcal{A}'}|\mathcal{A}'\subseteq\mathcal{K},|\mathcal{A}'|=l,k\in\mathcal{A}'\}$ are decoded at user-$k$ based on the decoding order $\pi_{l,k}$. We define  $\mathbf{s}_{\pi_{l,k}}=[s_{\pi_{l,k}{(1)}},\ldots,s_{\pi_{l,k}{(|\mathcal{S}_{l,k}|)}}]^H$ as the $l$-order stream vector decoded at user-$k$ with $s_{\pi_{l,k}{(i)}}$ being decoded before $s_{\pi_{l,k}{(j)}}$ if $i<j$.
	The rate of decoding $l$-order stream ${s}_{\pi_{l,k}{(i)}}$ at user-$k$ is 
	\begin{equation}
		R_k^{\pi_{l,k}{(i)}}=\log_{2}\left(1+\frac{|\mathbf{{h}}_{k}^{H}\mathbf{{p}}_{\pi_{l,k}{(i)}}|^{2}}{I_{\pi_{l,k}{(i)}}+\sigma_{n,k}^2}\right),
	\end{equation}
	where
	\[
	\begin{aligned}
		I_{\pi_{l,k}{(i)}}&=\sum_{j>i}|\mathbf{{h}}_{k}^{H}\mathbf{{p}}_{\pi_{l,k}(j)}|^{2}+\sum_{l'=1}^{l-1}\sum_{j=1}^{|\mathcal{S}_{l',k}|}|\mathbf{{h}}_{k}^{H}\mathbf{{p}}_{\pi_{l',k}(j)}|^{2}\\
		&+\sum_{\mathcal{A}'\subseteq\mathcal{K},k\notin\mathcal{A}'}|\mathbf{{h}}_{k}^{H}\mathbf{{p}}_{{\mathcal{A}'}}|^{2}
	\end{aligned}\]  
	is the interference received at user-$k$ when decoding  ${s}_{\pi_{l,k}{(i)}}$ with the first term $\sum_{j>i}|\mathbf{{h}}_{k}^{H}\mathbf{{p}}_{\pi_{l,k}(j)}|^{2}$ being the interference from the remaining non-decoded $l$-order streams in $\mathbf{s}_{{\pi_{l,k}}}$, the second term $\sum_{l'=1}^{l-1}\sum_{j=1}^{|\mathcal{S}_{l',k}|}|\mathbf{{h}}_{k}^{H}\mathbf{{p}}_{\pi_{l',k}(j)}|^{2}$ being the interference from lower order streams $\{\mathbf{s}_{{\pi_{l',k}}}| l'<l\}$ to be decoded at user-$k$, and the third term $\sum_{\mathcal{A}'\subseteq\mathcal{K},k\notin\mathcal{A}'}|\mathbf{{h}}_{k}^{H}\mathbf{{p}}_{{\mathcal{A}'}}|^{2}$ being the interference from the unintended streams. 
	%%%
	Following  (\ref{eq:commonstream1}) and  (\ref{eq:commonstream2}),  the  rate of the $|\mathcal{A}|$-order stream $s_{\mathcal{A}}$ 
	% ( $\mathcal{A}\in\mathcal{K},2\leq|\mathcal{A}|\leq K$)
	is
	\begin{equation}
		\label{eq:min rate}
		\sum_{k\in \mathcal{A}}C_{k}^{\mathcal{A}}=\min_{k'}\left\{ R_{k'}^{\mathcal{A}}\mid k'\in\mathcal{A}\right\},
	\end{equation} 
	where $C_k^{\mathcal{A}}$ is the portion of the rate of stream $s_{\mathcal{A}}$ allocated to user-$k$ $(k\in \mathcal{A})$ for $W_k^{\mathcal{A}}$.
	%%%
	The total achievable rate of user-$k$ is
	\begin{equation}
		R_{k,tot}=\sum_{\mathcal{A}'\subseteq\mathcal{K},k\in \mathcal{A}'}C_k^{\mathcal{A}'}+R_k.
	\end{equation}
	%%%
	\par 
	A 3-user generalized RS example is illustrated in Fig. \ref{fig: GRS}. 
	%%%
	Each user message is split into 4 sub-messages, i.e., $W_1$ is split into $\{W_{1}^{123}$, $W_{1}^{12}$, $W_{1}^{13}$, $W_{1}^{1}\}$. The total 12 sub-messages are recombined and encoded into one $3$-order stream $s_{123}$, three 2-order streams $\mathbf{s}_{2}=[s_{12},s_{13},s_{23}]^T$ and three $1$-order streams $\mathbf{s}_{1}=[s_{1},s_{2},s_{3}]^T$.
	%%%
	Each user employs 3 layers of SIC to decode the intended $4$ streams in order. Fig. \ref{fig: GRS} depicts the receiver of user-$1$ when the decoding order of the $2$-order streams is  $\pi_{2,1}=12\rightarrow13$. We have $s_{\pi_{2,1}{(1)}}=s_{12}$ and $s_{\pi_{2,1}{(2)}}=s_{13}$.
	%%%
	\begin{figure}[t!]
		\centering
		\includegraphics[width=3.0in]{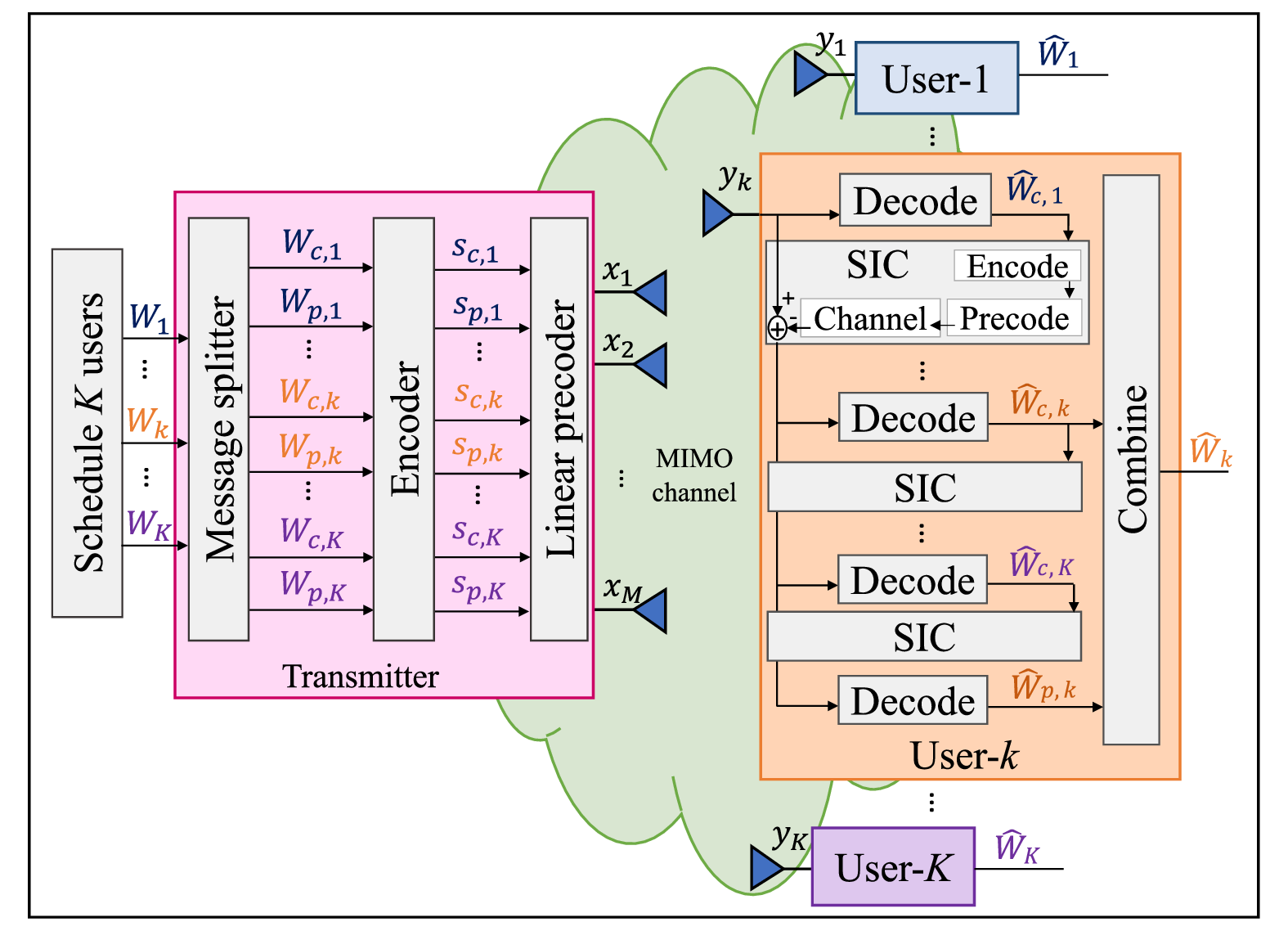}%
		\caption{Transceiver architecture of $K$-user RS-CMD.}
		\label{fig:RSCMD}
		\vspace{-2mm}
	\end{figure}
	%%%
	\subsubsection{RS-CMD}
	\label{sec:RSCMD}
	%%%
	RS-CMD, initially proposed in \cite{Ala2019IEEEAccess} for cloud-radio access networks (C-RAN), is a RSMA scheme that splits and encodes user messages without the use of message combiners, i.e., the message combiners at the RSMA transmitter in Fig. \ref{fig:RSMAtx} and the RSMA receiver in Fig. \ref{fig:RSMArx} are both turned off. 
	% is considered as an extension of the HK scheme to BC as C-RAN becomes a BC when the fronthaul capacity tends to infinity.
	%%%
	Fig. \ref{fig:RSCMD} illustrates a $K$-user RS-CMD transceiver for the MISO BC.
	%%%
	\par At the transmitter, message $W_k$ intended for user-$k$  is split into two sub-messages  $W_{c,k}$ and $W_{p,k}$. The resulting $2K$ sub-messages $\{W_{c,k}, W_{p,k} \mid k\in\mathcal{K}\}$ are independently encoded into $2K$  streams  $\{s_{c,k},s_{p,k}\mid k\in\mathcal{K}\}$ with $\{s_{c,k}\mid k\in\mathcal{K}\}$ being common streams decoded by all users\footnote{Here, we consider the case when the $K$ common streams $\{s_{c,k}\mid k\in\mathcal{K}\}$ are decoded by all users. The group of users that decodes each of the common streams could be further optimized to further enhance the performance of RS-CMD \cite{Alaa2019uavCRAN}.} and $\{s_{p,k}\mid k\in\mathcal{K}\}$ being private streams  decoded by the corresponding users only. The encoded streams are linearly precoded via precoders $\{\mathbf{p}_{c,k}, \mathbf{p}_{p,k} \mid k\in\mathcal{K}\}$. The resulting transmit signal of the $K$-user RS-CMD is given as follows
	%%%
	\begin{equation}
		\mathbf{{x}}=\sum_{k\in\mathcal{\mathcal{K}}}\left(\mathbf{{p}}_{c,k}s_{c,k}+\mathbf{{p}}_{p,k}s_{p,k}\right).
	\end{equation}
	%%%
	\par
	User-$k$ employs $K$ layers of SIC to decode the $K$ common streams $\{s_{c,k'} \mid k'\in\mathcal{K}\}$ based on decoding order $\pi_k$ before decoding the intended private stream $s_{p,k}$. Stream $s_{c,\pi_k(i)}$ is decoded before $s_{c,\pi_k(j)}$ at user-$k$ if $i<j$. The rates for decoding common stream $s_{c,\pi_k(i)}$ and private stream $s_{p,k}$ at user-$k$ are obtained as follows
	\begin{equation}
		%\small
		\resizebox{0.493\textwidth}{!}{$
			\begin{aligned}
				R_{k,\pi_k(i)}^{c}&=\log_{2}\left(1+\frac{\left|\mathbf{{h}}_{k}^{H}\mathbf{{p}}_{c,\pi_k(i)}\right|^{2}}{\sum\limits_{j>i}\left|\mathbf{{h}}_{k}^{H}\mathbf{{p}}_{c,\pi_k(j)}\right|^{2}+\sum\limits_{j\in\mathcal{K}}\left|\mathbf{{h}}_{k}^{H}\mathbf{{p}}_{p,j}\right|^{2}+\sigma_{n,k}^2}\right),
				\\
				R_{k}^{p}&=\log_{2}\left(1+\frac{\left|\mathbf{{h}}_{k}^{H}\mathbf{{p}}_{p,k}\right|^{2}}{\sum_{j\in\mathcal{K},j\neq k}\left|\mathbf{{h}}_{k}^{H}\mathbf{{p}}_{p,j}\right|^{2}+\sigma_{n,k}^2}\right)
			\end{aligned}
			$}
	\end{equation}
	The overall achievable rate of user-$k$ is given by
	\begin{equation}
		R_{k,tot}=R_{k}^{c}+R_{k}^{p},
	\end{equation}
	where $R_{k}^{c}=\min_{i}\left\{R_{i,k}^{c}\mid i\in\mathcal{K}\right\}$ is the achievable rate of common stream $s_{c,k}$ guaranteeing $s_{c,k}$  is successfully decoded at all users. 
	%%%
	Fig. \ref{fig:RSCMD} illustrates an example where the decoding order of the common streams at user-$k$ is as follows $s_{c,1}\rightarrow s_{c,2}\rightarrow\ldots\rightarrow s_{c,K}$.
	%%%
	\begin{figure}
		\centering
		\begin{subfigure}[b]{0.5\textwidth}
			\centering
			\includegraphics[width=\textwidth]{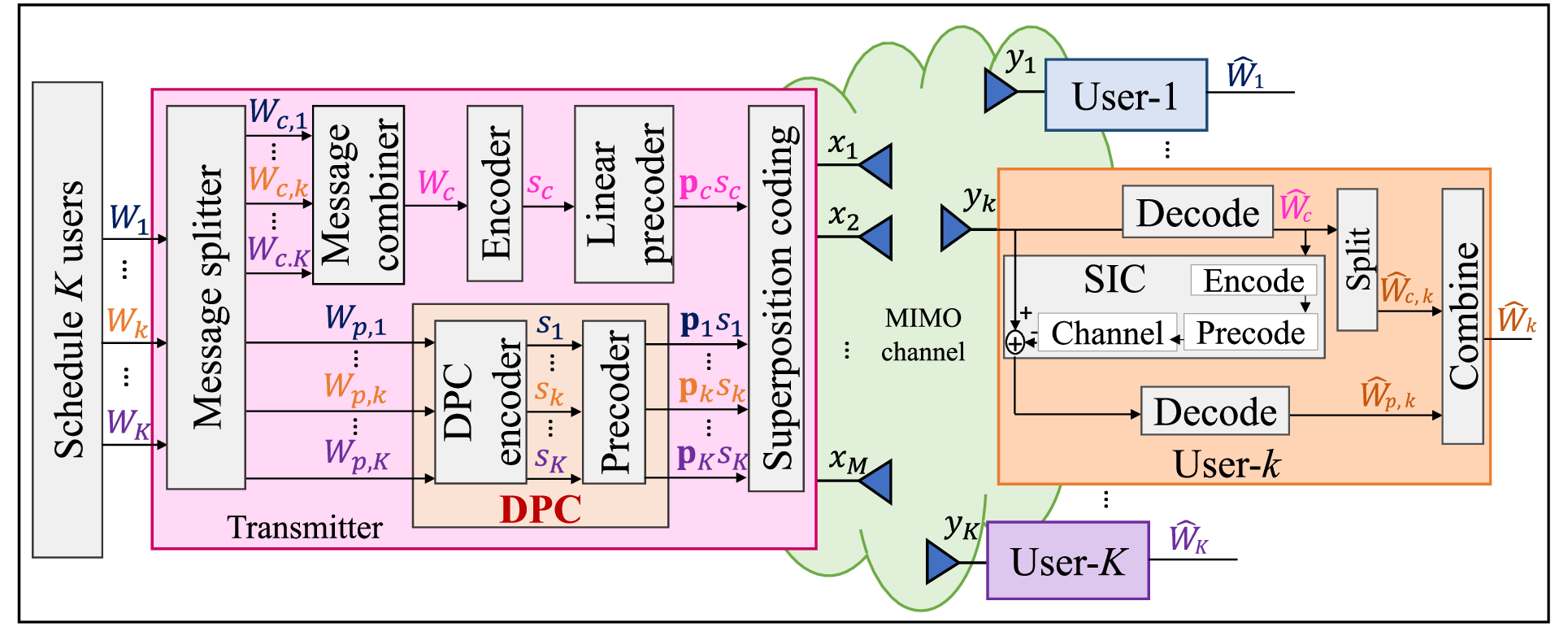}%
			\caption{ $K$-user 1-DPCRS}	
		\end{subfigure}%
		~\\
		\centering
		\begin{subfigure}[b]{0.48\textwidth}
			\centering
			\includegraphics[width=\textwidth]{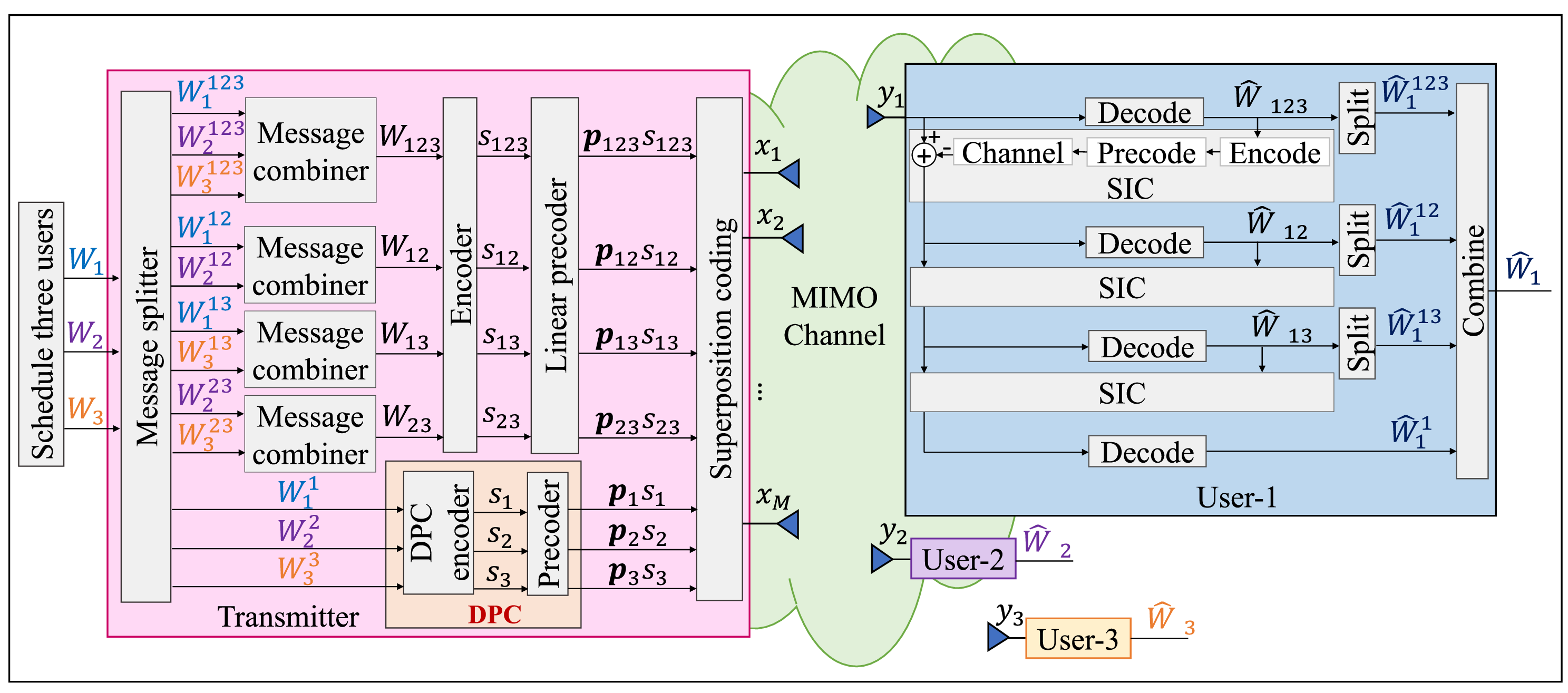}%
			%		\vspace{-0.4cm}
			\caption{$3$-user M-DPCRS}
		\end{subfigure}
		\caption{Transceiver architecture for  DPCRS \cite{mao2019beyondDPC}.}
		\label{fig: DPCRS}
	\end{figure}
	%%%
	\subsubsection{DPCRS}
	%%%
	\par Although DPC is the only known capacity-achieving strategy for the multi-antenna BC with perfect CSIT  \cite{jindal2002duality, duality2003Andrew, DPCrateRegion03Goldsmith}, it is vulnerable to CSIT estimation errors {\cite{DPC2012NanLee, yang2005impact}}. 
	%%%
	This limitation of DPC can be successfully compensated by DPCRS proposed in \cite{mao2019beyondDPC}, which marries RS and DPC to overcome the performance losses of DPC in imperfect CSIT.
	%%%
	The two DPCRS schemes proposed  in \cite{mao2019beyondDPC}  are ``1-layer Dirty Paper Coded RS (1-DPCRS)" and  ``multi-layer dirty paper coded rate-splitting (M-DPCRS)", as illustrated in Fig. \ref{fig: DPCRS}.
	%%%
	Compared with the corresponding linearly precoded  1-layer RS and generalized RS schemes, the main difference of DPCRS is that the private streams are generated using non-linear DPC. The common streams are still linearly precoded and superimposed on top of the non-linear private streams. 
	%%%
	The receiver architecture for DPCRS is the same as that for linearly precoded RS.
	%%%
	Readers are referred to \cite{mao2019beyondDPC} for more details on DPCRS. 
	%%%
	Note that besides 1-layer RS and generalized RS, other linearly precoded strategies, such as 2-layer HRS and RS-CMD, also can be extended to the corresponding DPC-based RS counterparts.

	% %%%
	% \par 
	% Define ${\pi}\triangleq[\pi(1),\ldots,\pi(K)]$ as one permutation of \{$1,\ldots,K$\} such that the message $W_{\pi(i)}$ is encoded before $W_{\pi(j)}$ if $i<j$.
	% %%%
	% For a given encoding order $\pi$, the private sub-message $W_{\pi(1)}$  for user-${\pi(1)}$  is encoded first until the sub-message $W_{\pi(K)}$ for user-${\pi(K)}$. 
	% %%%
	% The encoded data vector $\mathbf{s}\triangleq[s_c,s_{\pi(1)},\ldots,s_{\pi(K)}]^T$ is precoded by  $\mathbf{P}\triangleq[\mathbf{p}_c,\mathbf{p}_1,\ldots,\mathbf{p}_K]$ and the  transmit signal  of 1-DPCRS is 
	% \begin{equation}
	% \label{eq: transmit signal DPCRS}
	% \mathbf{x}=\mathbf{P}\mathbf{s}={{\mathbf{p}_c{s}_c}}+{{\sum_{k\in\mathcal{K}}\mathbf{p}_{\pi(k)}{s}_{\pi(k)}}}.
	% %	\vspace{-1mm}
	% \end{equation}
	% %%%
	% At user sides, the intended common and private streams are decoded in sequence as in 1-layer RS. 
	% The resulting rates of decoding $s_c$ and $s_{\pi(k)}$ at user-${\pi(k)}$  are \begin{equation}
	% \resizebox{0.49\textwidth}{!}{$
	% \begin{aligned}
	%  &{R_{c,\pi(k)}=\log_2\left(1+\frac{|{\mathbf{h}}_{\pi(k)}^H\mathbf{p}_{c}|^2}{\sum_{j\in\mathcal{K}}|\mathbf{h}_{\pi(k)}^H\mathbf{p}_{\pi(j)}|^2+\sigma_{n,k}^2}\right)}, 
	%  \\
	% &R_{\pi(k)}=\log_2\left(1+\frac{|{\mathbf{h}}_{\pi(k)}^H\mathbf{p}_{\pi(k)}|^2}{\sum_{i<k}|\widetilde{\mathbf{h}}_{\pi(k)}^H\mathbf{p}_{\pi(i)}|^2+\sum_{j> k}|\mathbf{h}_{\pi(k)}^H\mathbf{p}_{\pi(j)}|^2+\sigma_{n,k}^2}\right). 
	% \end{aligned}
	% $}
	% \end{equation}

	%%%
	\subsubsection{THPRS}
	%%%
	THPRS is proposed in \cite{Flores2018ISWCS} and combines the benefits of 1-layer RS and THP. Though THP  does not achieve as good a performance as DPC, it is less complex and considered to be a practical and suboptimal implementation of DPC. 
	%%%
	Compared with 1-DPCRS, the main difference of THPRS is the use of THP  to encode the private messages. The interference caused by the previously encoded private streams  can be eliminated using THP while the interference caused by the subsequently encoded private streams  is removed via beamforming, i.e., ZFBF based on LQ decomposition, see \cite{Flores2018ISWCS}. The common message $W_c$ is independently encoded into $s_c$ and superimposed on top of the THP encoded private streams. 
	%%%
	At the user sides, each user sequentially decodes the intended common and private streams. As THP requires a modulo operator at the transmitter to ensure that the transmit power does not exceed the power constraint, each user also applies the modulo operation to decode the intended private stream. 
	%%%
	The use of the modulo operation at the transmitter and receiver incurs power and modulo losses, which therefore make THPRS a suboptimal DPCRS implementation. 
	%%%
	Readers are referred to \cite{Flores2018ISWCS, Andre2021THP} for more details on THPRS. 
	%%%
	\subsection{Uplink RSMA}
	\label{sec:RSMAUL}
\begin{figure}[t!]
		\centering
		\includegraphics[width=3.0in]{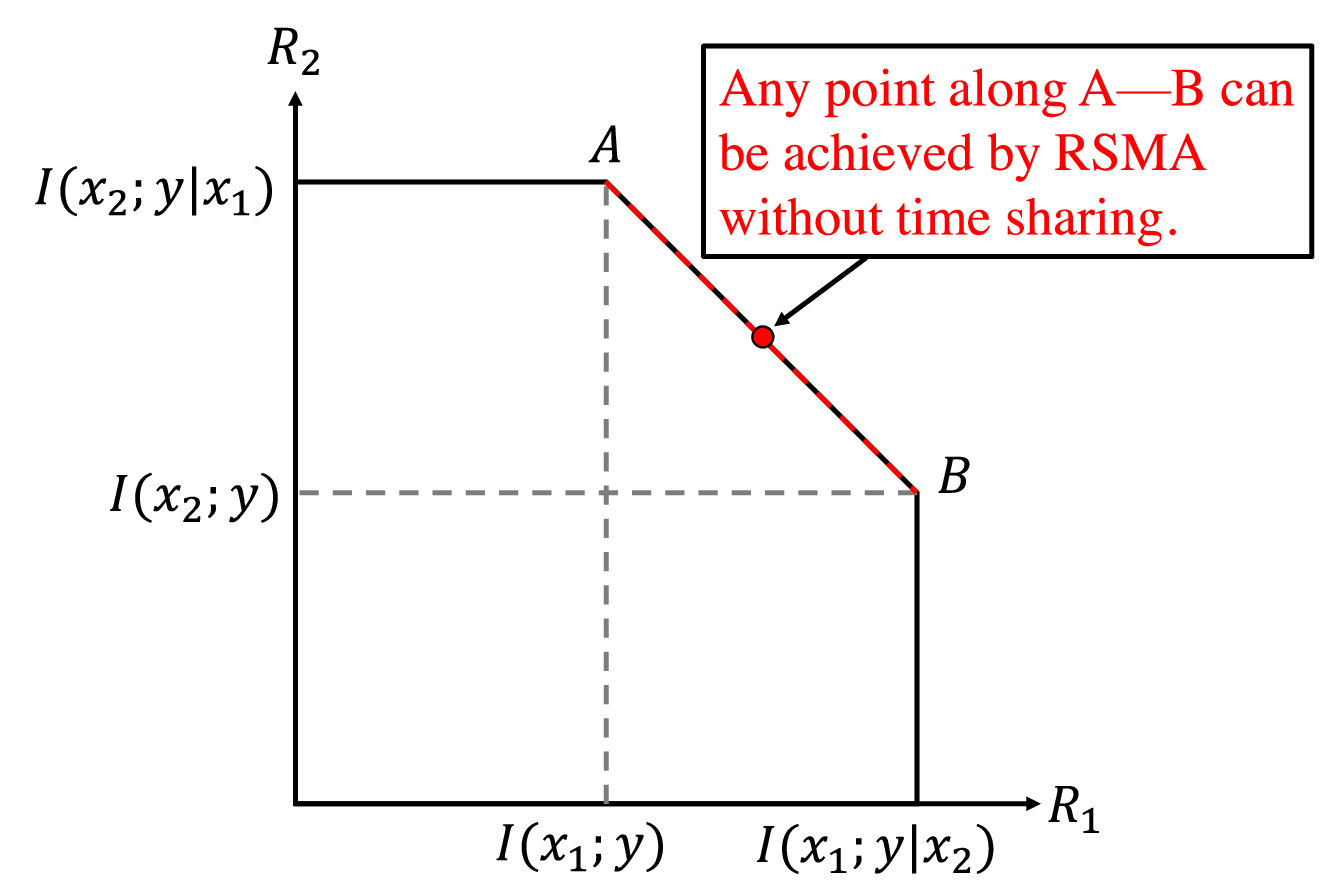}%
		\caption{Two-user Gaussian multiple-access capacity region. $x_1$ and $x_2$ are the information symbols of user-$1$ and user-$2$, respectively. $y$ is the received signal. $I(x_k;y)$ is the mutual information between $x_k$ and $y$. $I(x_k;y|x_j)$ is the conditional mutual information between $x_k$ and $y$ given $x_j$ \cite{tsefundamentalWC2005}.}
		\label{fig:RegionUplinkRS}
		\vspace{-2mm}
	\end{figure}

	\labelsubseccounter{IV-C}
	RSMA has been studied not only for the downlink, but also for the uplink \cite{Rimo1996,Grant2001,Medard2004Aloha,Cao2007distributeRS,zhaohui2020ULRS, Hongwu2020ULRS}. It was first proposed  in \cite{Rimo1996} for the  SISO MAC and was shown to achieve the capacity region of the Gaussian MAC without the need for time sharing among users. 
	%%%
	The capacity region of the two-user SISO Gaussian MAC is a pentagon as shown in Fig. \ref{fig:RegionUplinkRS}. 
	%%%
	The two corner points A and B can be achieved by SIC with two reverse decoding orders, which however does not apply to the rate points  along the line segment A--B. 
	%%%
	There are three methods that can achieve the points along the line segment A--B, namely, joint encoding/decoding, time sharing, and rate splitting. 
	%%%
	Joint encoding/decoding is not practical due to the high complexity of code construction and joint decoding at the receiver. The second method, time sharing, induces communication overhead and stringent synchronization requirements due to the coordination of the transmissions of all users \cite{Rimo1996}. Thus, it is not applicable for services requiring grant-free access, which allow collisions to reduce the access latency stemming from the channel grant procedure. 
	%%%
	%%%
	\begin{figure}[t!]
		\centering
		\includegraphics[width=3.0in]{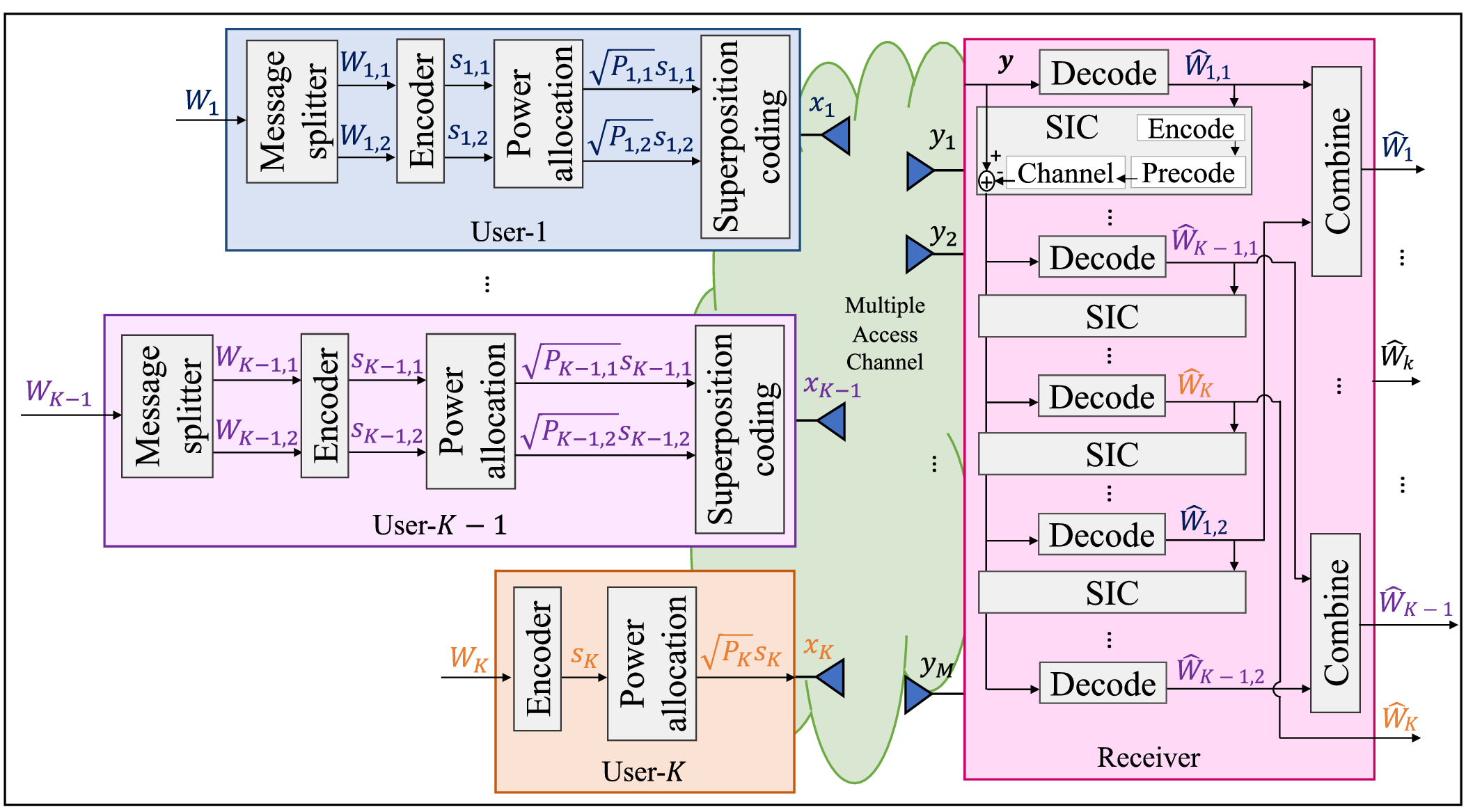}%
		\caption{Transceiver architecture of $K$-user uplink RSMA.}
		\label{fig:uplinkRS}
		\vspace{-2mm}
	\end{figure}
	%%%
	\par In comparison to the other methods, RSMA is more attractive as it achieves every point of the capacity region (including line segment A--B) via SC and SIC. 
	%%%
	Though RSMA introduces a coordination overhead for common and private streams’ rate selections, the complexity of selecting a versatile set of the parameters for RSMA is arguably lower than time sharing.
	%%%
	Therefore, RSMA can be a low-complexity enabler for services with grant-free access and user intermittency, in which users are not always active and there is always an element of unpredictability in their activation, such as URLLC and mMTC \cite{petar2018userIntermittency}.  
	%%%
	Fig. \ref{fig:uplinkRS}  illustrates $K$-user uplink RSMA based on SC  and SIC. 
	%%%
	In the $K$-user case, splitting the messages of $K-1$ users is sufficient to avoid time sharing and achieves every point of the capacity region \cite{Rimo1996}. Without loss of generality, we assume the messages of all users except user-$K$ are split.
	%%%
	At user-$k,k\in\{1,\ldots,K-1\}$, the message $W_k$ to be transmitted is split into two sub-messages $W_{k,1}$ and $W_{k,2}$. This can be interpreted as creating 2 virtual users. 
	%%%
	The messages $W_{k,1}$ and $W_{k,2}$ of the two virtual users   are independently encoded into  streams $s_{k,1}$ and $s_{k,2}$ with unit variance, i.e., $\mathbb{E}[\left |  s_{k,i}\right |^2]=1, i=1,2$. 
	%%%
	The two streams are then respectively allocated with certain powers $P_{k,1}$ and  $P_{k,2}$, and superposed at user-$k$. At user-$K$, the message $W_K$ is directly encoded into $s_K$ and allocated with power $P_{K}$. The transmit signal is given by
	\begin{equation}
		x_k=\begin{cases}
			\sqrt{P_{k,1}}s_{k,1}+\sqrt{P_{k,2}}s_{k,2}, & \text{ if } k\in\{1,\ldots,K-1\}, \\
			\sqrt{P_{K}}s_{K}, & \text{ if } k=K.
		\end{cases}
	\end{equation}
	% We use the same notation as in other downlink schemes, but there is no common and private sub-messages for uplink RSMA since all messages are decoded at the receiver (i.e., a BS). 
	
	%%%
	\par The signal observed at the receiver is
	\begin{equation}
		\mathbf{y}=\sum_{k\in\mathcal{K}}\mathbf{h}_kx_k+\mathbf{n}_r,
	\end{equation}
	where 
	$\mathbf{h}_k\in \mathbb{C}^{M\times1}$ is the channel vector between  user-$k$  and the receiver, and $\mathbf{n}_r\in\mathbb{C}^{M\times1}$ is the AWGN vector whose elements have zero-mean and variance $\sigma_{n_r}^2$.
	%%%
	The receiver employs filters $\mathbf{w}_{k,1}, \mathbf{w}_{k,2}$ to detect the two streams of user-$k, k\in\{1,\ldots,K-1\}$ and $\mathbf{w}_{K}$ to detect the stream of user-$K$.  
	%%%
	Let $\pi$ denote the decoding order of the $2K-1$ received streams $\{s_{k,i}, s_K \mid k\in\{1,\ldots,K-1\}, i\in\{1,2\}\}$, and let $\pi_{k,i}$
	denote the decoding order of $s_{k,i}$ with stream $s_{k,i}$ being decoded before stream $s_{k',i'}$ if $\pi_{k,i}<\pi_{k',i'}$. 
	%%%
	For user-$K$, $\pi_{K,i}$ is simplified to $\pi_{K}$. 
	%%%
	Assuming Gaussian signaling and infinite block length, the rate of decoding $s_{ki}$ at the receiver is given as follows
	\begin{equation}
		R_{k,i}=\log_{2}\left(1+\frac{P_{k,i}\left|\mathbf{{w}}_{k,i}^H\mathbf{{h}}_{k}\right|^{2}}{\sum\limits_{\pi_{k',i'}>\pi_{k,i}}P_{k',i'}\left|\mathbf{{w}}_{k,i}^H\mathbf{{h}}_{k'}\right|^{2}+\sigma_{n_r}^2}\right),
	\end{equation}
	\begin{equation}
		R_{K}=\log_{2}\left(1+\frac{P_{K}\left|\mathbf{{w}}_{K}^H\mathbf{{h}}_{K}\right|^{2}}{\sum\limits_{\pi_{k',i'}>\pi_{K}}P_{k',i'}\left|\mathbf{{w}}_{K}^H\mathbf{{h}}_{k'}\right|^{2}+\sigma_{n_r}^2}\right).
	\end{equation}
	%%%
	Optimizing the power allocation $\{P_{k,i},P_{K}\mid k\in\{1,\ldots,K-1\}, i\in\{1,2\}\}$ to maximize the sum rate has been investigated in \cite{zhaohui2020ULRS}.
	In order to decode all $2K-1$ streams, the receiver requires $2K-2$ layers of SIC. Therefore, higher receiver complexity is introduced in order to avoid time sharing.
	%%%
	\par The described system model for uplink RSMA is generic, and can be adapted to enable uplink multiple-access for both homogeneous and heterogeneous services. More specifically, RSMA can be employed to enable a specific service in a network, such as URLLC or mMTC; or it can be employed to enable non-orthogonal multiplexing of different services in heterogeneous networks, as in heterogeneous non-orthogonal multiple access (H-NOMA) \cite{petar2018userIntermittency}. The application of RSMA in both use cases require further study.
	
	%%%
	\subsection{Multi-cell RSMA}
	\label{sec:RSMAmulticell}
\begin{figure}[t!]
		\centering
		\begin{subfigure}[b]{0.36\textwidth}
			\centering
			\includegraphics[width=\textwidth]{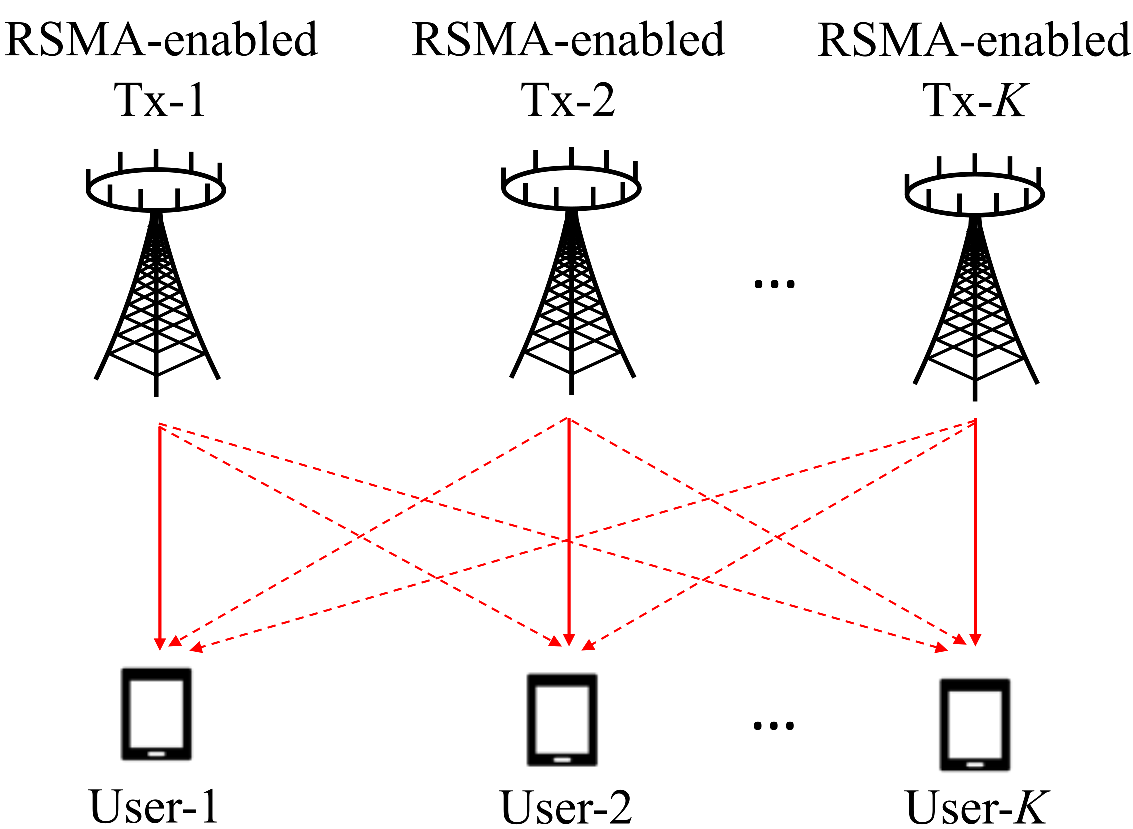}%
			\caption{Coordinated transmission}	
		\end{subfigure}%
		~\\
		\centering
		\vspace{0.2cm}
		\begin{subfigure}[b]{0.3\textwidth}
			\centering
			\includegraphics[width=\textwidth]{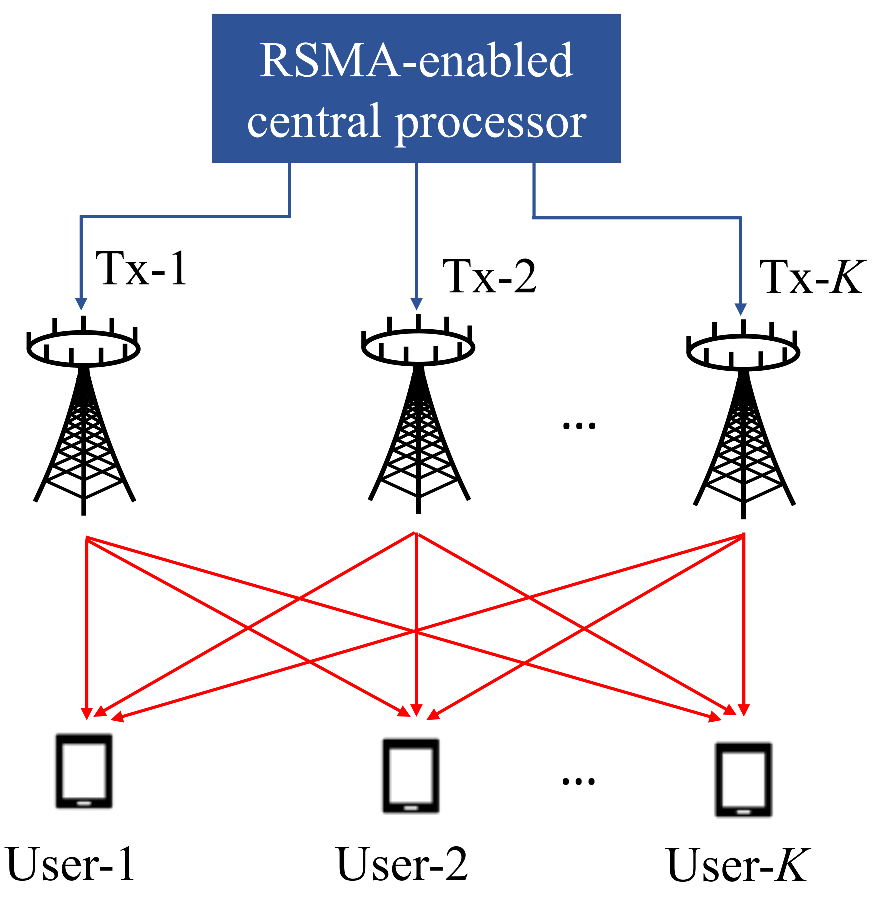}%
			\caption{Cooperative transmission}
		\end{subfigure}
		\caption{Multi-cell RSMA-enabled transmission.}
		\label{fig:multiCell}
	\end{figure}

	\labelsubseccounter{IV-D}
	Next, we increase our scope of RSMA by considering a network comprising multiple cells, each hosting one or multiple users. 
	%%%
	There are two categories of multi-cell networks, namely, ``coordination" and ``cooperation" networks \cite{clerckx2013mimo}. 
	%%%
	The former terminology refers to schemes that do not rely on data sharing while the latter refers to schemes relying on data sharing, as illustrated in Fig. \ref{fig:multiCell}, but both (normally) require CSI sharing among the BSs. 
	% \par
	The benefits of RSMA have been explored for both coordinated multi-cell networks \cite{carleial1978RS, TeHan1981,chenxi2017brunotopology,Tse2008,Tervo2018SPAWC,Jia2020SEEEtradeoff,wonjae2020multicell} and cooperative multi-cell networks \cite{mao2018networkmimo, Alaa2019uavCRAN,Ala2019IEEEAccess,cran2019wcl,alaa2020EECRAN,alaa2020gRS,alaa2020powerMini,alaa2020cranimperfectCSIT,alaa2020IRScran}. RSMA has been shown to achieve enhanced SE in both categories of multi-cell networks due to its ability to tackle not only  the intra-cell interference but also the inter-cell interference. 
	%%%
	%%%
	%%%
	\subsubsection{Coordinated transmission} Coordination techniques are motivated by interference channel (IC) concepts from information theory. The data of each user is  sent from  one cell while  user scheduling, beamforming, and power control are  designed based on coordination among cells. By using RS among cell, the inter-cell interference can be  partially decoded and partially treated as noise, which therefore improves the interference management between cells. 
	%%%
	Fig. \ref{fig:multiCell}(a) illustrates a $K$-cell MISO IC where each $M$-antenna transmitter (Tx)  serves one single-antenna user.
	%%%
	By employing RSMA at each transmitter-$k$, the  message $W_k$ of user-$k$ is split into two parts $W_{c,k}$ and $W_{p,k}$, which are then  encoded into two streams $s_{c,k}, s_{p,k}$ and linearly precoded via precoders $\mathbf{p}_{c,k}, \mathbf{p}_{p,k}\in \mathbb{C}^{M\times 1}$. The resulting transmit signal at  transmitter-$k$ is given by
	%%%
	\begin{equation}
		\mathbf{{x}}_k=\mathbf{{p}}_{c,k}s_{c,k}+\mathbf{{p}}_{p,k}s_{p,k}.
	\end{equation}
	%%%
	\par The signal received at the receiver is given by 
	\begin{equation}
		y_k=\sum_{j=1}^K\mathbf{h}_{kj}^H(\mathbf{{p}}_{c,j}s_{c,j}+\mathbf{{p}}_{p,j}s_{p,j})+n_k,
	\end{equation}
	where $\mathbf{h}_{kj}\in \mathbb{C}^{M\times 1}$ is the channel between transmitter-$j$ and user-$k$. 
	%%%
	The receiver design follows the  RS-CMD concept illustrated in Fig. \ref{fig:RSCMD}.
	%%%
	Each user-$k$ successively decodes and removes all  common streams $s_{c,1},\ldots,s_{c,L}$ using SIC, and then decodes the desired private stream $s_{p,k}$.
	%%%
	Based on the decoding order $\pi_k$, where stream $s_{c,\pi_k(i)}$ is decoded before $s_{c,\pi_k(j)}$ at user-$k$ if $i<j$, the rates for decoding $s_{c,\pi_k(i)}$ and $s_{p,k}$ at user-$k$ are obtained as follows
	\begin{equation}
		%\small
		\resizebox{0.493\textwidth}{!}{$
			\begin{aligned}
				R_{k,\pi_k(i)}^{c}&=\log_{2}\left(1+\frac{\left|\mathbf{{h}}_{k\pi_k(i)}^{H}\mathbf{{p}}_{c,\pi_k(i)}\right|^{2}}{\sum\limits_{j>i}\left|\mathbf{{h}}_{k\pi_k(j)}^{H}\mathbf{{p}}_{c,\pi_k(j)}\right|^{2}+\sum\limits_{j\in\mathcal{K}}\left|\mathbf{{h}}_{kj}^{H}\mathbf{{p}}_{p,j}\right|^{2}+\sigma_{n,k}^2}\right),
				\\
				R_{k}^{p}&=\log_{2}\left(1+\frac{\left|\mathbf{{h}}_{kk}^{H}\mathbf{{p}}_{p,k}\right|^{2}}{\sum_{j\in\mathcal{K},j\neq k}\left|\mathbf{{h}}_{kj}^{H}\mathbf{{p}}_{p,j}\right|^{2}+\sigma_{n,k}^2}\right).
			\end{aligned}
			$}
	\end{equation}
	The overall achievable rate of user-$k$ follows
	\begin{equation}
		R_{k,tot}=R_{k}^{c}+R_{k}^{p},
	\end{equation}
	where $R_{k}^{c}=\min_{i}\left\{R_{i,k}^{c}\mid i\in\mathcal{K}\right\}$.
	%%%
	
	\par 
	As discussed in \cite{chenxi2017brunotopology}, we could further split the message of each user into $N>2$ parts  with each part being decoded by a different group of users based
	on the specific instantaneous CSIT quality. By such means, the DoF of the MISO BC can be further boosted.
	%%%
	\subsubsection{Cooperative transmission} 
	Cooperative transmission (a.k.a. network MIMO/joint transmission) on the other hand is motivated by the MIMO BC. It requires both user data and CSI to be shared among cells.
	%%%
	Fig. \ref{fig:multiCell}(b) illustrates a $K$-cell network MIMO where each transmitter is equipped with $M$ antennas and all transmitters serve  $K$ users jointly. The $K$ transmitters are assumed to be connected with  a central processor via high speed and large bandwidth fronthaul links. The central processor has access to the messages of all $K$ users and to the CSI of all channels between the BSs and the users.
	%%%
	Therefore, cooperative transmission allows the transmitters to operate as a virtual joint transmitter and all downlink RSMA strategies illustrated in Section \ref{sec:RSMADL} can be applied for cooperative transmission.
	%%%
	The major differences compared with the single-cell MIMO BC lie in the transmit power constraint and the fronthaul capacity constraint.
	%%%
	Contrary to the single-cell MIMO BC where transmit signals are not subject to a sum-power constraint,  cooperative multi-cell networks require  per BS power constraints. 
	%%%
	If each transmitter is equipped with a single antenna, i.e., $M=1$, 
	the cooperative multi-cell network  is equivalent to a single-cell MIMO BC under a per antenna power constraint \cite{mao2018networkmimo}.
	%%%
	Moreover, contrary to the single-cell MIMO BC which does not include fronthaul capacity limitations,  cooperative multi-cell networks such as C-RAN or F-RAN have practical  fronthaul capacity constraints, which  influences  RSMA design \cite{alaa2020cranimperfectCSIT}. 
	%%%
	
	\section{Information-Theoretic Background and Milestones of RSMA}
	\label{sec:literatureReview}
	%%%
	RSMA has its roots in network information theory,
	which lays a solid foundation to the communication theory and emerging applications of RSMA. 
	%%%
	In this section, we provide a comprehensive survey of RSMA from an information theoretic perspective for both single-antenna  and multi-antenna networks, followed by a summary of the main milestones in the history of RSMA.  
	%%%
	\subsection{RSMA in Single-Antenna Networks}
	\label{sec:reviewSISO}
\begin{figure}[t!]
		\centering
		%	\hspace*{0.2cm}
		\begin{subfigure}[b]{0.4\textwidth}
			\centering
			\includegraphics[width=\textwidth]{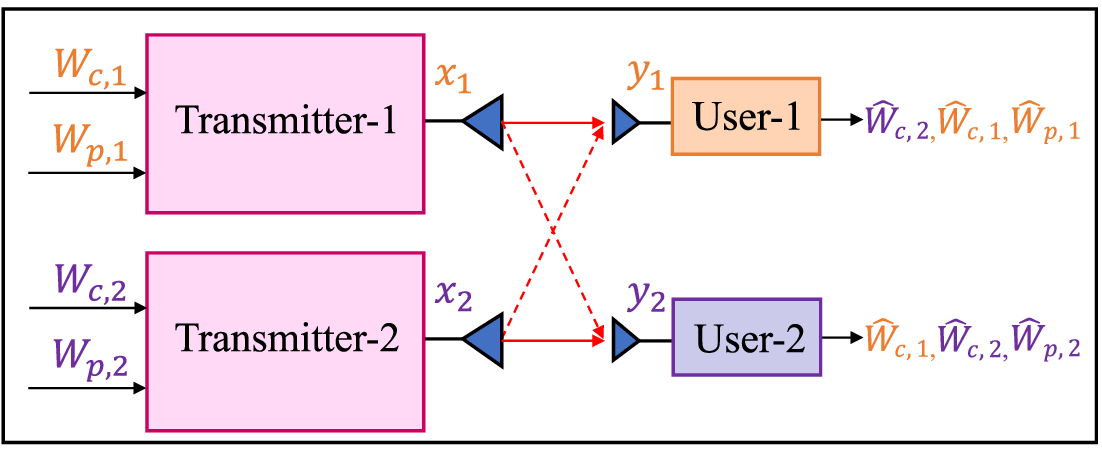}%
			\caption{RSMA for two-user SISO IC (HK scheme) \cite{TeHan1981}}
		\end{subfigure}%
		\\
		\vspace{0.2cm}
		\begin{subfigure}[b]{0.4\textwidth}
			\centering
			\includegraphics[width=\textwidth]{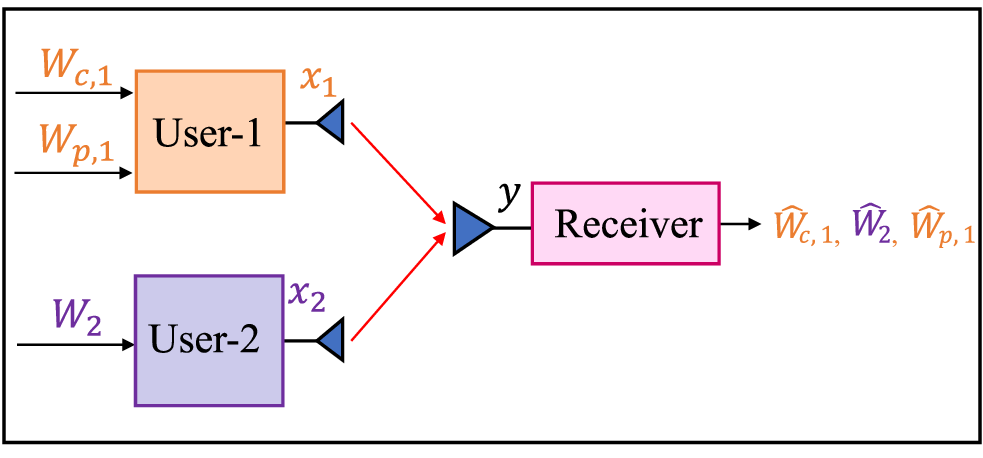}%
			\caption{RSMA for two-user SISO MAC \cite{Rimo1996}}
		\end{subfigure}%
		\caption{Two-user RSMA transmission framework for SISO IC and SISO MAC in information theory.}
		\label{fig:2UserRSMA}
		\vspace{-0.2cm}
	\end{figure}

	\labelsubseccounter{III-A}
	%%%
	The idea of RS was introduced by Carleial in \cite{carleial1978RS} for the two-user SISO IC where an inner bound on the capacity region based on RS and  successive cancellation decoding was proposed. This inner bound was further improved by Han and Kobayashi (HK) in  \cite{TeHan1981} via RS and simultaneous decoding. 
	%%%
	In \cite{Tse2008}, a special case of the HK scheme was further shown to achieve the capacity region of the two-user SISO IC within one bit.
	%%%
	An example of the HK scheme  is illustrated in Fig. \ref{fig:2UserRSMA}(a).
	%%%
	The  message $W_k$ of user-$k$  is split into  a common part $W_{c,k}$ and a private part $W_{p,k}$, which are independently encoded at the corresponding transmitter. Each user decodes the common part of the other user so as to cancel part of the interference and the remaining private part of the other user is treated as noise.
	%%%
	Fig. \ref{fig:2UserRSMA}(a) illustrates the case when user-1 decodes $W_{c,2}$, $W_{c,1}$, and $W_{p,1}$ while  user-2 decodes $W_{c,1}$, $W_{c,2}$, and $W_{p,2}$.
	%%%
	Such HK scheme enables flexible and powerful interference management by {partially decoding the interference and partially treating the remaining interference as noise}. This is the fundamental principle of RS on which all applications of RS in wireless networks are based. It allows RS to bridge the two extreme strategies of fully decoding the interference and fully treating the interference as noise.
	%%%
	Therefore, the HK scheme provides the basic concept of RS, and it can be seen as an RS strategy for the two-user SISO IC, which is a special case of the $K$-cell MISO IC illustrated in Fig. \ref{fig:multiCell}(a). 
	%%%
	%%%
	\begin{table*}[t!]
		\caption{Information-theoretic literature on RSMA}
		\label{tab:ITsurvey}
		%	\addtolength\tabcolsep{-2pt}
		\begin{tabular}{@{}|C{0.7cm}C{0.6cm}p{3.5cm}p{1.8cm}p{1.9cm}p{7.1cm}|@{}}
			\hline 	
			\textbf{Timeline}	&	\textbf{Ref.}     &  \textbf{Scenario}    & \textbf{CSI Condition}   &\textbf{Metric}  & \textbf{Main Discovery}    
			\\ \hline
			1978 &     \cite{carleial1978RS}   & $2$-cell SISO IC & Perfect CSIT  &Rate region &   Proposed the idea of RS for the $2$-cell SISO IC.
			\\  \hline
			1981 &     \cite{TeHan1981}   & $2$-cell SISO IC & Perfect CSIT  &Rate region &   Proposed the HK scheme for the $2$-cell SISO IC. 
			\\  \hline
			1996 &        \cite{Rimo1996}   & $K$-user SISO MAC   & Perfect CSIT  & Capacity region & Proposed uplink RSMA which achieves the capacity region of the $K$-user  Gaussian MAC without time sharing. 
			\\  \hline
			2007  &      \cite{Liang2007RBC}   & $2$-user SISO BC with cooperative relaying & Perfect CSIT    &Rate region &   Showed that RS achieves an inner bound on the capacity region of the two-user SISO BC with cooperative relaying.
			\\  \hline
			2008  &      \cite{Tse2008}   & $2$-cell SISO IC & Perfect CSIT    &Rate region &   Showed that a simplified HK scheme achieves rates within 1 bit/s/Hz of the capacity of the $2$-cell SISO (Gaussian) IC. 
			\\  \hline
			2009  &      \cite{Xu2009confidential}   & $2$-user SISO BC with a multicast message & Perfect CSIT    & Capacity region &   Showed that RS achieves the proposed inner bound on the capacity region of the two-user SISO BC with a multicast message.
			\\  \hline
			2009  &      \cite{Nair2009SISOBC}   & $3$-user SISO BC with multicast messages & Perfect CSIT    & Capacity region &   Showed that RS achieves the capacity region of the three-user SISO BC with two-degraded messages sets. 
			\\  \hline
			2013&        \cite{DoF2013SYang}  &  $2$-user underloaded MISO BC & Imperfect CSIT  &Sum-DoF &  Showed that RS achieves the optimum sum-DoF of the $2$-user underloaded MISO BC with imperfect CSIT. 
			\\  \hline
			2016 &        \cite{RS2016joudeh}   & $K$-user underloaded MISO BC & Imperfect CSIT    & Symmetric-DoF  &  Showed that RS achieves a higher symmetric-DoF  than SDMA  for the $K$-user underloaded MISO BC with imperfect CSIT.
			\\ \hline
			2016 &       \cite{enrico2016bruno}   & $K$-user overloaded MISO BC & Imperfect CSIT  &DoF region &  Showed that RS achieves the entire DoF region of the $K$-user overloaded MISO BC with heterogeneous CSIT qualities. 
			\\ \hline
			2016 &       \cite{RS2016hamdi}   & $K$-user underloaded MISO BC & Imperfect CSIT  &Sum-DoF &  Showed that RS achieves the optimum sum-DoF of the $K$-user underloaded MISO BC with imperfect CSIT. 
			\\ \hline
			2017 &       \cite{enrico2016bruno}   & $K$-user underloaded MISO BC & Imperfect CSIT   &DoF region &  Showed that RS achieves the entire DoF region of the $K$-user underloaded MISO BC with imperfect CSIT. \\ \hline
			2017 &       \cite{hamdi2017bruno}   & $K$-user MISO BC with overloaded multigroup multicast & Perfect CSIT  & Symmetric-DoF & Showed that RS achieves higher symmetric-DoF  than SDMA and multi-antenna NOMA with one user group for the $K$-user overloaded multigroup multicast with perfect CSIT. \\\hline
			2017 &         \cite{chenxi2017brunotopology}   & $K$-cell MISO IC  & Imperfect CSIT  &DoF region & Showed that RS achieves the best known DoF region of the $K$-cell MISO IC with imperfect CSIT. \\\hline
			2017 &         \cite{chenxi2017bruno}   & $2$-cell MIMO IC   & Imperfect CSIT  &DoF region & Showed that RS achieves the optimum DoF region of the $2$-cell  MIMO IC with imperfect CSIT under certain antenna-configurations and CSIT qualities. 
			\\\hline 
			2017 &  \cite{AG2017Gdof}   & $2$-user underloaded MISO BC & Imperfect CSIT  & GDoF & Showed that RS-assisted approach achieves the entire GDoF region of the $2$-user underloaded MISO BC with imperfect CSIT.  
			\\\hline 
			2018 &  \cite{SYang2018SPAWC}   &  $2$-user underloaded MISO BC & Perfect CSIT & Sum-rate &  Showed that RS achieves the sum capacity within a constant gap for the $2$-user MISO BC with perfect CSIT. 
			\\\hline 
			2018 &  \begin{tabular}[c]{@{}c@{}}\cite{yang2018itw}\\\cite{Zheng2021TIT}\end{tabular}   &  $2$-user underloaded MIMO BC & Perfect CSIT & Rate region & Showed that RS achieves the whole capacity region within a constant gap for the $2$-user MIMO BC with perfect CSIT.
			\\\hline 
			2020 &  \cite{romero2020rate}   & $K$-user SISO BC with multicast messages  & Perfect CSIT  & Rate region &  Showed that RS achieves a general inner bound for the discrete memoryless BC with an arbitrary number of users and an arbitrary set of message demands. 
			\\\hline 
			2021 &  \cite{weiyu2021FDrelay}   & $2$-user SISO BC with cooperative relaying    & Perfect CSIT  & Capacity region &  Showed that RS
			attains the capacity region within a constant gap for a scalar Gaussian full-duplex cellular network with device-to-device (D2D) messages.
			\\\hline 
			2021 &  \cite{longfei2020satellite}   & $K$-user MISO BC with underloaded and overloaded multigroup multicast  & Imperfect CSIT  & Symmetric-DoF &  Showed that RS achieves a symmetric-DoF  gain over SDMA  for the $K$-user underloaded and overloaded multigroup multicast channel with imperfect CSIT.
			\\\hline 
			2021 &  \cite{bruno2021MISONOMA}   & $K$-user underloaded and overloaded MISO BC & Perfect CSIT, \newline Imperfect CSIT & Sum-DoF, \newline Symmetric-DoF &  Showed that RS achieves both a sum-DoF and a symmetric-DoF gain over SDMA and multi-antenna NOMA  in both the overloaded and underloaded MISO BC with perfect and imperfect CSIT. 
			\\ \hline 
		\end{tabular}
	\end{table*}
	%%%
	\begin{figure*}[!t]
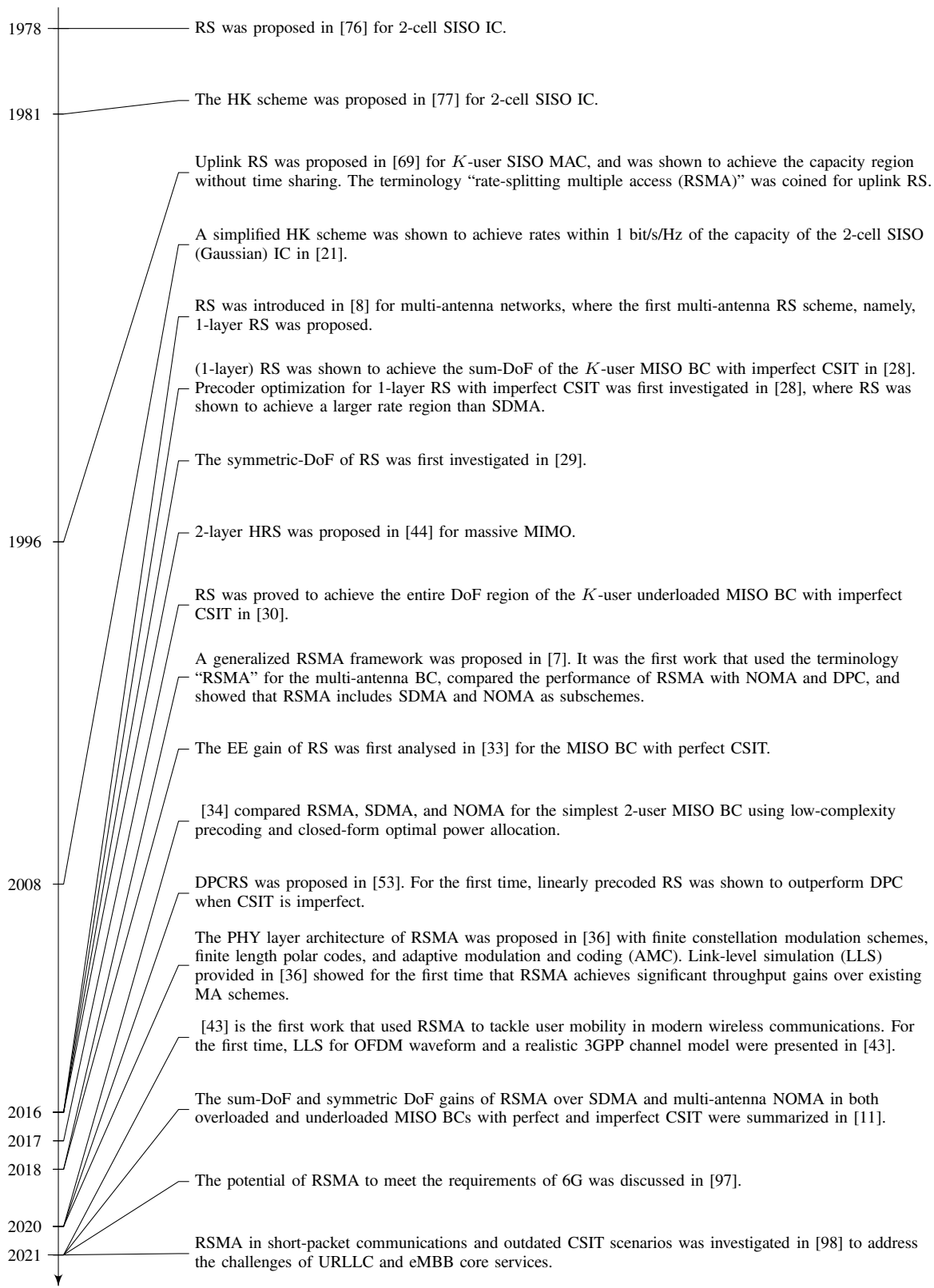

		\centering
		\begin{footnotesize}
			\begin{timeline}{1978}{2021}{2cm}{2cm}{1.4\columnwidth}{21.3cm}
				\entry{1978}{RS was  proposed in \cite{carleial1978RS} for $2$-cell SISO IC.}
				\entry{1981}{The HK scheme was proposed in \cite{TeHan1981} for $2$-cell SISO IC.}
				\entry{1996}{Uplink RS was proposed in  \cite{Rimo1996} for $K$-user SISO MAC, and was shown to achieve the capacity region without time sharing. The terminology ``rate-splitting multiple access (RSMA)" was coined for uplink RS.}
				\entry{2008}{A simplified HK scheme was shown to achieve rates within 1 bit/s/Hz of the capacity of the $2$-cell SISO (Gaussian) IC in \cite{Tse2008}.}
				\entry{2016}{RS was introduced in \cite{RSintro16bruno} for multi-antenna networks, where the first multi-antenna RS scheme, namely, 1-layer RS was proposed.}
				\plainentry{2016}{(1-layer) RS was shown to achieve the sum-DoF of the $K$-user MISO BC with imperfect CSIT in \cite{RS2016hamdi}.
					Precoder optimization for 1-layer RS with imperfect CSIT was first investigated in \cite{RS2016hamdi}, where RS was shown to achieve a larger rate region than SDMA. }
				\plainentry{2016}{The symmetric-DoF of RS was first investigated in \cite{RS2016joudeh}. }
				\plainentry{2016}{2-layer HRS was proposed in \cite{Minbo2016MassiveMIMO} for massive MIMO.}
				\entry{2017}{RS was proved to achieve the entire DoF region of the $K$-user underloaded MISO BC with imperfect CSIT in \cite{enrico2016bruno}.}
				\entry{2018}{A generalized RSMA framework was proposed in \cite{mao2017rate}.  It was  the first work that used the terminology ``RSMA" for the multi-antenna BC, compared the performance of RSMA with NOMA and DPC, and showed that RSMA includes SDMA and NOMA as subschemes.}
				\plainentry{2018}{The EE gain of RS was first analysed in \cite{mao2018EE} for the MISO BC with perfect CSIT.}
				\entry{2020}{\cite{bruno2019wcl} compared RSMA, SDMA, and NOMA for the simplest 2-user MISO BC using low-complexity precoding and closed-form optimal power allocation.}
				\plainentry{2020}{DPCRS was proposed in \cite{mao2019beyondDPC}. For the first time, linearly precoded RS was shown to outperform DPC when CSIT is imperfect.}
				\plainentry{2020}{The PHY layer architecture of RSMA was proposed in \cite{Onur2020LLS} with finite constellation modulation schemes, finite length polar codes, and adaptive modulation and coding (AMC). Link-level simulation (LLS) provided in \cite{Onur2020LLS} showed for the first time that RSMA achieves significant throughput gains over existing MA schemes.}
				\entry{2021}{\cite{onur2021mobility} is the first work that used RSMA to tackle user mobility in modern wireless communications. For the first time, LLS for OFDM waveform and a realistic 3GPP channel model were presented in \cite{onur2021mobility}. }
				\plainentry{2021}{The sum-DoF and symmetric DoF gains of RSMA over SDMA and multi-antenna NOMA  in both overloaded and underloaded MISO BCs with perfect and imperfect CSIT were summarized in \cite{bruno2021MISONOMA}.}
				\plainentry{2021}{The potential of RSMA to meet the requirements of 6G was discussed in \cite{onur2020sixG}. }
				\plainentry{2021}{RSMA in short-packet communications and outdated CSIT scenarios was investigated in \cite{onur2021sixG} to address the challenges of URLLC and eMBB core services.}
			\end{timeline}
		\end{footnotesize}
		\caption{Milestones in the history of RSMA.}
		\label{tab:milesone}
	\end{figure*}
	
	%%%
	\par
	The terminology ``rate-splitting multiple access (RSMA)" was first introduced in \cite{Rimo1996} for the SISO MAC.
	% where multiple users are simultaneously transmitting their messages to the same base station (BS). 
	%%%
	Relying on splitting user messages at the transmitters (i.e., users) and successive cancellation  at the receiver (i.e., base station--BS), RSMA achieves all points on the boundary of the SISO MAC capacity region  without  the need for time sharing and hence synchronization among users. 
	%%%
	This contrasts with conventional capacity-achieving schemes for the SISO BC (i.e., DPC and SC--SIC), which need time sharing to attain part of the capacity region boundary.
	%%%
	% Specifically in the $K$-user SISO MAC, by allowing $K-1$ users to split each of their messages into two sub-messages, the original $K$-user SISO MAC is transformed into a virtual $2K-1$-user MAC where each any rate tuple on the boundary of the capacity region of the original $K$-user MAC can be achieved via successive single-user decoding of $2K-1$ virtual users with  appropriate ordering and power allocation.
	%%%
	RSMA was further studied in \cite{Medard2004Aloha}  for the two-user Gaussian MAC where  each user independently transmits  data packets with predefined collusion probability. It has been shown that the capacity region of such systems is the same as the capacity region of MA systems where the users continuously transmit. 
	%%%
	In \cite{petar2007cognitiveRS,petar2011RS}, RSMA was used for opportunistic interference cancellation in cognitive networks and was shown to achieve all points on the boundary of the rate region without 
	time sharing between primary and secondary users.
	%%%
	The  optimal decoding order  and power allocation of uplink RSMA when the BS is equipped with multiple antennas was further investigated in \cite{zhaohui2020ULRS}. 
	%%%
	Fig. \ref{fig:2UserRSMA}(b) illustrates a two-user RSMA example for the SISO MAC, which is a special case of the $K$-user uplink RSMA in Fig. \ref{fig:uplinkRS}.
	%%%
	User-1 splits its message $W_1$ into $W_{c,1}$ and $W_{p,1}$ and the receiver (i.e., the BS) decodes $W_{c,1}, W_{p,1}$, and $W_2$ via successive cancellation. With appropriate decoding order and power allocation, the entire capacity region of the two-user SISO MAC can be attained.
	%%%
	Though uplink RSMA  has a different structure than the HK scheme, they both lean on the concept of RS to split user messages at the transmitters,  and partially decode interference and partially treat interference as noise at the receiver(s). 
	%%%
	However, the motivations for the two schemes are distinct. RS for the MAC was proposed to avoid time sharing while RS for the  IC was advocated to enhance the rate region over existing schemes that fully treat interference  as noise or fully decode interference.
	%%%
	\par
	Besides the SISO IC and MAC, RS has been studied for the SISO BC with degraded message sets (i.e., with one or multiple multicast messages intended for multiple users)  \cite{Liang2007RBC, Nair2009SISOBC, EL2011networkIT, romero2020rate}. In this context, RS has been shown to achieve the general inner bound on the capacity region  of the discrete memoryless BC with an arbitrary number of users and arbitrary sets of message demands \cite{romero2020rate}. 
	%%%
	
	\subsection{RSMA in Multi-Antenna Networks}
	\label{sec:reviewMIMO}
	\labelsubseccounter{III-B}
	\par
	The success of multi-antenna networks has shifted the research focus of RSMA from single-antenna networks to multi-antenna networks. 
	%%%
	The existing  information-theoretic  literature on RSMA in multi-antenna networks mainly has studied four metrics, namely, the sum-DoF and DoF region, max-min fairness/symmetric-DoF, generalized DoF (GDoF), and capacity region as elaborated in the following.
	%%%

	%%%
	\par
	\textit{Sum-DoF and DoF region:}
	As the capacity region of  the $K$-user MISO BC with partial CSIT remains an open problem hitherto, the research has focused on identifying the corresponding DoF and GDoF region (within a constant gap). 
	%%%
	RS was first studied for the two-user MISO BC in \cite{DoF2013SYang} where RS was shown to achieve the optimum sum-DoF of the two-user underloaded MISO BC with partial CSIT.
	%%%
	% The result is then extended to the $K$-user case in \cite{RS2016hamdi} where the strategy based on linearly precoded RS and SIC receivers achieves the optimum sum-DoF of the $K$-user underloaded MISO BC with imperfect CSIT.
	%%%
	The authors of \cite{AG2015} further discovered a novel sum-DoF upper bound based on aligned image sets for the $K$-user underloaded MISO BC with partial CSIT. Surprisingly, the sum-DoF achieved by RS was shown in \cite{RS2016hamdi} to match the DoF upper bound obtained in \cite{AG2015}.
	%%%
	This sharply contrasts with the sum-DoF of SDMA and NOMA for the MISO BC with partial CSIT, which are clearly suboptimal \cite{bruno2021MISONOMA}. 
	%%%
	The authors of \cite{enrico2017bruno} then proved that the entire DoF region of the underloaded MISO BC with partial CSIT is achieved by RS. 
	%%%
	Besides the conventional MISO BC, the DoF gain of RS has also been brought to light in more complicated underloaded networks with partial CSIT considering multiple transmitters \cite{chenxi2017brunotopology}  and multiple antennas at each receiver (MIMO BC and MIMO IC) \cite{chenxi2017bruno}.
	%%%
	For the MIMO BC, RS was again shown to achieve the optimal DoF region in the general case of  asymmetric numbers of receive antennas (i.e., different numbers of receive antennas deployed at different users), when the DoF region was achieved by RS matching the DoF-region upper bound \cite{chenxi2017bruno,Davoodi2021DoF}.
	%%%
	For the overloaded MISO BC, where users have heterogeneous CSIT qualities (namely, no-CSIT users with statistical CSIT only and partial-CSIT users with imperfect instantaneous CSIT), it was shown in \cite{enrico2016bruno} that the entire optimal DoF region could be achieved by adopting SC to transmit the degraded symbols for the no-CSIT users and linearly precoded RS symbols for the partial-CSIT users.
	
	\par
	\textit{Max-min fairness/symmetric-DoF:}
	Taking user fairness into consideration, the authors of  \cite{RS2016joudeh} derived the optimum max-min fairness (MMF)-DoF (a.k.a. symmetric DoF, i.e., the DoF that can be simultaneously achieved by all users) of RS   for the underloaded MISO BC with imperfect CSIT, which outperforms that of SDMA.
	%%%
	For the $K$-user overloaded multigroup multicast channel, the achievable MMF-DoF of RS were characterized for perfect CSIT in \cite{hamdi2017bruno} and for imperfect CSIT in \cite{longfei2020satellite}. Explicit symmetric DoF gains over SDMA and NOMA are obtained in both settings.
	%%%
	In \cite{bruno2021MISONOMA}, the sum-DoF and MMF-DoF of RS, SDMA, and multi-antenna NOMA for the underloaded and overloaded MISO BC with perfect and imperfect CSIT are summarized comprehensively. This shows that RS fully exploits  the multi-antenna DoF and the benefits of SIC receivers, and  outperforms SDMA and NOMA in both underloaded and overloaded regimes relying on perfect and imperfect CSIT.
	
	%%%
	\par 
	{\textit{GDoF:}}
	Conventional DoF metrics reflect how fast the user rates increase with SNR in the high SNR regime. However, they are not able to capture the diversity of the channel strengths among the users. The GDoF introduced in \cite{Tse2008} overcome this limitation of DoF. The authors of \cite{AG2016Gdof,AG2017Gdof} characterized the entire GDoF region of the two-user underloaded MISO BC with imperfect CSIT, and the GDoF optimal achievable scheme is built upon RS.
	
	%%%
	\par 
	{\textit{Capacity region:}}
	Besides the DoF metrics,  RS has been shown to achieve the sum capacity \cite{SYang2018SPAWC} and further the entire capacity region \cite{yang2018itw, Zheng2021TIT} within a constant gap for the two-user MIMO BC with perfect CSIT. 
	%%%
	The reviewed information-theoretic works on RSMA are summarized in Table \ref{tab:ITsurvey}.

	\subsection{Milestones of RSMA}
	\label{sec:milestone}
	\labelsubseccounter{III-C}
	Looking back at the history of RSMA, there are a number of significant milestones paving the way  towards a novel PHY layer design for future generations of wireless systems based on RSMA, which are summarized in Fig. \ref{tab:milesone}. As time goes on, RSMA will continue to evolve and motivate research for  next generation multiple access aiming at better performance in terms of SE, EE, throughput, latency, reliability, robustness, coverage, cost, among others.
	%%%
	
	\begin{table*}[]
		\caption{Communication-theoretic literature on RSMA for the multi-antenna BC.}
		\label{tab:commuSurveyMISOBC}
		%	\addtolength\tabcolsep{-2pt}
		\begin{tabular}{@{}|C{0.6cm}|C{0.6cm}|C{1.1cm}|C{1.25cm}|C{1.1cm}|C{1.25cm}|C{1.0cm}|C{1.25cm}|C{0.7cm}|C{0.7cm}|C{0.7cm}|C{0.9cm}|C{0.9cm}|@{}}
			\hline
			\multicolumn{1}{|c|}{\multirow{2}{*}{\textbf{Year}}} &
			\multicolumn{1}{c|}{\multirow{2}{*}{\textbf{Ref.}}} &
			\multicolumn{2}{c|}{\textbf{RSMA Scheme}} &
			\multicolumn{2}{c|}{\textbf{Precoding Scheme}} &
			\multicolumn{2}{c|}{\textbf{CSIT Condition}} &
			\multicolumn{5}{c|}{\textbf{KPIs}} \\ \cline{3-13} 
			\multicolumn{1}{|c|}{} &
			\multicolumn{1}{c|}{} &
			\multicolumn{1}{c|}{\textbf{1-layer}} &
			\multicolumn{1}{c|}{\textbf{multi-layer}} &
			\multicolumn{1}{c|}{\textbf{Linear}} &
			\multicolumn{1}{c|}{\textbf{Non-linear}} &
			\multicolumn{1}{c|}{\textbf{Perfect}} &
			\multicolumn{1}{c|}{\textbf{Imperfect}} &
			\multicolumn{1}{c|}{\textbf{WSR}} &
			\multicolumn{1}{c|}{\textbf{MMF}} &
			\multicolumn{1}{c|}{\textbf{EE}}  &
			\multicolumn{1}{c|}{\textbf{Mobility}} &
			\multicolumn{1}{c|}{\textbf{Latency}} \\ \hline
			2015 &
			\cite{RS2015bruno} &
			$\surd$ &
			&
			$\triangle$ &
			&
			&
			$\bigodot$ &
			$\surd$ &
			&
			&
			&
			\\ \hline
			\multirow{3}{*}{2016} &
			\cite{RS2016hamdi} &
			$\surd$ &
			&
			$\blacktriangle$ &
			&
			& $\bigoplus$ &
			$\surd$ &
			&
			&
			&
			\\ \cline{2-13} 
			&
			\cite{RS2016joudeh} &
			$\surd$ &
			&
			$\blacktriangle$ &
			&
			&
			$\bigodot$ &
			&
			$\surd$ &
			&
			&
			\\ \cline{2-13} 
			&
			\cite{enrico2016bruno} &
			$\surd$ &
			&
			$\triangle$ &
			&
			$\surd$ &
			$\bigoplus$ &
			$\surd$ &
			&
			&
			&
			\\ \hline
			\multirow{5}{*}{2018} &
			\cite{Lu2018MMSERS} &
			$\surd$ &
			&
			$\triangle$ &
			&
			&
			$\bigodot$ &
			$\surd$ &
			&
			&
			&
			\\ \cline{2-13} 
			&
			\cite{mao2017rate} &
			$\surd$ &
			$\surd$ &
			$\blacktriangle$ &
			&
			$\surd$ &
			$\bigoplus$ &
			$\surd$ &
			&
			&
			&
			\\ \cline{2-13} 
			&
			\cite{Medra2018SPAWC} &
			$\surd$ &
			&
			$\blacktriangle$ &
			&
			&
			$\bigodot$ &
			&
			$\surd$ &
			&
			&
			\\ \cline{2-13} 
			&
			\cite{mao2018EE} &
			$\surd$ &
			&
			$\blacktriangle$ &
			&
			$\surd$ &
			&
			&
			&
			$\surd$ &
			&
			\\ \cline{2-13} 
			&
			\cite{Flores2018ISWCS} &
			$\surd$ &
			&
			&
			$\triangle$ &
			&
			$\bigodot$ &
			$\surd$ &
			&
			&
			&
			\\ \hline
			\multirow{7}{*}{2020} &
			\cite{bruno2019wcl} &
			$\surd$ &
			&
			$\triangle$ &
			&
			$\surd$ &
			&
			$\surd$ &
			&
			&
			&
			\\ \cline{2-13} 
			&
			\cite{mao2019beyondDPC} &
			$\surd$ &
			$\surd$ &
			&
			$\blacktriangle$ &
			&
			$\bigoplus$ &
			$\surd$ &
			&
			&
			&
			\\ \cline{2-13} 
			&
			\cite{hongzhi2020LLS} &
			$\surd$   &
			&
			$\blacktriangle$ &
			&
			$\surd$ &
			&
			&
			$\surd$ &
			&
			&
			\\ \cline{2-13} 
			&
			\cite{Onur2020LLS} &
			$\surd$  &
			&
			$\blacktriangle$ &
			&
			&
			$\bigoplus$ &
			$\surd$ &
			&
			&
			&
			\\ \cline{2-13} 
			&
			\cite{bruno2020MUMIMO} &
			$\surd$ &
			&
			&
			&
			&
			$\bigodot$ &
			$\surd$ &
			&
			&
			&
			\\ \cline{2-13} 
			&
			\cite{Gui2020EESEtradeoff} &
			$\surd$ &
			&
			$\blacktriangle$ &
			&
			$\surd$ &
			&
			$\surd$ &
			&
			$\surd$ &
			&
			\\ \cline{2-13} 
			&
			\cite{Zheng2020JSAC} &
			$\surd$   & $\surd$
			&
			$\blacktriangle$ &
			&
			&
			$\bigcirc$ &
			$\surd$ &
			&
			&
			&
			\\ \hline
			\multirow{8}{*}{2021} &
			\cite{mao2021IoT} &
			$\surd$ &
			&
			$\blacktriangle$ &
			&
			&
			$\bigoplus$ &
			$\surd$ &
			&
			&
			&
			\\ \cline{2-13} 
			&
			\cite{bho2021globalEE} &
			$\surd$ &
			&
			$\blacktriangle$ &
			&
			$\surd$ &
			&
			&
			&
			$\surd$ &
			&
			\\ \cline{2-13} 
			&
			\cite{Andre2021THP} &
			$\surd$ &
			&
			&
			$\triangle$ &
			&
			$\bigoplus$ &
			$\surd$ &
			&
			&
			&
			\\ \cline{2-13} 
			&
			\cite{wonjae2021imperfectCSIR} &
			$\surd$ &
			&
			$\blacktriangle$ &
			&
			&
			$\bigodot$ &
			$\surd$ &
			&
			&
			&
			\\ \cline{2-13} 
			&
			\cite{longfei2021statisticalCSIT} &
			$\surd$ &
			&
			$\blacktriangle$ &
			&
			$\surd$ &
			$\bigcirc$ &
			&
			$\surd$ &
			&
			&
			\\ \cline{2-13} 
			&
			\cite{onur2021mobility} &
			$\surd$  &
			&
			$\blacktriangle$ &
			&
			&
			$\bigoplus$ &
			$\surd$ &
			&
			&
			$\surd$ &   $\surd$
			\\ \cline{2-13} 
			&
			\cite{yunnuo2021FBL} &
			$\surd$ &
			&
			$\blacktriangle$ &
			&
			$\surd$ &
			&
			$\surd$ &
			&
			&
			&
			$\surd$ \\ \cline{2-13} 
			&
			\cite{anup2021MIMO} &
			$\surd$ &
			&
			$\blacktriangle$ &
			&
			&
			$\bigoplus$  &
			$\surd$&
			&
			&
			&
			\\ \hline
		\end{tabular}
		\vspace{0.1cm}
		
		Notations:  $\blacktriangle$: Optimized precoding;  $\triangle$: Low-complex precoding;
		$\bigoplus$: Partial instantaneous CSIT with unbounded channel estimation error; $\bigodot$: Partial instantaneous CSIT with bounded channel estimation error; $\bigcirc$: Statistical CSIT.
	\end{table*}

	\section{Communication-Theoretic Background of RSMA}
	\label{sec:resourceAllocation}
	%%%
	In recent years, the superiority of RSMA, unveiled by the information-theoretic results discussed in Section \ref{sec:literatureReview}, has motivated the study of RSMA from the communication-theoretic perspective in the moderate SNR regime.
	%%%
	The main research focus in this regime is on the resource allocation for RSMA including 
	precoder design, power control, common rate allocation, user scheduling, and subcarrier allocation, which are crucial elements to reap the benefits of RSMA in modern wireless communication networks.
	%%%
	The communication-theoretic literature on RSMA for the conventional multi-antenna BC is summarized in Table \ref{tab:commuSurveyMISOBC}.
	%%%
	As most of the existing works focus on a single carrier (even if multiple carriers are presented), in this section, we first elaborate the resource allocation for single-carrier RSMA, and then extend to scenarios with multiple carriers.
	\subsection{Resource Allocation for Single-Carrier RSMA}
	\label{sec:SC-RA}
	\labelsubseccounter{V-A}
	%%%
	Resource allocation policies for single-carrier RSMA mainly consider precoder design, power control, and common rate allocation, which all influence the performance of RSMA, especially in the finite SNR regime. 
	%%%
	There are two lines of research as far as RSMA resource allocation is concerned, namely, joint resource optimization  and low-complexity resource allocation. 
	%%%
	The former aims to unveil the maximum achievable performance of RSMA by jointly optimizing the precoders, powers, and common rate, while the latter tries to achieve a favorable trade-off between performance and complexity for practical implementation by designing low-complexity resource allocation schemes. 
	%%%
	Both lines of research are recapped in this subsection.
	%%%
	
	\subsubsection{Joint Resource Optimization}
	\label{sec:precoderOpt}
	The availability of CSI at the transmitter plays a significant role for joint optimization of precoders, power, and common rate allocation. If instantaneous CSIT is available, the aforementioned wireless resources can be optimized for each instantaneous channel realization and the instantaneous rate is guaranteed to be achievable. 
	%%%
	However, such method is not applicable when the instantaneous CSI is imperfect or not available at all at the transmitter.
	%%%
	Two methods have been proposed for single-carrier RSMA resource allocation with imperfect/statistical CSIT, namely, 
	long-term resource optimization \cite{RS2016hamdi} and worst-case resource optimization \cite{RS2016joudeh}.
	%%%
	The former averages out the impact of CSIT estimation errors by designing precoders (including power allocation) and common rate allocation based on the ergodic performance over a long sequence of fading states while the latter considers bounded CSIT estimation errors and the wireless resources are designed to guarantee the performance for all possible channels (a.k.a. the worst-case performance) in the corresponding uncertainty regions. 
	%%%
	In this subsection, the RSMA  resources optimization problems for both perfect and imperfect CSIT are formulated followed by a review of optimization algorithms adopted to solve the corresponding problems. 
	%%%
	\par
	\underline{\textit{Perfect CSIT}}:
	The joint precoder (including power allocation) and common rate optimization for RSMA has been widely studied in perfect CSIT for different design objectives, such as maximizing the WSR \cite{ mao2017rate, yunnuo2021FBL, mao2018networkmimo, Ala2019IEEEAccess, Alaa2019uavCRAN, mao2019swipt, jian2019crs,lihua2020multicarrier,alaa2020gRS,xu2021rate,jaafar2020UAV,jaafar2020UAV2,siyu2020vlc,naser2020vlc, Liping2020secrecyCRS, onur2021DAC,Park2021WCNC},  maximizing the minimum rate among users (max-min rate) \cite{hamdi2017bruno,cran2019wcl,mao2019maxmin,yalcin2020RSmultigroup,hongzhi2021RSLDPC}, maximizing the EE \cite{mao2018EE, bho2021globalEE, AP2017cognitive, Tervo2018SPAWC, UAVRS2019ICC,alaa2020EECRAN,Jia2020SEEEtradeoff,zhaohui2020IRS}, and minimizing the sum-power consumption \cite{alaa2020powerMini,Camana2020swiptRS,alaa2020IRScran}. 
	%%%
	For different design purposes, though the objective functions are different, the supplementary constraints introduced by RSMA are identical. Therefore, a general problem   that accommodates different optimization criteria can be formulated.
	%%%
	For the sake of simplicity,  we summarize the resource optimization for 1-layer RS for perfect CSIT. The formulated problem can be easily extended to other RSMA schemes based on the system model specified in Section \ref{sec:principle}.
	%%%
	\par
	For a given transmit power constraint $P$ and QoS rate constraint $R_k^{th}$ of each user-$k$, the joint precoder  $\mathbf{{P}}=[\mathbf{{p}}_{c}, \mathbf{{p}}_{1},\ldots, \mathbf{{p}}_{K}]$ and  common rate allocation $\mathbf{c}=[C_1,\ldots,C_K]$ optimization problem for maximization of a certain utility function $U\left ( \mathbf{{P}},\mathbf{c} \right )$ for 1-layer RS can be formulated as follows
	\vspace{0.5mm}
	\begin{center}
		\fbox{
			\begin{minipage}{0.8\linewidth}
				\begin{subequations}
					\label{eq:perfectCSIopt}
					\begin{align}
						\mathcal{P}_1:  \quad
						&\max_{\mathbf{{P}}, \mathbf{c}} \,\,U\left ( \mathbf{{P}},\mathbf{c} \right )\\
						\mbox{s.t.}\,\,
						& \sum_{k=1}^K C_k\leq \min \{R_{c,1},\ldots,R_{c,K}\} \label{const:1}\\
						&	\text{tr}(\mathbf{P}\mathbf{P}^{H})\leq P \label{const:2}\\
						&   {R}_{k} + {C}_{k}\geq R_k^{th}, \forall k\in\mathcal{K} \label{const:3}\\
						& C_k\geq 0, \forall k\in\mathcal{K} \label{const:4}
					\end{align}
				\end{subequations}
			\end{minipage}
		}
	\end{center}
	\vspace{0.6mm}
	\vspace{0.4mm}
	%%%
	where $R_k$ and $R_{c,k}$, $k\in\mathcal{K}$ are functions of $\mathbf{P}$ as specified in (\ref{eq:rate1RS}). Depending on the design objective, the utility function is given by
	\begin{equation}
		\begin{aligned}
			U\left ( \mathbf{{P}},\mathbf{c} \right )=&\begin{cases}
				\sum_{k=1}^Ku_{k}({R}_{k} + {C}_{k}), & \textrm{WSR}\\ 
				\underset{k \in \mathcal{K}}{\min}
				\ ({R}_{k} + {C}_{k}),& \textrm{MMF-rate}\\ 
				\frac{\sum_{k=1}^K({R}_{k} + {C}_{k})}{	\frac{1}{\eta}\text{tr}(\mathbf{P}\mathbf{P}^{H})+P_{\textrm{cir}}}, & \textrm{EE}\\ 
				-\text{tr}(\mathbf{P}\mathbf{P}^{H}),& \textrm{sum-power}.
			\end{cases}    \nonumber
		\end{aligned}
	\end{equation}
	In $\mathcal{P}_1$, $u_k$ for WSR optimization is the weight allocated to user-$k$. For EE maximization,  $\eta\in(0,1]$ and $P_{\textrm{cir}}$ are respectively the power amplifier efficiency and the circuit power consumption. 
	%%%
	Constraint (\ref{const:1}) guarantees the decodability of the common stream at all users. 
	%%%
	Constraint (\ref{const:2}) is the sum-power constraint at the transmitter, which can be omitted if the sum-power consumption is considered as the objective function.
	%%%
	Constraint (\ref{const:3}) is the QoS constraint, guaranteeing that the rate of each user is no less than a certain threshold. If $R_k^{th}=0$, there is no QoS rate constraint for user-$k$. This is commonly used when exploring the largest achievable rate region of 1-layer RS.
	%%%
	Constraint (\ref{const:4}) guarantees that the rate of the common stream allocated to each user is non-negative. 
	%%%
	\par By turning off the rate allocation to the common stream, i.e., $C_k=0, \forall k\in\mathcal{K}$, $\mathcal{P}_1$ reduces to the problem of MU--LP design for SDMA.
	%%%
	This reveals that RSMA is more general than SDMA and enlarges the optimization space. 
	%%%
	Optimization algorithms adopted in existing works to solve $\mathcal{P}_1$ for different utility functions are discussed in Section \ref{sec:optAlgo} and corresponding numerical results obtained by solving $\mathcal{P}_1$ are illustrated  in part in Section \ref{sec:performanceCompare}. 
	%%%
	\par 
	\underline{\textit{Imperfect CSIT}}:
	Resource optimization for RSMA has also been studied extensively for imperfect  CSIT with the objective of maximizing the WSR \cite{RS2016hamdi, mao2019beyondDPC, Zheng2020JSAC, mao2021IoT, wonjae2021imperfectCSIR , anup2021MIMO,mao2019TCOM,mao2020DPCNOUM,alaa2020cranimperfectCSIT,onur2021jamming,rafael2021radarsensing}, maximizing the minimum rate among users \cite{ RS2016joudeh, longfei2021statisticalCSIT, fuhao2020secrecyRS,longfei2020satellite,si2021imperfectSatellite}, maximizing the EE \cite{Lin2021Cognitive}, and minimizing the sum-power consumption \cite{Medra2018SPAWC, RSswiptIC2019CL}. 
	%%%
	For imperfect CSIT, the BS only has the knowledge of channel estimate $\widehat{\mathbf{h}}_{k}$ for each instantaneous CSI vector $\mathbf{h}_{k}$ of user-$k$. 
	%%%
	A typical imperfect CSIT model is given as follows
	\begin{equation}
		\label{eq:imperfectCSIT}
		\mathbf{h}_k=\widehat{\mathbf{h}}_k+\widetilde{\mathbf{h}}_k,
	\end{equation}
	where $\widetilde{\mathbf{h}}_{k}$  is the CSIT  estimation error.
	%%%
	Different sources of CSIT impairment would lead to different models for $\widehat{\mathbf{h}}_{k}$ and  $\widetilde{\mathbf{h}}_{k}$. 
	%%%
	For example, when the CSIT imperfection results from user mobility and delay in CSI feedback, it is common to model  $\widehat{\mathbf{h}}_k[t]$ at time instance $t$ as the exact channel  $\mathbf{h}_k[t-1]$ at time instance $t-1$ while  
	$\mathbb{E}\{\widehat{\mathbf{h}}_k\widehat{\mathbf{h}}_k^H\}=\epsilon^2\mathbf{I}$, $\mathbb{E}\{\mathbf{h}_k\mathbf{h}_k^H\}=\mathbf{I}$,  and $\mathbb{E}\{\widetilde{\mathbf{h}}_k\widetilde{\mathbf{h}}_k^H\}=(1-\epsilon^2)\mathbf{I}$. 
	%%%
	Here, $\epsilon$ is the time correlation coefficient obeying the Jakes' model \cite{onur2021mobility}.  
	%%%
	Depending on the strength of $\widetilde{\mathbf{h}}_{k}$,  the existing robust precoding designs for RSMA  can be generally classified into worst-case resource optimization and long-term resource optimization.
	%%%
	Specifically, the former assumes that $\widetilde{\mathbf{h}}_{k}$ is bounded, while the latter typically assumes that $\widetilde{\mathbf{h}}_{k}$ is unbounded. The resource optimization problems resulting for both cases are discussed in the following.
	%%%
	\par 
	\textit{Worst-case resource optimization:}
	Worst-case resource optimization has been investigated in \cite{RS2016joudeh,fuhao2020secrecyRS, Lin2021Cognitive,RSswiptIC2019CL}. For each instantaneous CSI vector $\mathbf{h}_{k}$ of user-$k$, the CSIT  estimation error $\widetilde{\mathbf{h}}_{k}$ is assumed to be bounded by an origin-centered sphere with radius $\delta_k$. The instantaneous CSI $\mathbf{h}_{k}$ of each user-$k$ is therefore confined within an uncertainty region \cite{RS2016joudeh}:
	\begin{equation}
		\label{eq:boundedCSI}
		\mathbb{H}_{k} = \left\{ \mathbf{h}_{k} \mid \mathbf{h}_{k} = \widehat{\mathbf{h}}_{k} +  \widetilde{\mathbf{h}}_{k},
		\left \| \widetilde{\mathbf{h}}_{k} \right \| \leq \delta_{k}  \right\}.
	\end{equation}
	This is a typical imperfect CSIT model for quantized feedback where the properties of the quantization codebook are used to bound the CSIT error. 
	%%%
	The bounded imperfect CSIT leads to a bounded uncertainty of the users' achievable rates. To guarantee successful decoding at the users for all possible channels within the
	uncertainty region, robust optimization is carried out where the resources are allocated with respect to the worst-case achievable rates defined as follows
	\begin{equation}
		\label{eq:worstCaseRate}
		\widehat{R}_{\mathrm{c},k} =  \min_{\mathbf{h}_{k} \in \mathbb{H}_{k}} R_{\mathrm{c},k}(\mathbf{h}_{k})
		\quad \text{and} \quad
		\widehat{R}_{k} =  \min_{\mathbf{h}_{k} \in \mathbb{H}_{k}} R_{k}(\mathbf{h}_{k}).
	\end{equation}
	$R_{\mathrm{c},k}(\mathbf{h}_{k})$ and $R_{k}(\mathbf{h}_{k})$ are the instantaneous rates of user-$k$ based on the instantaneous CSIT, as specified in (\ref{eq:rate1RS}). The worst-case common rate is therefore defined as \mbox{$\widehat{R}_{\mathrm{c}}=\min\{\widehat{R}_{\mathrm{c},1},\ldots, \widehat{R}_{\mathrm{c},K}\}$}.
	%%%
	Denote the portion of the worst-case common rate allocated to user-$k$ as $\widehat{C}_{k}$, such that $\sum_{k=1}^K\widehat{C}_k=\widehat{R}_{\mathrm{c}}$, then the worst-case achievable rate of each user is $\widehat{R}_{k,tot}=\widehat{R}_k+\widehat{C}_k$.
	%%%
	The  resource optimization problem for imperfect CSIT is formulated in problem $\mathcal{P}_2$ as follows
	\vspace{0.5mm}
	\begin{center}
		\fbox{
			\begin{minipage}{0.8\linewidth}
				\begin{subequations}
					\label{eq:imperfectCSIopt}
					\begin{align}
						\mathcal{P}_2:  \quad
						&\max_{\mathbf{{P}}, \widehat{\mathbf{c}}} \,\,U\left ( \mathbf{{P}},\widehat{\mathbf{c}}\right )\\
						\mbox{s.t.}\,\,
						& \sum_{k=1}^K \widehat{C}_{k}\leq \min\{\widehat{R}_{c,1},\ldots,\widehat{R}_{c,K}\} \\
						&	\text{tr}(\mathbf{P}\mathbf{P}^{H})\leq P\\
						&   \widehat{R}_{k} + \widehat{C}_{k}\geq R_k^{th}, \forall k\in\mathcal{K} \\
						& \widehat{C}_{k}\geq 0, \forall k\in\mathcal{K}
					\end{align}
				\end{subequations}
			\end{minipage}
		}
	\end{center}
	\vspace{0.6mm}
	\vspace{0.4mm}
	where  $\widehat{\mathbf{c}}=\{\widehat{C}_1,\ldots,\widehat{C}_K\}$. 
	%%%
	The corresponding objective function is given as follows
	\begin{equation}
		\begin{aligned}
			U\left ( \mathbf{{P}},\widehat{\mathbf{c}} \right )=&\begin{cases}
				\sum_{k=1}^Ku_{k}(\widehat{R}_{k} + \widehat{C}_{k}), & \textrm{WSR}\\ 
				\underset{k \in \mathcal{K}}{\min}
				\ (\widehat{R}_{k} + \widehat{C}_{k}),& \textrm{MMF-rate}\\ 
				\frac{\sum_{k=1}^K(\widehat{R}_{k} + \widehat{C}_{k})}{	\frac{1}{\eta}\text{tr}(\mathbf{P}\mathbf{P}^{H})+P_{\textrm{cir}}}, & \textrm{EE}\\ 
				-\text{tr}(\mathbf{P}\mathbf{P}^{H}),& \textrm{sum-power}.
			\end{cases}    \nonumber
		\end{aligned}
	\end{equation}
	%%%
	\par 
	{
		%%%
		Optimization algorithms adopted to solve $\mathcal{P}_2$ for different utility functions are discussed in Section \ref{sec:optAlgo}. 
	}
	%%%
	\par 
	\textit{Long-term resource optimization:}
	When the CSIT  estimation error $\widetilde{\mathbf{h}}_{k}$ is unbounded, for example, when the transmitter only knows the Gaussian distribution of the CSIT estimation error, a worst-case precoder optimization is not possible any more. 
	%%%
	Instead of designing resource allocation for each instantaneous channel, in this case, long-term RSMA resource optimization over a long sequence of fading states is more commonly used, see, e.g.,   \cite{RS2016hamdi,mao2019beyondDPC,mao2021IoT,anup2021MIMO,mao2019TCOM,mao2020DPCNOUM,alaa2020cranimperfectCSIT,onur2021jamming,rafael2021radarsensing,longfei2021statisticalCSIT,si2021imperfectSatellite}. 
	%%%
	By defining $\mathbf{H}=[\mathbf{h}_1,\ldots,\mathbf{h}_K]$, $\widehat{\mathbf{H}}=[\widehat{\mathbf{h}}_1,\ldots,\widehat{\mathbf{h}}_K]$,  and $\widetilde{\mathbf{H}}=[\widetilde{\mathbf{h}}_1,\ldots,\widetilde{\mathbf{h}}_K]$ respectively as the actual CSI,  the estimated CSIT, and the corresponding channel estimation errors of all users, we have
	$\mathbf{H}=\widehat{\mathbf{H}}+\widetilde{\mathbf{H}}$. 
	%%%
	Assuming the fading process to be ergodic and stationary,  the ergodic rates (ER) characterizing the long-term rate performance of the streams over all possible joint
	fading states $\{\mathbf{H},\widehat{\mathbf{H}}\}$  are defined as
	\begin{equation}
		\label{eq:ER}
		\overline{R}_{\mathrm{c},k} =  \mathbb{E}_{\{\mathbf{H},\widehat{\mathbf{H}}\}}\{R_{c,k}\}
		\quad \text{and} \quad   \overline{R}_{k} =  \mathbb{E}_{\{\mathbf{H},\widehat{\mathbf{H}}\}}\{R_{k}\}.
	\end{equation}
	%%%
	Sending the common stream and  the private stream of user-$k$ at respectively ERs $\overline{R}_{\mathrm{c},k}$ and $\overline{R}_{k}$ increases the robustness of the system.
	%%%
	To further ensure that the common stream is decodable at all users, the ER of the common stream cannot exceed $\overline{R}_{\mathrm{c}}=\min\{\overline{R}_{\mathrm{c},1},\ldots, \overline{R}_{\mathrm{c},K}\}$ and the ergodic common rate allocated to user-$k$, $\overline{C}_k$, has to satisfy  $\sum_{k=1}^K\overline{C}_k=\overline{R}_{\mathrm{c}}$. The resulting ergodic achievable rate of each user is $\textrm{ER}_{k,tot}=\overline{R}_k+\overline{C}_k$.
	%%%
	The corresponding long-term resource optimization problem $\mathcal{P}_3$ has a similar form as $\mathcal{P}_2$,  when $\widehat{R}_{\mathrm{c},k}$, $\widehat{R}_{k}$, $\widehat{C}_{k}$ are simply replaced by
	$\overline{R}_{\mathrm{c},k}$, $\overline{R}_{k}$, $\overline{C}_{k}$, respectively.
	%%%
	\begin{figure}
		\centering
		\includegraphics[width=3.5in]{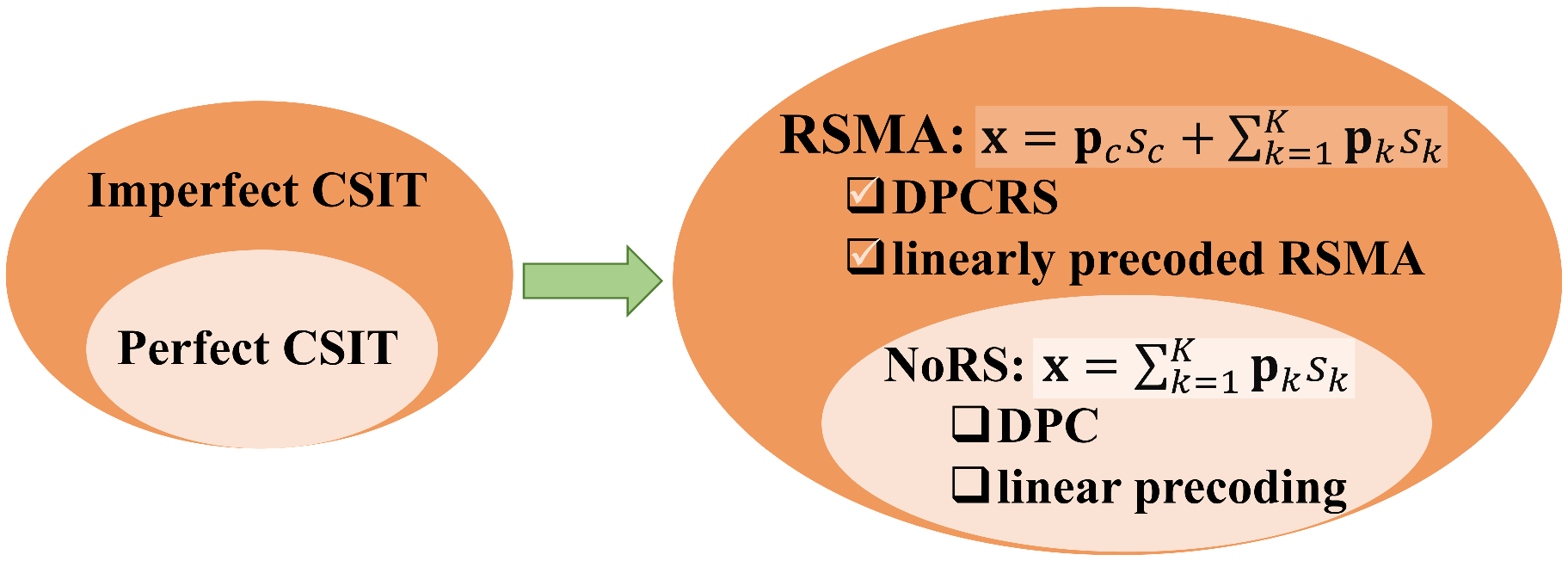}%
		\caption{Relationship between the optimization problems for perfect and imperfect CSIT, and relationship between the problems for RSMA and SDMA (a.k.a. NoRS). RSMA  for  imperfect  CSIT  introduces  a  more  general class  of  problems  for  MIMO  networks. }
		\label{fig:biggermap}
	\end{figure}
	%%%
	\par
	{
		\begin{remark}
			\label{remark:biggerMap}
			When the strength of the channel estimation error is zero (i.e., $\|\widetilde{\mathbf{h}}_{k}\|=0$), the CSIT becomes perfect (i.e., $\mathbf{h}_{k} = \widehat{\mathbf{h}}_{k}$). In this case, the worst-case resource optimization problem, $\mathcal{P}_2$, reduces to the problem for perfect CSIT,  $\mathcal{P}_1$,  and the long-term resource optimization problem, $\mathcal{P}_3$, can be equivalently decomposed into a $\mathcal{P}_1$ subproblem for each instantaneous channel realization.
			%%%
			Therefore, the problem to be solved for perfect CSIT is equivalent to the problems to be solved for imperfect CSIT when the channel estimation error is zero. 
			%%%
			Fig. \ref{fig:biggermap} illustrates the relationship between the problems for perfect and imperfect CSIT. 
			%%%
			The relation between the problems obtained for  RSMA and existing linearly precoded and non-linearly precoded strategies are also illustrated.
			%%%
			Since the simplest RSMA schemes (i.e., linearly precoded 1-layer RS and 1-DPCRS) are more general than the  conventional precoded schemes, RSMA for imperfect CSIT  further enlarges the optimization space.
			%%%
			Therefore, RSMA for imperfect CSIT  introduces a more general class of optimization problems for MIMO networks and the solutions to  these problems lead to a more general class of transmission strategies.
		\end{remark}
	}
	%%%
	\par
	\underline{\textit{Optimization Algorithms}}:
	\label{sec:optAlgo}
	%%%
	Problems $\mathcal{P}_1$, $\mathcal{P}_2$, and $\mathcal{P}_3$ are  challenging to solve due to the non-convexity of the coupled rate expressions (and the  fractional objective function if EE is employed as utility function).
	%%%
	Following Remark \ref{remark:biggerMap}, the algorithms proposed to solve $\mathcal{P}_2$ and  $\mathcal{P}_3$ can also be applied for solving $\mathcal{P}_1$. 
	%%%
	In general, the resource optimization algorithms proposed to solve $\mathcal{P}_1$, $\mathcal{P}_2$, and $\mathcal{P}_3$ can be categorized into globally optimal algorithms and suboptimal algorithms.
	%%%
	In \cite{bho2021globalEE}, a globally optimal algorithm, namely, the successive incumbent transcending (SIT)  branch and bound (BB) algorithm was proposed to solve $\mathcal{P}_1$ optimally, when the WSR, EE, and sum-power objective functions were considered. 
	%%%
	To reduce the computational complexity, a large number of works on RSMA have focused on developing suboptimal algorithms to solve $\mathcal{P}_1$, $\mathcal{P}_2$, and $\mathcal{P}_3$ with affordable and tractable complexities, such as weighted minimum mean square error (WMMSE)-based algorithms \cite{RS2016hamdi,mao2017rate,mao2018networkmimo,jian2019crs,jaafar2020UAV2,longfei2021statisticalCSIT ,si2021imperfectSatellite,alaa2020cranimperfectCSIT,mao2020DPCNOUM,anup2021MIMO,mao2021IoT,mao2019beyondDPC}, successive convex approximation (SCA)-based algorithms \cite{mao2018EE,mao2019maxmin,Tervo2018SPAWC, Zheng2020JSAC,Jia2020SEEEtradeoff,yunnuo2021FBL,Ala2019IEEEAccess,mao2019swipt,alaa2020gRS,Liping2020secrecyCRS,onur2021jamming,mao2019TCOM,alaa2020powerMini,zhaohui2020IRS,alaa2020IRScran,alaa2020EECRAN,UAVRS2019ICC}, alternating direction method of multipliers-based algorithms \cite{xu2021rate,onur2021DAC,onur2022DFRC,rafael2021radarsensing,JRCLoli2022}, and semidefinite relaxation-based  algorithms \cite{fuhao2020secrecyRS,Medra2018SPAWC,RSswiptIC2019CL,Camana2020swiptRS,wonjae2021imperfectCSIR}. 
	%%%
	These algorithms are able to converge to the Karush–Kuhn–Tucker (KKT) points of the original problem. 
	%%%
	Numerical results in \cite{bho2021globalEE} show that the WSR performance achieved by a WMMSE-based algorithm and the EE performance of a SCA-based algorithm almost coincide with  the corresponding globally optimal performance. 
	%%%
	\par
	Though the computational complexity of the suboptimal algorithms is tractable,  it is still unfavorable for real-world applications.
	%%%
	For example, SCA requires solving a series of  convex subproblems, and  in each subproblem the non-convex objective and constraints are approximated by locally tight lower bounds. 
	%%%
	%Readers are referred to the corresponding papers for more details of the algorithms. 
	%%%
	
	\subsubsection{Low-Complexity Resource Allocation}
	\label{sec:lowComplexPrecoding}
	To further reduce the computational complexity, another line of RSMA research focuses on closed-form and low-complexity resource allocation \cite{RS2015bruno,enrico2016bruno,onur2021mobility,Lu2018MMSERS,bruno2019wcl,Gui2020EESEtradeoff,bruno2020MUMIMO,Minbo2016MassiveMIMO,AP2017bruno,Asos2018MultiRelay,bansal2021IRS},  which is summarized in this subsection.
	%%%
	\par 
	The precoder $\mathbf{p}_i$ for stream-$i$ can be written as $\mathbf{p}_i=\sqrt{P_i}\bar{\mathbf{p}}_i$, where $\bar{\mathbf{p}}_i=\frac{\mathbf{p}_i}{\|\mathbf{p}_i\|}$ is the direction of the precoder for stream-$i$ and  $P_i=\|\mathbf{p}_i\|^2$ is the power allocated to that precoder.
	%%%
	Different from resource optimization, where the precoders $\mathbf{p}_i, \forall i$, (including their direction and power)  of all streams are jointly optimized at the transmitter, for low-complexity resouce allocation, usually the design of the precoder directions $\bar{\mathbf{p}}_i, \forall i$, and the optimization of  power $P_i, \forall i$, are separated.
	%%%
	For 1-layer RS, there are $K+1$ streams indexed by $i\in\{c,1,\ldots, K\}$. The common stream $s_c$ is required to be decoded by all users and its precoder direction $\bar{\mathbf{p}}_c$ is usually designed in a multicast manner based on the following typical methods:
	\begin{itemize}
		\item Random precoding \cite{RS2015bruno,enrico2016bruno,onur2021mobility,Lu2018MMSERS}:
		Random precoding, such as setting $\mathbf{p}_c=\mathbf{e}_1$ (where $\mathbf{e}_1$ is a null vector with one entry 1 and all other entries 0), is commonly used in the literature of RSMA for DoF analysis. 
		%%%
		Using such random precoding for the common stream is sufficient for RSMA to achieve the entire DoF region, but is not efficient in terms of improving the SE performance. 
		%%%
		\item Weighted matched beamforming (MBF) \cite{Minbo2016MassiveMIMO,AP2017bruno,Asos2018MultiRelay,bansal2021IRS,Lu2018MMSERS,bruno2019wcl,Gui2020EESEtradeoff}: The precoder direction  $\bar{\mathbf{p}}_c$ is designed to maximize the achievable rate of the common stream, which can be formulated as follows
		\begin{subequations}
			\label{eq:optcommonprecoder}
			\begin{align}
				\quad
				\max_{\bar{\mathbf{p}}_c} \,\min_k \,\,& \omega_k|{\mathbf{h}}_k^H\bar{\mathbf{p}}_c|^2\\
				\mbox{s.t.}\,\,
				&	\text{tr}(\bar{\mathbf{p}}_c\bar{\mathbf{p}}_c^{H})\leq 1.
			\end{align}
		\end{subequations}
		The optimal precoder direction  $\bar{\mathbf{p}}_c^*$ to solve (\ref{eq:optcommonprecoder}) is the weighted MBF, which is given by
		\begin{equation}
			\bar{\mathbf{p}}_c=\sum_{k\in\mathcal{K}}\varpi_k\mathbf{h}_k,
		\end{equation}
		where the optimal weight $\varpi_k$ for the two-user case is obtained in \cite{bruno2019wcl}, and the  asymptotically optimal weight $\varpi_k$ as $M\rightarrow\infty$ is obtained in  \cite{Minbo2016MassiveMIMO,AP2017bruno}. In both cases, $\varpi_k$ is a function of $\omega_k$.  To obtain a more  tractable precoder design, equal weighted MBF with $\varpi_k=\frac{1}{\sqrt{MK}}$ is commonly employed \cite{Minbo2016MassiveMIMO,Lu2018MMSERS}.
		\item  Singular value decomposition (SVD) \cite{RS2016hamdi,bansal2021IRS}: SVD based designs choose the dominant left singular vector of $\mathbf{H}=[\mathbf{h}_1,\ldots,\mathbf{h}_K]$ for $\bar{\mathbf{p}}_c$. Such approach is also used to initialize the precoder of the common stream for the WMMSE and SCA algorithms in \cite{RS2016hamdi}.   
	\end{itemize}
	%%%
	\par
	The private stream $s_k$ is only decoded by user-$k$ and is treated as noise by the other users. Therefore, popular linear precoding methods such as ZFBF, regularized ZFBF (RZF)/MMSE, and block-diagonalization (BD) can be used to design the precoder direction of the private streams:
	\begin{itemize}
		\item ZFBF \cite{RS2015bruno,enrico2016bruno,onur2021mobility}: ZFBF
		steers  the precoder direction of each user $\bar{\mathbf{p}}_k$ to the  space orthogonal to the space  spanned by the other users' channel vectors, i.e., $\bar{\mathbf{p}}_k\in \mathrm{null}\left( \left[ \mathbf{h}_{1},\ldots,\mathbf{h}_{k-1},\mathbf{h}_{k+1},\ldots,\mathbf{h}_{K} \right]^{H} \right)$. It is therefore limited to the underloaded MISO BC, i.e.,  $M\geq K$.  Specifically, ZF designs $\bar{\mathbf{P}}_p=[\bar{\mathbf{p}}_1,\ldots,\bar{\mathbf{p}}_K]$ as
		\begin{equation}
			\label{eq:ZFBF}
			\bar{\mathbf{P}}_p={{\left(\mathbf{H}\mathbf{H}^H\right)^{-1}}}\mathbf{H}.
		\end{equation}
		Such ZFBF for the private streams together with random beamforming for the common stream  achieves the optimal  DoF region for the MISO BC with imperfect CSIT \cite{RS2016hamdi,enrico2016bruno,enrico2017bruno}.
		%%%
		\item  RZF/MMSE \cite{Minbo2016MassiveMIMO,AP2017bruno,Asos2018MultiRelay,Lu2018MMSERS}:
		RZF precoding aims to tackle the ill-conditioned behavior of the largest eigenvalue of $\left(\mathbf{H}\mathbf{H}^H\right)^{-1}$ via a regularized form of channel inversion.
		RZF designs the precoder direction $\bar{\mathbf{P}}_p$ as follows
		\begin{equation}
			\label{eq:RZF}
			\bar{\mathbf{P}}_p=\underset{\mathbf{F}}{\underbrace{\left(\mathbf{H}\mathbf{H}^H+\kappa\mathbf{I}\right)^{-1}}}\mathbf{H},
		\end{equation}
		where $\kappa=K\sigma_{n}^2$ is a regularization parameter (assuming the noise power  is the same at all users, i.e., $\sigma_{n}^2=\sigma_{n,k}^2, \forall k\in\mathcal{K}$). RZF (\ref{eq:RZF}) reduces to ZFBF (\ref{eq:ZFBF}) if $\kappa=0$.

		\item (Regularized-)BD \cite{bruno2020MUMIMO,onur2021mobility}:
		%%%
		(Regularized-)BD extends (Regularized-)ZFBF to the MIMO BC with $Q \,(2\leq Q\leq M)$ receive antennas at each user. The precoder direction $\bar{\mathbf{P}}_k \in \mathcal{C}^{M\times Q}$ based on regularized-BD is designed as a cascade of two precoding matrices
		\begin{equation}
			\bar{\mathbf{P}}_k=\bar{\mathbf{V}}_k\bar{\mathbf{W}}_k,
		\end{equation}
		%%%
		where the first filter $\bar{\mathbf{V}}_k$ is designed to partially remove multi-user interference.  Define $\bar{\mathbf{H}}_k=[{\mathbf{H}}_1, \ldots,{\mathbf{H}}_{k-1},{\mathbf{H}}_{k+1},\ldots,{\mathbf{H}}_K]$ as the channel matrix excluding ${\mathbf{H}}_k$. Applying SVD to $\bar{\mathbf{H}}_k$, we have $\bar{\mathbf{H}}_k={\mathbf{U}}_k{\mathbf{\Lambda }}_k{\mathbf{V}}_k$. $\bar{\mathbf{V}}_k$ is chosen as $\bar{\mathbf{V}}_k={\mathbf{V}}_k({\mathbf{\Lambda }}_k^T{\mathbf{\Lambda }}_k+Q\sigma_{n,k}^2\mathbf{I})^{-\frac{1}{2}}$. 
		%%%
		The second filter $\bar{\mathbf{W}}_k$ is designed to enable parallel symbol stream detection at user-$k$. Defining the effective channel matrix as $\tilde{\mathbf{H}}_k=\mathbf{H}_k\bar{\mathbf{V}}_k^H$ and applying SVD to $\tilde{\mathbf{H}}_k$, we have $\tilde{\mathbf{H}}_k=\tilde{\mathbf{U}}_k\tilde{\mathbf{\Lambda }}_k\tilde{\mathbf{V}}_k^H$. $\bar{\mathbf{W}}_k$ is designed as $\bar{\mathbf{W}}_k=\tilde{\mathbf{V}}_k$. $\mathbf{U}_k^H$ is used to design  the receive filter at user-$k$.
	\end{itemize}
	%%%
	\par
	Besides the precoder direction, the power allocation among the common and private precoders, i.e., selecting the values of $P_{c}, P_{1}, \ldots, P_{K}$, is also crucial for the performance of 1-layer RS, especially the division of power between the common stream $P_c$ and the private streams $P_1+\ldots+P_K$.
	%%%
	Defining the  fraction  of the  transmit power $P$  allocated to the private streams as $\tau \,(0\leq \tau\leq1)$, we have $P_c=(1-\tau)P$ as $P_c+P_1+\ldots+P_K=P$.
	%%%
	Optimal closed-form expressions for $\tau$ have been reported in \cite{bruno2019wcl} for perfect CSIT and in \cite{Minbo2016MassiveMIMO,onur2021mobility} for imperfect CSIT. 
	%%%
	The powers allocated to the different private streams can be chosen  equal \cite{onur2021mobility} or obtained via   water-filling (WF) \cite{bruno2019wcl}.
	%%%
	\subsubsection{Resource Allocation by Machine Learning} 
	Apart from optimization based resource allocation designs, machine learning methods have also been used to optimize the power allocation strategies for given precoders \cite{AI20206G}. In \cite{hieu2021}, the authors propose a deep reinforcement learning algorithm to design the power allocation for each transmit stream, and show that RSMA achieves a significant performance gain compared to SDMA for imperfect CSIT.
	
	\subsection{Resource Allocation for Multicarrier RSMA}
	\label{sec:MC-RA}
	\labelsubseccounter{V-B}
	The pivotal role of resource allocation in single-carrier RSMA further motivates the study of RSMA in multicarrier systems.
	%%%
	In multicarrier RSMA, each subcarrier is occupied by multiple users and each user can use multiple subcarriers. 
	%%%
	Therefore, besides the precoder design, power control, and common rate allocation as in single-carrier RSMA, user scheduling, and subcarrier assignment are also key variables for resource allocation in multicarrier RSMA. 
	%%%
	\par
	The study of multicarrier RSMA is still in its infancy. 
	%%%
	The authors in \cite{HongzhiChen2020subcarrierRS} first investigate a joint subcarrier allocation, precoder design, power allocation, and common rate allocation problem to maximize the minimum rate among users in multicarrier multigroup multicast systems.
	%%%
	The optimization problem is a mixed integer non-linear programming problem due to the applied subcarrier allocation indicators. To solve this problem, the binary subcarrier allocation indicators are relaxed to continuous variables and the non-convex constraints are reformulated and solved via semidefinite programming and SCA methods.
	%%%
	The authors further investigated the optimal MMF-rate performance of RSMA by exhaustively searching over all possible subcarrier allocation strategies and optimizing the precoder, power, and common rate allocation for each possible subcarrier allocation scheme \cite{hongzhi2021RSLDPC}.
	%%%
	One major observation in \cite{HongzhiChen2020subcarrierRS,hongzhi2021RSLDPC} is that RSMA without subcarrier optimization can even outperform SDMA with subcarrier optimization. Unlike SDMA which requires users with (semi-)orthogonal channel conditions in each subcarrier, RSMA is capable of serving users with arbitrary channel conditions in each subcarrier. Therefore, it can ease overhead introduced by subcarrier allocation and user scheduling at the transmitter. A similar observation has also been made in \cite{mao2019TCOM} for user scheduling in single-carrier RSMA. In Section \ref{sec:RSMAvsOtherMA}, the scheduling complexity of RSMA is further discussed.

	\par
	The aforementioned works  \cite{HongzhiChen2020subcarrierRS,hongzhi2021RSLDPC} only considered resource optimization in perfect CSIT.
	%%%
	RSMA for multicarrier cognitive radio systems with imperfect CSIT was recently investigated in \cite{onur2021jamming, onur2021CR} under the assumption that all legitimate users are served in all subcarriers. 
	%%%
	Hence, the subcarrier allocation and user scheduling for multicarrier RSMA with imperfect CSIT remains an open problem. 
	%%%
	Further work on low-complexity subcarrier allocation strategies for both perfect and imperfect CSIT is also needed.

	\section{RSMA PHY layer Design}
	\label{sec:RSMAphyLayer}
	%All results provided in Section \ref{sec:RSMAvsOtherMA} are obtained under the idealistic assumptions
	%of Gaussian signaling and infinite block lengths.
	%%%
	The discussions on RSMA so far have been in terms of abstract message definitions and Shannon capacity, which assumes Gaussian signalling and infinite block length coding. A practical PHY layer architecture for RSMA has been proposed in \cite{Onur2020LLS} including finite constellation modulation schemes, finite length polar codes, and AMC. This architecture has been further extended  in \cite{onur2021mobility} to the MISO/MIMO BC with CSI feedback delay and user mobility, in \cite{anup2021MIMO} to the MIMO BC employing a V-BLAST receiver, in \cite{longfei2021LLS} to satellite communications, and in \cite{JRCLoli2022} to joint communications and sensing. 
	%%%
	In this section, we first provide some toy examples to give an intuition about a practical PHY layer architecture of RSMA. Then, we thoroughly delineate the PHY layer transmitter and receiver architectures.
	
	%%%
	\subsection{A Quick Introduction to RSMA PHY Layer}
	\label{sec:quickintro}
	The examples in this section are based on the RSMA PHY layer architecture described in \cite{Onur2020LLS}. In particular, we explain how the user messages are split and combined at the transmitter based on the transmission rates and how the received signals are processed by demonstrating the bit-level processing of each user's message. 
	
	We consider a scenario where a multi-antenna transmitter serves two single-antenna receivers. The transmitter aims to send two independent messages to the users, with each message being intended only for one user. According to the RSMA framework, the transmitter forms one common message and two private messages to transmit in such a setting.  
	
	%(or a back-off is applied to the calculated actual transmission rates to account for the performance degradation due to the employed FBL coding and BICM schemes).

	\begin{figure}[t!]
		\begin{subfigure}[b]{\columnwidth}
			\centering
			\includegraphics[width=0.99\textwidth]{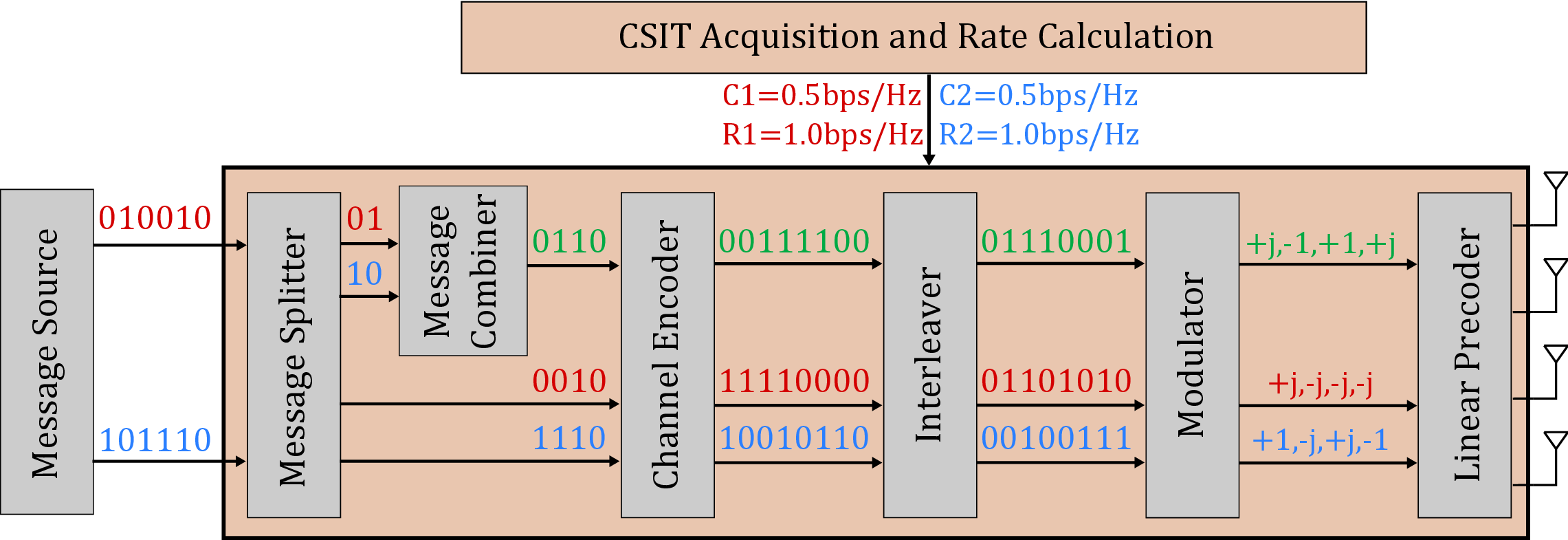}%
			\caption{Example 1: $C_{1}=C_{2}=0.5$ bps/Hz, $R_{1}=R_{2}=1.0$ bps/Hz. The coding rate is $1/2$ and the modulation scheme is QPSK for the common and private streams, respectively.}
			\label{fig:tx_example1}
			\vspace{0.3cm}	
		\end{subfigure}
		%	~\\
		\begin{subfigure}[b]{\columnwidth}
			\centering
			\includegraphics[width=0.99\textwidth]{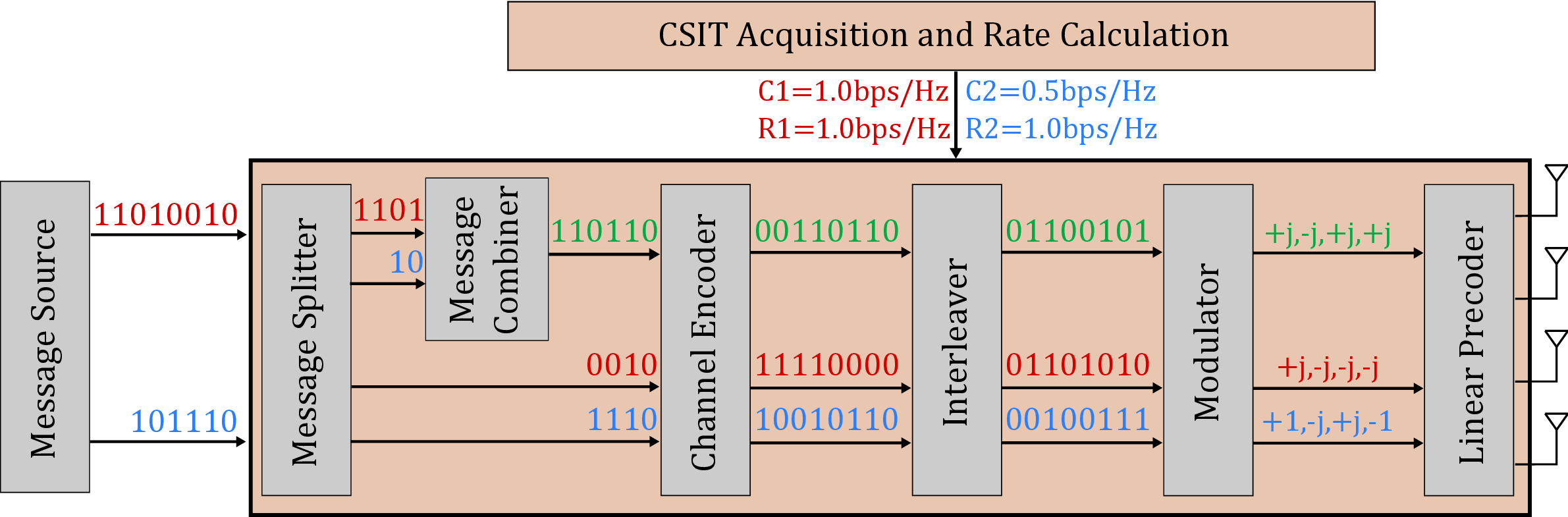}%
			\caption{Example 2: $C_{1}=1.0$ bps/Hz, $C_{2}=0.5$ bps/Hz, $R_{1}=R_{2}=1.0$ bps/Hz. The coding rate is $3/4$ for the common stream and $1/2$ for the private streams. The modulation scheme is QPSK for the common and private streams, respectively.}
			\label{fig:tx_example2}
			\vspace{0.3cm}
		\end{subfigure}
		% ~\\
		\begin{subfigure}[b]{\columnwidth}
			\centering
			\includegraphics[width=0.99\textwidth]{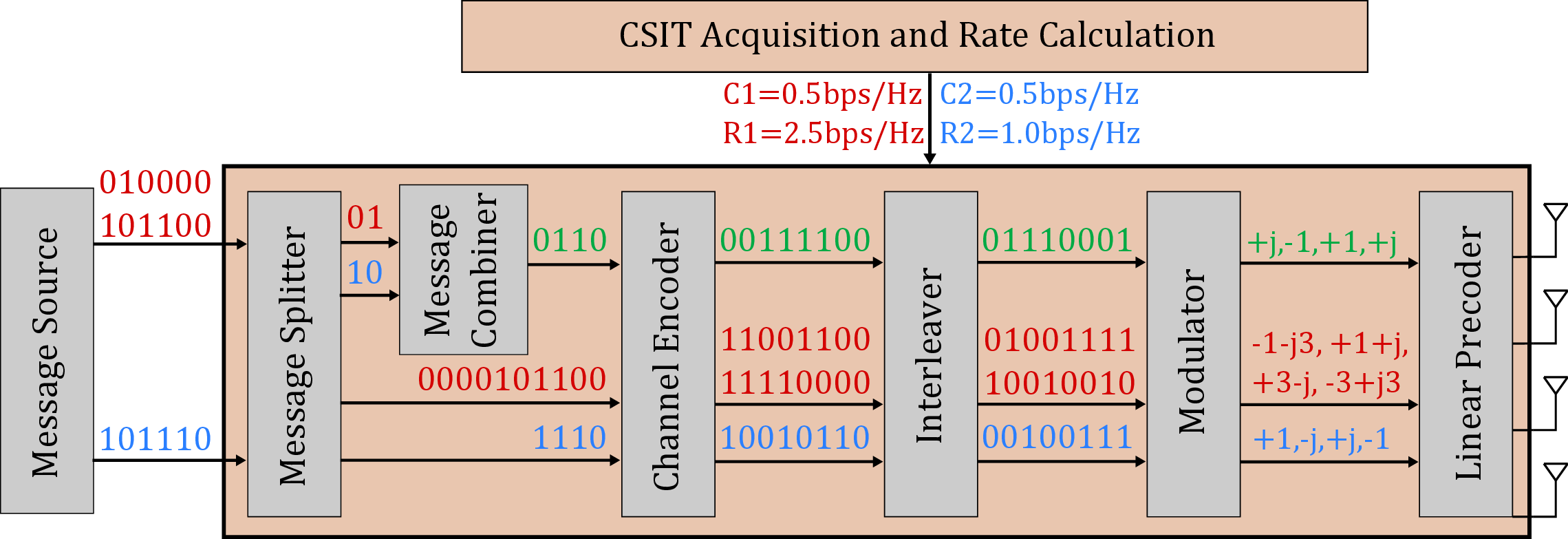}%
			\caption{Example 3: $C_{1}=C_{2}=0.5$ bps/Hz, $R_{1}=2.5$ bps/Hz, $R_{2}=1.0$ bps/Hz. The coding rates is $1/2$ for the common stream and the private stream of user-$2$, respectively, and $5/8$ for the private stream of user-$1$. The modulation scheme is QPSK for the common stream and the private stream for user-$2$, respectively, and 16-QAM for the private stream of user-$1$.}
			\label{fig:tx_example3}
		\end{subfigure}%
		\caption{Examples for RSMA transmitters. Two separate messages are transmitted to two users over $4$ consecutive symbols. The red bits represent the message intended for user-$1$, the blue bits represent the message intended for user-$2$, and the green bits represent the formed common message after message splitting and combining. $R_{1}$ and $R_{2}$ denote the rates of the private streams and $C_{1}$ and $C_{2}$ denote the portions of the rate of the common stream allocated to user-$1$ and user-$2$, respectively. Appropriate coding rates ($1/2$, $3/4$, $5/8$) and modulation schemes (QPSK, $16$-QAM) are chosen to support the calculated transmission rates.} 
		\label{fig:example_rsma_tx}
	\end{figure}
	
	\subsubsection{RSMA transmitter operation}
	\label{sec:rsmaTx}
	\labelsubseccounter{II-A}
	Fig.~\ref{fig:example_rsma_tx} illustrates the operations at the transmitter for different transmission rates. In the figures, $R_{1}$ and $R_{2}$ denote the transmission rates of the private streams and $C_{1}$ and $C_{2}$ denote the portions of the rate of the common stream allocated to user-$1$ and user-$2$, respectively. The transmission rate calculations for RSMA are explained in detail in Section~\ref{sec:principle}. The rates can be calculated either at the transmitter using the available CSIT, or obtained as a feedback from the users to be served. In the following examples, we assume that the transmission rates provided to the transmitter are achievable by means of finite blocklength
	(FBL) coding and bit-interleaved coded modulation (BICM).

	The transmission rates are input to the modules which perform message splitting, message combining, modulation and coding; and they are used to determine the number of message bits, the modulation scheme, and the channel coding parameters. The transmitter is allocated $4$ consecutive symbols (channel uses) to transmit two independent messages, each intended for only one of the users. In the figures, the message intended for user-$1$ is depicted in red color and the message for user-$2$ is depicted in blue. We assume that the bits of the messages are independent. 
	%(the messages cannot be compressed without any losses). 
	%The transmission rates for the common stream and the private streams of each user are calculated according to the CSIT to determine the length of the user messages, the coding rates and the modulation schemes. 
	The transmission rates dictate the number of message bits transmitted in one symbol. Thus, the number of message bits that can be carried by the common and private streams is $4$ times the value of the corresponding transmission rates. 
	
	For Example 1 (Fig.~\ref{fig:tx_example1}), the transmission rates for the common and private streams are all $1$ bps/Hz. These rates allow $4$ message bits to be carried over each stream. We consider the case where the modulation scheme for transmission is chosen as QPSK and the coding rate is set as $1/2$. 
	%Note that any other MCS that achieves a SE of $1$ bps/Hz can be chosen for transmission as well ({\sl e.g.,} 16-QAM and rate $1/4$). 
	Next, we check the portions of the common rate of the users, which are both equal to $0.5$ bps/Hz. This implies that $2$ bits per user can be transmitted over the common stream, which, together with the messages carried by the private streams, sums up to $6$ message bits for each user. One can also verify that the sum-rate for a single user is $C_{1}+R_{1}=1.5$ bps/Hz, which corresponds to transmission of $6$ bits during $4$ channel uses. Next, we extract $6$ message bits from the message source buffer for each user. As the message bits are independent, one can split the $6$ message bits of each user into the common part of $2$ bits and the private part of $4$ bits by means of any splitting pattern. 
	
	After the message splitting is performed, we move to the process of message combining to form the single common message to be transmitted over the common stream. As the common parts of the messages from the two users are independent, one can simply concatenate the $2$-bit common message parts to form the $4$-bit common message by means of any concatenation pattern.  
	
	\begin{remark}
		In the considered example, the splitting operation is performed by assigning the first $2$ bits to the common stream and the remaining bits to the private stream. However, any other splitting pattern can be considered (such as, selecting $2$ bits randomly for the common stream), as long as the message bits are independent and a matching message desplitting operation is performed at the receivers. The splitting pattern can also depend on the message structure (for example, if the message bits from the message source are concatenated messages for two separate applications, splitting can be performed to separate these messages into common and private messages), the protocols on the upper layers (MAC and above), or any other criterion imposed by the system design.
		
		Similarly, the combining operation is performed by appending the common message parts from the two users directly, with the common part from user-$1$ representing the first $2$ bits of the new message and the part from user-$2$ representing the last $2$ bits.
		However, any other combining pattern can be employed as long as the message bits are independent and a matching message decomposing operation is performed at the receivers. 
		For example, a comb-like combining operation can be performed, where each bit from one user's message is followed by a bit from the other user's message, or a 1-1 mapping can be performed with a new message set consisting of all $4$-bit messages.
	\end{remark}
	
	After the message splitting and the message combining operations are performed, the resulting $4$ bit messages are independently encoded by channel codes of rates $1/2$ to obtain codewords of $8$-bits. 
	The channel coding applied on each stream can be performed over the same codebook or different codebooks\footnote{Here, the terminology codebook refers to the set which contains all codewords of a channel code.}, as long as the codebook for the common stream is known by both users and the codebooks for the private streams are known by the corresponding users. Similarly, the employed interleavers can be identical or different as long as the interleaver for the common stream is known by both users and the interleavers for the private streams are known by the corresponding users.
	The codewords are then interleaved and modulated by QPSK modulation to obtain the $4$ symbols to be transmitted. 
	%We note here that since the common message is intended for both users, the polar code codebook for the common stream should be designed for the compound setting consisting of two channels for optimal error performance. 
	%However, any practical code design method can be employed to achieve an improved performance, as we demonstrate by numerical results in Section~\ref{sec:LLS}.

	\begin{figure*}[t!]
		\begin{subfigure}[b]{0.64\columnwidth}
			\centering
			\includegraphics[width=0.99\textwidth]{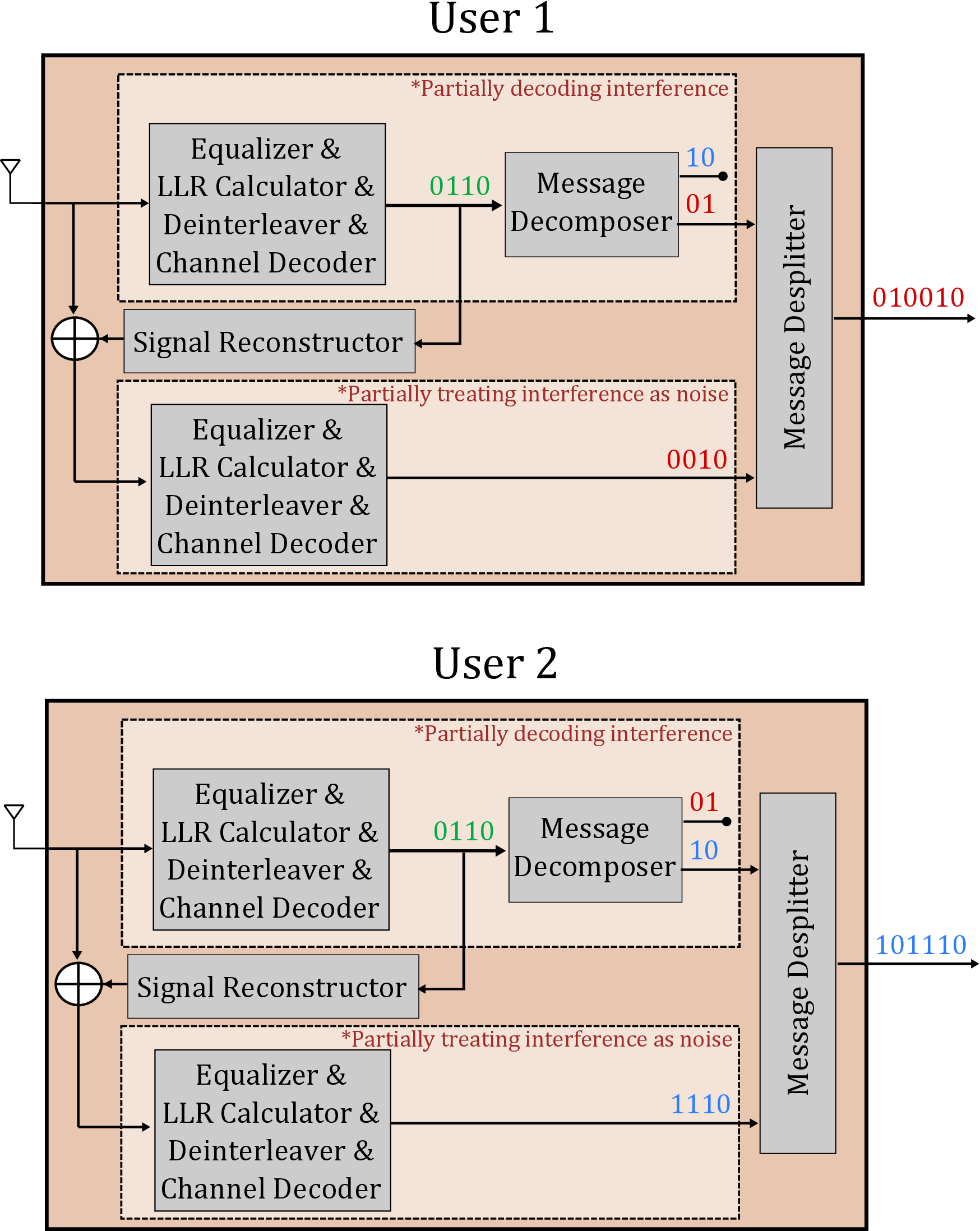}%
			\caption{Example 1.}
			\label{fig:rx_example1}
			%\vspace{0.3cm}	
		\end{subfigure}
		%	~\\
		\begin{subfigure}[b]{0.64\columnwidth}
			\centering
			\includegraphics[width=1.005\textwidth]{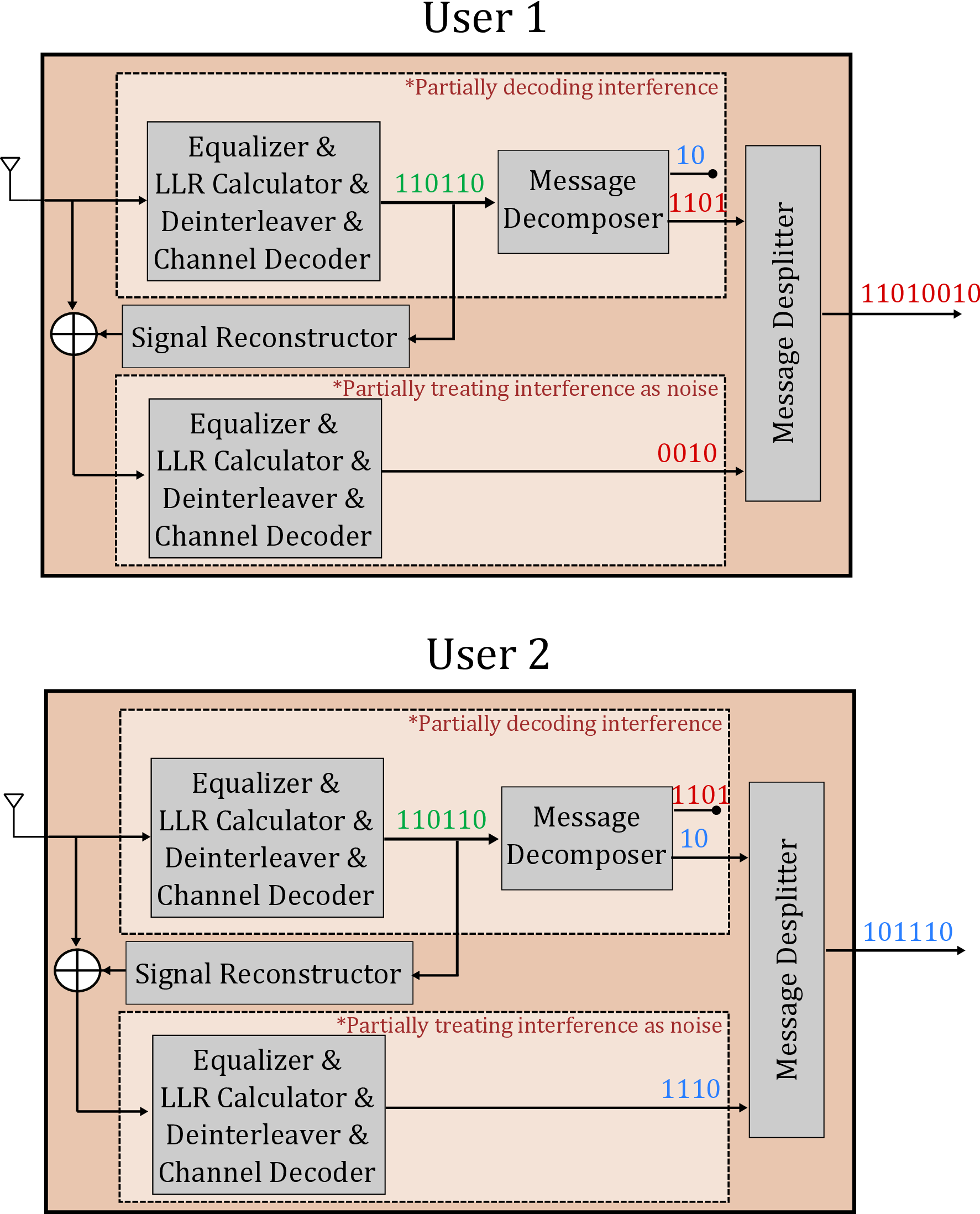}%
			\caption{Example 2.}
			\label{fig:rx_example2}
			%\vspace{0.3cm}
		\end{subfigure}
		% ~\\
		\begin{subfigure}[b]{0.64\columnwidth}
			\centering
			\includegraphics[width=0.99\textwidth]{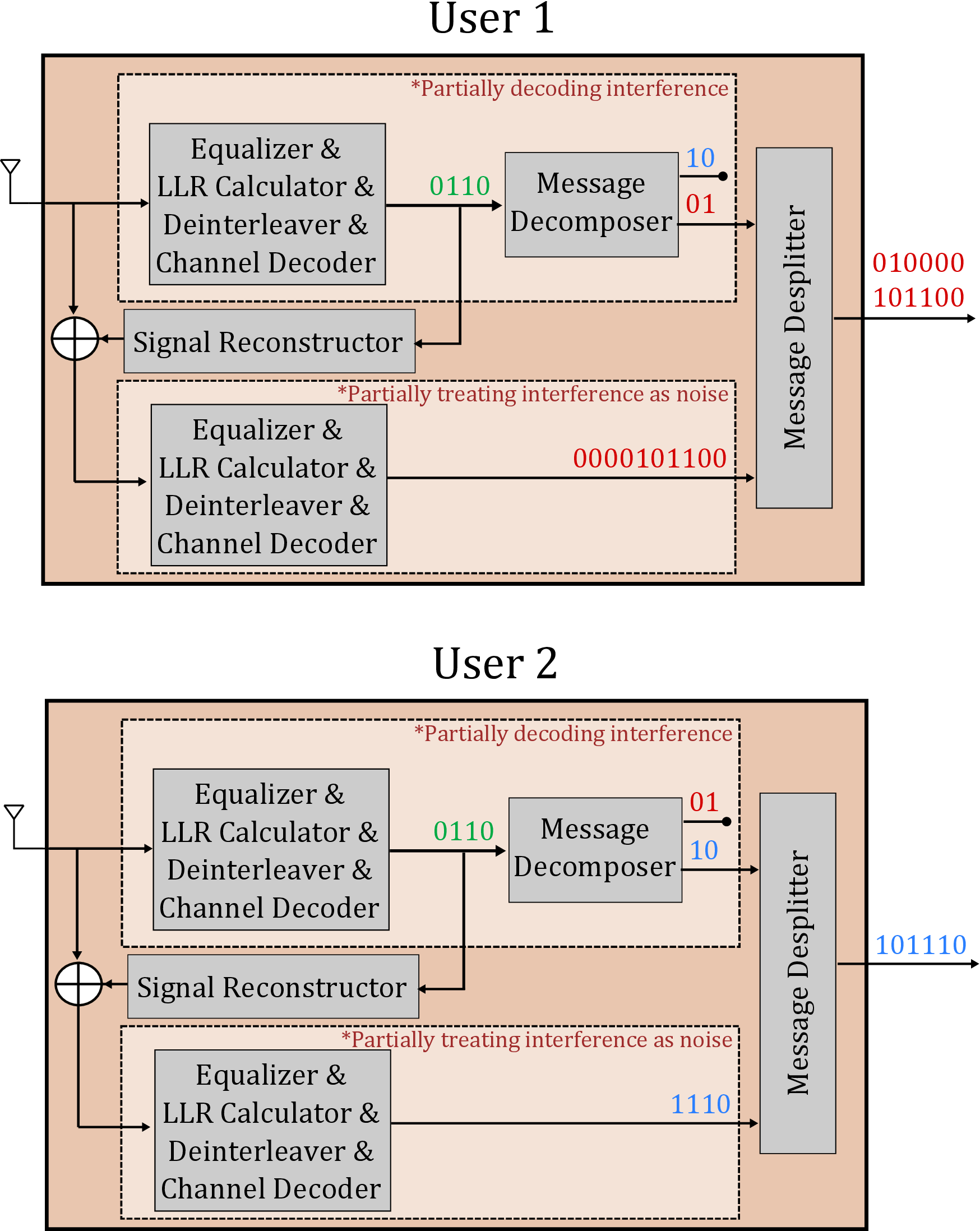}%
			\caption{Example 3.}
			\label{fig:rx_example3}
		\end{subfigure}%
		\caption{Receiver operations for RSMA for the examples in Fig. \ref{fig:example_rsma_tx}. Each receiver performs message decomposing and desplitting to obtain the transmitted message bits.}
		\label{fig:example_rsma_rx}
	\end{figure*}

	In Example 2 (Fig.~\ref{fig:tx_example2}), the transmission rate of the common stream and the portion of user-$1$ are increased compared to Example 1. Again, the transmitter is allocated $4$ consecutive symbols for transmission and the modulation scheme for transmission is chosen as QPSK. In this case, $4$ bits can be transmitted over the common stream to user-$1$, while the number of bits that can be transmitted to user-$2$ stays the same ($2$ bits). Consequently, the total number of message bits intended for user-$1$ increases to $8$ bits (this can be verified by multiplying the total sum-rate of user-$1$, $C_{1}+R_{1}=2$ bps/Hz, by the number of transmit symbols). 
	
	Similar to the case in Example 1, the message splitting operation for user-$1$ is performed by assigning the first $4$ bits of the message of user-$1$ to the common part, and the splitting operation for user-$2$ is performed in the exact same fashion as in Example 1. The message combiner concatenates the $4$-bit message that is the common message part of user-$1$ and the $2$-bit message that is the common message part of user-$2$ by appending them, resulting in a common message of $6$ bits. Since the transmission is performed over $4$ QPSK symbols and a total of $8$ bits can be transmitted over a stream, the coding rate is calculated as $3/4$ (which can also be calculated by dividing the total common rate $C_{1}+C_{2}=1.5$ bps/Hz by the SE of QPSK modulation). Finally, the encoding, interleaving, and modulation operations are performed as done in Example 1. 
	
	In Example 3 (Fig.~\ref{fig:tx_example3}), the transmission rate of the private stream for user-$1$ is increased compared to the one in Example 1. Again, the transmitter is allocated $4$ consecutive symbols for transmission. Contrary to the previous examples, the transmission rate of the private stream of user-$1$ ($2.5$ bps/Hz) is larger than the SE of QPSK modulation ($2.0$ bps/Hz). Therefore, we choose 16-QAM modulation for transmission, which has a larger SE than QPSK. The number of message bits that can be transmitted over the private stream of user-$1$ is calculated as $10$ bits and the number of codeword bits that can be transmitted using 16-QAM modulation is $16$ bits, resulting in a coding rate of $5/8$ for the considered stream (again, the coding rate can also be calculated by dividing $R_{1}=2.5$ bps/Hz by the SE of 16-QAM modulation). The number of message bits that can be transmitted over the common stream and the private stream of user-$2$ are the same as in Example 1, yielding an additional $2$ bits to be transmitted to user-$1$ over the common stream. Consequently, the total number of message bits intended for user-$1$ becomes $12$, while the total number of message bits intended for user-$2$ is $6$. 
	
	The message splitting and message combining operations are performed in the exact same fashion as in Example 1 (first, $2$ message bits are split for the common parts, and then combined by appending them). Then, the resulting common and private messages are encoded and interleaved separately. Note that, in this case, the codebook for the channel code applied to the private stream of user-$1$ cannot be the same as those for the common stream and the private stream of user-$2$, as opposed to the cases in Examples 1 and 2. Similarly, the interleaver should be longer for the private stream of user-$1$. After interleaving, QPSK modulation is applied to the common stream and the private stream of user-$2$ and 16-QAM modulation is applied for the private stream of user-$1$, all resulting in $4$ symbols for transmission.
	
	\begin{remark}
		It is worth noting that, for the considered two-user scenarios, removing the message splitting operation of the considered transmitter structure would result in the conventional transmitter architecture for other MA schemes, such as SDMA and NOMA. More specifically, SDMA transmitters do not apply message splitting and combining, and directly encode the data signals for the two users into two separate private streams. Similarly, NOMA transmitters do not apply message splitting and combining, and directly encode the data signal for one user into the common stream and that for the other user into one private stream (the reasoning behind using the common stream terminology here for NOMA will become clearer after the receiver operations are described in the next section).
	\end{remark}
	
	%%%
	\begin{figure*}[t!]
		\centering
		\includegraphics[width=\textwidth]{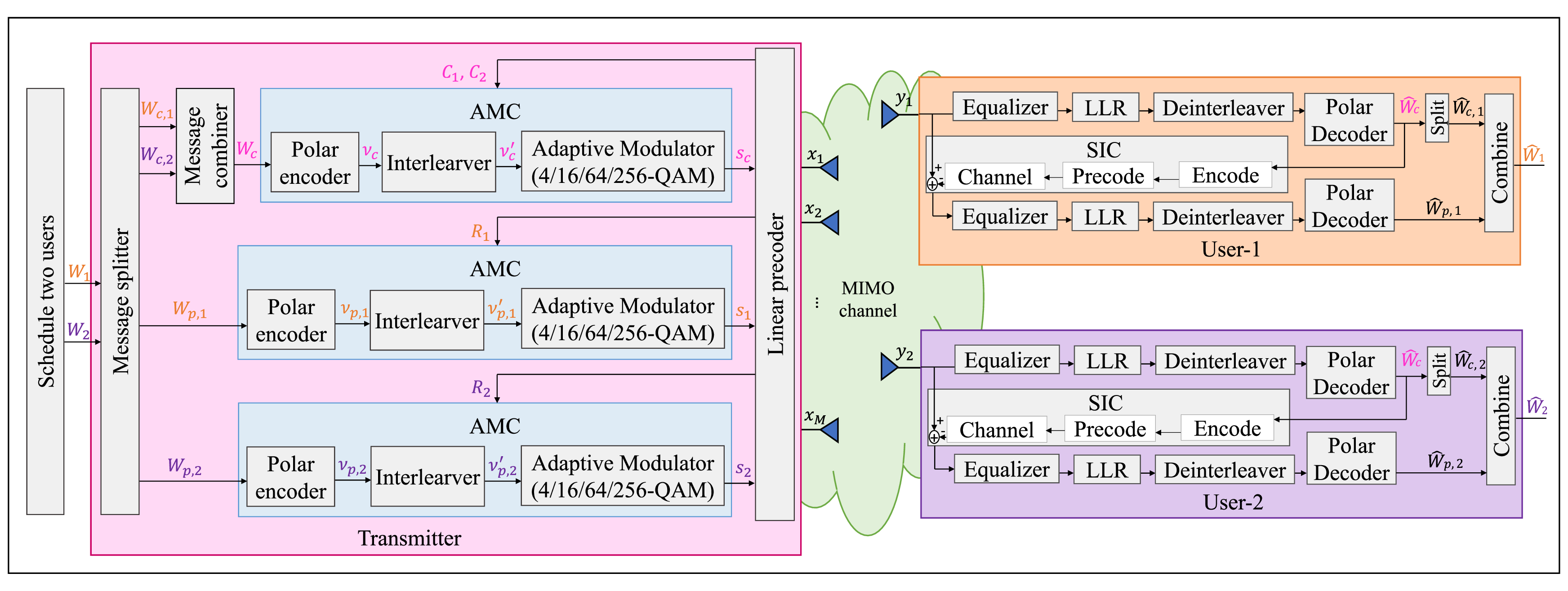}%
		\caption{PHY layer transmitter and receiver structures of RSMA with finite constellation modulation schemes, finite length polar codes, and adaptive modulation and coding (AMC) \cite{Onur2020LLS}. }
		\label{fig:PHYdesign}
	\end{figure*}
	%%%
	
	\subsubsection{RSMA receiver operation}
	\label{sec:rsmaRx}
	\labelsubseccounter{II-B}
	Next, we explain the operations performed at the receiver for the examples considered above. Fig.~\ref{fig:example_rsma_rx} depicts the operations at the receivers for each example. Each receiver starts by processing the common stream by performing equalization, Log-Likelihood Ratio (LLR) calculation, deinterleaving, and decoding. Assuming correct detection and decoding, the outputs of the decoders at the users are the message bits carried over the common stream. For Examples 1 and 3 (Figs.~\ref{fig:rx_example1}~and~\ref{fig:rx_example3}, respectively), the outputs are the $4$-bit common message and for Example 2 (Fig.~\ref{fig:rx_example2}), the outputs are the $6$-bit common message. The decoded common messages are then input to the message decomposers, which reverse the operation done by the message combiner at the transmitter. The resulting decomposed messages are the common parts of the user messages. Each user proceeds with the corresponding common parts of their messages and discards the other user's common message part, as illustrated in Fig.~\ref{fig:example_rsma_rx} (the green message is decomposed into the red and blue messages, user-$1$ proceeds with the red message and discards the blue, while it is the other way around for user-$2$). 
	
	After signal reconstruction and interference cancellation, the corresponding private streams are processed at each user following the same steps as used for decoding the common stream. Assuming correct detection and decoding, the outputs of the decoders are the private message bits intended for the corresponding users. For Examples 1 and 2 (Figs.~\ref{fig:rx_example1}~and~\ref{fig:rx_example2}, respectively), the decoder outputs at each user are the $4$-bit private messages and for Example 3 (Fig.~\ref{fig:rx_example2}), the output is the $10$-bit private message at user-$1$ and the $4$-bit private message at user-$2$. 
	
	One can notice that processing the common stream at any user involves decoding a portion of the message of the other user. As the multi-user interference at one user stems from the message of the other user, such a process can be interpreted as partially decoding the interference. 
	On the other hand, processing the private stream at any user is done by treating a portion of the message of the other user as noise. As the multi-user interference at one user stems from the message of the other user, such a process can be interpreted as partially treating the interference as noise. 
	
	Finally, the decoded common and private messages at each user are input to the message desplitters, which reverse the operation done by the message splitter at the transmitter.  The outputs of the message desplitters at each user are the intended messages for the corresponding users.
	\begin{remark}
		As a final note in this section, we would like to comment on the relation between the receiver operations needed for RSMA, SDMA and NOMA. A conventional SDMA receiver does not detect or decode a common stream, and performs detection of its intended private signal only by treating all interference as noise. On the other hand,  a conventional NOMA receiver in a two-user scenario performs decoding of the common stream (recall from Remark 4 that the common stream in NOMA includes the full message of one user and no contributions from the message of the other user). For the user whose message is carried in the common stream, this can be interpreted as treating the interference as noise, as it does not continue with any decoding operations after the common message. For the other user (whose message is carried in one private stream), this can be interpreted as fully decoding the interference, as it performs interference cancellation to continue with decoding its own message. This underlines a key feature of RSMA - the capability to bridge the extreme regimes of decoding interference and treating it as a noise.
	\end{remark}
	
	\subsection{RSMA PHY Layer Architecture}
	\label{sec:PHYdesign} 
	\labelsubseccounter{VII-A}
	Fig. \ref{fig:PHYdesign} illustrates the two-user PHY layer transmitter and receiver architecture of  1-layer RS for the two-user MISO BC including message splitting, finite-alphabet modulation,  finite-length polar coding, AMC, and SIC.  The architecture can be extended in a straightforward manner to the PHY layer of the RSMA schemes discussed in Section \ref{sec:principle}. In the following, the transmitter and receiver structures are explained in detail.
	%and readers are referred to  \cite{Onur2020LLS} for more details on each module. 
	%%%
	\subsubsection{Transmitter}
	%%%
	At the transmitter,  message $W_k$ intended for user-$k$ is split into two sub-messages $W_{c,k}, W_{p,k}, k\in\{1,2\}$. The two common sub-messages $W_{c,1},W_{c,2}$ are combined into one common message $W_c$.
	%%%
	The three obtained messages $W_c, W_{p,1}, W_{p,2}$ contain $K_c, K_{p,1}, K_{p,2}$ information bits, respectively, which are assumed to be independently and uniformly distributed in  binary fields of dimensions $K_c, K_{p,1}, K_{p,2}$, respectively.
	%%%
	The three obtained messages  $W_c, W_{p,1}, W_{p,2}$ are independently encoded into codewords $\nu_c, \nu_{p,1}, \nu_{p,2}$ with code blocklengths $N_c, N_{p, 1}, N_{p,2}$, respectively.
	%%%
	%Polar coding is considered here due to its strong error correction capabilities. 
	%Readers are referred to \cite{arikan2009polar} for more information of polar codes.
	%%%
	%Note that channel coding of RSMA is not limited to polar coding, and other types of channel codes such as low-density parity-check (LDPC) \cite{hongzhi2021RSLDPC} have also been applied to RSMA. 
	
	%%%
	\par
	Before passing the codewords to the modulators, the bits in 
	$\nu_c, \nu_{p,1}, \nu_{p,2}$ are interleaved 
	based on the BICM concept  in order to further improve the coding performance especially for high-order modulation \cite{guseppe1998BICM}.
	%%%
	The interleaved bit vectors $\nu'_c, \nu'_{p,1}, \nu'_{p,2}$ are respectively modulated into common symbol stream $s_c$ and private symbol streams $s_1, s_2$.
	%%%
	%The modulation schemes for $s_c, s_1, s_2$  are adaptively selected from  four candidate modulation schemes 4-QAM, 16-QAM, 64-QAM and 256-QAM based on the AMC algorithm.
	%%%
	Aiming at enhancing the system throughput, an AMC algorithm uses the rates $C_1, C_2, R_1, R_2$ calculated based on the designed precoders in Section \ref{sec:resourceAllocation} 
	and the Shannon bound in (\ref{eq:rate1RS})--(\ref{eq:commonstream2}) 
	as link quality metrics to determine the corresponding modulation schemes and the coding rates for the symbol streams. 
	%%%
	%Note that the modulation scheme is not limited to QAM.
	%Besides QAM, PSK modulation (other than QPSK) has also been applied to RSMA in \cite{hongzhi2020LLS,hongzhi2021RSLDPC}.
	%%%
	After modulation, the modulated symbol streams are then mapped to transmit antennas via precoding matrix $\mathbf{P}$.
	
	\begin{remark}
		The PHY layer transceiver architecture described in \cite{Onur2020LLS} employs polar codes \cite{arikan2009polar} for channel coding and the modulation formats 4-QAM, 16-QAM, 64-QAM and 256-QAM. However, any other types of coding and modulation techniques can be applied in an RSMA transceiver. For example, Low-Density Parity-Check (LDPC) codes and Phase-Shift Keying (PSK) modulation have been employed in the RSMA transceiver architectures described in \cite{hongzhi2020LLS,hongzhi2021RSLDPC}.
	\end{remark}

	%\textcolor{magenta}{.}
	% \par
	% \underline{\textit{Polar encoder}}:
	% %%%
	% \par
	% \underline{\textit{Adaptive modulator}}:
	% %%%
	% \par
	% \underline{\textit{Linear precoder}}:
	% %%%
	% \par
	% \underline{\textit{AMC algorithm}}:
	%%%
	\subsubsection{Receiver}
	At  user-$k$,  MMSE equalizer $g_{c,k}^{\textrm{MMSE}}$ is first employed to detect the common symbol stream $s_c$. The MMSE equalizer is calculated by minimizing mean square error (MSE) metric $\mathbb{E}\{|g_{c,k}y_k-s_c|^2\}$, and obtained as follows
	\begin{equation}
		g_{c,k}^{\textrm{MMSE}}=\frac{\mathbf{p}_c^H\mathbf{h}_k}{\left|\mathbf{{h}}_{k}^{H}\mathbf{{p}}_{c}\right|^{2}+\sum_{l=1}^K\left|\mathbf{{h}}_{k}^{H}\mathbf{{p}}_{l}\right|^{2}+\sigma_{n,k}^2}.
	\end{equation}
	%%%
	\par 
	Log-likelihood ratios (LLRs) are  then calculated for soft decision (SD) decoding from the equalized symbols. 
	%%%
	The bit LLRs are sent to the bit deinterleaver and then to the channel decoder to obtain common message bits $\widehat{W}_c$. Finally, the intended common part $\widehat{W}_{c,k}$ is  extracted from $\widehat{W}_c$. 
	%To enhance the error performance of the polar codes, cyclic redundancy check (CRC)-aided successive cancellation list (SCL) decoder is employed \cite{Tal2015SCL}.
	%%%
	After obtaining $\widehat{W}_c$, user-$k$  employs hard decision (HD) SIC to reconstruct 
	$s_c$ from $\widehat{W}_c$  by replicating the operations at the transmitter. $s_c$ is then removed from the received signal as $y'_k=y_k-\mathbf{h}_k^H\mathbf{p}_cs_c$. MMSE equalizer $g_{k}^{\textrm{MMSE}}$ is employed to detect the private symbol stream $s_k$. The MMSE equalizer is calculated by minimizing MSE $\mathbb{E}\{|g_{k}y'_k-s_k|^2\}$, and obtained as follows 
	%The resulting MMSE equalizer $g_{k}^{\textrm{MMSE}}$ is
	\begin{equation}
		g_{k}^{\textrm{MMSE}}=\frac{\mathbf{p}_k^H\mathbf{h}_k}{\sum_{\substack{l=1, \\ l \neq k}}^K\left|\mathbf{{h}}_{k}^{H}\mathbf{{p}}_{l}\right|^{2}+\sigma_{n,k}^2}.
	\end{equation}
	%%%
	Following the same decoding procedure, $\widehat{W}_{c}$ and  $\widehat{W}_{p,k}$ are obtained. By merging  $\widehat{W}_{p,k}$ and $\widehat{W}_{c,k}$, $\widehat{W}_{k}$ is recovered. 
	% %%%
	% \par
	% \underline{\textit{Equalizer}}:
	% %%%
	% \par
	% \underline{\textit{LLR calculator}}:
	% %%%
	% \par
	% \underline{\textit{Decoder}}:

	\section{Comparison of Multiple Access Schemes}
	\label{sec:RSMAvsOtherMA}
	In this section, we comprehensively compare downlink RSMA with existing MA schemes in terms of their design principles and frameworks, DoF performance, SE performance, LLS performance, and their complexities. The advantages and disadvantages of RSMA are summarized at the end of this section.
	\subsection{Framework Comparison}
	\label{sec:FrameworkCompare}
\begin{figure}[t!]
		\centering
		\includegraphics[width=2.4in]{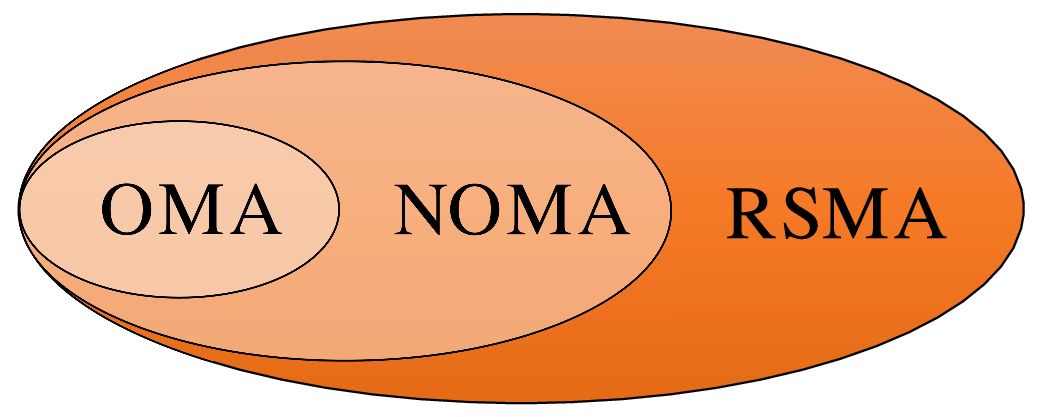}%
		\caption{Comparison of MA schemes for $K$-user uplink transmission.}
		\label{fig:RSMAvsOtherMAUL}
		\vspace{-2mm}
	\end{figure}

\begin{figure}[t!]
		\centering
		\includegraphics[width=3.3in]{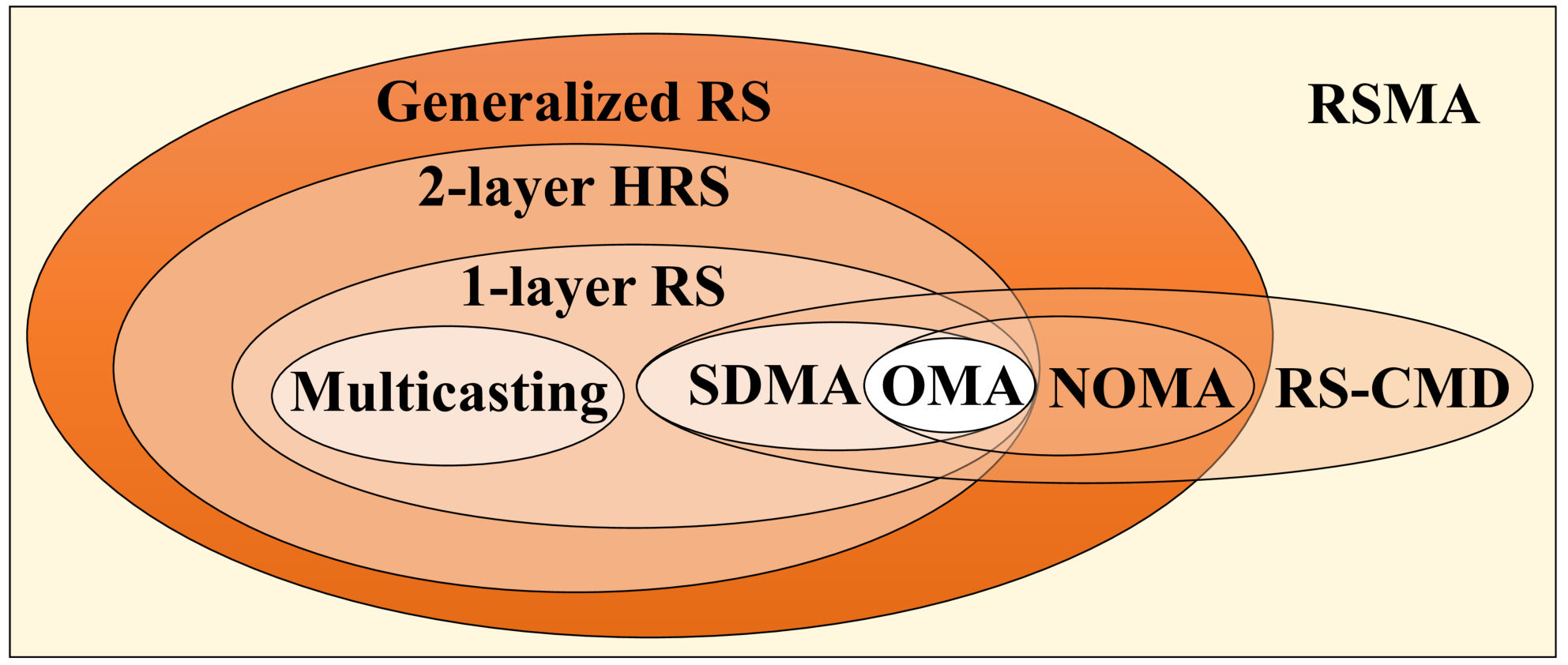}%
		\caption{Comparison of  linearly precoded MA schemes for $K$-user downlink transmission.}
		\label{fig:RSMAvsOtherMA}
		\vspace{-2mm}
	\end{figure}

	\labelsubseccounter{VI-A}
	Fig. \ref{fig:RSMAvsOtherMAUL} and Fig. \ref{fig:RSMAvsOtherMA}  illustrate the relationship between different MA schemes for uplink and downlink transmission, respectively. RSMA is more general than NOMA and OMA in both cases. 
	%%%
	For uplink transmissions, RSMA in Fig. \ref{fig:uplinkRS} reduces to uplink NOMA if the transmit power  at each user-$k$ is fully allocated to one stream, i.e., $s_{k1}$,  such that $W_k$ is encoded directly into $s_{k1}$ directly. When only one user is active in each time-frequency resource block, uplink NOMA further reduces to OMA.
	%%%
	\par
	%%%
	For downlink transmission, Fig. \ref{fig:RSMAvsOtherMA} compares linearly precoded MA schemes including multicasting, SDMA, OMA, NOMA, and four linearly precoded RSMA schemes, namely, generalized RS, 2-layer HRS, 1-layer RS, and RS--CMD.
	%%%
	Generalized RS is a universal linearly precoded scheme that comprises all other schemes except RS-CMD as sub-schemes.
	By allocating non-zero power to the $K$-order stream, $|\mathcal{K}_i|$-order streams, and $1$-order streams while turning off all other streams, generalized RS reduces to 2-layer HRS.
	%%%
	2-layer HRS  reduces to 1-layer RS when the $|\mathcal{K}_i|$-order streams of 2-layer HRS are turned off (with zero transmit power allocated to them). 
	%%%
	RS-CMD is not a special instance of other RSMA schemes as RS-CMD does not combine and jointly encode user messages. Though the common streams in RS-CMD are decoded by all or at least several users, each stream is  intended only for a single user.
	%%%
	\par
	SDMA illustrated in Fig. \ref{fig:twoUserBaseline}(b) is a sub-scheme of all linearly precoded RSMA schemes. By turning off the power allocated to all common streams, RSMA reduces to SDMA. 
	%%%
	In contrast, NOMA based on SC--SIC  in Fig. \ref{fig:twoUserBaseline}(b) or SC--SIC per group in  Fig. \ref{fig:SCSICperGroup} utilizes each common stream to fully encode the entire message of a single user.  It is a sub-scheme of generalized RS. 
	%%%
	By optimizing the group of users to decode each common stream, RS-CMD can be made more general than NOMA. Therefore, NOMA is shown as a special case of RS-CMD in Fig.  \ref{fig:RSMAvsOtherMA}.
	%%%
	For OMA, illustrated in  Fig. \ref{fig:twoUserBaseline}(a),  only a single user can be active in each time-frequency resource block, i.e., the transmit power is fully allocated to a single private stream. It is the simplest MA scheme and is therefore a  sub-scheme of all other MA schemes except for PHY layer multicasting. 
	%%%
	PHY layer multicasting allows the transmitter to  serve multiple users simultaneously by jointly encoding  the messages intended for different users into one common stream  decoded by all users. In other words, multicasting is a sub-scheme of 1-layer RS/2-layer RS/generalized RS when the transmit power is fully allocated to the $K$-order common stream.
	%%%
	%%%
	%%%
	%%%
	%%%
	%%%
	\begin{table*}
		\centering
		\caption{Comparison of different MA schemes.}
		\label{tab: design comparison}
		% 		\addtolength\tabcolsep{-3pt}
		\begin{tabular}{|L{1.5cm}|L{2.2cm}|L{2.2cm}|L{3.3cm}|L{3.3cm}|L{2.8cm}|}
			\hline
			\centering\textbf{Multiple Access}    & \centering \textbf{OMA}      & \centering \textbf{SDMA}            & \multicolumn{2}{c|}{ \textbf{NOMA}}                                                                                                                                                                                                                                                                        &                               \multicolumn{1}{c|}{ \textbf{RSMA}  }                                                \\ \hline
			\centering\textbf{Strategy}    &\centering \textbf{LP}     &\centering \textbf{MU--LP}             & \centering\textbf{Multi-antenna NOMA ($G=1$)}                                                                       & \centering\textbf{Multi-antenna NOMA ($G>1$)}                                                                                                                                                                                        &\multicolumn{1}{c|}{ \textbf{All forms of RS} }                                               \\ \hline
			\textbf{Design principle}          & Orthogonal resource allocation to avoid interference          & Fully treat interference as noise              & Fully decode interference                                                              & Fully decode interference in each group and treat interference between groups as noise                                                                                                & Partially decode interference and partially treat interference as noise \\ \hline
			\textbf{Decoder architecture}    & Treat interference as noise   & Treat interference as noise    & SIC at receivers                                                                       & SIC at receivers                                                                                                                                                                                      & SIC at receivers                                                        \\ \hline
			\textbf{Ideal user deployment scenario} & Any angle between user channels and any disparity in channel strengths   & Users channels are (semi-) orthogonal with similar channel strengths. & Users experience aligned channel directions and a large disparity in channel strengths. & Users in the same group experience aligned channel directions and a large disparity in channel strengths. Users in different groups experience orthogonal channels. & Any angle between channels and any disparity in channel strengths       \\ \hline
			\textbf{Network load}       & Only one active user (in each radio resource)                     & Better suited for underloaded network                                                       & Better suited for overloaded network                                                                                                                             & Better suited for underloaded network                                    & Suited to any network load                                              \\ \hline
		\end{tabular}
	\end{table*}
	%%%
	\par Table \ref{tab: design comparison} summarizes the comparison of the different MA schemes in terms of their design principles as well as the user deployment scenarios and network loads they are best suited for. 
	%%%
	Apparently, the most important characteristic  that makes RSMA distinct from the other MA schemes lies in its flexible interference management policy of partially decoding interference and partially treating residual interference as noise, which  allows RSMA to generalize and encompass multi-antenna NOMA, SDMA, OMA, and multicasting as sub-schemes. Most importantly, RSMA smoothly bridges different sub-schemes without hard switching between them. 
	%%%
	\begin{table}
		\caption{Message-to-stream\! mapping\! in\! the two-user\! MISO\! BC \cite{bruno2019wcl}.}
		\centering
		\begin{tabular}{|p{.2\textwidth}|p{.6\textwidth}|p{2.3cm}|p{2.5cm}|}
			\hline
			\multicolumn{1}{|c|}{}             & \multicolumn{1}{c|}{$s_1$}     & \multicolumn{1}{c|}{$s_2$}     & \multicolumn{1}{c|}{$s_c$}             \\ \hline
			\multicolumn{1}{|c|}{ SDMA}         & \multicolumn{1}{c|}{\color{blue} $\,\,\quad \quad W_1\quad\quad\,\,$}     & \multicolumn{1}{c|}{\color{blue} $W_2$}     & \multicolumn{1}{c|}{\color{red} --}                \\ \hline
			\multicolumn{1}{|c|}{ NOMA}         & \multicolumn{1}{c|}{\color{blue} $W_1$}     & \multicolumn{1}{c|}{\color{blue} --}        & \multicolumn{1}{c|}{\color{red} $W_2$}             \\ \hline
			\multicolumn{1}{|c|}{ OMA}          & \multicolumn{1}{c|}{\color{blue} $W_1$}     & \multicolumn{1}{c|}{\color{blue} --}        & \multicolumn{1}{c|}{\color{red} --}                \\ \hline
			\multicolumn{1}{|c|}{Multicasting} & \multicolumn{1}{c|}{\color{blue} --}        & \multicolumn{1}{c|}{\color{blue} --}        & \multicolumn{1}{c|}{\color{red} $W_1,W_2$}         \\ \hline
			\multicolumn{1}{|c|}{1-layer RS}           & \multicolumn{1}{c|}{\color{blue} $W_{p,1}$} & \multicolumn{1}{c|}{\color{blue} $W_{p,2}$} & \multicolumn{1}{c|}{\color{red} $W_{c,1},W_{c,2}$} \\ \hline
			\multicolumn{1}{c}{}  & \multicolumn{2}{c}{\color{blue} decoded by its intended user and}            &  \multicolumn{1}{c}{\color{red} decoded by}                         \\
			\multicolumn{1}{c}{}  &\multicolumn{2}{c}{\color{blue} treated as noise by the other user}          & \multicolumn{1}{c}{ \color{red} both users   }                         
		\end{tabular}
		\label{fig_mapping}
		%\vspace{-0.3cm}
	\end{table}
	%%%
	\begin{table*}
		\centering
		\caption{Qualitative comparison of the complexity of different MA schemes.}
		\label{tab: complexity}
		\addtolength\tabcolsep{-5pt}
		\begin{tabular}{|L{1.36cm}|L{1.6cm}|L{1.7cm}|L{2.3cm}|L{2.5cm}|L{1.7cm}|L{1.9cm}|L{1.95cm}|L{1.8cm}|}
			\hline
			\centering\textbf{Multiple Access}   &  \centering\textbf{OMA}  & \centering\textbf{SDMA}     & \multicolumn{2}{c|}{\textbf{NOMA}}                                                  & \multicolumn{4}{c|}{\textbf{RSMA}}  \\ \hline
			\centering\textbf{Strategy}        & \centering \textbf{LP}      & \centering\textbf{MU--LP}        & \centering\textbf{Multi-antenna NOMA ($G=1$)}  & \centering\textbf{Multi-antenna NOMA ($G>1$)}  & \centering \textbf{1-layer RS}       & \centering\textbf{2-layer HRS}                    & \centering \textbf{Generalized RS}    &  \multicolumn{1}{c|}{\textbf{RS-CMD}}   \\ \hline
			\textbf{Encoder complexity}& Encode $K$ streams & Encode $K$ streams & Encode $K$ streams & Encode $K$ streams
			& Encode $K+1$ streams                
			& Encode $K+G+1$ streams	
			& Encode $2^K-1$ streams 
			& Encode $2K$ streams\\ \hline
			\textbf{Scheduler complexity} & Complex as OMA requires subcarrier/time-slot allocation for all users.  & Complex as MU--LP requires to pair semi-orthogonal users with similar channel gains.  & Very complex as it requires finding aligned users and decide between $K!$ user orderings. & Very complex as it requires to divide users into orthogonal groups, with aligned users in each group. Decide between  $\sum_{k=1}^KS(K,k)$ grouping method and  at most $K!$  decoding orders  for each grouping method. 	& Simple user scheduling as it can cope with any user deployment scenario, and does not rely on user grouping or user ordering.  &	Decide between  $\sum_{k=1}^KS(K,k)$ groupings without decoding order. 
			& Decide between $\prod_{k=2}^{K-1} \binom{K}{k}! $  decoding orders.                  
			& Decide upon $\sum_{k=1}^KS(K,k)$ grouping methods and at most  $(K!)^K$ decoding orders.  \\ \hline
			\textbf{Receiver complexity} & Does not require SIC. & Does not require SIC.  & Requires $K-1$ layers of SIC at each user.  Sensitive to error propagation.                                                        & Requires $|\mathcal{K}_i|-1$ layers of SIC at each user in  group-$g$.    Sensitive to error propagation.          & Requires 1 layer of SIC at each user.  Less sensitive to error propagation (compared to NOMA, and other more complex RS schemes).            & Requires 2 layers of SIC at each user.  Less sensitive to error propagation (compared to NOMA, and other more complex RS schemes).                                                                                & Requires $2^{K-1}-1$ layers of SIC at each user. Sensitive to error propagation.             & Requires $K$ layers of SIC at each user. Sensitive to error propagation.                           \\ \hline
		\end{tabular}
	\end{table*}
	%%%
	\par 
	In the simple two-user case, 1-layer RS (see Fig. \ref{fig:twoUserBaseline}(d)) becomes a super-set of not only OMA (see Fig. \ref{fig:twoUserBaseline}(a)), SDMA (see Fig. \ref{fig:twoUserBaseline}(b)), and multicasting, but also NOMA (see Fig. \ref{fig:twoUserBaseline}(c)).
	%%%
	As illustrated in Table \ref{fig_mapping}, OMA, SDMA, NOMA, and multicasting are particular instances of 1-layer RS. 
	%%%
	Specifically, SDMA is a special case of 1-layer RS by forcing $\left \| \mathbf{p}_c \right \|^2=0$.
	In this way, $W_k$ is directly encoded into $s_k$.
	%%%
	By encoding message $W_2$ entirely into $s_c$ (i.e., $W_c=W_2$) and $W_1$ into $s_1$ while turning off  $s_2$ ($\left \| \mathbf{p}_2\right \|^2=0$), we obtain  NOMA. 
	%%%
	% In such case, user-1 is required to fully decodes the interference from the message of user-2. 
	%%%
	When only user-1 is scheduled (i.e., $\left \| \mathbf{p}_c \right \|^2=\left \| \mathbf{p}_2 \right \|^2=0$), we obtain OMA.  
	%%%
	When messages  $W_1, W_2$ are both encoded into $s_c$ (i.e., $W_c=\{W_1,W_2\}$) and the private streams are turned off  ($\left \| \mathbf{p}_1 \right \|^2=\left \| \mathbf{p}_2 \right \|^2=0$), we obtain multicasting. 
	%%%
	Note that the multicasting specified in  Table \ref{fig_mapping} is from a  perspective  of  PHY layer transmission regardless of the content that $s_c$ conveys, as discussed in Remark \ref{remark:multicast}.
	%%%
	
	\subsection{Complexity Comparison}
	\label{sec:complexitycompare}
	\labelsubseccounter{VI-B}
	%%%
	\par The  complexity of different linearly precoded strategies are qualitatively compared in Table \ref{tab: complexity} in terms of their encoder complexity, scheduler complexity, and receiver complexity. 
	%%%
	\subsubsection{Encoder complexity}
	Due to the use of message splitting, more streams need to be encoded at the RSMA transmitter. The encoder complexity of RSMA is therefore higher than those of other MA schemes. Among the linearly precoded RSMA schemes, 1-layer RS has the lowest encoding complexity with only one additional stream (compared to the SDMA and NOMA baselines where $K$ messages are encoded into $K$ streams) encoded at the transmitter in the $K$-user case.  For 2-layer  HRS,  as there is one group-specific common stream for each user group,  in total $G$  additional common streams are encoded at the transmitter besides the  common stream for all users. 
	%%%
	Therefore, the encoder complexity difference between  1-layer RS/2-layer HRS and  SDMA/NOMA\footnote{Here, the encoder complexity is measured in terms of the number of encoded streams.}  is invariant to the number of users $K$ while the encoder complexity difference between  RS-CMD/generalized RS and  SDMA/NOMA increases with $K$ due to the large number of  sub-messages split from each message, as illustrated in Table \ref{tab:RSschemesCompare}.
	%%%
	\subsubsection{Scheduler complexity}
	Even though SDMA has the lowest  encoder and receiver complexities, it entails a relatively high scheduler complexity due to the requirement of pairing users with semi-orthogonal channels and similar channel strengths to achieve high performance. 
	%%%
	Fig. \ref{fig: operationRegion} in Section \ref{sec:performanceCompare} further shows that SDMA is mainly beneficial when the users' channels are sufficiently orthogonal.
	%%%
	Moreover, as discussed in Section \ref{sec:motivations}, well-designed user scheduling requires accurate CSIT and leads to high signaling overhead and latency. 
	%%%
	The SIC receivers used in  NOMA and RSMA introduce additional scheduler complexity compared with OMA and SDMA.
	%%%
	Two additional aspects have to be considered at the transmitter due to the use of  SIC, namely, the decoding order  and the user grouping. The coupled nature of user grouping,  decoding order, precoders,  and user selection  requires a joint optimization at the transmitter \cite{ding2017survey}.
	%%%
	Among all strategies using SIC, multi-antenna NOMA ($G>1$), 2-layer HRS, and RS-CMD bear the highest user grouping complexity. As discussed in \cite{mao2017rate}, the total number of possible user groupings for these three schemes is $\sum_{k=1}^KS(K,k)$, where  $S(K,k)=\frac{1}{k!}\sum_{i=0}^k(-1)^i\binom{k}{i}(k-i)^K$ is a Stirling set number \cite{riordan2012introduction} representing the total number of partitions of a set of $K$ elements into $k$ non-empty sets.
	%%%
	\par 
	In terms of the decoding order, multi-antenna NOMA, generalized RS, and RS-CMD all entail inevitably high decoding order complexity. 
	%%%
	For multi-antenna NOMA ($G=1$), all users employ the same decoding order to decode the user messages and the transmitter should select between  $K!$ possible decoding orders for the $K$ streams. 
	%%%
	Note that for multi-antenna NOMA ($G=1$), it is advantageous for the scheduler to  select users with aligned channels and significant channel strength differences, which however increases the scheduler complexity.
	%%%
	For multi-antenna NOMA ($G>1$), as the user grouping is jointly optimized with the decoding order, at most $K!$ decoding orders have to be considered in each user group. 
	%%%
	For generalized RS, though the decoding order starts from the $K$-order streams downwards to the $1$-order streams (see Section \ref{sec: generalized RS}), there are $\binom{K}{k}$ $k$-order streams for $2 \leq k\leq K-1$, which results in $\binom{K}{k}!$ possible decoding orders. Therefore, the transmitter has to select between $\prod_{k=2}^{K-1} \binom{K}{k}! $ possible decoding orders for users to decode all intended streams \cite{mao2017rate}. 
	%%%
	For RS-CMD, if it is assumed that the $K$ common streams are decoded at all users as in Section \ref{sec:RSCMD},  in total $(K!)^K$ decoding orders need to be considered at the transmitter without the issue of user grouping. 
	%%%
	Otherwise, the transmitter has to jointly decide the decoding order and the group of common streams decoded at each user. 
	%%%
	\par
	In summary, both multi-antenna NOMA ($G>1$) and RS-CMD have relatively the highest scheduler complexity  since the decoding order and user grouping have to be jointly decided. 
	%%%
	For multi-antenna NOMA ($G>1$), users with semi-orthogonal channels should be assigned to different user groups and the user channels within the same  user group should be aligned.
	%%%
	In contrast, 1-layer RS has relatively the lowest scheduler complexity among all aforementioned MA schemes (including OMA, SDMA, NOMA, and all RSMA schemes) as it does not require grouping and ordering, and is less sensitive to user pairing. Hence, 1-layer RS does not require complex user scheduling and pairing.
	%%%
	Generally, all RSMA schemes including 1-layer RS, 2-layer HRS, generalized RS, and RS-CMD are suited for users with any channel strength disparity and channel directions.
	%%%
	As per Fig. 13 in \cite{mao2019TCOM}, surprisingly generalized RS and 1-layer RS without user scheduling outperform SDMA with user scheduling when the CSIT is sufficiently inaccurate.

	\subsubsection{Receiver complexity}
	\par Both OMA and SDMA have comparatively the lowest receiver complexity as they do not need SIC.
	%%%
	Among the  schemes using SIC (including NOMA and all RSMA schemes), 1-layer RS and 2-layer HRS entail the lowest receiver complexity. In the $K$-user scenario, the number of SIC operations required at each user is $1$ for 1-layer RS and  $2$ for 2-layer HRS, which is relatively low and  independent of the number of users $K$. Hence, both 1-layer RS and 2-layer HRS are not severely affected by error propagation.
	%%%
	All remaining schemes, including NOMA (for a fixed user group $G$),  generalized RS, and RS-CMD, require several SIC operations depending on $K$, which  not only leads to  an increased transceiver complexity  but also increases the susceptibility to SIC error propagation. 
	%%%
	Specifically, each user  requires $K-1$ layers of SIC in multi-antenna NOMA ($G=1$) while each user from user group-$i$  requires $|\mathcal{K}_i|-1$ layers of SIC in multi-antenna NOMA ($G>1$)\footnote{For a certain decoding order, the number of SIC layers used at different users for decoding the data streams of other users is different. However, as the decoding order depends on the users' instantaneous CSI, all users are required to implement the maximum number of SIC layers as each of them may have to decode the streams of all other users (in the same user group) for some channel conditions.}, where $|\mathcal{K}_i|\leq K$ is the number of users in  group-$i$\footnote{Readers are referred to \cite{bruno2021MISONOMA} for more details on the system model of multi-antenna NOMA.}.
	%%%
	% Though SC--SIC simplifies the scheduler complexity at the transmitter (without user grouping), it increases the receiver complexity.
	%%%
	The receivers for RS-CMD and generalized RS are both complex as $K$ and $2^{K-1}-1$ layers of SIC are required, respectively. Though generalized RS provides a higher flexibility in interference management compared with the other schemes, it also entails a higher receiver complexity.
	%%%
	In view of the high transmitter and receiver complexities of RS-CMD and generalized RS, attempts towards low-complexity versions have been made \cite{Zheng2020JSAC,cran2019wcl,Ala2019IEEEAccess,alaa2020cranimperfectCSIT} by controlling the number of common streams  decoded at each user and simplifying the decoding order design.
	%%%
	Moreover, readers are reminded that there are two motivations for introducing  generalized RS, one is to explore the best achievable performance at the expense of a higher complexity at the transceivers, and the other one is to identify which common stream(s) lead(s) to the highest performance gains, and from there identify which stream(s) could be turned off so as to reduce the encoding and decoding complexities.
	%%%
	\subsection{Performance Comparison}
	\label{sec:performanceCompare}
	\labelsubseccounter{VI-C}
	Next, we evaluate the performance  of the considered MA schemes in terms of their DoF region, rate region,  operational region, user fairness, and LLS performance.
	%%%
	Here, the operational region characterizes the channel conditions of the users which are favorable for a given MA scheme \cite{bruno2019wcl}.
	%%%
	\subsubsection{DoF region}
	The degrees-of-freedom (DoF)  (a.k.a. multiplexing gain) of user-$k$ achieved by communication scheme $j, j\in\{\textrm{N, M, R}\}$, (N, M, R  respectively stand for  NOMA, MU--LP, RSMA) are defined as follows
	\begin{equation}
		d_k^{(j)}=\lim_{P\rightarrow \infty}\frac{R_k^{(j)}(P)}{\log_2(P)},
	\end{equation}
	where $R_k^{(j)}(P)$ is the rate of user-$k$ for scheme $j$ under transmit power constraint $P$.
	%%%
	Mathematically, DoF $d_k^{(j)}$ is the first-order approximation of the rate of user-$k$  at high SNR and is therefore the pre-log factor of $R_k^{(j)}(P)$ at high SNR. It can  be also interpreted as the number or fraction of interference-free streams that can be simultaneously transmitted from the transmitter to user-$k$ using scheme $j$. The larger $d_k^{(j)}$ is, the faster the rate of user-$k$ increases with SNR. 
	%%%
	\par
\begin{figure}[t!]
		\centering
		\includegraphics[width=0.9\columnwidth]{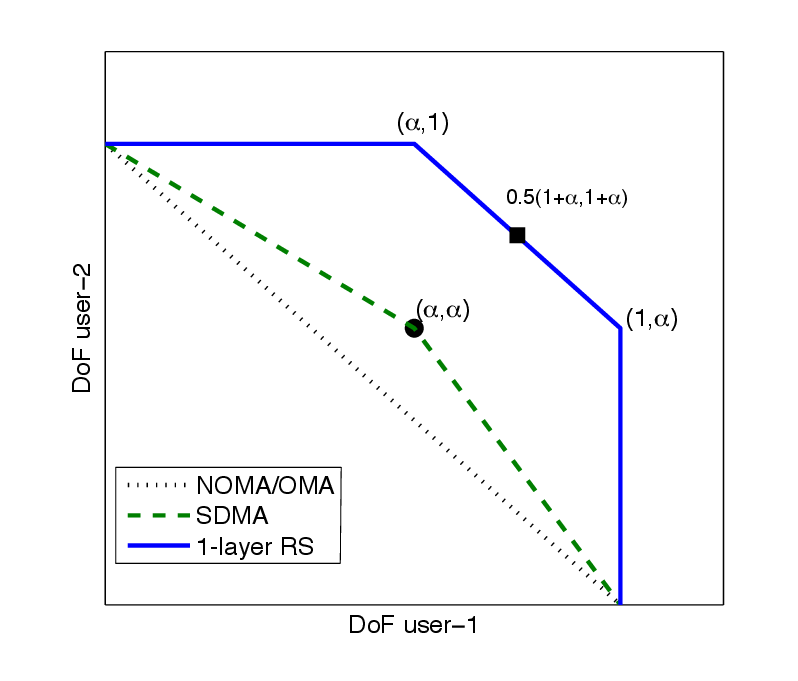}
		\caption{The DoF region for the two-user MISO BC with imperfect CSIT, $\alpha=0.6$ \cite{RSintro16bruno}.}
		\label{fig:DoFregion}
	\end{figure}

 
	A DoF region is defined as the enclosure of all possible DoF tuples ($d_1^{(j)},\ldots,d_K^{(j)}$).
	We recall here that 1-layer RS, which is the simplest RSMA scheme, already achieves the optimal DoF region in the multi-antenna BC with imperfect CSIT \cite{enrico2017bruno,hamdi2019spawc,chenxi2017bruno,Davoodi2021DoF}. All other MA strategies therefore can only achieve the same or worse DoF compared to RSMA.
	%%%
	The  DoF region illustrated in Fig. \ref{fig:DoFregion} for the two-user MISO BC confirms the explicit DoF gain achieved by RSMA.
	%%%
	Each DoF point in Fig. \ref{fig:DoFregion} is calculated based on the imperfect CSIT model in (\ref{eq:imperfectCSIT}), where the estimated channel of user-$k$ at the transmitter $\widehat{\mathbf{h}}_k$ and the corresponding channel estimation error $\widetilde{\mathbf{h}}_k$ have i.i.d. complex Gaussian entries following distributions $\mathcal{CN}(0,\sigma_k^2)$ and $\mathcal{CN}(0,\sigma_{e,k}^2)$, respectively. The error variance scales  with SNR as $\sigma_{e,k}^2\sim \mathcal{O}(P^{-\alpha})$, where  $\alpha\in[0,\infty)$ is the CSIT scaling factor representing the quality of the CSIT  in the high SNR regime \cite{AG2015,NJindalMIMO2006,doppler2010Caire,DoF2013SYang,RS2016hamdi}. 
	%%%
	In limited feedback systems, where users send quantized versions of their channels back to the BS, $\alpha$ can be interpreted in terms of the number of feedback bits.
	%%%
	$\alpha=0$ represents partial CSIT with finite precision, i.e., a constant number of feedback bits, while $\alpha=\infty$ represents perfect CSIT. Normally, we choose $\alpha\in [0,1]$ since  $\alpha=1$ corresponds to perfect CSIT in the DoF sense \cite{DoF2013SYang}.   
	%%%
	%In Fig. \ref{fig:DoFregion}, $\alpha=0.6$ is considered. In such imperfect CSIT scenarios, the optimal sum-DoF achieved by NOMA/OMA is 1, and that achieved by SDMA is $2\alpha$ with a DoF of $\alpha$ achieved by each user.
	%%%
	In Fig. \ref{fig:DoFregion}, $\alpha=0.6$ is considered. 
	In such imperfect CSIT scenarios, SDMA can only achieve a DoF of $\alpha$ for each user.
	%%%
	In comparison, 1-layer RS leverages the common stream to achieve an extra DoF of $1-\alpha$ while keeping an achievable DoF of $2\alpha$ for the two private streams.
	%%%
	By respectively allocating the entire common stream to transmit the sub-messages of user-$1$ and user-$2$, the  corner points of $(1,\alpha)$ and $(\alpha,1)$ of 1-layer RS are achieved. 
	%%%
	The line segment between the two corner points can be achieved by adjusting the amount of the common stream allocated to transmit the sub-messages of the two users. 
	%%%
	%%%
	%%%
	\begin{table*}
		\centering
		\caption{Comparison of sum-DoF  and MMF-DoF of different MA schemes for  perfect and imperfect CSIT \cite{bruno2021MISONOMA}.}
		\label{tab:DoF}
		\addtolength\tabcolsep{-2pt}
		\begin{tabular}{|C{2.1cm}|C{2.3cm}|C{3.2cm}|C{5.6cm}|}
			\hline   \vspace{0.2cm}
			\textbf{Scheme}    \vspace{0.2cm}                 & \textbf{Sum/MMF DoF}                                               & \textbf{Perfect CSIT}                                                                                                                                                  & \textbf{Imperfect CSIT}                                                                                                                                                                                                                                                                                              \\ \hline 
			\multirow{2}{*}{ \textbf{NOMA}} & \vspace{0.1cm}$d_{\textnormal{s}}^{(\textnormal{N})}$  \vspace{0.1cm}          & $\min\left(M,G\right)$               & $\max\left(1,\min\left(M,G\right)\alpha\right)$              \\ \cline{2-4} 
			&
			\textbf{$d_{\textnormal{mmf}}^{(\textnormal{N})}$}
			& \vspace{0.1cm} \begin{tabular}[c]{@{}l@{}}
				$ \left\{\begin{matrix}
					\frac{1}{g}, &M\geq K-g+1 \\ 
					\\
					0,& M < K-g+1
				\end{matrix}\right.$ \end{tabular} \vspace{0.1cm}& \vspace{0.1cm}
			\begin{tabular}[c]{@{}l@{}}$\left\{\begin{matrix}
					\frac{\alpha}{g},  & G>1 \:\textnormal{and}\: M\geq K-g+1 \\ 
					0,&G>1 \:\textnormal{and}\: M < K-g+1\\
					\frac{1}{K}, & \hspace{-2.55cm} G=1 
				\end{matrix}\right.$\end{tabular}                                \vspace{0.1cm}                                                                                                          \\ \hline
			\multirow{2}{*}{\textbf{SDMA}}    & \vspace{0.1cm}\textbf{$d_{\textnormal{s}}^{(\textnormal{M})}$} \vspace{0.1cm}  
			& $\min\left(M,K\right)$                       & $\max\left(1,\min\left(M,K\right)\alpha\right)$                \\ \cline{2-4} 
			& \textbf{$d_{\textnormal{mmf}}^{(\textnormal{M})}$} &
			\vspace{0.1cm} \begin{tabular}[c]{@{}l@{}}
				$ \left\{\begin{matrix}
					1, &M\geq K \\ 
					\\
					0,& M < K
				\end{matrix}\right.$ \end{tabular} \vspace{0.1cm}               & 
			\vspace{0.1cm} \begin{tabular}[c]{@{}l@{}}
				$ \left\{\begin{matrix}
					\alpha, &M\geq K \\ 
					\\
					0,& M < K
				\end{matrix}\right.$ \end{tabular}     \vspace{0.1cm}                          \\ \hline
			\multirow{2}{*}{\textbf{RSMA}}        & \vspace{0.1cm}\textbf{$d_{\textnormal{s}}^{(\textnormal{R})}$} \vspace{0.1cm} 
			& $\min\left(M,K\right)$                       & $1+\left(\min\left(M,K\right)-1\right)\alpha$                  \\ \cline{2-4} 
			& \textbf{$d_{\textnormal{mmf}}^{(\textnormal{R})}$} & 
			\vspace{0.1cm} \begin{tabular}[c]{@{}l@{}}
				$ \left\{\begin{matrix}
					1, &M\geq K \\ 
					\\
					\frac{1}{1+K-M},& M < K
				\end{matrix}\right.$ \end{tabular} \vspace{0.1cm}           
			& 
			\vspace{0.1cm} \begin{tabular}[c]{@{}l@{}}
				$ \left\{\begin{matrix}
					\frac{1+(K-1)\alpha}{K},&\hspace{-2.2cm}M\geq K \\ 
					\frac{1+(M-1)\alpha}{K}, & M < K \:\textnormal{and}\: \alpha\leq\frac{1}{1+K-M}\\ 
					\frac{1}{1+K-M}, &  M < K \:\textnormal{and}\: \alpha > \frac{1}{1+K-M}
				\end{matrix}\right.$ \end{tabular}
			\vspace{0.1cm}
			\\ \hline
		\end{tabular}
	\end{table*}
	%%%
	%%%
	\begin{figure}[t!]
		\centering
		\includegraphics[width=0.9\columnwidth]{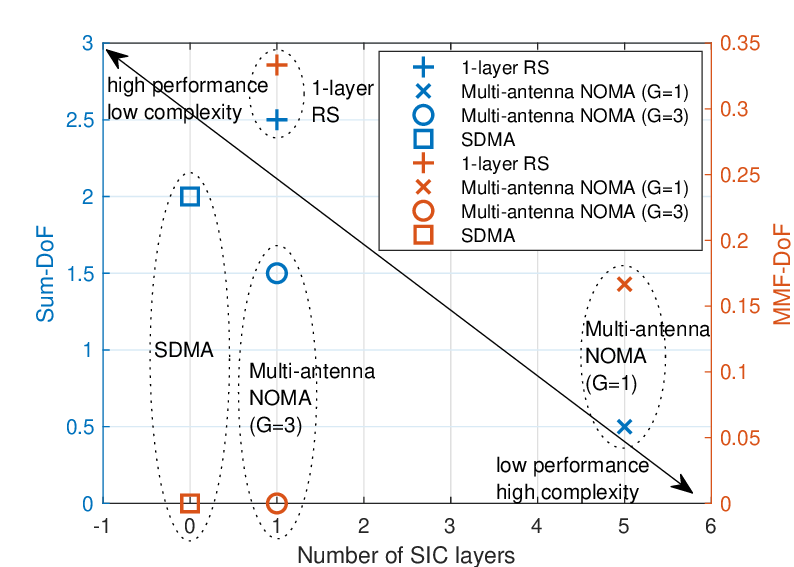}
		\caption{Sum/MMF-DoF  vs. number of SIC layers for MISO BC with imperfect CSIT, $M=4$, $K=6$, $\alpha=0.5$ \cite{bruno2021MISONOMA}.}
		\label{DoF_vs_SIC_imperfect}
	\end{figure}
	%%%
	\par
	Besides the DoF region, sum-DoF and MMF-DoF have been used in the MIMO literature to assess the capability of a strategy to exploit multiple antennas. These performance metrics are defined as follows
	\begin{equation}
		d_{\textnormal{s}}^{(j)}=\lim_{P\rightarrow \infty}\frac{R_{\textnormal{s}}^{(j)}(P)}{\log_2(P)}=\sum_{k=1}^K d_k^{(j)},
	\end{equation}
	\begin{equation}
		d_{\textnormal{mmf}}^{(j)}=\lim_{P\rightarrow \infty}\frac{R_{\textnormal{mmf}}^{(j)}(P)}{\log_2(P)}=\min_{k=1,...,K} d_k^{(j)},
	\end{equation}
	where   $R_{\textnormal{s}}^{(j)}=\sum_{k=1}^K R_k^{(j)}$ is the sum-rate and $R_{\textnormal{mmf}}^{(j)}=\min_{k} \{R_k^{(j)}|k\in\mathcal{K}\}$ is the MMF rate.
	Sum-DoF $d_{\textnormal{s}}^{(j)}$ is  the total  number  of  interference-free  data  streams  that  can  be   transmitted  to  all users  by  employing   scheme $j$.  MMF-DoF  $d_{\textnormal{mmf}}^{(j)}$ (a.k.a. symmetric DoF) are the DoF that can be simultaneously achieved by all users.
	%%%
	The sum-DoF and MMF-DoF of NOMA, SDMA, and RSMA for the MISO BC with perfect and imperfect CSIT have been calculated in \cite{bruno2021MISONOMA} and are summarized in Table \ref{tab:DoF} for convenience.  Without loss of generality, we assume the numbers of users in different user groups are  all equal to $g=|\mathcal{K}_i|, i\in\mathcal{G}$, for SC--SIC per group.  
	%%%
	\par In Fig. \ref{DoF_vs_SIC_imperfect}, the tradeoff between the sum/MMF-DoF and the number of SIC layers for the MISO BC with $M=4$ and $K=6$ is illustrated. Imperfect CSIT is considered and the CSIT scaling factor is $\alpha=0.5$.
	%%%
	As can be observed, multi-antenna NOMA ($G=3$) achieves better sum-DoF but  worse MMF-DoF compared to multi-antenna NOMA ($G=1$). 
	%%%
	1-layer RS achieves the highest sum/MMF-DoF among all schemes and entails a much lower  complexity  compared to NOMA. 
	\begin{figure}[t!]
		\centering
		\begin{subfigure}[b]{\columnwidth}
			\centering
			\includegraphics[width=0.9\textwidth]{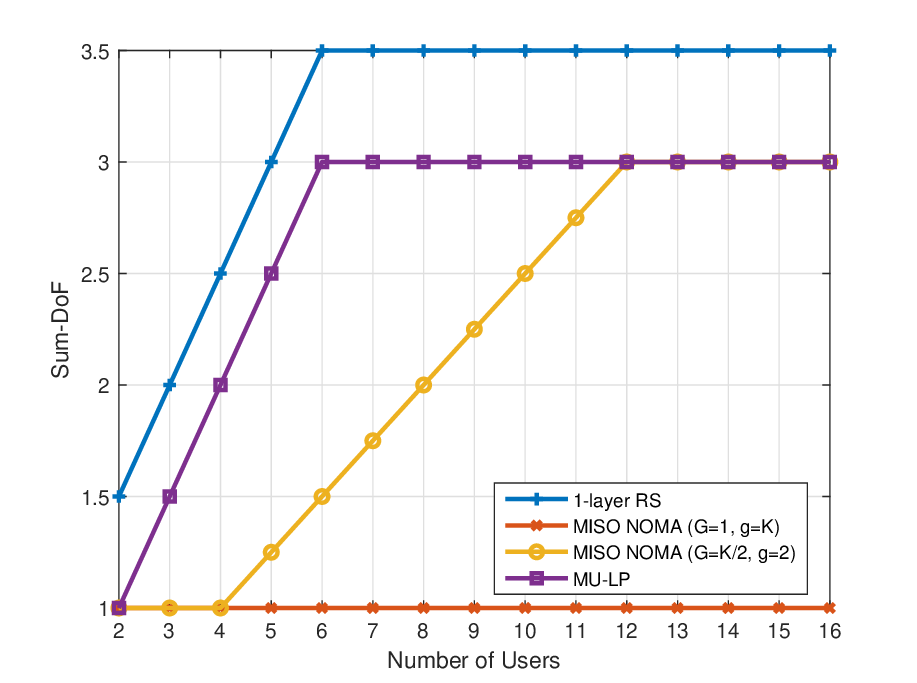}%
			\caption{Sum-DoF vs. number of users $K$}
		\end{subfigure}
		~\\
		\begin{subfigure}[b]{\columnwidth}
			\centering
			\includegraphics[width=0.9\textwidth]{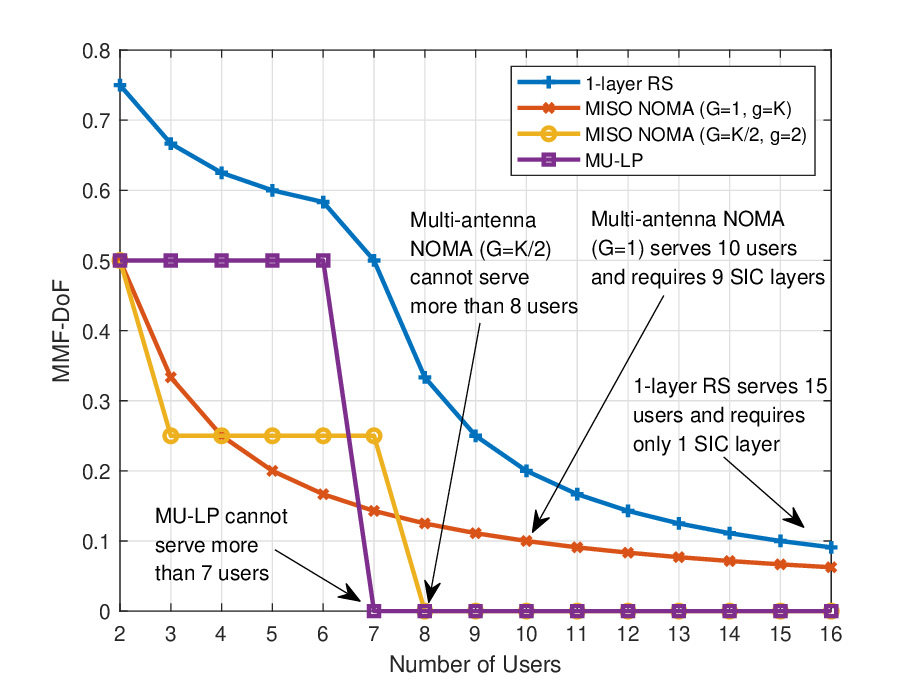}%
			\caption{MMF-DoF vs. number of users $K$}
		\end{subfigure}%
		\caption{Sum/MMF-DoF vs. number of users $K$ for MISO BC with imperfect  CSIT, $\alpha=0.5$, $M=6$ \cite{bruno2021MISONOMA}.}
		\label{DoF_vs_K_M6}
	\end{figure}
	%%%
	%%%
	\begin{figure}
		\centering
		%	\hspace*{0.2cm}
		\begin{subfigure}[b]{0.245\textwidth}
			\centering
			\includegraphics[width=\textwidth]{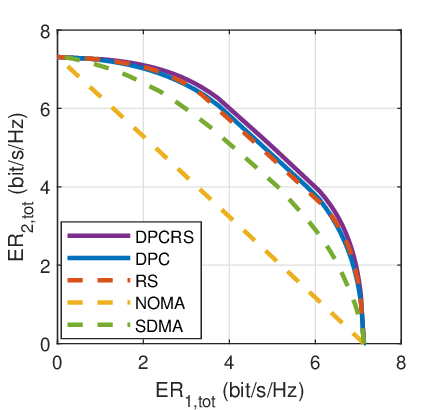}%
			\caption{$\alpha=0.9, \sigma_2^2=1$}
		\end{subfigure}%
		~
		\begin{subfigure}[b]{0.245\textwidth}
			\centering
			\includegraphics[width=\textwidth]{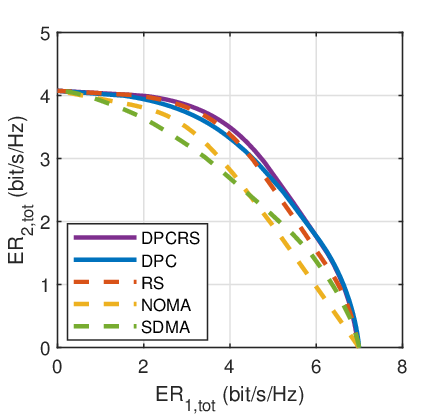}%
			\caption{$\alpha=0.9, \sigma_2^2=0.09$}
		\end{subfigure}%
		\\
		\begin{subfigure}[b]{0.245\textwidth}
			\centering
			\includegraphics[width=\textwidth]{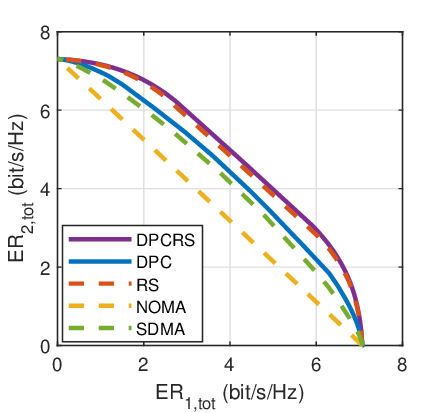}%
			\caption{$\alpha=0.6, \sigma_2^2=1$}
		\end{subfigure}%
		~
		\begin{subfigure}[b]{0.245\textwidth}
			\centering
			\includegraphics[width=\textwidth]{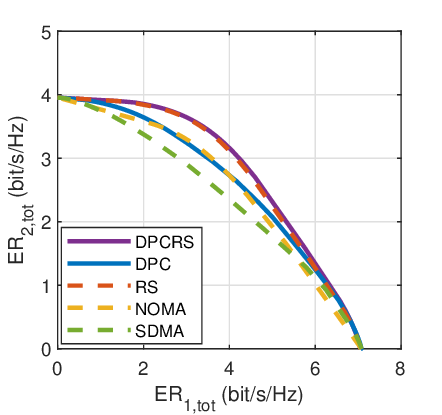}%
			\caption{$\alpha=0.6, \sigma_2^2=0.09$}
		\end{subfigure}%
		\caption{Ergodic rate region comparison of different multiple access schemes with imperfect CSIT, averaged over 100 random channel realizations, $K=2$, SNR = 20 dB \cite{mao2019beyondDPC}.}
		\label{fig: rateRegion}
	\end{figure}
	%%%
	%%%
	\begin{figure*}
		\centering
		%	\hspace*{0.2cm}
		\begin{subfigure}[b]{0.33\textwidth}
			\centering
			\includegraphics[width=\textwidth]{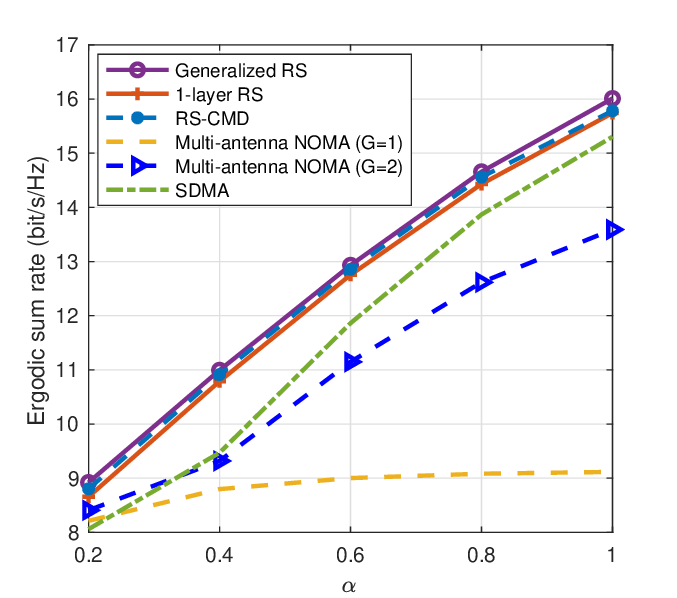}%
			\caption{$\sigma_1^2=\sigma_2^2=\sigma_3^2=1, M=4$}
		\end{subfigure}%
		~
		\begin{subfigure}[b]{0.33\textwidth}
			\centering
			\includegraphics[width=\textwidth]{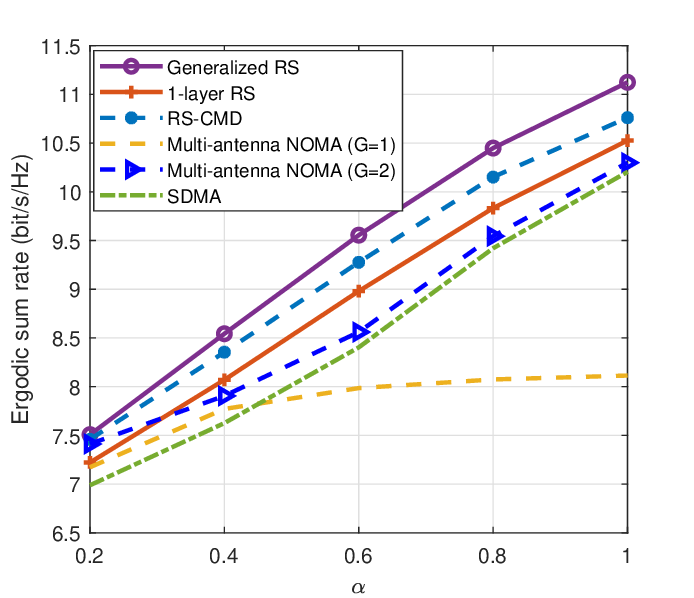}%
			\caption{$\sigma_1^2=\sigma_2^2=\sigma_3^2=1, M=2$}
		\end{subfigure}%
		~
		\begin{subfigure}[b]{0.33\textwidth}
			\centering
			\includegraphics[width=\textwidth]{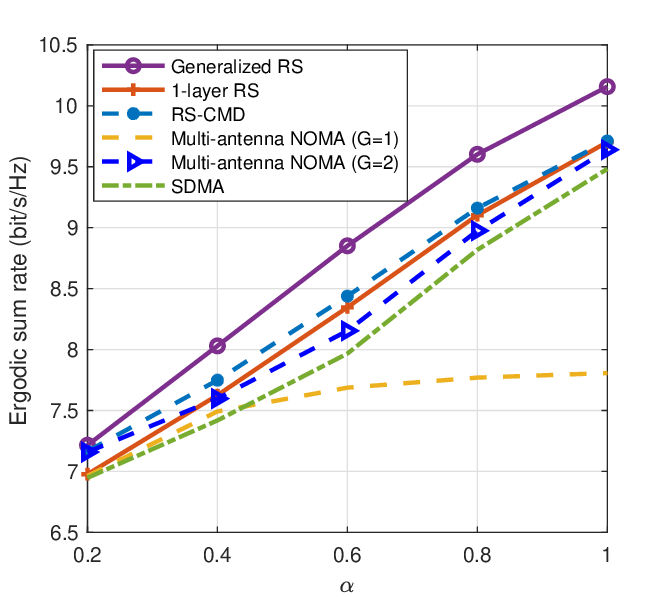}%
			\caption{$\sigma_1^2=\sigma_2^2=1, \sigma_3^2=0.09, M=2$}
		\end{subfigure}%	
		\caption{Ergodic sum rate versus CSIT quality comparison of different MA schemes, averaged over 100 random channel realizations, $K=3$, SNR = 20 dB.}
		\label{fig: ESR vs alpha}
	\end{figure*}
	%%%
	\par 
	Fig. \ref{DoF_vs_K_M6} illustrates the sum- and MMF-DoF for the MISO BC with imperfect CSIT versus the number of users $K$  when the number of transmit antennas is $M=6$ and the CSIT scaling factor is $\alpha=0.5$.
	%%%
	1-layer RS shows an explicit sum- and MMF-DoF gain compared to all other schemes for all $K$. 
	In Fig. \ref{DoF_vs_K_M6}(b), the MMF-DoF of SDMA  and multi-antenna NOMA ($G=1$) drops to 0 when $K>6$ and $K>7$, respectively. This implies that the rate of the worst user does not scale with the SNR and the rate saturates at high SNR. 
	%%%
	To maintain multiple access and fairness among users, multi-antenna NOMA ($G=1$) is more suitable when $K>M$ but it requires $K-1$ layers of SIC at each user. 
	%%%
	Taking 0.1 as the  MMF-DoF threshold for all users in Fig. \ref{DoF_vs_K_M6}(b), we observe that 1-layer RS can serve around 15 users using 1 layer of SIC while multi-antenna NOMA ($G=1$) can only serve 10 users at most using 9 layers of SIC.
	%%%
	Hence, RSMA is significantly more efficient than NOMA since RS with only one layer of SIC can support a much larger number of users than NOMA with multiple layers of  SIC. 
	%%%
	\subsubsection{Rate region}
	\par Motivated by the DoF optimality of RSMA in the high SNR regime, attempts have been made to explore the achievable rate (or SE) of RSMA in the finite SNR regime \cite{RS2016hamdi,mao2019beyondDPC}. Fig. \ref{fig: rateRegion}  evaluates the ER region of different MA schemes averaged over 100 random channel realizations for the MISO BC with imperfect CSIT. 
	%%%
	The BS has $M=2$ antennas and serves two single-antenna users.
	%%%
	The imperfect CSIT model follows (\ref{eq:imperfectCSIT}) 
	and the channel estimates of each user, $\widehat{\mathbf{h}}_k$, have i.i.d. complex Gaussian entries following distribution $\mathcal{CN}(0,\sigma_k^2)$. 
	%%%
	The variance of $\widehat{\mathbf{h}}_1$ is fixed to $\sigma_1^2=1$ while the variance of $\widehat{\mathbf{h}}_2$ varies between $\sigma_2^2=1$ and $\sigma_2^2=0.09$  representing equal channel strength and 10 dB channel strength disparity between the two user channels, respectively. 
	%%%
	ER  is defined in the same way as in (\ref{eq:ER}) and
	$\textrm{ER}_{k,tot}=\overline{R}_{k} + \overline{C}_{k}$.
	%%%
	Each rate pair $\textrm{ER}_{1,tot}, \textrm{ER}_{2,tot}$ at the boundary of the ER region is obtained by solving $\mathcal{P}_3$ in Section \ref{sec:precoderOpt} for a certain weight pair $u_1, u_2$ with $U_n( \mathbf{{P}},\overline{\mathbf{c}})= \sum_{k=1}^Ku_{k}(\overline{R}_{k} + \overline{C}_{k})$  and $R_k^{th}=0$ bit/s/Hz. By changing the weight pair and optimizing the rate pair accordingly, we obtain the entire ER region. The  ER region of DPC/DPCRS is the convex hull of the rate regions for all possible encoding orders. Readers are referred to \cite{mao2019beyondDPC} for details on the optimization.  
	%%%
	In the two-user case, M-DPCRS and generalized RS  respectively reduce to 1-DPCRS and 1-layer RS. In Fig. \ref{fig: rateRegion}, we use the terminology ``DPCRS" to represent ``M-DPCRS" and ``1-DPCRS", and the terminology ``RS" to represent ``Generalized RS" and ``1-layer RS", respectively. 
	%%%
	\begin{figure}
		\centering
		\includegraphics[width=3.0in]{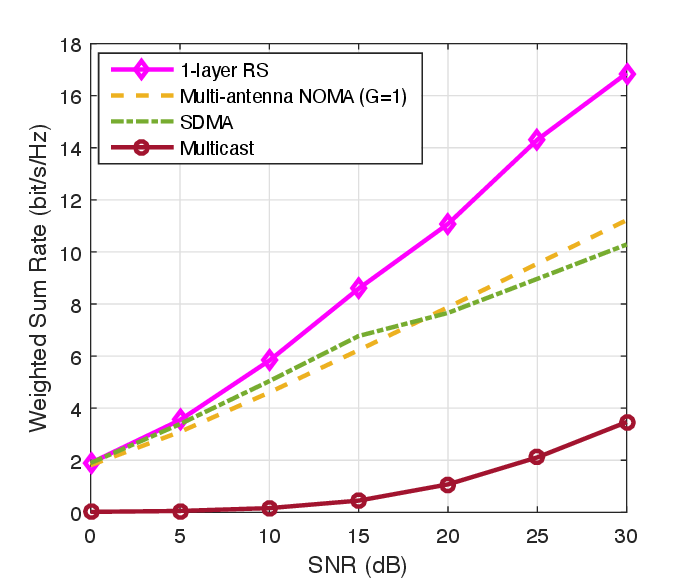}%
		\caption{Weighted sum rate versus SNR comparison of different MA schemes for an overloaded ten-user deployment with perfect CSIT, $M=2$ \cite{mao2017rate}. }
		\label{fig: 10UErate}
	\end{figure}
	%%%
	\par
	From Fig. \ref{fig: rateRegion}, we observe that   DPCRS enlarges the ER region of  all other schemes and linearly precoded RS even outperforms DPC when the CSIT quality is poor.
	%%%
	DPC and SDMA (based on MU--LP) can be seen as two extreme schemes that highly rely on accurate CSIT to fully manage interference  at the transmitter side, and both of them are therefore  sensitive to CSIT inaccuracy. Hence, the ER regions of DPC and SDMA (based on MU--LP)  quickly shrink as the CSIT quality decreases from $\alpha=0.9$ (in Fig. \ref{fig: rateRegion}(a) and (b)) to  $\alpha=0.6$ (in Fig. \ref{fig: rateRegion}(c) and (d)).
	%%%
	On the other hand, NOMA can be seen as an extreme scheme that fully manages/cancels interference  at the receiver side. However, it cannot fully exploit the available DoF and results in poor performance \cite{bruno2021MISONOMA}.
	%%%
	DPCRS and RS bridge the two extremes and use the common streams to smartly create a mix of transmitter-side and receiver-side interference cancellation. 
	%%%
	By adjusting the contribution of the common stream, it can dynamically decide the level of interference that needs to be canceled at the transmitter or the receiver side. 
	%%%
	Therefore, we conclude that in practical deployments with partial CSIT, RSMA with a combination of transmitter-side and receiver-side interference cancellation outperforms  full transmitter-side interference cancellation schemes such as SDMA (based on MU--LP or DPC) and  full receiver-side interference cancellation schemes such as NOMA (based on SC--SIC). RSMA is more flexible, and hence, more robust to CSIT uncertainties and everchanging user deployments. 
	%%%
	\par In Fig. \ref{fig: ESR vs alpha}, we further evaluate the ergodic sum rate (ESR) performance of different linearly precoded schemes for a three-user MISO BC with imperfect CSIT where generalized RS does not reduce to  1-layer RS.  
	%%%
	RS--CMD is also considered where each user is required to decode three common streams resulting in $6^3$ possible decoding orders to be considered at the transmitter. To ease the computational burden, we only consider one decoding order, namely, the ascending order of the  users' channel strengths.  
	%%%
	The precoders of all schemes are designed to maximize the ESR (when users have equal weights). 
	%%%
	In all figures, generalized RS outperforms all other schemes, especially when the network is overloaded, see Fig. \ref{fig: ESR vs alpha} (b) and (c). 
	%%%
	Generalized RS achieves significant gains compared to the other MA schemes when multi-user interference is strong and there is enough power allocated to all common streams.
	%%%
	The ESRs of multi-antenna NOMA ($G=1$) and SDMA drop dramatically as $\alpha$ decreases from 1 to 0.2.
	%%%
	Note that in the three-user case,  1-layer RS yields an explicit performance gain over multi-antenna NOMA ($G=1$) and is robust to CSIT uncertainties. Most importantly, 1-layer RS   does not require  transmitter scheduling and only needs one layer of SIC at each user, while multi-antenna NOMA ($G=1$) requires the transmitter to select from 6 decoding orders and each user has to be capable of performing two layers of SIC. The joint transmitter-side and receiver-side interference cancellation enables 1-layer RS to simultaneously enhance  SE and reduce transceiver complexity.  
	%%%
	\par
	To further emphasize the capability of 1-layer RS to jointly boost the SE and reduce the transceiver complexity, we consider  an extremely overloaded scenario with perfect CSIT in Fig. \ref{fig: 10UErate}  where the BS has $M=2$ antennas and serves  ten single-antenna users. 
	%%%
	The user channels have i.i.d. complex Gaussian entries following distribution $\mathcal{CN}(0,\sigma_k^2)$. The variances of the ten user channels are $\sigma_1^2=1,  \sigma_2^2=0.9, \ldots, \sigma_{10}^2=0.1$.
	%%%
	To  enable service to multiple users in such an extremely overloaded scenario, we set the QoS rate requirement of each user for SNR $=[0,5,\ldots,30]$ dB to be $[0, 0.001, 0.004, 0.01, 0.03, 0.06, 0.1]$ bit/s/Hz, respectively.
	%%%
	The precoders and message splits are designed by solving $\mathcal{P}_1$ in Section \ref{sec:precoderOpt}.
	%%%
	The rate of each user is averaged over 10 randomly generated channels and the weight of each user is assumed to be equal to 1.
	%%%
	Fig. \ref{fig: 10UErate} shows an explicit WSR gain of 1-layer RS over multi-antenna NOMA ($G=1$), SDMA, and multicast.  RS exploits the maximum DoF of 2 in the considered deployment (that is limited by the two transmit antennas) by using the common stream to pack messages from eight users while using the two private streams to serve the remaining two users.
	%%%
	In contrast,  the DoF achieved by  MU--LP is 1 because it cannot coordinate
	the inter-user interference when there is a non-negligible QoS rate requirement for each user, and its achieved DoF drops to 1. 
	%%%
	Multi-antenna NOMA ($G=1$) also achieves a DoF of 1. %%%
	It makes inefficient use of the transmit antennas and 9 layers of SIC (deployed but not always used) at each user.
	%%%
	Therefore, the WSR improvement of 1-layer RS comes with a much lower receiver complexity compared with multi-antenna NOMA ($G=1$). 
	
	\subsubsection{Operational region}
	As defined in \cite{bruno2019wcl}, the operational region of an MA scheme is  the  region of  channel conditions that it prefers.
	%%%
	%In the two-user case, as SDMA, NOMA, and OMA are sub-schemes of RSMA, their WSRs are always less than or equal to that of RSMA. 
	%%%
	We obey the following rules to select the relevant MA scheme from RSMA, SDMA, NOMA, OMA: 
	\begin{enumerate}[i)]
		\item if $\textrm{WSR}_{\textrm{RSMA}}-\textrm{WSR}_{\textrm{OMA}}<\epsilon' $,  RSMA boils down to OMA.	
		\item if $\textrm{WSR}_{\textrm{SDMA}}-\textrm{WSR}_{\textrm{OMA}}>\epsilon' $ and $\textrm{WSR}_{\textrm{RSMA}}-\textrm{WSR}_{\textrm{SDMA}}<\epsilon' $,  RSMA boils down to SDMA.	
		\item if $\textrm{WSR}_{\textrm{NOMA}}-\textrm{WSR}_{\textrm{SDMA}}>\epsilon' $ and $\textrm{WSR}_{\textrm{RSMA}}-\textrm{WSR}_{\textrm{NOMA}}<\epsilon' $,  RSMA boils down to NOMA.	
		\item if $\textrm{WSR}_{\textrm{RSMA}}-\textrm{WSR}_{\textrm{SDMA}}>\epsilon' $ and $\textrm{WSR}_{\textrm{RSMA}}-\textrm{WSR}_{\textrm{NOMA}}>\epsilon' $,   RSMA is superior to all other MA schemes.
	\end{enumerate}
	$\epsilon'$ is the tolerance for selection. RSMA is selected as the preferred scheme when it is not equivalent to any other MA scheme. 
	%%%
	%%%
	\begin{figure}[t!]
		\centering
		%	\hspace*{0.2cm}
		\begin{subfigure}[b]{0.25\textwidth}
			\centering
			\includegraphics[width=\textwidth]{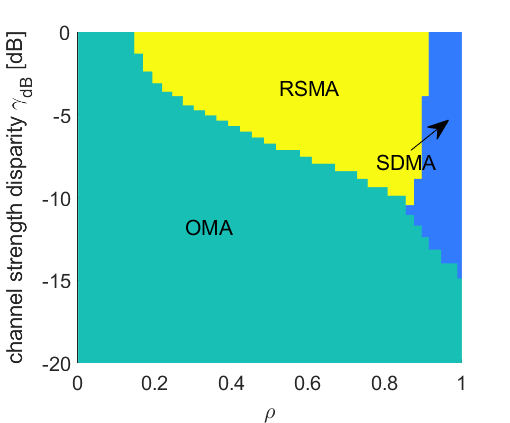}%
			\caption{$u_1=10^{0.5}, u_2=1$}
		\end{subfigure}%
		~
		\begin{subfigure}[b]{0.25\textwidth}
			\centering
			\includegraphics[width=\textwidth]{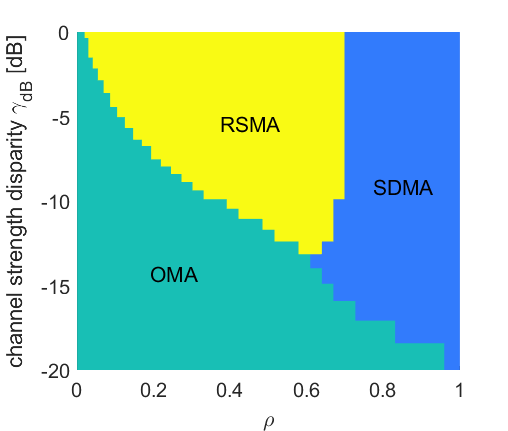}%
			\caption{$u_1=1, u_2=1$}
		\end{subfigure}%
		\\
		\begin{subfigure}[b]{0.25\textwidth}
			\centering
			\includegraphics[width=\textwidth]{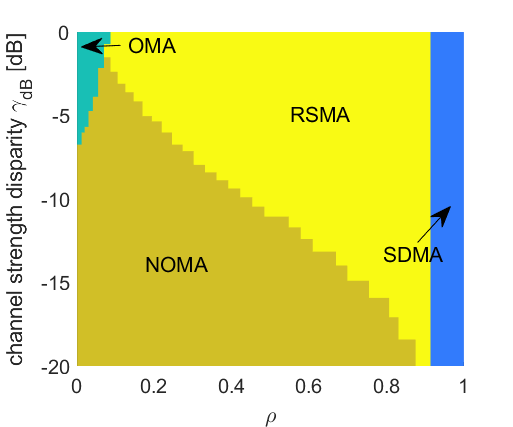}%
			\caption{$u_1=1, u_2=10^{0.5}$}
		\end{subfigure}%
		\caption{Regions of operation for different MA schemes, $K=2$, SNR = 20 dB. }
		\label{fig: operationRegion}
	\end{figure}
	%%%
	\par
	Following the evaluations in \cite{bruno2019wcl}, we illustrate the preferred operational regions for RSMA, SDMA, NOMA, and OMA for the two-user MISO BC with perfect CSIT in Fig. \ref{fig: operationRegion}. The transmitter is equipped with $M=2$ antennas and the channel vectors  are  given by $\mathbf{h}_1=1/\sqrt{2}[1, 1]^H$ and $\mathbf{h}_2=\gamma/\sqrt{2}[1, e^{j\theta}]^H$.
	%%%
	The precoders of all MA schemes are designed using the WMMSE optimization framework developed in \cite{RS2016hamdi,mao2017rate} for solving $\mathcal{P}_1$ in Section \ref{sec:precoderOpt} for a certain weight pair $u_1, u_2$ with $U_n( \mathbf{{P}},\overline{\mathbf{c}})= \sum_{k=1}^Ku_{k}(\overline{R}_{k} + \overline{C}_{k})$  and $R_k^{th}=0$ bit/s/Hz.  Readers are referred to  \cite{mao2017rate}  for more details on the optimization of all MA schemes. 
	%%%
	The colors in  Fig. \ref{fig: operationRegion} illustrate the  preferred MA schemes (RSMA, SDMA, NOMA, OMA) that maximize the WSR as a function of $\rho=1-\frac{|\mathbf{h}_1^H\mathbf{h}_2|^2}{\left\|\mathbf{h}_1\right\|^2\left\|\mathbf{h}_1\right\|^2}$ ($0\leq\rho\leq1$) and $\gamma_{\mathrm{dB}}=20\log_{10}(\gamma)$  ($-20 \textrm{dB}\leq\gamma_{\mathrm{dB}}\leq 0$). Therefore, user-1 and user-2 have a long-term SNR of 20dB and $ 0\textrm{dB}\leq 20\textrm{dB}+\gamma_{\mathrm{dB}}\leq 20\textrm{dB}$, respectively. 
	
	\par 
	From Fig. \ref{fig: operationRegion}(a) and (b), we obtain that NOMA has no benefit at all when an equal or higher weight is allocated to the user with the stronger channel strength. Only when user fairness is considered, i.e.,  a higher weight is allocated to the weaker user as in Fig. \ref{fig: operationRegion}(c), RSMA boils down to NOMA, especially for user  deployments where the users are closely aligned with a large channel strength disparity (i.e., small $\rho$ and small $\gamma_{\mathrm{dB}}$). 
	%%%
	In all subfigures, we observe that RSMA boils down to SDMA when $\rho$ is sufficiently large. SDMA is sufficient to manage inter-user interference when the user channels are orthogonal. 
	%%%
	For different user weights, there is always an explicit operational region for RSMA where it  provides a larger WSR than SDMA, NOMA, and OMA. 
	%%%
	In this critically loaded scenario, RSMA offers high flexibility in all user deployments and  is capable of enhancing user fairness as well.
	%%%
	\subsubsection{User fairness}
	%%%
	\begin{figure}[t!]
		\begin{subfigure}[b]{\columnwidth}
			\centering
			\includegraphics[width=0.89\textwidth]{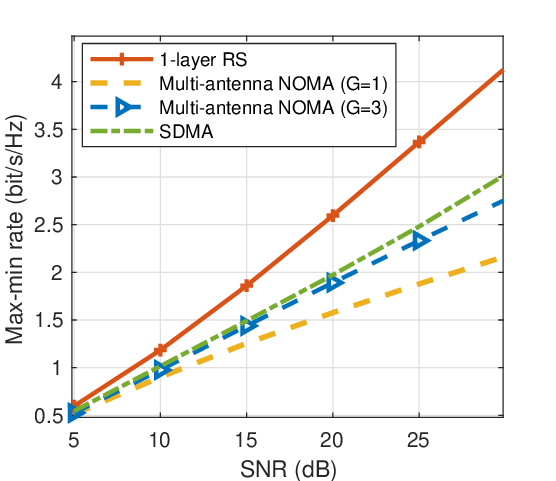}%
			\caption{Max-min rate vs. SNR in dB.}
		\end{subfigure}
		~\\
		\begin{subfigure}[b]{\columnwidth}
			\centering
			\includegraphics[width=0.9\textwidth]{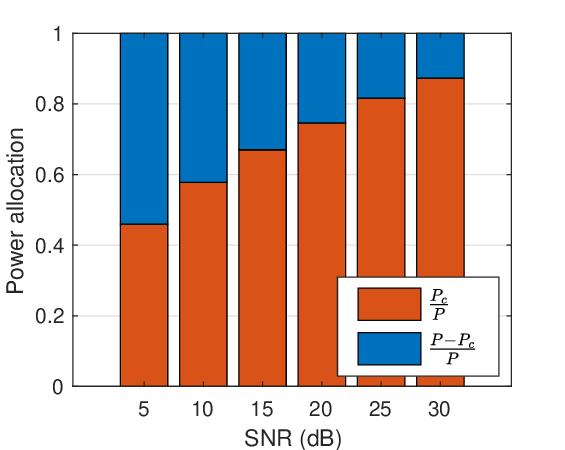}%
			\caption{Power allocation to the common and private streams for 1-layer RS.}
		\end{subfigure}%
		\caption{Max-min rate performance for the MISO BC with imperfect  CSIT $\alpha=0.5$, $M=K=6$.  $P_c=\left\|\mathbf{p}_{c}\right\|^2$ is the power allocated to the common stream. $P-P_c$ is the sum power allocated to all private streams. }
		\label{eg:MMFrate}
	\end{figure}
	%%%
	In practice, due to the disparate path losses experienced by  users in different locations, the issue of user fairness surfaces where users closer to the transmitter are allocated with more (time/frequency/power) resources, while users far from the transmitter are starving. 
	%%%
	To cope with  this issue, the  max-min rate  utility function is preferred for  precoder design such that a certain rate performance is guaranteed for all users. 
	%%%
	Next, we evaluate the max-min rate performance of the users when the precoders are designed by solving $\mathcal{P}_2$ in Section \ref{sec:precoderOpt}
	with $U( \mathbf{{P}},\overline{\mathbf{c}})= {\min}_{k \in \mathcal{K}}(\overline{R}_{k} + \overline{C}_{k})$ and $R_k^{th}=0$ bit/s/Hz.
	%%%
	\par 
	Fig. \ref{eg:MMFrate} illustrates the max-min rate performance and the corresponding power allocation between the common and private streams for RSMA in the MISO BC with imperfect CSIT when the transmitter is equipped with $M=6$ antennas and serves $K=6$ single-antenna users.  The imperfect CSIT model follows (\ref{eq:imperfectCSIT}) with a CSIT scaling factor of $\alpha=0.5$ and the results are averaged over 100 random channel realizations. 
	%%%
	1-layer RS shows superior max-min rate  performance compared to multi-antenna NOMA  and SDMA in Fig. \ref{eg:MMFrate}(a).  
	%%%
	As specified in Table \ref{tab: complexity}, the scheduler complexity and the receiver complexity of multi-antenna NOMA are much higher than those for 1-layer RS and SDMA. 
	%%%
	At the transmitter, for  multi-antenna NOMA ($G=3$), the user grouping and decoding order have to be jointly optimized with the precoders. When $K=6$, each user requires (but not always uses all) 5 layers of SIC for multi-antenna NOMA ($G=1$). Such high complexity, however, yields worse performance than 1-layer RS  with a single layer of SIC at each user or MU--LP without SIC. 
	%%%
	Therefore,  multi-antenna NOMA makes inefficient use of SIC as  highlighted in \cite{bruno2021MISONOMA}.
	%%%
	%The significant performance gain of RS underscores its  capability to enhance user fairness while maintaining a relatively low complexity.
	%%%
	\par
	To shed more light on the reasons behind the large max-min rate gain achieved by 1-layer RS, the power allocation between the common and private streams is illustrated in Fig. \ref{eg:MMFrate}(b). The sum power allocated to all $K+1$ streams for 1-layer RS is  $P=\left\|\mathbf{p}_{c}\right\|^2+\sum_{k=1}^K\left\|\mathbf{p}_{k}\right\|^2$.
	%%%
	Let $P_c=\left\|\mathbf{p}_{c}\right\|^2$ be the power allocated to the common stream $s_c$, then $P-P_c$ becomes the power allocated to all private streams $s_1,\ldots,s_K$. $\frac{P_c}{P}$ is therefore the fraction of transmit power allocated to the common stream and $\frac{P-P_c}{P}$ is the fraction of transmit power allocated to all private  streams. 
	%%%
	Fig. \ref{eg:MMFrate}(b) shows that for all transmit SNR, 1-layer RS allocates a large portion of power to the common stream and
	$P_c$ increases with the SNR as the interference becomes stronger (compared to the noise) in the high SNR regime. 
	%%%
	The common stream therefore plays an essential role in improving user fairness in 1-layer RS.
	%%%
	By  adjusting the message split and the power allocation to the common stream based on the level of interference that can be canceled by the receiver, 1-layer RS  realizes a
	powerful interference management/cancellation strategy.
	%%%
	\begin{figure}
		\centering
		\includegraphics[width=3.5in]{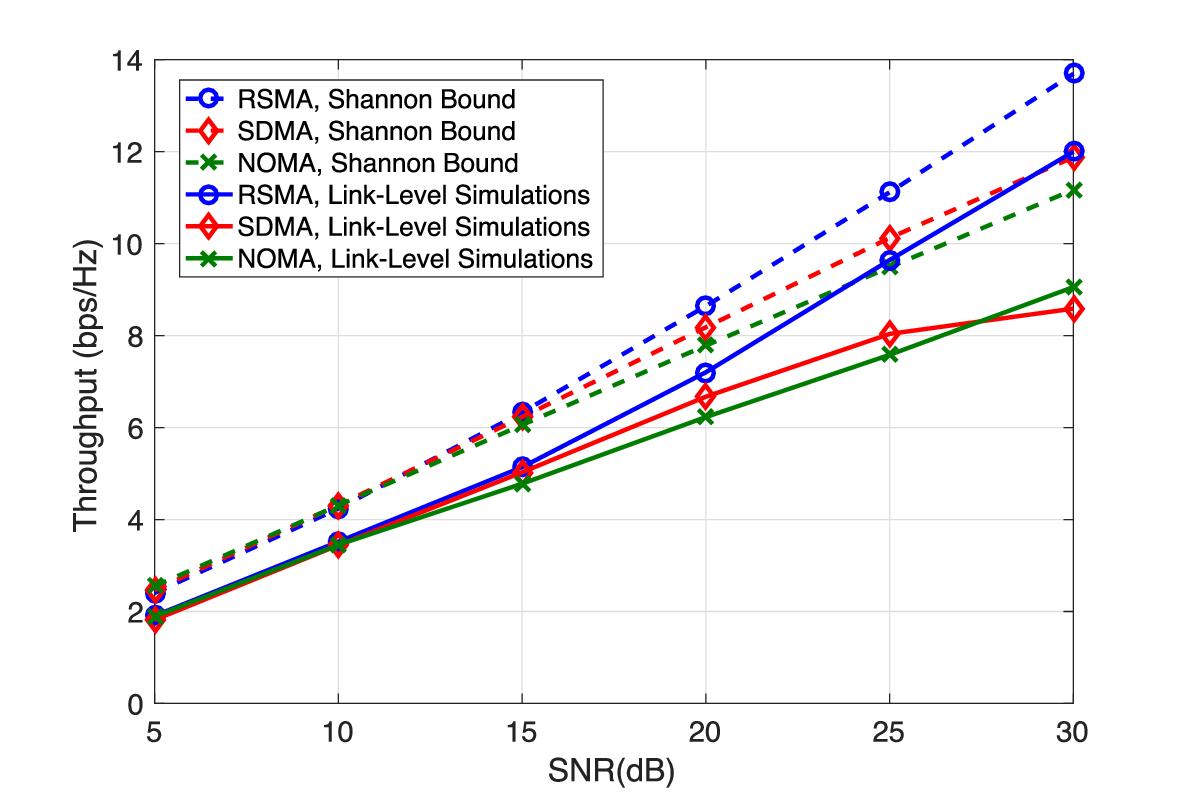}%
		\caption{Throughput vs. SNR for different MA schemes for the MISO BC with imperfect CSIT, $M=K=2$, $\alpha=0.6$ \cite{Onur2020LLS}.}
		\label{fig:LLS1}
	\end{figure}
	%%%
	\subsubsection{LLS Performance}
	\label{sec:LLS}
	\labelsubseccounter{VII-B}
	%%%
	The PHY layer architecture illustrated in  Section \ref{sec:PHYdesign} has been used to establish an LLS platform for RSMA, {\sl e.g.,} \cite{Onur2020LLS,onur2021mobility,anup2021MIMO,longfei2021LLS,JRCLoli2022,DLloli2022}. 
	%%%
	In this section, we study the throughput performance of RSMA obtained from LLS for a two-user scenario, and compare the obtained throughput with the Shannon bounds. 
	%%%
	\par
\begin{figure}[t!]
		\begin{subfigure}{.5\textwidth}
			\centering
			\includegraphics[width=3.3in]{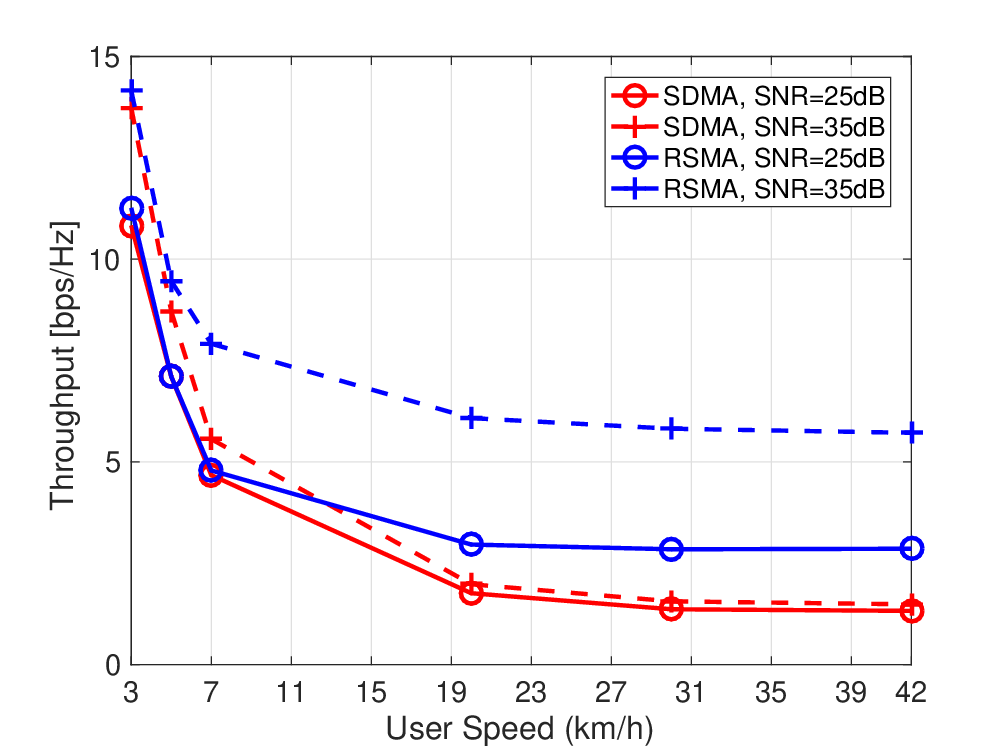}%
			\caption{$M=32, K=8, N_r=1, Q_c=Q_p=1$.}   
			\label{fig:LLSmobility_MISO}
		\end{subfigure}	
		~\\
		\begin{subfigure}{.5\textwidth}
			\centering
			\includegraphics[width=3.3in]{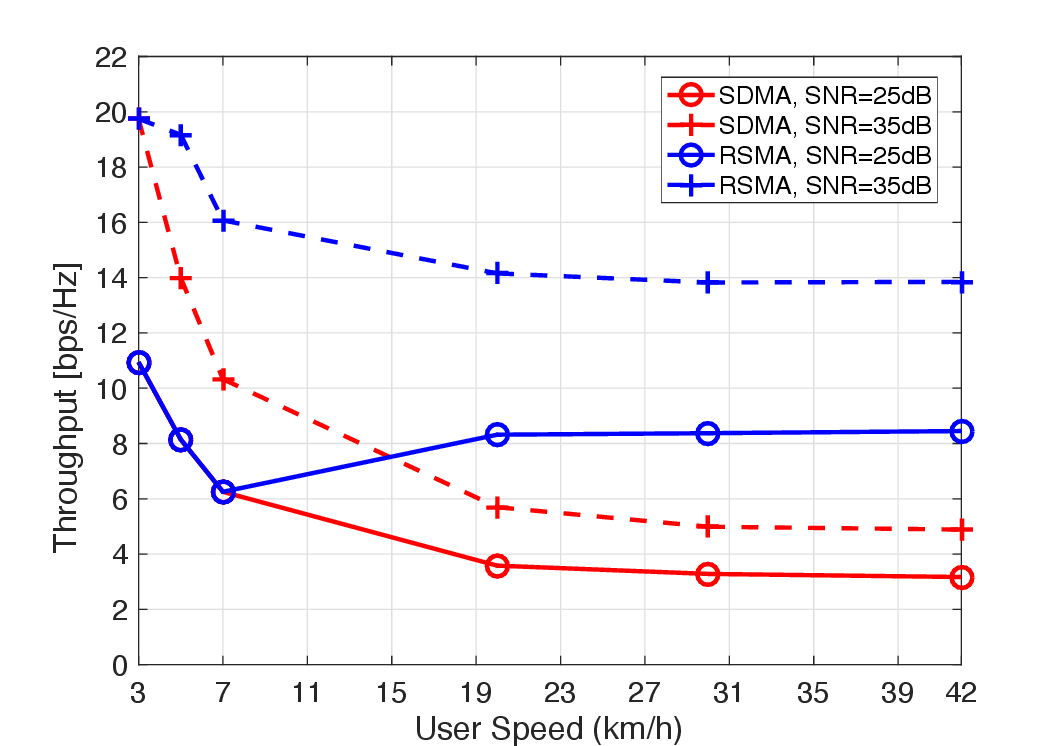}%
			\caption{$M=64, K=8, N_r=4, Q_c=Q_p=3$.}   
			\label{fig:LLSmobility_MIMO}
		\end{subfigure}
		\caption{Throughput vs. user speed for different MA schemes for (massive) MIMO BC with outdated CSIT employing OFDM and a 3GPP channel model, and $10$ ms CSI feedback delay \cite{onur2021mobility}.}   
		\label{fig:LLSmobility}
	\end{figure}

 
	The system throughput is defined as the total  number of successfully recovered information bits at all users per channel use. 
	%%%
	Denote $S^{(l)}$ and $D_{s,k}^{(l)}$ respectively  as the number of channel uses in the $l$-th Monte-Carlo realization and  the number of information bits successfully recovered at user-$k$ for the intended common sub-message $\widehat{W}_{c,k}$ and private sub-message $\widehat{W}_{p,k}$. Then, the system throughput is given as follows
	\begin{equation}
		\textrm{Throughput (bps/Hz)}=\frac{\sum_l\left(D_{s,1}^{(l)}+D_{s,2}^{(l)}\right)}{\sum_l S^{(l)}}.
	\end{equation}
	In total 100 Monte-Carlo realizations were used to obtain the throughput performance results shown in Figs. \ref{fig:LLS1} and \ref{fig:LLSmobility}, and $S^{(l)}=256, \forall l\in \{1,2,\ldots, 100\}$.
	%%%
	\par In Fig. \ref{fig:LLS1}, both the Shannon bounds and throughput levels achieved by RSMA, SDMA, and NOMA for the MISO BC with imperfect CSIT are illustrated. The transmitter has $M=2$ antennas and serves $K=2$ single-antenna users.
	%%%
	The imperfect CSIT model follows (\ref{eq:imperfectCSIT}) with CSIT scaling factor $\alpha=0.6$.
	%%%
	Generally, the trend of the throughput performance obtained by the LLS is in line with the corresponding Shannon bounds with a significant throughput gain of RSMA compared to SDMA and NOMA. 
	%%%
	The throughput gain of RSMA over SDMA is even larger than expected based on the Shannon bound.
	%%%
	Furthermore, the throughput of SDMA is observed to saturate in the high SNR regime. 
	%%%
	Due to  imperfect CSIT, SDMA frequently switches to single-user transmission (i.e., OMA) during the Monte-Carlo simulations of different channel realizations.  This  results in a large transmission rate for a single user, which exceeds the maximum SE achieved by the largest modulation order and code rate for a single stream considered in the design. AMC therefore cannot assign an appropriate modulation and coding scheme in such cases, resulting in a throughput loss.
	%%%
	%%%
	%%%
	\par 
	Fig. \ref{fig:LLSmobility} further extends the LLS to a (massive) MIMO BC with imperfect CSIT where the transmitter has $M=32$ or $M=64$ antennas and serves $K=8$ users, each equipped with $N_r=1$ or $N_r=4$ receive antennas. 
	%%%
	In Fig. \ref{fig:LLSmobility}, the imperfect CSIT is caused by user mobility and delay in CSI feedback from the users to the transmitter. 
	%%%
	The instantaneous channel matrix $\mathbf{H}_{k}[t]\in\mathbb{C}^{N_r\times M}$ at time instant $t$ is modelled as follows \cite{onur2021mobility}
	\begin{align}
		\mathbf{H}_{k}[t]=\sqrt{\epsilon^{2}} \mathbf{H}_{k}[t-1] + \sqrt{1-\epsilon^{2}} \mathbf{E}_{k}[t],
		\label{eq:mobilechannel}
	\end{align}where $\epsilon=J_{0}(2\pi f_{D}T)$ is the time correlation coefficient obeying the Jakes' model, $J_{0}(\cdot)$ is the modified Bessel function of the first kind and order zero, $T$ is the channel instantiation interval, and $f_{D}=vf_{c}/c$ is the maximum Doppler frequency for given user speed $v$, speed of light $c$, and carrier frequency $f_{c}$.
	%%%
	Both $\mathbf{H}_{k}[t]$ and $\mathbf{E}_{k}[t]\in\mathbb{C}^{N_r\times M}$ have i.i.d. entries distributed according to $\mathcal{CN}(0,1)$.
	%%%
	We assume that the transmitter at time $t$ only has knowledge of the channel matrix observed at the previous time instant  $\mathbf{H}_{k}[t-1]$ due to the latency in CSI feedback.
	%%%
	To simplify the RSMA transmitter design, low-complexity ZFBF and BD methods are used to design the precoders of the private streams for MISO and MIMO channels, respectively, while the leftmost eigenvector is used as the precoder for the common streams. The system employs an Orthogonal Frequency Division Multiplexing (OFDM) waveform in a frequency-selective channel obtained according to the Clustered Delay Line (CDL) model of 3GPP. The employed precoders are calculated for each subcarrier of the waveform for optimal performance. The closed-form power allocation scheme proposed in  \cite{onur2021mobility} for MISO channels is used to perform the power allocation between the common and the private streams. 
	% which however results in the the non-monotonic behavior of RSMA in Fig. \ref{fig:LLSmobility}. 
	%%%
	\par
	Fig. \ref{fig:LLSmobility} reveals a significant throughput gain of  RSMA over SDMA without and with multi-antenna receivers.
	With the same QoS constraint (i.e., 8 bps/Hz), RSMA supports a much higher user speed (i.e., 40 km/h) than SDMA (i.e., 5 km/h). 
	In Fig. \ref{fig:LLSmobility}(b), one can note the non-monotonic behaviour of the RSMA throughput for SNR$=25$ dB, which occurs as the employed power allocation algorithm is suboptimal for MIMO channels. The results clearly show that RSMA is indeed more robust to user mobility  than SDMA owing to its ability to partially decode interference and partially treat interference as noise.

	\subsection{Advantages of RSMA}
	\label{sec:advantageRSMA}
	\labelsubseccounter{VI-D}
	\par The above comparisons between RSMA and existing MA schemes including SDMA, NOMA, OMA, and multicasting in terms of their respective framework, complexity, and performance allow us to unveil the major advantages of RSMA, which are summarized in the following:
	\begin{itemize}
		\item \textbf{Universality}: 
		RSMA is a universal MA scheme that subsumes  SDMA, NOMA, OMA, and multicasting as sub-schemes.
		%%%
		In other words, RSMA is a superset of existing MA schemes and therefore always achieves equal  or better performance compared to SDMA, NOMA, OMA, and multicasting. 
		%%%
		Henceforth, there does not seem to be a need to investigate other MA schemes but rather use a single unified RSMA framework.
		
		\item \textbf{Flexibility}: 
		RSMA is suited for different network loads  (underloaded/overloaded) and  user deployments (diverse channel directions/strengths). 
		%%%
		It is a powerful strategy that can manage multi-user interference originating from different sources flexibly. 
		%%%
		Different from SDMA that fully mitigates interference at the transmitter side or NOMA that fully mitigates/cancels interference at the receiver side, RSMA creates common streams to enable a smart combination of transmitter-side and receiver-side interference mitigation. By adjusting the amount of resources allocated to the common stream, the level of interference canceled at the transmitter and receiver can be  adjusted flexibly. 
		%%%
		In fact, RSMA  employs a flexible  interference management strategy by partially decoding the interference and partially treating the interference as noise, which generalizes the design principle of fully treating interference as noise (as in SDMA), fully decoding interference (as in NOMA), and single-user transmission to avoid interference (as in OMA).
		%%%
		By optimizing the power allocated to the common and private streams, RSMA automatically simplifies to SDMA when  the network is underloaded and user channels are orthogonal with perfect CSIT and boils down to NOMA when the network is extremely overloaded and user channels are aligned with certain channel strength disparities. 
		For other channel conditions, i.e., when the user channels are neither orthogonal nor aligned, RS takes advantage of the common streams to better manage the interference and  outperforms all other MA schemes.  
		\item \textbf{Robustness}: 
		%%%
		As the application of RSMA in multi-antenna networks is motivated by its DoF optimality in the multi-antenna BC with imperfect CSIT, RSMA is ideally suited to manage multi-user interference originating from imperfect CSIT and is robust to CSIT uncertainties resulting from different sources of impairment, such as quantized feedback \cite{Minbo2016MassiveMIMO}, pilot contamination \cite{anup2022Pilot,anup2022CFMIMO}, channel estimation errors \cite{RS2016hamdi}, and user mobility \cite{onur2021mobility}. 
		%%%
		User mobility is one of the major sources of interference in NR and is expected to become more severe and to occur more frequently in 6G systems. 
		%%%
		Numerical results in \cite{onur2021mobility, onur2021sixG} have shown that RSMA has the capability to cope with the high mobility envisioned in 6G. Contrary to SDMA, for which the  sum-rate performance  quickly saturates  for increasing SNR and drops quickly with the user speed, RSMA ensures robust multi-user connectivity and achieves significant sum-rate and throughput gains under user mobility.
		%%%
		This is in contrast to the other MA schemes (OMA, SDMA, NOMA), which are primarily designed for perfect CSIT and are vulnerable to imperfect CSIT \cite{mao2019beyondDPC}.
		%%%
		
		\item \textbf{Higher spectrally efficiency}:
		The SE of RSMA always exceeds or is equal to that of the other MA schemes for both perfect and imperfect CSIT.
		%%%
		When the CSIT is perfect, the  rate region of linearly precoded RSMA is larger than that of MU--LP and NOMA, and is closer to the capacity region achieved by DPC. 
		%%%
		When the CSIT is imperfect, linearly precoded RSMA can even achieve a larger rate region than DPC.  By marrying the benefits of RS and DPC, nonlinearly precoded DPCRS further enlarges the rate region of the MISO BC with imperfect CSIT beyond that of DPC and that of linearly precoded RS, as illustrated in Fig. \ref{fig: rateRegion}. 
		%%%
		RSMA optimally exploits the  available  CSIT and the spatial dimension of the multi-antenna BC. Furthermore, it achieves the optimum sum-DoF in both perfect CSIT and imperfect CSIT (as illustrated Table \ref{tab:DoF} and Fig. \ref{fig:DoFregion}--\ref{DoF_vs_K_M6}). 
		%%%
		\item \textbf{Higher energy efficiency}: 
		Since RSMA is a superset of OMA, SDMA, NOMA, and multicasting, 
		the performance gain of RSMA is not limited to SE, but also extends to EE  and their trade-off \cite{Gui2020EESEtradeoff,Jia2020SEEEtradeoff,Hassan2021FRAN}. 
		RSMA has been shown to outperform existing MA schemes in terms of EE for various applications  with a wide range of network loads and user deployments  \cite{mao2018EE, bho2021globalEE,AP2017cognitive, Tervo2018SPAWC, UAVRS2019ICC, mao2019TCOM, alaa2020EECRAN, zhaohui2020IRS,Lin2021Cognitive}. 
		\item \textbf{Enhanced QoS and user fairness}: 
		RSMA not only better exploits the sum-DoF but also the MMF-DoF in the multi-antenna BC with perfect and imperfect CSIT, as illustrated in Table \ref{tab:DoF} and Figs. \ref{DoF_vs_SIC_imperfect}, \ref{DoF_vs_K_M6}.
		%%%
		In the finite SNR regime, RSMA also exhibits a substantial max-min rate gain over the other MA schemes, as shown in Fig. \ref{eg:MMFrate}(a) and  a number of recent works \cite{RS2016joudeh,hongzhi2020LLS,longfei2021statisticalCSIT,cran2019wcl,mao2019maxmin,HongzhiChen2020subcarrierRS,yalcin2020RSmultigroup,hongzhi2020LLSmultigroup,fuhao2020secrecyRS,longfei2020satellite,si2021imperfectSatellite}.
		%%%
		Moreover, the performance gain of RSMA becomes more pronounced when the user rates are subject to QoS constraints or when a larger weight is assigned to users with weaker channel strengths  \cite{mao2017rate}. 
		%%%

		\item \textbf{Lower complexity}:
		RSMA entails a lower transceiver complexity than many existing MA schemes.
		%%%
		Although linearly-precoded RSMA achieves a larger rate region than DPC, see RS versus DPC in Fig. \ref{fig: rateRegion}, this gain comes with a simpler transmitter design. 
		%%%
		Figs. \ref{DoF_vs_SIC_imperfect}, \ref{DoF_vs_K_M6} and Figs. \ref{fig: ESR vs alpha}, \ref{fig: 10UErate} show that the sum/MMF-DoF gain and the sum-rate gain of 1-layer RS over multi-antenna NOMA comes with lower scheduling and receiver complexity.
		%%%
		Different from multi-antenna NOMA, which requires joint design of user grouping and ordering at the transmitter and layer(s) of SIC at each user, 1-layer RS does not require user ordering and grouping at the transmitter, and it only requires one SIC layer at each user.   
		%%%
		Moreover, 1-layer RS is less sensitive to difference channel conditions, which as a result, can further simplify the scheduler. 
		%%%
		Besides the merits of reducing the scheduler and receiver complexity,  RSMA is also capable of  reducing the CSI feedback overhead  in the presence of quantized feedback \cite{RS2015bruno, minbo2017mmWave}.  
		\item \textbf{Coverage extension}: 
		It has been shown in \cite{jian2019crs,mao2019maxmin} that cooperative rate-splitting (CRS), which incorporates cooperative user relaying with 1-layer RS,  and allows one user to decode and forward the common stream to the other users, significantly improves the max-min rate when users with large channel strength disparities are jointly served. 
		%%%
		Therefore, RSMA with coverage extension techniques such as unmanned aerial vehicles (UAV)/intelligent reconfigurable surface (IRS)/relay stations  can further boost the rate of the users at the cell edge and  offers substantial coverage extension benefits. 
		\item \textbf{Low latency}: 
		URLLC is one major use case in NR and 6G. One approach to achieve low-latency communication is reducing the transmitted packet size. RSMA has been shown to improve throughput over SDMA and NOMA with finite length polar codes in \cite{Onur2020LLS}, and to attain the same transmission rate as SDMA and NOMA but with shorter blocklength and therefore lower latency in \cite{yunnuo2021FBL,onur2021sixG}. Hence, RSMA is a promising enabling technology to reduce the latency for URLLC service in 6G.
	\end{itemize}
	%%%
	\subsection{Disadvantages of RSMA}
	\label{sec:disadvantageRSMA}
	\labelsubseccounter{VI-E}
	Despite of all the aforementioned advantages, RSMA inevitably also introduces some disadvantages, which are summarized as follows.  
	%%%
	\begin{itemize}
		%%%
		\item \textbf{SIC  requirement at receivers}:
		As summarized in Table \ref{tab:RSschemesCompare}, RSMA requires at least one layer of SIC at each receiver. 
		%%%
		For some RSMA schemes, i.e., 1-layer RS, 2-layer HRS, 1-DPCRS, THPRS, the number of SIC layers at each user does not depend on the number of user $K$. 
		%%%
		However, for generalized RS and M-DPCRS, the number of SIC layers increases exponentially with the number of user $K$, which imposes high receiver complexity. 
		%%%
		\item \textbf{Higher encoding complexity}: 
		For existing MA schemes including SDMA, NOMA, and OMA, the number of data streams encoded at the transmitter is equal to the number of users. 
		%%%
		In comparison,  RSMA requires more data streams  to be encoded due to the message splitting (and recombining) at the transmitter as per Table \ref{tab:RSschemesCompare}.
		%%%
		The encoding complexity of RSMA is therefore higher than that of other MA schemes.
		%%%
		\item \textbf{Higher signaling burden}:
		As RSMA splits (and combines) user messages at the transmitter, each receiver needs to know how to (split and) combine the decoded messages in order to recover the intended message from the decoded common and private messages.
		%%%
		Therefore, a higher downlink signaling burden might be imposed by RSMA-enabled transmission networks compared with other MA-enabled networks so as to guarantee the transmitter and the receivers can coordinate message splitting and combining.  
		%%%
		\item \textbf{Higher optimization burden}:
		In terms of precoder optimization, RSMA requires the precoders of the common and private streams to be jointly optimized  with the common rate allocation as per Section  \ref{sec:precoderOpt}. 
		%%%
		Though the optimization space is enlarged and more DoFs for precoder design are achieved, a higher optimization burden  is exerted by RSMA compared with the conventional SDMA schemes. 
		%%%
		To reduce the computation time,  efforts have been made to design low-complexity precoders. However, as discussed in Section \ref{sec:lowComplexPrecoding}, the power allocation between  the common and private streams is crucial for the performance of RSMA. The fraction of the transmit power allocated to the private streams $\tau$ has to be optimized for different operational scenarios in order to realize the potential performance gain of RSMA.
		%%%
	\end{itemize}
	Although RSMA inevitably introduces the above problems, the appealing advantages of RSMA far outweigh its disadvantages. 
	%%%
	As the study of RSMA is still in its infancy, these problems merit future exploration.
	%%%
	\begin{table*}
		\caption{Communication-theoretic literature on the applications of RSMA.}
		\label{tab:commuSurveyApp}
		%	\addtolength\tabcolsep{-2pt}
		\begin{tabular}{@{}|C{0.6cm}|C{1.47cm}|p{4.9cm}|C{1.4cm}|C{0.8cm}|C{1cm}|C{0.8cm}|C{1.2cm}|C{0.6cm}|C{0.6cm}|C{0.6cm}|@{}}
			\hline
			\multirow{2}{*}{\textbf{Year}} &
			\multirow{2}{*}{\textbf{Ref.}} &
			\multicolumn{1}{c|}{\multirow{2}{*}{\textbf{Application}}} &
			\multirow{2}{*}{\textbf{\begin{tabular}[c]{@{}c@{}}RSMA \\ Scheme\end{tabular}}} &
			\multicolumn{2}{c|}{\textbf{Precoding Scheme}} &
			\multicolumn{2}{c|}{\textbf{CSIT Condition}} &
			\multicolumn{3}{c|}{\textbf{KPIs}} \\ \cline{5-11} 
			&
			&
			\multicolumn{1}{c|}{} &
			&
			\textbf{Linear} &
			\textbf{\begin{tabular}[c]{@{}c@{}}Non-\\ linear\end{tabular}} &
			\textbf{Perfect} &
			\textbf{Imperfect} &
			\textbf{WSR} &
			\textbf{MMF} &
			\textbf{EE} \\ \hline
			2016 &
			\cite{Minbo2016MassiveMIMO} &
			Massive MIMO &
			\circled{1}, \circled{2} &
			$\triangle$ &
			&
			&
			$\surd$ &
			$\surd$ &
			&
			\\ \hline
			\multirow{4}{*}{2017} &
			\cite{AP2017bruno} &
			Massive MIMO with hardware impairment &
			\circled{1} &
			$\triangle$ &
			&
			&
			$\surd$ &
			$\surd$ &
			&
			\\ \cline{2-11} 
			&
			\cite{AP2017cognitive} &
			Cognitive D2D &
			\circled{1} &
			$\blacktriangle$ &
			&
			$\surd$ &
			&
			&
			&
			$\surd$ \\ \cline{2-11} 
			&
			\cite{hamdi2017bruno} &
			Multigroup multicast &
			\circled{1} &
			$\blacktriangle$ &
			&
			$\surd$ &
			&
			&
			$\surd$ &
			\\ \cline{2-11} 
			&
			\cite{minbo2017mmWave} &
			mmWave MISO BC &
			\circled{1} &
			$\blacktriangle$ &
			&
			&
			$\surd$ &
			$\surd$ &
			&
			\\ \hline
			\multirow{3}{*}{2018} &
			\cite{Tervo2018SPAWC} &
			Multi-cell  multigroup multicast &
			\circled{1} &
			$\blacktriangle$ &
			&
			$\surd$ &
			&
			&
			&
			$\surd$ \\ \cline{2-11} 
			&
			\cite{Asos2018MultiRelay} &
			Multi-pair massive MIMO relaying &
			\circled{1} &
			$\triangle$ &
			&
			&
			$\surd$ &
			$\surd$ &
			&
			\\ \cline{2-11} 
			&
			\cite{RSSatellite2018ISWCS} &
			Multi-beam  satellite networks &
			\circled{1} &
			$\blacktriangle$ &
			&
			&
			$\surd$ &
			$\surd$ &
			&
			\\ \hline
			\multirow{9}{*}{2019} &
			\cite{mao2018networkmimo} &
			Cooperative multi-cell joint transmission &
			\circled{1}, \circled{2}, \circled{3} &
			$\blacktriangle$ &
			&
			$\surd$ &
			&
			$\surd$ &
			&
			\\ \cline{2-11} 
			&
			\cite{Ala2019IEEEAccess} &
			C-RAN &
			\circled{6} &
			$\blacktriangle$ &
			&
			$\surd$ &
			&
			$\surd$ &
			&
			\\ \cline{2-11} 
			&
			\cite{Alaa2019uavCRAN} &
			UAV-assisted  C-RAN &
			\circled{6} &
			$\blacktriangle$ &
			&
			$\surd$ &
			&
			$\surd$ &
			&
			\\ \cline{2-11} 
			&
			\cite{UAVRS2019ICC} &
			MmWave UAV-assisted MISO BC &
			\circled{1} &
			$\blacktriangle$ &
			&
			$\surd$ &
			&
			&
			&
			$\surd$ \\ \cline{2-11} 
			&
			\cite{cran2019wcl} &
			C-RAN &
			\circled{3} &
			$\blacktriangle$ &
			&
			$\surd$ &
			&
			&
			$\surd$ &
			\\ \cline{2-11} 
			&
			\cite{mao2019swipt} &
			SWIPT for MISO BC &
			\circled{1} &
			$\blacktriangle$ &
			&
			$\surd$ &
			&
			$\surd$ &
			&
			\\ \cline{2-11} 
			&
			\cite{RSswiptIC2019CL} &
			SWIPT for MISO IC &
			\circled{6} &
			$\blacktriangle$ &
			&
			&
			$\surd$ &
			$\surd$ &
			&
			\\ \cline{2-11} 
			&
			\cite{jian2019crs} &
			Cooperative user relaying for MISO BC &
			\circled{1} &
			$\blacktriangle$ &
			&
			$\surd$ &
			&
			$\surd$ &
			&
			\\ \cline{2-11} 
			&
			\cite{mao2019TCOM} &
			Non-orthogonal unicast and multicast &
			\circled{1}, \circled{2}, \circled{3} &
			$\blacktriangle$ &
			&
			$\surd$ &
			$\surd$ &
			$\surd$ &
			&
			$\surd$ \\ \hline
			\multirow{18}{*}{2020} &
			\cite{mao2019maxmin} &
			Cooperative user relaying for MISO BC &
			\circled{1} &
			$\blacktriangle$ &
			&
			$\surd$ &
			&
			$\surd$ &
			&
			\\ \cline{2-11} 
			&
			\cite{mao2020DPCNOUM} &
			Non-orthogonal unicast and multicast &
			\circled{5} &
			&
			$\blacktriangle$ &
			&
			$\surd$ &
			$\surd$ &
			&
			\\ \cline{2-11} 
			&
			\cite{HongzhiChen2020subcarrierRS} &
			Multicarrier multigroup multicast &
			\circled{1} &
			$\blacktriangle$ &
			&
			$\surd$ &
			&
			&
			$\surd$ &
			\\ \cline{2-11} 
			&
			\cite{yalcin2020RSmultigroup} &
			Multigroup multicast &
			\circled{1} &
			$\blacktriangle$ &
			&
			$\surd$ &
			&
			&
			$\surd$ &
			\\ \cline{2-11} 
			&
			\cite{lihua2020multicarrier} &
			Multicarrier MISO BC &
			\circled{1} &
			$\blacktriangle$ &
			&
			$\surd$ &
			&
			$\surd$ &
			&
			\\ \cline{2-11} 
			&
			\cite{alaa2020EECRAN} &
			C-RAN &
			\circled{6} &
			$\blacktriangle$ &
			&
			$\surd$ &
			&
			&
			&
			$\surd$ \\ \cline{2-11} 
			&
			\cite{alaa2020gRS} &
			C-RAN &
			\circled{3} &
			$\blacktriangle$ &
			&
			$\surd$ &
			&
			$\surd$ &
			&
			\\ \cline{2-11} 
			&
			\cite{alaa2020powerMini} &
			C-RAN &
			\circled{6} &
			$\blacktriangle$ &
			&
			$\surd$ &
			&
			$\surd$ &
			&
			\\ \cline{2-11} 
			&
			\cite{xu2021rate} &
			Joint radar and communication &
			\circled{1} &
			$\blacktriangle$ &
			&
			$\surd$ &
			&
			$\surd$ &
			&
			\\ \cline{2-11} 
			&
			\begin{tabular}[c]{@{}c@{}}\cite{jaafar2020UAV,jaafar2020UAV2}\end{tabular} &
			UAV-aided MISO BC &
			\circled{1} &
			$\blacktriangle$ &
			&
			$\surd$ &
			&
			$\surd$ &
			&
			\\ \cline{2-11} 
			&
			\cite{siyu2020vlc} &
			Visible light communication &
			\circled{1} &
			$\blacktriangle$ &
			&
			$\surd$ &
			&
			$\surd$ &
			&
			\\ \cline{2-11} 
			&
			\cite{naser2020vlc} &
			Multi-cell visible light communication &
			\circled{1} &
			$\blacktriangle$ &
			&
			$\surd$ &
			&
			$\surd$ &
			&
			\\ \cline{2-11} 
			&
			\cite{Jia2020SEEEtradeoff} &
			Multi-cell coordination &
			\circled{1} &
			$\blacktriangle$ &
			&
			$\surd$ &
			&
			$\surd$ &
			&
			$\surd$ \\ \cline{2-11} 
			&
			\cite{Liping2020secrecyCRS} &
			PHY layer security with  user relaying &
			\circled{1} &
			$\blacktriangle$ &
			&
			$\surd$ &
			&
			$\surd$ &
			&
			\\ \cline{2-11} 
			&
			\cite{Camana2020swiptRS} &
			Cognitive radio system with SWIPT &
			\circled{1} &
			$\blacktriangle$ &
			&
			$\surd$ &
			&
			$\surd$ &
			&
			\\ \cline{2-11} 
			&
			\cite{fuhao2020secrecyRS} &
			PHY layer security &
			\circled{1} &
			$\blacktriangle$ &
			&
			&
			$\surd$ &
			&
			$\surd$ &
			\\ \cline{2-11} 
			&
			\cite{zhaohui2020IRS} &
			IRS-assisted MISO BC &
			\circled{1} &
			$\blacktriangle$ &
			&
			$\surd$ &
			&
			&
			&
			$\surd$ \\ \hline
			\multirow{17}{*}{2021} &
			\cite{alaa2020IRScran} &
			IRS-assisted C-RAN &
			\circled{6} &
			$\blacktriangle$ &
			&
			$\surd$ &
			&
			$\surd$ &
			&
			\\ \cline{2-11} 
			&
			\cite{alaa2020cranimperfectCSIT} &
			C-RAN &
			\circled{6} &
			$\blacktriangle$ &
			&
			&
			$\surd$ &
			$\surd$ &
			&
			\\ \cline{2-11} 
			&
			\cite{hassan2020fogRAN} &
			D2D in F-RAN &
			\circled{6} &
			$\triangle$ &
			&
			$\surd$ &
			&
			$\surd$ &
			&
			\\ \cline{2-11} 
			&
			\cite{longfei2020satellite} &
			Multi-beam satellite networks &
			\circled{1} &
			$\blacktriangle$ &
			&
			&
			$\surd$ &
			&
			$\surd$ &
			\\ \cline{2-11} 
			&
			\cite{abdel2019finite} &
			Finite constellation in MISO BC &
			\circled{1} &
			$\triangle$ &
			&
			$\surd$ &
			&
			$\surd$ &
			&
			\\ \cline{2-11} 
			&
			\cite{Lin2021Cognitive} &
			Cognitive satellite-terrestrial networks &
			\circled{1} &
			$\blacktriangle$ &
			&
			&
			$\surd$ &
			&
			&
			$\surd$ \\ \cline{2-11} 
			&
			\cite{hongzhi2021RSLDPC} &
			Multicarrier multigroup multicast &
			\circled{1} &
			$\blacktriangle$ &
			&
			$\surd$ &
			&
			&
			$\surd$ &
			\\ \cline{2-11} 
			&
			\cite{Hassan2021FRAN} &
			F-RAN &
			\circled{1} &
			$\triangle$ &
			&
			$\surd$ &
			&
			$\surd$ &
			$\surd$ &
			\\ \cline{2-11} 
			&
			\cite{bansal2021IRS} &
			IRS-assisted MISO BC &
			\circled{1} &
			$\triangle$ &
			&
			$\surd$ &
			&
			$\surd$ &
			&
			\\ \cline{2-11} 
			&
			\cite{onur2021jamming,onur2021CR} &
			Joint communication and jamming &
			\circled{1} &
			$\blacktriangle$ &
			&
			&
			$\surd$ &
			$\surd$ &
			&
			\\ \cline{2-11} 
			&
			\cite{si2021imperfectSatellite} &
			Multi-beam satellite networks &
			\circled{1} &
			$\blacktriangle$ &
			&
			&
			$\surd$ &
			&
			$\surd$ &
			\\ \cline{2-11} 
			&
			\cite{longfei2021LLS} &
			Multi-beam satellite networks &
			\circled{1} &
			$\blacktriangle$ &
			&
			&
			$\surd$ &
			&
			$\surd$ &
			\\ \cline{2-11} 
			&
			\cite{onur2021DAC} &
			Joint radar and communication &
			\circled{1} &
			$\blacktriangle$ &
			&
			$\surd$ &
			&
			$\surd$ &
			&
			\\ \cline{2-11} 
			&
			\cite{onur2022DFRC} &
			Joint radar and communication &
			\circled{1} &
			$\blacktriangle$ &
			&
			&
			$\surd$ &
			&
			& $\surd$
			\\ \cline{2-11} 
			&
			\cite{xu2021rate} &
			Joint radar and communication &
			\circled{1} &
			$\blacktriangle$ &
			&
			$\surd$ &
			&
			$\surd$ &
			&
			\\ \cline{2-11} 
			&
			\cite{rafael2021radarsensing} &
			Joint radar  and communication &
			\circled{1} &
			$\blacktriangle$ &
			&
			&
			$\surd$ &
			$\surd$ &
			&
			\\ \cline{2-11} 
			&
			\cite{alaa2021cacheCRAN} &
			Cache-aided C-RAN &
			\circled{6} &
			$\blacktriangle$ &
			&
			&
			$\surd$ &
			&
			$\surd$ &
			\\ \cline{2-11} 
			&
			\cite{mashuai2021vlc} &
			Visible light communication &
			\circled{1} &
			$\blacktriangle$ &
			&
			$\surd$ &
			$\surd$ &
			$\surd$ &
			&
			\\ \hline
		\end{tabular}
		\vspace{0.1cm}
		
		Notations: \circled{1}: 1-layer RS; \circled{2}: 2-layer hierarchical RS (HRS); \circled{3}: Generalized RS; \circled{4}: THPRS; \circled{5}: DPCRS; \circled{6}: RS-CMD;  $\blacktriangle$: Optimized precoding;  $\triangle$: Low-complex precoding.
	\end{table*}
	\section{Emerging Applications, Challenges, and Future Research Trends of RSMA}
	\label{sec:emergingApp}
	%%%
	\par 
	As summarized in Table \ref{tab:commuSurveyApp}, 
	the appealing benefits of RSMA discovered for the multi-antenna BC have spawned an explosion of RSMA studies investigating its applicability to and interplay with multifarious 5G/6G enabling communication techniques such as massive MIMO, IRS, visible light communication (VLC), UAV,  joint communication and sensing, and satellite communications networks.
	%%%
	Despite the substantial interest,
	the study of RSMA is still in its infancy. 
	%%%
	% For all applications summarized in Table \ref{tab:commuSurveyApp}, 
	For all applications listed in Section \ref{sec:literatureReview},
	there are numerous open problems in the categories of  PHY layer design (such as coding and modulation, precoder design, receiver design), cross-layer design (such as joint scheduling and precoder design, subcarrier allocation, large-scale networks, stochastic geometry analysis), performance analysis (such as  throughput, bit error rate, outage probability), and so on.
	%%%
	In this section, we summarize the applications of RSMA to enabling technologies in 6G, and the research challenges for each application, followed by some potential research directions.
	%%%
	Subsequently, we discuss the standardization and implementation of RSMA.  
	
	\subsection{Technical Aspects of RSMA -- The Road Ahead}
	\label{sec:roadahead}
	\labelsubseccounter{VIII-A}
	\subsubsection{MIMO BC}
	The studies of RSMA  for the multi-antenna broadcast channel heretofore have mainly considered a single antenna at each receiver (a.k.a. MISO BC). 
	%%%
	The extension to multi-antenna receivers, however, is not well-understood in information or communication theoretic terms. When each user has multiple receive antennas, it is capable of receiving vectors of common streams and private streams, and the receive filter can be further designed for  interference management. Therefore, the number of  common streams transmitted in parallel, the DoFs, and the transceiver design of RSMA have to be reconsidered for the MIMO BC. 
	%%%
	\par 
	In the information-theoretic literature, an achievable DoF region of RSMA in the two-user  MIMO BC with asymmetric numbers of antennas and asymmetric levels of partial CSIT has been established in  \cite{chenxi2017bruno}, which is further shown to be the optimal DoF region in \cite{Davoodi2021DoF}. 
	%%%
	The authors of \cite{anup2021MIMO} further establish the achievable sum-DoF of RSMA  in the $K$-user symmetric MIMO BC with $M$ transmit antennas,
	$Q$ receive antennas at each user, and an arbitrary number of common streams in the range of $[1, \min(M,Q)]$. 
	%%%
	In \cite{Zheng2021TIT}, RSMA was shown to achieve the capacity region of the two-user MIMO BC with perfect CSIT up to a constant gap, but such constant-gap optimality does not extend to the three-user case.
	%%%
	\par
	As for the communication-theoretic literature, \cite{bruno2020MUMIMO} was the first work that studied the precoder design for RS in the MIMO BC but with only a single common stream. Each user therefore receives $Q$ replicated common streams and a vector of private streams. 
	%%%
	Several practical stream combining techniques  along with regularized block diagonalization  linear precoders are proposed in \cite{bruno2020MUMIMO}.
	%%%
	Even though the combining techniques for RS yield a sum-rate gain over conventional multi-user MIMO (with linear precoding), the DoFs of RS in the MIMO BC are not fully exploited as multiple receive antennas at each user enable a vector of common streams to be transmitted.
	%%%
	Recent works \cite{tuan2019, schober2021MUMIMORS, anup2021MIMO}  study precoder optimization for RSMA. Both \cite{tuan2019} and \cite{schober2021MUMIMORS} are limited to the underloaded MIMO BC with perfect CSIT and a fixed number of common streams, while \cite{anup2021MIMO} generalizes to both the underloaded and overloaded MIMO BC with perfect and imperfect CSIT. The impact of transmitting different numbers of common streams in the MIMO BC is also investigated in \cite{anup2021MIMO}.
	%%%
	Numerical results show a substantial ESR and sum-DoF improvement for RSMA in all settings and that this gain increases with the number of transmitted common streams when CSIT is imperfect. 
	%%%
	Besides the theoretical results under the assumptions of Gaussian signaling and infinite block lengths, the PHY layer architecture and the corresponding LLS of RSMA for realistic finite constellation modulation, finite-length polar codes, and AMC are studied in \cite{onur2021mobility, anup2021MIMO}. The observed significant throughput gain of RSMA over MU--MIMO and MIMO NOMA is consistent with its Shannon ESR gain.   
	%%%
	Therefore, RSMA has significant potential for communication over the MIMO BC. 
	%%%
	\par
	The capacity region of the $K$-user MIMO BC with imperfect CSIT remains an open problem. 
	%%%
	As (linearly-precoded) RSMA was shown to achieve the optimal DoF region in the $2$-user MIMO BC with imperfect CSIT as well as the  optimal DoF region of the $K$-user MISO BC with imperfect CSIT, there is a good chance that RSMA will also be an efficient solution for the $K$-user MIMO BC with imperfect CSIT. Further studies of the fundamental limits of the MIMO BC with imperfect CSIT and the role played by RSMA to achieve those limits need to be conducted. Even though the performance gain of RSMA increases with the number of common streams, the receiver complexity also increases. 
	%%%
	Therefore, in view of a practical implementation, the number of common streams that RSMA   requires to achieve a favorable tradeoff between complexity and performance is worth further investigation.
	%%%
	\subsubsection{MIMO IC}
	%%%
	The entire capacity region and the  corresponding capacity-achieving strategy for the MIMO IC remain unknown for both perfect and imperfect CSIT.
	%%%
	The HK scheme proposed for the SISO IC has been  proved to achieve the capacity region of the two-cell MIMO IC within a constant gap \cite{MIMOIC2013capacityRegion}. 
	%%%
	The authors of \cite{chenxi2017brunotopology} further show that RSMA achieves the best known DoF region for the $K$-cell MISO IC with imperfect CSIT and the result has also been extended to the (two-cell) MIMO IC with an arbitrary number of antennas at each node in \cite{chenxi2017bruno}.
	%%%
	In \cite{Jia2020SEEEtradeoff}, coordinated  beamforming for RSMA for the multi-cell MISO IC with perfect CSIT is studied and is  shown to achieve a better SE--EE tradeoff than SDMA and NOMA.
	%%%
	A comprehensive performance analysis for the $K$-cell RSMA-aided MIMO IC is provided in \cite{su2021MIMOICRS}, where the benefits of integrating RSMA with interference alignment are investigated in terms of the average sum rate, outage probability, and symbol error rate.
	%%%
	\par
	Motivated by the DoF, SE, and EE performance enhancement realized by RSMA in the MIMO IC, it is of interest to further explore its potential for achieving the entire DoF and capacity region of the MIMO IC for both perfect and imperfect CSIT. 
	%%%
	In terms of practical designs, the PHY layer design and LLS of the RSMA-aided MIMO IC have not been studied, yet. Existing works on the MIMO IC mainly resort to interference alignment to confine undesired interference at each receiver into a lower-dimensional subspace. This method however is very sensitive to CSIT uncertainties.
	%%%
	The integration of RSMA and interference alignment has great potential to marry the advantages of both techniques so as to further enhance the SE of transmission. 
	%%%
	The resulting open problems such as transceiver design and cross-layer resource allocation for different objectives are worth studying.
	%%%
	\subsubsection{Alternative Receiver Design}
	SIC at the receivers is a major component of RSMA for interference management. A well-known problem of SIC is error propagation, which affects the decoding reliability of messages after interference cancellation. It has been shown in \cite{Onur2020LLS,onur2021mobility,longfei2021LLS,anup2021MIMO,JRCLoli2022,DLloli2022} that RSMA performs well with receivers employing SIC for perfect CSIR even with error propagation due to incorrectly decoded common messages. However, performance degradation occurs for imperfect CSIR, even if the message to be cancelled is decoded correctly \cite{DLloli2022}. A deep learning based receiver algorithm is proposed in \cite{DLloli2022} to address this problem. The results show that the proposed receiver mitigates the effects of interference cancellation under imperfect CSIR and enhances the error rate performance by improving the detection of modulated symbols before and after interference cancellation. The results in \cite{DLloli2022} highlight that the error rate performance of RSMA under imperfect CSIR is an important subject that merits further investigation. 
	\subsubsection{Hybrid Automatic Repeat Request (HARQ) Design}
	HARQ is an important mechanism to improve the efficiency of packet-based transmission in wireless networks. HARQ combines the MAC-layer ACK-NACK feedback mechanism with channel coding to obtain an efficient re-transmission scheme for recovering messages with missing packets. Therefore, HARQ is an important mechanism in modern wireless communication standards. The design of HARQ schemes for RSMA has not been considered in the literature yet, to the best of the authors' knowledge. 
	\par
	For SDMA, HARQ is realized by encoding the packets intended for different users separately. Under the RSMA framework, this task is less straightforward due to the message splitting and combining operations, and the variety of scenarios where common and/or private messages are incorrectly decoded at each user. Therefore, how the user messages are to be encoded and how the re-transmissions are performed for each user when RSMA is combined with HARQ requires further study.
	\subsubsection{Waveform Design}
	The vast majority of studies on RSMA neglected the aspect of waveform design. The authors in \cite{lihua2020multicarrier, hongzhi2021RSLDPC, HongzhiChen2020subcarrierRS, onur2021jamming, onur2021CR} considered resource allocation and precoder optimization for multi-carrier waveforms, however, without considering any imperfections related to the waveform structure. In practice, waveforms may be susceptible to impairments for various reasons, as in the case of OFDM suffering from intercarrier interference under Doppler shift.
	RSMA can help to improve the robustness of waveforms by addressing such vulnerabilities and provide flexibility in waveform design to achieve higher SE.
	\subsection{RSMA for Enabling Technologies in 6G}
	\label{sec:enablingTech}
	\labelsubseccounter{VIII-B}
	\subsubsection{Massive MIMO}
	Massive MIMO is a MU--MIMO technique where the transmitter is equipped with a large number of antennas and serves multiple  users in the same resource block. 
	%%%
	It has been considered as the most promising air interface technique for achieving high SE and EE for 5G NR.
	%%%
	To realize its substantial potential gains, massive MIMO heavily relies on accurate CSIT,  which however is particularly challenging to attain  when the number of transmit antennas is large. 
	%%%
	The unaffordable feedback overhead in FDD massive MIMO, and  the imperfect channel estimation during the training phase and pilot contamination in TDD massive MIMO, make imperfect CSIT a major bottleneck for realizing the benefits of massive MIMO.
	%%%
	\par 
	As has been highlighted in Section~\ref{sec:advantageRSMA}, RSMA is more robust to CSIT uncertainties (resulting from different kinds of impairment sources) than existing MA schemes, which therefore makes the integration of RSMA and massive MIMO a promising approach to tackle the aforementioned bottleneck of massive MIMO. 
	%%%
	The authors of \cite{Minbo2016MassiveMIMO} pioneered the integration of RSMA and (FDD) massive MIMO.
	%%%
	To reduce the CSI feedback, a commonly used approach in FDD massive MIMO is two-tier precoding,  where the outer precoder controls the inter-group interference based on long-term CSIT and the inner precoder controls intra-group interference based on a short-term effective channel \cite{Chen2014twoTierMassiveMIMO,jPark2015twoTierMassiveMIMO,Kim2015twoLayerMassiveMIMO}. 
	%%%
	Motivated by the two-tier precoding, 2-layer HRS, as illustrated in Fig. \ref{fig: HRS}, was proposed in \cite{Minbo2016MassiveMIMO}. By utilizing two layers of common streams to respectively manage the inter-group and inner-group interference, 2-layer HRS reaps the rate saturation at high SNR and achieves a significant sum-rate gain over existing two-tier precoding strategies. 
	%%%
	For TDD massive MIMO, RSMA has been studied in  a more realistic scenario where the channel estimation at the transmitter is hampered by hardware impairments in \cite{AP2017bruno}. The authors of \cite{anup2022Pilot} further investigate the potential of RSMA to mitigate pilot contamination. 1-layer RS has been shown to achieve a substantial gain over conventional massive MIMO based on MU--LP when CSIT is compromised by two different impairment sources (hardware impairments and pilot contamination).
	%%%
	The impact of user mobility on (TDD/FDD) massive MIMO was studied in \cite{onur2021mobility}. As illustrated in Fig. \ref{fig:LLSmobility}, RSMA yields a significant throughput gain over conventional massive MIMO based on MU--LP and it is able to maintain multi-user connectivity in mobile deployments.
	%%%
	Besides CSIT imperfection, RSMA has also shown its capability of mitigating self-interference at relay stations equipped with massive antennas in multi-pair massive MIMO relay systems \cite{Asos2018MultiRelay}.
	%%%
	\par
	\textit{Challenges and future work:} 
	Encouraged by the appealing performance gain of RSMA in massive MIMO, the study of RSMA in cell-free massive MIMO  is another promising research direction \cite{anup2022CFMIMO}. 
	%%%
	Cell-free massive MIMO extends single-cell massive MIMO to multi-cell scenarios without cell  boundaries, and therefore, enjoys the benefits of both network MIMO and massive MIMO \cite{Ngo2017cellFreeMIMO}. 
	%%%
	Cell-free massive MIMO usually assumes no instantaneous CSI sharing among users, and relies on the use of non-orthogonal uplink pilot sequences for channel estimation, which therefore results in pilot contamination.
	%%%
	This could be successfully resolved by RSMA as in conventional massive MIMO \cite{anup2022Pilot}. 
	%%%
	Another issue that hinders the wide use of  massive MIMO is its high power consumption dominated by the power used by the digital-to-analog converters (DACs) and analog-to-digital converters (ADCs) in each RF chain.
	%%%
	To alleviate the power consumption burden, low-resolution ADCs/DACs could be used, which however, distort the transmit/received signals.
	%%%
	Whether RSMA is more robust against the impact of low-resolution ADCs/DACs, and whether RSMA can help reduce the resolution of ADCs/DACs while maintaining comparable performance as conventional approaches are topics worth investigating. 
	%%%
	\subsubsection{Millimeter-wave}
	%%%
	The explosive growth of mobile data traffic and the scarcity of the  spectrum resources available for cellular networks have motivated the investigation of mmWave communication which exploits the frequency band from 30 GHz to 300 GHz for wireless communication. 
	%%%
	However, mmWave communication suffers from many challenges, among which high propagation loss is the most significant one.  
	%%%
	To compensate such loss, one effective approach is to deploy massive antennas at the transmitter for providing high array gain.  
	%%%
	The small carrier wavelength of mmWave systems enables deploying massive antennas in a small physical space, but makes fully digital arrays impractical.
	% %%%
	% Small physical separation among antennas leaves insufficient room to accommodate all radio frequency (RF) chains.
	%%%
	Hybrid analog/digital processing, consisting of a high-dimensional analog beamformer cascaded with a reduced-dimensional digital precoder, has been shown to be a more practical solution \cite{Ayach2014hybridBFmmWave}. Its performance, however, is limited by imperfect CSIT and strong multi-user interference.
	%%%
	The conventional feedback procedure for mmWave massive MIMO relies on a two-stage approach to sequentially feedback the indices of the codewords for the analog precoders and the quantized effective channel, which entails high complexity and overhead \cite{twoStageBF2015TWC}. 
	%%%
	\par 
	Inspired by the advantages of RSMA in the multi-antenna BC, the interplay of RSMA and mmWave was first studied in \cite{minbo2017mmWave}.
	%%%
	By leveraging statistical CSIT to design digital precoders, RSMA further reduces the training and feedback complexities via a simple one-stage feedback while achieving a sum-rate comparable to that of conventional mmWave massive MIMO based on MU--LP and two-stage feedback.
	%%%
	The authors \cite{zhengli2019mmWaveRS} further study the joint analog and digital precoder optimization for the RSMA-aided mmWave model proposed in \cite{minbo2017mmWave}.
	%%%
	The extension to a UAV network was  studied in \cite{UAVRS2019ICC} where the RSMA-aided mmWave UAV network was shown to provide higher EE than NOMA for the practical 3GPP Standardized 5G  channel model. 
	%%%
	\par
	\textit{Challenges and future work:}
	%%%
	To further reduce the feedback overhead and to exploit the sparse nature of mmWave channels in the angle domain, compressive sensing or deep learning-based approaches could be applied in RSMA-aided mmWave massive MIMO networks to compress the pilot measurements into lower-dimensional matrices.
	%%%
	As quantization distortion is more tolerable with RSMA, the joint use of CSI compression techniques and RSMA in mmWave massive MIMO would be beneficial to ease the feedback overhead.
	%%%
	\subsubsection{Multigroup multicasting}
	PHY layer multicasting characterizes the point-to-multipoint transmission when a transmitter simultaneously sends one multicast content to a group of recipients. It has been included as multimedia broadcast/multicast service (MBMS) in 3GPP Release 9 for 4G, evolved MBMS (eMBMS) in 3GPP Release 11 for LTE-A \cite{eMBMS2012LTEA}, and NR multicast and broadcast services (MBS) in 3GPP Release 17 for 5G \cite{3gpp38413}.
	%%%
	Due to the diverse requirements of different user groups, multigroup multicasting has emerged as an extension of conventional multicasting where multiple multicast contents are simultaneously disseminated to  different  groups of users (a.k.a. multicast group). 
	%%%
	% It has been incorporated in 3GPP Release-17 of the NR
	% specification, which is known as eMBMS service [19].
	%%%
	Though such approach is promising to enhance SE, inter-group interference is introduced. 
	%%%
	\par Motivated by the superior interference management capability of RSMA, research efforts have been dedicated to its application in multigroup multicasting networks \cite{hamdi2017bruno,Tervo2018SPAWC,HongzhiChen2020subcarrierRS,longfei2020satellite,yalcin2020RSmultigroup,hongzhi2021RSLDPC,hongzhi2020LLSmultigroup,longfei2021LLS,alaa2021cacheCRAN}.
	%%%
	Reference \cite{hamdi2017bruno} is the first work that studied the integration of RSMA and multigroup multicasting. For the MISO BC with perfect CSIT,  RSMA was shown to achieve notable MMF-DoF gains at high SNR  and  a  max-min rate enhancement at low SNR by enabling partial decoding of the inter-group interference and partially treating the inter-group interference as noise.  
	%%%
	The authors of \cite{longfei2020satellite} further extended the DoF and rate analysis of RSMA to the imperfect CSIT setting for a more practical multibeam satellite communication application, and showed large  DoF and rate gains for RSMA when the CSIT quality is poor.
	%%%
	RSMA was also shown to achieve higher EE in multi-cell multigroup systems \cite{Tervo2018SPAWC} and  enhanced user fairness
	in multicarrier multigroup multicast systems \cite{HongzhiChen2020subcarrierRS}.
	%%%
	PHY layer design for multigroup multicasting was studied in \cite{longfei2021LLS} and LLS results were presented for multigroup multicast cellular and satellite communications, and in \cite{hongzhi2021RSLDPC}  for multicarrier multigroup multicast.
	%%%
	\par
	\textit{Challenges and future work:} 
	Each multicast stream in multigroup multicasting has to be decoded by all users in the corresponding user group, which therefore limits its achievable rate to the worst-case rate amongst the users. 
	%%%
	The large fluctuation of terrestrial channels often results in deep channel fades at the users, which therefore degrades the achievable rate of some multicast streams.
	%%%
	Contrary to terrestrial communications, satellite communications provide a larger coverage, and are well-suited for providing multicasting services for users distributed in a wider range. 
	%%%
	Integrated terrestrial-satellite  networks, aiming at providing ubiquitous services for ground users via cooperative transmission between BSs and satellites, have  therefore been introduced as a new paradigm for the next generation of communication networks \cite{kuang2018integratedsatellitete}. 
	%%%
	Inspired by the DoF, SE, and EE gains of RSMA-aided multigroup multicasting in both terrestrial and satellite networks, the application of RSMA in integrated terrestrial-satellite networks is interesting and promising to simultaneously reap its benefits in both component networks.
	%%%
	\subsubsection{Non-orthogonal unicast and multicast}
	%%%
	Recently, non-orthogonal multiplexing (NOM) was subject to intensive research focus in academia and industry \cite{3gpp38913, LDM2016ATSC, 3gpp5Gbroadcast, 2020TWCNOUM}. NOM of broadcast/multicast signals has been approved by 3GPP as a study item in Release 16 as LTE-based 5G Broadcast \cite{3gpp5Gbroadcast} and applied in the digital TV standard ATSC 3.0 under the name layer division multiplexing  \cite{LDM2016ATSC}. 
	%%%
	The research interest has been further extended to NOM of multiple services. 
	%%%
	Non-orthogonal unicast and multicast transmission (NOUM), which enables the concurrent delivery of both unicast and multicast services to users in the same time-frequency resource block, is one of the key research directions in multi-service NOM.
	%%%
	In multi-antenna NOUM, a conventional approach is to adopt MU--LP  at the transmitter to superpose the multicast stream on top of the unicast streams and  to use SIC at each user to decode and remove the multicast stream before decoding the intended private stream.  
	%%%
	Such MU--LP-aided approach does not fully exploit the advantages of SIC as SIC is only used to manage interference between unicast and multicast streams and the interference between unicast streams is fully treated as noise. 
	%%%
	\par  Motivated by the performance gains of RSMA for unicast-only and multicast-only transmissions, RSMA has been applied to NOUM.  The authors of \cite{mao2019TCOM} propose a novel RSMA-aided NOUM where the unicast messages are split into common and private parts at the transmitter, and the common parts are encoded with the multicast message into a super-common stream for all users. 
	%%%
	Each user can still use one layer of SIC to sequentially decode the super-common and the intended private stream, but the function of SIC is better  exploited in RSMA-aided NOUM as SIC is used not only  for managing the interference between the multicast and unicast streams, but also for managing the interference between the unicast streams.
	%%%
	The proposed RSMA-aided NOUM was shown to yield SE and EE gains over conventional NOUM without increasing the receiver complexity, which therefore makes RSMA well-suited for NOUM.
	%%%
	The authors of \cite{mao2020DPCNOUM} further propose a dirty paper coded RS-aided NOUM and showed its capability to enlarge the rate region achieved by 
	conventional DPC-aided NOUM and  linearly precoded RS-assisted NOUM when CSIT is imperfect.
	%%%
	\par
	\textit{Challenges and future work:} 
	The existing works \cite{mao2019TCOM,mao2020DPCNOUM} are limited to the case when the multicast message is intended for all users, i.e., a single multicast group. When multiple multicast groups are considered, each user receives three-dimensional interference, namely, interference among the multicast streams, interference between the multicast and unicast streams, and interference among the unicast streams. How to design the transceiver for RSMA for such non-orthogonal unicast and multigroup multicasting networks in order to deal with the three-dimensional interference remains unknown.

	\subsubsection{Multi-cell MIMO including coordinated multi-point (CoMP), cloud-radio access network (C-RAN), and fog-radio access network (F-RAN)}
	\label{sec:multicell}
	The advancement of RSMA in the single-cell MIMO BC has motivated its study for multi-cell networks for both coordinated transmission and cooperative transmission, as illustrated in Section \ref{sec:RSMAmulticell}.
	%%%
	In coordinated multi-cell networks with only CSI shared among all BSs, RSMA has been shown to achieve enhanced SE and EE performance when the inter-cell interference is fully treated as noise and RSMA is applied to only manage the intra-cell interference \cite{Tervo2018SPAWC,Jia2020SEEEtradeoff}, or when the inter-cell interference is partially decoded and partially treated as noise via RSMA \cite{carleial1978RS, TeHan1981,chenxi2017brunotopology,Tse2008,wonjae2020multicell}.
	%%%
	Apparently, schemes that manage inter-cell interference via RSMA outperform  their counterparts that fully treat inter-cell interference as noise. However, the former approach imposes higher decoding burdens at the receiver side.  
	%%%
	\par Recent research has investigated cooperative multi-cell networks \cite{mao2018networkmimo, Alaa2019uavCRAN,Ala2019IEEEAccess,cran2019wcl,alaa2020EECRAN,alaa2020gRS,alaa2020powerMini,alaa2020cranimperfectCSIT,alaa2020IRScran} where all BSs share the CSI and data of all users through fronthaul (a.k.a. backhaul in some papers) links  as this allows the users in different cells to share the same common stream(s), which therefore simplifies the receiver design. 
	%%%
	Reference \cite{mao2018networkmimo} was the first work that investigated RSMA in cooperative multi-cell networks under the assumption of unlimited fronthaul capacity, where RSMA was shown to achieve higher SE than MU--LP  and multi-antenna NOMA for various inter-user and inter-cell channel strength disparities. 
	%%%
	Considering practical cooperative multi-cell networks with non-ideal limited fronthaul capacity, C-RAN  has attracted intense research interests \cite{Peng2015CRAN}. 
	%%%
	The application of RSMA in C-RAN has been widely studied for both perfect CSIT \cite{Alaa2019uavCRAN,Ala2019IEEEAccess,cran2019wcl,alaa2020EECRAN,alaa2020gRS,alaa2020powerMini} and imperfect CSIT \cite{alaa2020cranimperfectCSIT}, where RSMA was shown to alleviate the fronthaul capacity limitation for a given QoS rate requirement.
	%%%
	To further ease the pressure of information exchange and signal processing at the central processor, F-RAN has been proposed as a new paradigm for 5G and beyond to exploit intelligence at the network edge. Specifically, F-RAN enables edge-caching to prefetch content with high reuse probability from the centralized processor at the remote radio heads  or end-users. 
	%%%
	RSMA in cache-aided C-RAN and F-RAN has been studied in   \cite{alaa2021cacheCRAN,hassan2020fogRAN,Hassan2021FRAN}, where RSMA based on RS--CMD was shown to achieve significant performance  gains over MU--LP and NOMA requiring a much smaller caching size and fronthaul capacity. 
	%%%
	Therefore, RSMA has great potential to enhance user performance and simplify the transceiver design in cache-aided C-RAN.
	%%%
	\par
	\textit{Challenges and future work:} 
	Existing works on RSMA-aided F-RAN assume that the cache placement  is determined in advance (without using RSMA), and RSMA is only used in the delivery phase. All messages, no matter whether they are cached or not, are split and transmitted as in conventional C-RAN and caching only influences the fronthaul capacity constraint. 
	%%%
	Though cache placement normally uses a much larger timescale than content delivery, it is still possible to jointly optimize content placement and delivery based on a mixed timescale to better enhance the system performance for a limited cache size in RSMA-aided F-RAN. 
	
	\subsubsection{Simultaneous wireless information and power transfer (SWIPT)}
	SWIPT unifies wireless information transfer (WIT) and wireless power transfer (WPT) by enabling the transmitter to deliver information and energy to the corresponding information receivers and  energy receivers at the same time.
	%%%
	One major issue in conventional multi-antenna SWIPT networks based on MU--LP is that a dedicated energy-carrying signal is needed for each energy receiver to guarantee its harvested power \cite{jiexu2014swipt} if energy receivers and information receivers are separated. 
	%%%
	\par
	Inspired by the appealing WIT performance of RSMA, RSMA for multi-antenna SWIPT was proposed in \cite{mao2019swipt}  where a multi-antenna transmitter is serving multiple separated  information receivers and  energy receivers.
	%%%
	Numerical results show that for a given lower bound on the power harvested at the energy receivers, the rate region of the information receivers for 1-layer RS outperforms that of MU--LP.
	%%%
	Moreover, the rate region performance for dedicated energy-carrying signals is almost the same as the one without dedicated energy-carrying signals. 
	%%%
	The common stream of 1-layer RS is capable of  improving the interference management among information receivers, and at the same time can guarantee the  energy harvested at the energy receivers.
	%%%
	When the information receivers and  energy receivers are co-located, i.e., each receiver requires both information and energy, RSMA can save more transmit power than SDMA as illustrated in \cite{RSswiptIC2019CL,Camana2020swiptRS}.
	%%%%
	\par
	\textit{Challenges and future work:} 
	One major challenge of SWIPT is the limited transmission range to the energy receivers, which is a common limitation of WPT. A potential approach to resolve this issue is to exploit other coverage extension techniques such as user relaying, IRS, etc. The performance of RSMA in these SWIPT scenarios remains unknown.
	%%%
	In addition, ensuring information security is more challenging  in SWIPT as energy receivers are normally located closer to the transmitter compared with information receivers. Energy receivers are therefore capable of  eavesdropping the information transmitted to the information receivers with relatively higher received signal strengths.
	%%%
	As RSMA encodes the information of each information receiver into common and private streams, only when both the intended common and private  streams  are successfully decoded at the energy receivers, the information of that information receiver can be wiretapped.
	%%%
	Therefore, RSMA has great potential to address the security issue of SWIPT, which merits further investigation in  future works.
	%%%
	\subsubsection{Cooperative user relaying}
	Cooperative user relaying is a promising technique to enhance system capacity, transmission reliability, and  coverage without requiring extra antennas at the transceivers. 
	%%%
	Motivated by the broadcast nature of wireless transmissions, cooperative user relaying allows users to overhear the information of other users emitted by  the transmitter and then forward what is received to the intended users. By creating spatially independent transmission paths from the transmitter and relaying user,  cooperative user relaying improves the spatial diversity and enhances  the received signal strength. 
	%%%
	Therefore, it has been included in several wireless standards such as IEEE 802.11, IEEE 802.15.4a, and IEEE 802.16 standards \cite{akyildiz2005wireless,Ghosh2005WiMax,Lee2006mesh}.
	%%%
	One of the most attractive strategies for cooperative user relaying is cooperative NOMA. As NOMA  requires multiple users to decode the messages of other users, it can be easily integrated with cooperative user relaying. 
	%%%
	However, as discussed in Section \ref{sec: NOMA}, NOMA suffers from  limitations imposed by the full decoding of the interference and this drawback also limits cooperative NOMA.
	%%%
	\par 
	Inspired by the advantages of 1-layer RS in the multi-antenna BC and the need of all users to decode the common stream, the authors of \cite{jian2019crs} first studied cooperative user relaying and RSMA in a  two-user MISO BC and proposed a novel RSMA strategy, namely, cooperative rate-splitting (CRS). 
	%%%
	The authors of \cite{mao2019maxmin} further extended the proposed transmission framework to the $K$-user case.
	%%%
	Specifically, the transmission requires two phases. In the first phase, the RSMA-enabled transmitter serves all users as in the conventional BC. In the second phase, the selected relaying user(s) opportunistically forward the decoded common stream to the remaining users so as to further enhance the rate of the common stream.
	%%%
	CRS is indeed a general  strategy that includes 1-layer RS as a special case when the second phase of CRS is turned off.
	%%%
	The numerical results in \cite{jian2019crs,mao2019maxmin} have shown substantial rate region and max-min fairness gains of CRS over cooperative NOMA, 1-layer RS, and MU--LP, even when the SNR is low. 
	%%%
	Furthermore, CRS is capable of  enhancing PHY layer security, as shown in \cite{Liping2020secrecyCRS}. 
	%%%
	\par
	\textit{Challenges and future work:} 
	Existing works on CRS all assume perfect CSIT. As imperfect CSIT is the major bottleneck in current wireless networks, more research attention should be dedicated to CRS with imperfect CSIT resulting from different sources of impairment. 
	%%%
	Besides the heuristic relaying user selection algorithm proposed in \cite{mao2019maxmin}, more sophisticated selection algorithms including a joint optimization of precoding and relaying user selection are worth investigating.
	%%%
	\subsubsection{Wireless caching}
	Wireless caching is an auspicious technology to ease  backhaul traffic loads and address the issues of spectrum scarcity and latency in future wireless networks.  
	%%%
	There are two phases involved in cache-aided transmission, namely, the placement and the delivery phases. Popular contents are prefetched to local nodes such as small-cell BSs and end users during the placement phase while the transmitter delivers the requested data, including the cached and uncached contents, to the users during the delivery phase. 
	%%%
	Caching has been shown to be beneficial in wireless networks to manage interference  by allowing multiple transmitters to cache and transmit the same content collaboratively \cite{naderializadeh2017caching}.
	\par 
	Motivated by the excellent interference management capability of RSMA, its application in cache-aided networks has received significant attention \cite{enrico2016bruno, Cache2017TIT, lampiris2017cache, alaa2021cacheCRAN,Demarchou2022}. By splitting each user message  into a cached part and an uncached part during the placement phase, and using 1-layer RS  to transmit the uncached parts during the delivery phase,  RSMA-aided caching  has been shown to boost the DoF and the coded-caching gain, and it further reduces the CSIT requirement at the transmitter \cite{enrico2016bruno}. 
	System level caching-aided RSMA is further investigated in \cite{Demarchou2022}, and the gains introduced by the co-design of wireless caching and RSMA are further exploited.
	%%%
	\par
	\textit{Challenges and future work:} 
	Existing works on RSMA-aided caching  mainly tackle information-theoretic problems.  
	%%%
	The optimization of  caching methods (including placement and delivery policies) to enhance SE/EE or to reduce the transmission delay are worth studying. 
	%%%
	Moreover, RSMA-aided caching in F-RAN with fronthaul constraints still needs more efforts as mentioned in Section~\ref{sec:multicell}.
	%%%
	\subsubsection{Unmanned aerial vehicles-aided communications}
	UAV (a.k.a. drones) are gaining prominence  in 5G NR and 6G for their merits of coverage extension, convenient deployment, and low cost and highly controllable 3D mobility.
	%%%
	\par 
	Applications of RSMA in UAV-aided communications have emerged recently \cite{Alaa2019uavCRAN,UAVRS2019ICC,jaafar2020UAV,jaafar2020UAV2,Lin2021Cognitive}. 
	%%%
	In \cite{Alaa2019uavCRAN}, the UAVs are used to replace  broken BSs in order to maintain service to mobile users. By applying RSMA at the UAVs and BSs, the sum-rate performance shows significant improvement over  conventional SDMA. 
	%%%
	The authors of \cite{UAVRS2019ICC} consider RSMA in UAV-aided mmWave transmission, where the UAVs are receivers and served by a terrestrial BS, and RSMA is shown to achieve superior EE compared to NOMA.
	%%%
	The transition of MA from OMA and NOMA to RSMA for aerial networks has been documented in \cite{jaafar2020UAV}.
	%%%
	The issue of UAV placement is addressed in \cite{jaafar2020UAV2}, and RSMA is shown to achieve WSR gains over SDMA and NOMA.
	%%%
	In \cite{Lin2021UAV}, RSMA is further shown to improve the sum-rate in satellite and aerial integrated networks.
	%%%
	\par
	\textit{Challenges and future work:} 
	One major challenge in UAV-aided communication is the UAV deployment and trajectory design based on the estimated channels between the UAVs and the ground nodes. Perfect tracking of the channels is impossible due to the 3D channel structure and the high mobility of UAVs, which therefore  results in severe co-channel interference. RSMA has great potential to address this issue thanks to its robustness towards  CSIT inaccuracy. However, the trajectory design problem for RSMA and UAV-aided communication systems and the influence of UAV mobility on the system performance have not been investigated, yet. 
	%%%
	Moreover, whether using RSMA can help overcome the lifetime limitation of UAVs merits investigation.
	%%%
	\subsubsection{Physical layer security}
	Privacy/security  is one of the most challenging and critical issues in wireless communications due to the broadcast nature of wireless channels.
	%%%
	One promising method  to enhance security is to protect the data transmission in the PHY layer  by preventing the eavesdroppers from decoding data while ensuring successful decoding at the legitimate users.
	%%%
	\par
	To bring RSMA on the map of the next generation communication networks, it is necessary to investigate the secrecy performance of RSMA. 
	%%%
	Reference \cite{fuhao2020secrecyRS} was the first work to study the secrecy performance of RSMA in a two-user MISO BC where the multi-antenna transmitter simultaneously  served two legitimate users and there was one eavesdropper. 
	%%%
	By maximizing the minimum secrecy rate of the two legitimate users, 1-layer RS was shown to achieve an  explicit max-min secrecy rate gain over MU--LP and NOMA. 
	%%%
	The common stream in 1-layer RS was not only used as a useful means to enhance the transmission rate of the legitimate users, but also  as  artificial noise to confound the eavesdropper.
	%%%
	The authors of \cite{Liping2020secrecyCRS} further investigate the secrecy rate of CRS when cooperative user relaying is enabled at the two legitimate users such that one user decodes and forwards the common stream to the other user. 
	%%%
	Again, CRS yields a secrecy sum-rate gain over cooperative NOMA and MU--LP.
	%%%
	Jamming is another important approach in PHY layer security, which aims at generating noise signals to confuse  potential eavesdroppers.
	%%%
	The authors of \cite{onur2021jamming} further investigate RSMA for joint communication and jamming where the transmitter simultaneously serves multiple information users (IUs) with imperfect CSIT and performs jamming on pilot subcarriers of adversarial users (AUs) based on statistical CSIT.
	%%%
	For the same amount of jamming power at the AU pilot subcarriers, RSMA achieves a significantly higher sum-rate for the IUs than SDMA.
	%%%
	\par
	\textit{Challenges and future work:} 
	Ensuring the secrecy in RSMA  is challenging when the eavesdroppers are internal legitimate users, i.e., one legitimate user decodes the intended streams, and then eavesdrops the private streams of the other legitimate users. After decoding the intended common and private streams, the interference impairing the private stream of the other legitimate users also becomes smaller.  Such problem, however, has not been studied, yet.
	%%%
	Moreover, in TDD massive MIMO, the uplink channel estimation could be affected by pilot-contamination attackers who are  actively sending the same pilot sequence as the legitimate users in order to deteriorate the estimated CSIT of the legitimate users.
	%%%
	Inspired by the enhanced robustness of RSMA against pilot contamination \cite{anup2022Pilot},  it is interesting to investigate the secrecy rate performance of RSMA in TDD massive MIMO to understand whether it is suitable for combating pilot contamination attacks.
	%%%
	\subsubsection{Satellite Communications}
	To further enhance the coverage and transmission reliability  especially in  areas where  terrestrial infrastructures are difficult to deploy, the development of satellite networks  as complements to terrestrial cellular networks has been investigated.
	%%%
	Satellite networks have evolved from single-beam to multibeam architectures, where the satellite is typically equipped with multiple feeds and serves multiple users groups within multiple co-channel beams. 
	%%%
	Such multibeam satellite communication  follows the cellular multigroup multicasting paradigm, which  introduces inter-beam interference.
	%%%
	\par The benefits of RSMA discovered for multigroup multicasting transmission in cellular networks have inspired the study of RSMA for multibeam satellite communications \cite{RSSatellite2018ISWCS,longfei2020satellite,longfei2020multibeam,longfei2021LLS,si2021imperfectSatellite,Lin2021UAV,Lin2021Cognitive,miguel2018RSsatellite}.
	%%%
	The authors of \cite{RSSatellite2018ISWCS} initiated the study of RSMA in a two-beam satellite communication network where the inter-beam interference was managed by RSMA while the inter-user interference within each beam was coordinated via TDMA. The rate region of the proposed RSMA-aided satellite communication framework was shown to be larger than that of other strategies which do not use RS or beam cooperation.
	%%%
	The authors of \cite{longfei2020multibeam,longfei2020satellite} further demonstrate MMF-DoF and max-min rate gains of RSMA over MU--LP in a generalized framework considering multiple beams in the same time-frequency resource blocks and multiple users in each beam with imperfect CSIT. 
	%%%
	RSMA is further applied to  satellite systems with multiple gateways and feeder link interference in \cite{si2021imperfectSatellite} and is shown to be powerful in handling inter-gateway interference. 
	%%%
	PHY-layer designs and LLS for satellite communications are provided in \cite{longfei2021LLS} with significant throughput gains achieved by RSMA over MU--LP.
	%%%
	To further guarantee seamless and ubiquitous service, integrated terrestrial-satellite networks have been proposed and are a promising approach.
	%%%
	RSMA has been studied in \cite{Lin2021UAV,Lin2021Cognitive,longfei2021satelliteTerres} for integrated terrestrial-satellite networks, where RSMA is shown to achieve significant sum-rate, secrecy rate, and EE gains over conventional MU--LP. 
	%%%
	\par
	\textit{Challenges and future work:} 
	Existing works on RSMA in integrated terrestrial-satellite networks are limited to single beam without inter-beam interference and the inter-user interference is completely managed by the BS on the ground. Furthermore, the cooperation between BSs and satellites has not been fully exploited yet. How satellites can efficiently act as a complement and cooperatively serve users with poor terrestrial channels is still unknown. 
	%%%
	\subsubsection{Integrated Radar/Sensing and Communication}
	The growing issue of radio spectrum congestion has catalyzed the electromagnetic RF convergence paradigm by enabling cooperative spectrum sharing among multiple RF systems.
	%%%
	Integrated   sensing and communication (ISAC) is one such approach that allows  both functionalities to share
	the same transmit signals and even the same hardware
	platforms.
	%%%
	However, such new paradigm incurs interference between communication and sensing, which has become one of the major challenges in ISAC \cite{RadCom2018TWC}.
	%%%
	\par 
	Inspired by the formidable interference management capability of RSMA, an RSMA-aided dual-functional radar-communication (DFRC) transmission framework was first proposed in \cite{chengcheng2020radCom,xu2021rate} considering one multi-antenna transmitter simultaneously detecting   radar targets and serving multiple communication users based on 1-layer RS. 
	%%%
	By designing the precoders to simultaneously maximize the WSR of the communication users and minimize the MSE for radar beampattern approximation, RSMA improves the tradeoff between WSR and MSE compared to conventional SDMA-based DFRC. 
	%%%
	RSMA not only better manages the interference among communication users, but also  successfully handles the interference between communication and radar.
	%%%
	The authors of \cite{rafael2021radarsensing, JRCLoli2022} show the tradeoff gain of  RSMA-aided DFRC for imperfect CSIT.  The authors of \cite{onur2021DAC,onur2022DFRC} further consider low resolution digital-to-analog converter (DAC) units and RF chain selection for RSMA-aided DFRC. RSMA is shown to achieve significant  sum-rate gains over SDMA for all considered numbers of quantization bits.
	%%%
	\par
	\textit{Challenges and future work:} 
	The spectrum sharing between communication and radar raises security concerns, especially for military applications. Whether RSMA-aided DFRC can enhance PHY layer security has not been investigated yet. The tradeoff between radar detection and  secrecy rate is indeed worth studying.
	%%%
	\subsubsection{Cognitive radio}
	Cognitive radio (CR) is a promising technique to overcome the scarcity of radio resources and enhance SE by allowing secondary users (SUs) to opportunistically use the frequency band shared by the licensed primary users (PUs) as long as the interference from SUs to PUs is controlled within a certain level.
	%%%
	One major challenge in CR is the interference management between SUs and PUs as PUs are normally unaware of the existence of the SUs and the burden of interference management is moved to the transmitter and SUs.
	%%%
	\par
	The powerful interference management capability of RSMA makes it a perfect fit for MIMO CR networks. 
	%%%
	The authors of \cite{Camana2020swiptRS} first investigated RSMA-aided MIMO CR networks. By minimizing the transmit power subject to a minimum data rate constraint at the SU and a maximum interference power level constraint at the PUs, RSMA was shown to reduce the power consumption at the transmitter compared with SDMA.
	%%%
	The authors of \cite{onur2021CR} further identify the performance benefits of RSMA for joint communications and jamming in a MIMO CR network with imperfect and statistical CSIT.
	%%%
	\par
	\textit{Challenges and future work:} 
	% Channel uncertainty is one major challenging  in CR, which can be potentially addressed by RSMA. However, there is  a lack of investigation of RSMA-aided MIMO CR networks with imperfect CSIT.
	%%%
	The spectrum sharing between SUs and PUs gives rise to security concerns in CR as SUs and PUs may wiretap each others' information. 
	%%%
	The secrecy performance of RSMA in such scenario, however, remains an open problem.
	%%%
	\subsubsection{Massive machine-type communication}
	IoT is essential for 5G and beyond as it provides advanced solutions for smart cities via a massive number of IoT devices. 
	%%%
	mMTC, which is tailored for massive IoT applications, has been included in  3GPP Release 13 \cite{3gpp45820} as an instrumental usage scenario for ultimately realizing massive connectivity among IoT devices with low complexity and low power. 
	How to serve massive devices with lean control signaling and receiver complexity has become the major challenge in massive IoT. 
	%%% 
	\par 
	Existing studies of RSMA have shown that RSMA is capable of 
	achieving higher SE, reducing  receiver complexities, and enhancing robustness with respect to CSIT inaccuracy and network loads, which therefore makes RSMA an appealing MA scheme for mMTC.
	%%%
	The authors of \cite{enrico2016bruno} first showed from an information-theoretic perspective that RSMA-aided power partitioning achieves the optimal DoF region in an overloaded MISO BC, where the BS has partial CSIT of some high-end users and statistical CSIT of some low-end IoT users. 
	%%%
	The DoF gain of the proposed RSMA-aided power partitioning approach in the high SNR regime motivates the study of its achievable sum-rate  in the finite SNR regime, where RSMA also facilitates significant SE enhancement \cite{mao2021IoT}.
	%%%
	\par
	\textit{Challenges and future work:} 
	IoT users in 6G are envisioned to have hybrid requirements, such as simultaneously enhancing the throughput and reducing the transmission delay. Such hybrid service is known as enhanced eMBB-URLLC-mMTC in 6G and has been identified as one of the core services in 6G. 
	%%%
	Whether RSMA is a promising  MA technique to accommodate the hybrid requirements of massive IoT users is not known, yet.
	%%%
	\subsubsection{Visible light communication}
	%%%
	VLC, which employs the visible light spectrum between 400 and 789 THz  for communication, is emerging as a compelling technique to complement the  RF-based mobile communication networks for high speed communications, especially in indoor environments.  
	%%%
	To further enhance the data rate of VLC, MU--MIMO VLC, which allows multiple light emitting diodes (LEDs) to simultaneously serve multiple users, has been developed and widely studied in the VLC literature  \cite{naser2020vlc}.
	%%%
	%Contrary to complex RF signals, the input signal in optical communications is real and positive and the user channels are also highly correlated.  
	However, multi-user interference is still a major bottleneck in MU--MIMO VLC.
	%%%
	\par Motivated by the strong interference management capability of RSMA in RF communications and the high operating SNR offered by VLC, the application of RSMA in MU--MIMO VLC is promising as RSMA achieves high DoF and SE gains over existing MA schemes. 
	%%%
	The authors of \cite{siyu2020vlc} and \cite{naser2020vlc}  were the first to study RSMA for MU--MIMO VLC. 
	%%%
	Both papers show clear sum-rate gains of RSMA over conventional MA schemes even when the user channels are highly correlated. Therefore, RSMA is  a suitable MA scheme for VLC networks. 
	%%%
	Both \cite{siyu2020vlc} and \cite{naser2020vlc} use the Shannon formula to calculate the sum-rate, which may thus not be achievable due to the unique characteristics of VLC. 
	%%%
	This motivated the authors of \cite{mashuai2021vlc} to derive the achievable rate for RSMA-aided VLC networks. Reference \cite{mashuai2021vlc} was also the first work that studied RSMA-aided VLC for imperfect CSIT.
	\par
	\textit{Challenges and future work:} 
	Open issues for RSMA in VLC systems have been thoroughly discussed in \cite{naser2020vlc}.  For example, practical limitations, such as imperfect CSIT, merit further investigation for RSMA-aided VLC. 
	
	\subsubsection{Intelligent reconfigurable surface-aided communications}
	IRS (a.k.a. reconfigurable intelligent surface--RIS, holographic MIMO surface--HMIMOS, software controlled metasurface) has emerged as a tempting technology for 6G. IRSs consist of a number of  meta-atoms or passive scatterers that are configurable and programmable in software.  
	%%%
	By dynamically tuning the reflection coefficients of the meta-atoms for an incoming signal, IRSs  have the potential to control  the propagation environment, thus  changing the design of wireless networks \cite{RIS2021ResourceManagement}. IRSs have  shown their benefits in improving PHY layer security, suppressing interference, improving wireless power transmission efficiency, improving PHY layer security, extending the cell coverage, etc.
	%%%
	\par
	The merits of RSMA in multi-antenna networks have motivated  the study of RSMA in IRS-aided communication \cite{zhaohui2020IRS,alaa2020IRScran, bansal2021IRS,jolly2021IRS,IRSRS2021TVT}.
	%%%
	The authors of \cite{zhaohui2020IRS} were the first to study RSMA for the MISO BC with multiple IRSs cooperatively assisting the downlink transmission from a BS to multiple users. By jointly optimizing the precoders, the message splits of 1-layer RS, and the phase shifts of the IRS to maximize EE, RSMA was shown to achieve a better EE performance than NOMA and OFDMA in the IRS aided MISO BC.
	%%%
	2-layer HRS and IRS-aided transmission strategies are further studied in \cite{bansal2021IRS, jolly2021IRS, IRSRS2021TVT}, where  the transmission from the BS to the users in each HRS user group is assisted by one independent IRS.
	%%%
	RSMA was shown to improve the WSR and outage performance compared to 1-layer RS/MU--LP/NOMA-based IRS transmission strategies.
	\par
	\textit{Challenges and future work:} 
	Due to the absence of RF chains, it is challenging to acquire the CSI at the IRS.
	%%%
	An alternative approach is to estimate the concatenated channel at the BS based on certain IRS reflection patterns, which however, may result in imperfect CSIT due to limited channel training resources \cite{IRSBeixiong2020WCL,IRS2020imperfectCSI}.
	%%%
	The integration of RSMA and IRS has a great potential to compensate such limitations of IRS and boost the system performance.
	%%%
	However, issues such as joint passive and active beamforming design, user scheduling, and IRS allocation have not been studied for RSMA based IRS-aided networks with imperfect CSIT, which thus merits investigation.
	%%%
	\subsubsection{Other open issues}
	The aforementioned 6G enabling technologies  are not always independent from each other. Some of them are complementary and the combination of those techniques may introduce new research directions for RSMA.
	%%%
	For example, both cooperative user relaying and NOUM have the capability of enhancing the SE.  
	%%%
	However, the performance of the multicast message in NOUM is restricted by the worst-case rate among the users.
	%%%
	By integrating CRS and NOUM, the system performance of NOUM could be potentially improved as both the common parts of the unicast messages and the multicast message are transmitted in the cooperative transmission phase. The second hop/phase of CRS is therefore used for the dual functions of improving the rate of the multicast message and enhancing the interference management among the unicast messages.
	%%%
	\par 
	In addition to the aforementioned research directions for RSMA, there are many other unplumbed and timely areas for RSMA research, such as space-air-ground integrated networks, vehicle-to-everything communications, 3D eMBB-URLLC-mMTC services, etc. More research efforts are needed in  each of these fields and the underlying motivation is discussed in the following: 
	\begin{itemize}
		\item \textit{Space-air-ground integrated networks (SAGIN)}: 
		6G is envisioned to be an all-coverage network that seamlessly integrates space, air, and ground communications. 
		%%%
		The concept of SAGIN has therefore been developed and has received considerable research attention recently. 
		%%%
		By offloading traffic among space, air, and ground network segments, the network load in each part could be balanced and alleviated. 
		%%%
		However, due to the spectrum sharing among the above segments, interference becomes one of the major challenges in SAGIN that significantly impacts the entire ecosystem of satellite-aerial-terrestrial communications. 
		%%%
		Motivated by the powerful interference management capability of RSMA demonstrated for satellite, air, and terrestrial cellular networks, RSMA has great potential to successfully address the interference in SAGIN. 
		%%%
		\item \textit{Vehicle-to-everything (V2X) communications}:
		V2X, which includes vehicle-to-network (V2N), vehicle-to-infrastructure (V2I), vehicle-to-pedestrian (V2P), and vehicle-to-vehicle (V2V) communications, has been considered as one platform for achieving intelligent transportation and is included in 3GPP release 14 as LTE-based V2X services \cite{3gppV2X,V2X2016magazine}.
		%%%
		Though appealing in its concept, V2X communication is susceptible to user mobility and there are rigorous requirements on transmission latency and reliability.
		%%%
		Recent studies of RSMA have shown that RSMA makes MU--MIMO systems robust against interference resulting from user mobility and feedback delay \cite{onur2021mobility}, and it outperforms existing MA schemes for finite block length coding \cite{yunnuo2021FBL}. It is therefore a promising option for V2X communication. 
		%%%
		\item \textit{3D eMBB-URLLC-mMTC services}:
		NR defines three core services for 5G, namely URLLC, mMTC, and eMBB. 
		%%%
		It is envisioned that 6G needs to provide three-dimensional services for satisfying the mixed demands of the new use cases. %%%
		%%%
		All possible combinations of the three core services in NR, such as the enhanced eMBB-URLLC-mMTC service,  enhanced eMBB-URLLC service, enhanced eMBB-mMTC service, and enhanced mMTC-URLLC service, have been identified as hybrid core services in 6G. 
		%%%
		As discussed in \cite{onur2021sixG}, RSMA provides a unified solution for facilitating the aforementioned hybrid core services in 6G thanks to its enhanced performance in terms of the 6G KPIs, i.e., data rate,  mobility, network density and dynamic topology, latency and reliability, and EE.
		%%%
		There are numerous open research topics in 6G that merit investigation in the context of RSMA.
		
	\end{itemize}
	\subsection{Standardization and Implementation of RSMA}
	\label{sec:standardRSMA}
	\labelsubseccounter{VIII-C}
	%%%
	\begin{figure}[t!]
		\centering
		\includegraphics[width=3.0 in]{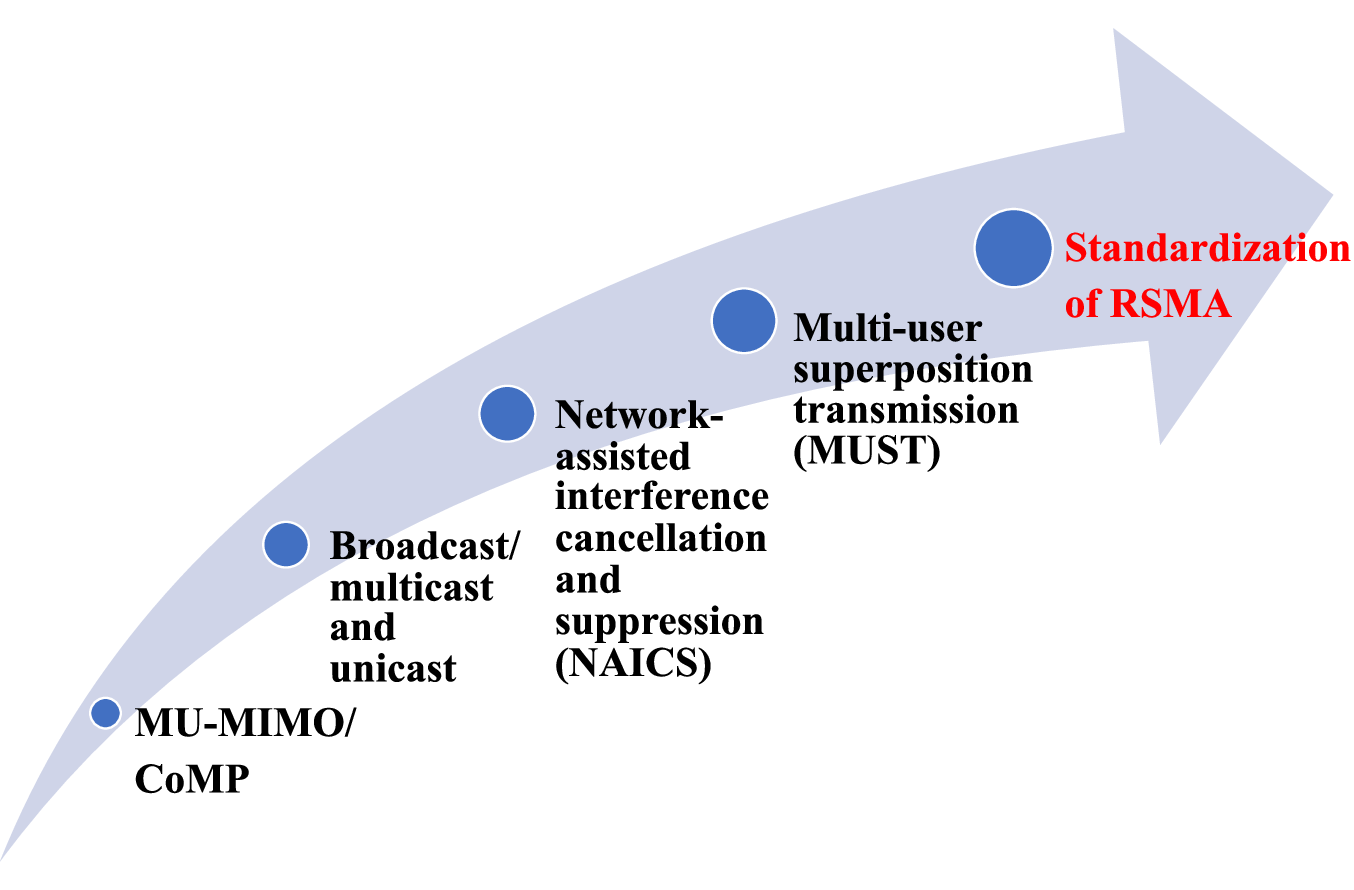}%
		\caption{Pathways to 3GPP}
		\label{fig: 3GPP}
	\end{figure}
	%%%
	The standardization of RSMA has yet to be considered by 3GPP. 
	Here, we provide details on several important topics to be studied in the context of standardization and implementation of RSMA as follows:
	\begin{itemize}
		\item \textit{HARQ}:
		As discussed in Section \ref{sec:roadahead},
		HARQ is crucial for efficient packet-based transmission in wireless networks, and it is less straightforward to design an HARQ scheme for RSMA. A low-complexity and compatible HARQ design is a major step for the adoption of RSMA in standards.
		\item \textit{Downlink and uplink signaling}:
		RSMA requires the knowledge of message splitting ratios at the transmitters and receivers. Additional downlink and uplink signaling is needed to guarantee the synchronization of such knowledge.  
		%%%
		\item \textit{System-level performance}: The PHY layer design and LLS performance of RSMA has been investigated in \cite{Onur2020LLS} and the performance gain of RSMA over the 5G NR design has been brought to light in \cite{onur2021mobility}. However,  system-level performance evaluations, which take into account the impact of the design of the higher layers (such as  HARQ, user scheduling, QoS provisioning in the application layer, etc.), are still not available.
	\end{itemize}
	It is clear that the abovementioned topics need to be addressed in detail for further progress with respect to standardization. 
	Fortunately, several key components of RSMA have been considered in previous 3GPP releases, which can provide a basis for RSMA mechanisms and reduce the workload for standardization and implementation studies.
	Fig. \ref{fig: 3GPP} illustrates four important study/work items, which pave the way to the inclusion of RSMA in future releases with minimal extra mechanisms to design and implement. Specifically, MU--MIMO (Release 8 \cite{3gppMIMO}) and CoMP (Release 11 \cite{3gpp36819}) use precoding and signalling mechanisms already implemented in the standards to enable multi-user transmission. Similarly, a work item on MBS (Release 17 \cite{3GPPNOUM1}) addresses broadcast/multicast transmission in NR, the signalling design for which can be modified to support RSMA transmission. SIC, which is an important mechanism for RSMA at the receiver side, is considered for network-assisted interference  cancellation and suppression (NAICS) (Release 12 \cite{3gppNAIC}) and multi-user superposition transmission (MUST) (Release 13 \cite{3gppMUST}), and thus, can readily be adopted for RSMA without any additional standardization and implementation burden.
	\par
	Although there is still work to be done before RSMA is employed in next generation networks, there are existing mechanisms in the standards that can be leveraged to make its integration easier.
	%%%
	
	\section{Conclusion}
	\label{sec:conclusion}
	In this paper, we have provided the first holistic tutorial on, and survey of, RSMA including 
	\begin{itemize}
		\item an exhaustive survey of the RSMA literature from both the information-theoretic and communication-theoretic perspectives, 
		\item a thorough elaboration of the principles and transmission frameworks of downlink/uplink and multi-cell RSMA, 
		\item a comprehensive summary of two strategies for precoder design, namely precoder optimization and low-complexity precoder design, 
		\item a detailed performance comparison between RSMA and existing MA schemes (including SDMA, NOMA, OMA, multicasting) in terms of their DoF, complexity, and performance,
		\item a summary of the major advantages and disadvantages of RSMA,
		\item an illustration of the PHY layer design and link-level simulation of RSMA,
		\item an extensive discussion of emerging applications of RSMA as well as corresponding research challenges and future directions including many uninvestigated areas,
		\item a synopsis of the standardization and implementation of RSMA.
	\end{itemize}
	The existing literature on RSMA has shown that non-orthogonal transmission in multi-antenna networks should be designed such that interference is partially decoded and partially treated as noise in order to fully exploit the benefits of multiple antennas at the transceivers and SIC at the receivers.
	%%%
	Thus, RSMA is a promising PHY layer transmission paradigm for interference management, non-orthogonal transmission, and multiple access in 6G, which will fundamentally reform the PHY layer and lower MAC layer design of wireless communication networks.
	%%%
	Many previous studies of communication design merit revisiting under the RSMA framework.
	%%%
	RSMA indeed opens up many interesting research problems for both industry and academia, and offers the prospect of boosting data rate, enhancing transmission reliability,  increasing energy efficiency, saving transmit power, reducing transmission latency, improving interference management, and providing robustness to CSIT uncertainties.
	%%%
	
	%\section*{Acknowledgement}The authors are deeply indebted to Prof. Petar Popovski for his helpful comments and suggestions.
	
	\bibliographystyle{IEEEtran}
	\bibliography{IEEEabrv,reference_MA}

\end{document}